\documentclass[11pt]{memoir}
\usepackage[T1]{fontenc}
\usepackage[utf8]{inputenc}
\usepackage[main=english,greek]{babel}

\usepackage{amsmath}
\usepackage{amssymb}
\usepackage{amsfonts}
\usepackage{mathrsfs}
\usepackage{ragged2e}
\usepackage{braket}
\usepackage{slashed}
\usepackage{graphicx}
\usepackage{subcaption}
\usepackage{empheq}
\usepackage[numbers,sort&compress]{natbib}
\usepackage[usenames,dvipsnames]{xcolor}
\usepackage{color}
\usepackage{float}
\usepackage{afterpage}
\usepackage{dcolumn}
\usepackage{indentfirst}
\usepackage{titletoc}
\usepackage[format=plain,indention=.5cm,labelfont={sc},textfont={sl}]{caption}
\usepackage[top=1in, bottom=0.8in, left=0.7in, right=0.7in, includefoot]{geometry}
\usepackage{xfrac}
\usepackage{perpage}
\MakePerPage{footnote}
\allowdisplaybreaks
\usepackage{dsfont}
\usepackage{multicol}
\usepackage{soul}
\usepackage{smartdiagram}
\usepackage{tikz}
\usetikzlibrary{matrix,calc,shapes}
\usetikzlibrary{arrows,automata}
\usepackage{booktabs}
\usepackage{siunitx}
\usepackage{wrapfig}

\usepackage[colorlinks=true
,urlcolor=MidnightBlue
,anchorcolor=blue
,citecolor=teal
,filecolor=blue
,linkcolor=blue
,menucolor=blue
,linktocpage=true
,pdfproducer=PDFLaTeX
,pdfa=true
]{hyperref}

\numberwithin{equation}{chapter}
\usepackage{pgfornament}

\newcommand\myastr{%
  \par\bigskip\noindent\hfill\pgfornament[width=17pt]{4}\hfill\null\par\bigskip
}

\newcommand\xquad
     {\hspace{0.7em plus.4em minus.4em}}
\titlecontents*{subsection}% <section>
  [3.5em]% <left>
  {\filright\small}% <above-code>
  { }% <numbered-entry-format>; you could also use {\thecontentslabel. } to show the numbers
  {}% <numberless-entry-format>
  {~(\thecontentspage)}% <filler-page-format>
  [\xquad\textbullet\xquad]% <separator>
  [\vspace*{0.25cm}]% <end>

\graphicspath{{images/}}

\usepackage{pifont}
\usepackage{enumitem}
\usepackage{rotating}

\begin{document}

\begin{titlingpage}
\selectlanguage{greek}
\centering
\begin{minipage}{0.15\textwidth}
\begin{flushright}
\includegraphics[scale=0.38]{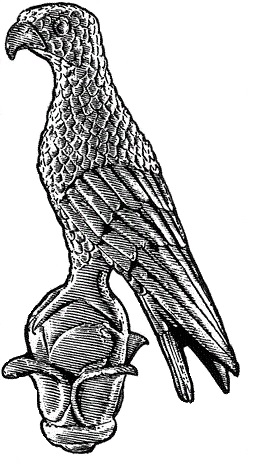}
\end{flushright}
\end{minipage}
\begin{minipage}{0.5 \textwidth}
\begin{center}
\textsc{\LARGE Πανεπιστήμιο Ιωαννίνων}\\[0.25cm]
\textsc{\Large Σχολή Θετικών Επιστημών}   \\[0.2cm] 
\textsc{\Large Τμήμα Φυσικής}
\end{center}
\end{minipage}

\vspace{3.25cm}

\LARGE{Φαινομενολογία του κοσμολογικού πληθωρισμού\\ της τετραγωνικής βαρύτητας\\ στο πλαίσιο του φορμαλισμού \textlatin{Palatini}}

\vspace{1.75cm}

\textit{\Large{Άγγελος Λύκκας}}

\vspace{2.75cm}

\Large{Διδακτορική διατριβή}
 
 \vfill

ΙΩΑΝΝΙΝΑ 2022
\clearpage
\end{titlingpage}

\begin{titlingpage}
\selectlanguage{english}
\centering
\begin{minipage}{0.15\textwidth}
\begin{flushright}
\includegraphics[scale=0.38]{uoi_logo.jpg}
\end{flushright}
\end{minipage}
\begin{minipage}{0.5 \textwidth}
\begin{center}
\textsc{\LARGE University of Ioannina}\\[0.25cm]
\textsc{\Large School of Natural Sciences}   \\[0.2cm] 
\textsc{\Large Department of Physics}
\end{center}
\end{minipage}

\vspace{3.25cm}

\LARGE{Inflationary phenomenology of quadratic gravity\\ in the Palatini formulation}

\vspace{1.75cm}

\Large{\textit{Angelos Lykkas}}

\vspace{2.75cm}

PhD Thesis

\vfill

IOANNINA 2022
\clearpage
\end{titlingpage}

\frontmatter

\thispagestyle{empty}

\vspace*{1.5cm}

\begin{center}
   {\textsc{\LARGE{Doctoral Committee}}}
\end{center}

\vspace*{1.25cm}

\begin{center}
   \large{\textbf{Three-member advisory committee}}
\end{center}

\begin{enumerate}[align=left]
    \item[\textbf{Kyriakos Tamvakis} (Advisor)] -- Emeritus Professor, Department of Physics, Ioannina U.
    \item[\textbf{Panagiota Kanti}] -- Professor, Department of Physics, Ioannina U.
    \item[\textbf{John Rizos}] -- Professor, Department of Physics, Ioannina U.
\end{enumerate}

\vspace{2.5cm}

\begin{center}
   \large{\textbf{Seven-member PhD examination committee}}
\end{center}

\begin{enumerate}[align=left]
    \item[\textbf{Athanasios Dedes}] -- Professor, Department of Physics, Ioannina U.
    \item[\textbf{Ioannis Florakis}] -- Assistant Professor, Department of Physics, Ioannina U.
    \item[\textbf{Panagiota Kanti}] -- Professor, Department of Physics, Ioannina U.
    \item[\textbf{Alexandros Kehagias}] -- Professor, Department of Physics, Natl. Tech. U., Athens
    \item[\textbf{John Rizos}] -- Professor, Department of Physics, Ioannina U.
    \item[\textbf{Vassilis Spanos}] -- Associate Professor, Department of Physics, Nat. and Kapodistrian U., Athens
    \item[\textbf{Kyriakos Tamvakis} (Advisor)] -- Emeritus Professor, Department of Physics, Ioannina U.
\end{enumerate}

\clearpage

\thispagestyle{empty}

\vspace*{5cm}

\begin{flushright}
... \emph{\large{dedicated to my family}}
\end{flushright}

\clearpage

\thispagestyle{empty}
\addcontentsline{toc}{chapter}{\protect\numberline{}\emph{Abstract}}

\vspace*{3.5cm}

\begin{center}
    \LARGE{\textbf{--- \emph{Abstract} ---}}
\end{center}

\vspace{0.5cm}

\noindent The theory of General Relativity was established on a spacetime manifold equipped with a metric tensor, $(\mathcal{M}_4,\text{g})$, and the connection on $\mathcal{M}_4$ identified with the Levi-Civita one. Even though there are valid reasons to assume a torsionless manifold that preserves the metric, it was shown that dealing away with these assumptions the Levi-Civita condition can be reproduced at the level of equations of motion of GR for a metric-affine connection. It was not long before the equivalence of General Relativity between the two descriptions, known as the Palatini or first-order formalism in which the connection is independent of the metric, and the conventional metric or second-order formalism, was broken for more complicated action functionals involving higher-order curvature invariants and/or nonminimal couplings between the gravitational and matter sector. Nowadays these types of theories are prominent in modeling inflation where they have found major success. Since the paradigm of inflation is fused with the gravitational degrees of freedom and thus their parametrisation, it is interesting to understand how the predictions of these models differ between the two formulations. For example, one of the outstanding models of inflation is the Starobinsky or quadratic gravity model, $R+R^2$, with continued success since its conception. However in the Palatini formalism the scalar degree of freedom sourced by the $R^2$ term is actually nonpropagating and therefore is unable to drive an inflationary phase. Then in order for inflation to be realised in the first-order formalism the Starobinsky model has to be coupled with a fundamental scalar field that will assume the role of the inflaton field. In this thesis we investigate different inflationary scenarios, starting with previously ruled-out models such as the free massive scalar, natural inflation, etc, where we find that the $R^2$ term has a significant role in flattening the Einstein-frame inflaton potential and thus giving the opportunity for these models to come in contact with observations in that specific program. Of particular interest is the study of Higgs inflation in this context and a possible comparison with results obtained in the usual metric formalism, as well as proposing a case of minimal Higgs inflation with the $R^2$ term. Contrary to their second-order formulation in which the field space of the models is two-dimensional, here we show that the models are actually one-dimensional in the field space and can be readily studied analytically leading to interesting results.

\clearpage

\thispagestyle{plain}

\addcontentsline{toc}{chapter}{\protect\numberline{}\emph{Extended Summary in Greek}}

\vspace*{3.5cm}

\selectlanguage{greek}
\begin{center}
    \LARGE{\underline{Περίληψη}}
    
    \vspace{0.5cm}
    
    \selectlanguage{english}
    \large{\emph{Extended summary in Greek}}
\end{center}

\vspace{0.75cm}

\selectlanguage{greek}

Η Γενική θεωρία της Σχετικότητας (ΓΣ) και το Καθιερωμένο Πρότυπο (ΚΠ) αποτελούν τις πιο ικανές θεωρίες να περιγράψουν με ακρίβεια τη φυσική γύρω μας. Το ΚΠ είναι μια κβαντική θεωρία πεδίου που περιγράφει τις αλληλεπιδράσεις των στοιχειωδών σωματιδίων και τις ιδιότητές τους, βασισμένη στη συμμετρία βαθμίδας $SU(3)_C\times SU(2)_L\times U(1)_Y$. Η τελευταία επιβεβαίωση του ΚΠ ήρθε με την πρόσφατη ανακάλυψη του μποζονίου \textlatin{Higgs} (\textlatin{Higgs boson}) στο Μεγάλο Επιταχυντή Αδρονίων (\textlatin{LHC}) στο Ευρωπαϊκό Πυρηνικό Κέντρο Ερευνών (\textlatin{CERN}), ολοκληρώνοντας έτσι την ενοποίηση των θεμελιώδων (μη-βαρυτικών) δυνάμεων της φύσης. Παρόλες τις επιτυχίες της θεωρίας υπάρχουν ακόμα αναπάντητα ερωτήματα και διάφορα πρόβληματα, μεταξύ άλλων είναι το πρόβλημα των μαζών των νετρίνων, η φύση της σκοτεινής ύλης του σύμπαντος και η αδυναμία του ΚΠ να ενοποιηθεί με την τέταρτη δύναμη, τη βαρύτητα. Η ΓΣ περιγράφει αποκλειστικά τις βαρυτικές αλληλεπιδράσεις και βρίσκει επιτυχία σε μια εντελώς διαφορετική περιοχή ισχύος από αυτή του ΚΠ και έχει αντέξει το πέρασμα του χρόνου, έχοντας επιβεβαιωθεί από πειράματα σε διάφορες περιπτώσεις με τελευταίο και ίσως πιο σημαντικό την παρατήρηση στον ανιχνευτή του \textlatin{LIGO} βαρυτικών κυμάτων από συγχωνευόμενες μελανές οπές. 

Η ασυμβατότητα του ΚΠ με την ΓΣ καταδεικνύει ότι η φιλόδοξη ιδέα της ενοποίησης των θεμελιωδών δυνάμεων κάτω από ένα μοναδικό θεωρητικό πλαίσιο απέχει ακόμα αρκετά. Ωστόσο, τα τελευταία χρόνια η κοσμολογία και η σωματιδιακή φυσική πλησιάζουν ολοένα και περισσότερο, το οποίο αποδεικνύεται καλύτερα από τις σύγχρονες εξελίξεις στη θεωρία του κοσμολογικού πληθωρισμού (\textlatin{inflation}). Επεκτείνοντας τη θεωρία της Γενικής Σχετικότητας με ένα επιπλέον θεμελιώδες βαθμωτό πεδίο, ικανό να περιγράψει μια διαστολή \textlatin{de-Sitter} τις πρώτες στιγμές του σύμπαντος, μπορούμε να εξηγήσουμε την παρατηρούμενη επιπεδότητα, ομοιογένεια και ισοτροπία που έχει παρατηρηθεί να ισχύει σε μεγάλες αποστάσεις στο σύμπαν. Η θεωρία του κοσμολογικού πληθωρισμού προτάθηκε αρχικά ακριβώς για να λύσει αυτά τα ζητήματα της κοσμολογίας, συγκεκριμένα θέματα συνδεδεμένα με τις αρχικές συνθήκες του σύμπαντος, που είναι ουσιαστικά το μεγαλύτερο μέρος της Φυσικής, λαμβάνει την κεντρική θέση. Διατυπωμένη στα τέλη της δεκαετίας του '70 με αρχές της δεκαετίας του '80, εφάρμοσε διάφορες γνώσεις, θεωρίες και αποτελέσματα από πολυάριθμα πεδία της σωματιδιακής φυσικής, αλλά ο ίδιος ο μηχανισμός της θεωρίας του κοσμολογικού πληθωρισμού είναι ακόμα άγνωστος. Ως εκ τούτου, προτείνονται διάφορα μοντέλα, ικανά να περιγράψουν το σενάριο του κοσμολογικού πληθωρισμού, που οδηγούν σε αντίστοιχες προβλέψεις, οι οποίες με τη σειρά τους συνδέονται με παρατηρήσιμα γεγονότα και έτσι έχουν την δυνατότητα να επαληθευτούν ή απορριφθούν. Ο κοσμολογικός πληθωρισμός έχει αποδειχθεί ότι είναι η απλούστερη πραγματοποίηση, όσον αφορά την εφαρμογή και τις υποθέσεις, μιας τέτοιας προσπάθειας κατανόησης της εξέλιξης του πρώιμου σύμπαντος.

Ενώ ο κοσμολογικός πληθωρισμός κατασκευάστηκε έτσι ώστε να μπορεί φυσικά να αντιμετωπίσει τα ζητήματα που συζητήθηκαν παραπάνω, περιλαμβάνει μια χαρακτηριστική πρόβλεψη -- την ικανότητα να παράγει τους «σπόρους» για  την υλοποίηση του σχηματισμού δομών μεγάλης κλίμακας του γνώριμου σύμπαντος (όπως άστρα, γαλαξίες κ.α.) μέσω των διαστελλόμενων κβαντικών διακυμάνσεων του πεδίου που υλοποιεί το σενάριο του κοσμολογικού πληθωρισμού. Το χαρακτηριστικό αυτό υποδηλώνει την προβλεπτική δύναμη της θεωρίας και την επιτυχία της να οδηγήσει σε ένα παρατηρήσιμο αποτέλεσμα της αλληλεπίδρασης μεταξύ της ΓΣ και Κβαντικής Θεωρίας, πιθανώς μία από τις λίγες γνωστές περιπτώσεις στη φυσική σήμερα. Αυτό το φαινόμενο μπορεί να επιβεβαιωθεί πειραματικά στις ανισοτροπίες που παρατηρούνται στο \textlatin{CMB} και, με αυξανόμενη ακρίβεια (οι αποστολές προγραμματίζονται για τις επόμενες δεκαετίες), μας επιτρέπει να περιορίσουμε τον μεγάλο αριθμό προτεινόμενων μοντέλων για το σενάριο του πληθωρισμού.

Ο κοσμολογικός πληθωρισμός χρησιμεύει επίσης για την «αραίωση» των ανεπιθύμητων υπολειμμάτων (π.χ. \textlatin{topological defects}). Ως αποτέλεσμα, στο τέλος παραμένουν μόνο το \textlatin{zero mode} του πεδίου \textlatin{inflaton} και οι μικροσκοπικές διακυμάνσεις της μετρικής. Τότε, είναι φυσικό να υποθέσουμε την πιθανή ύπαρξη μιας περιόδου κατά την οποία το σύμπαν θερμαίνεται, από κενό και κρύο αμέσως μετά το τέλος του πληθωρισμού έως τα μεγάλα επίπεδα ενέργειας και εντροπίας που παρατηρούνται στον ορίζοντα σήμερα. Κατά τη διάρκεια αυτής της περιόδου, γνωστής ως αναθέρμανση (\textlatin{reheating}), η πυκνότητα ενέργειας του \textlatin{inflaton} μετατρέπεται σε ακτινοβολία (ή άλλα έμμαζα σωματίδια) στο τέλος του κοσμολογικού πληθωρισμού μέσω διαφόρων μηχανισμών. Δυστυχώς, κατά την περίοδο της αναθέρμανσης, οι κινούμενες κλίμακες (\textlatin{comoving scales}) εισέρχονται ξανά στον ορίζοντα, καθιστώντας την έμμεση ανίχνευση σχεδόν αδύνατη, σε αντίθεση με την περίπτωση του πληθωρισμού όπου αυτές «παγώνουν» και αφήνουν ένα «αποτύπωμα» στο \textlatin{CMB}. Επομένως, η περίοδος αναθέρμανσης περιορίζεται ελάχιστα από κοσμολογικές παρατηρήσεις. Ένας τρόπος για να περιοριστεί τουλάχιστον η εξέλιξή της είναι να συσχετίσουμε τις κλίμακες που επανέρχονται στον ορίζοντα με εκείνες που εξέρχονται από τον ορίζοντα κατα τον πληθωρισμό. Αυτό το είδος παραμετροποίησης μπορεί να παρέχει έμμεσες ενδείξεις για την περίοδο της αναθέρμανσης βοηθώντας στην πιθανή ελαχιστοποίηση του χώρου παραμέτρων του αντίστοιχου πληθωριστικού μοντέλου.

Η εκπληκτική διορατικότητα του \textlatin{Albert Einstein} πρέπει να αναφερθεί όταν κάποιος συζητά τη Γενική θεωρία της Σχετικότητας, η οποία βασίζεται στη συνειδητοποίηση ότι ένας παρατηρητής που βρίσκεται σε ελεύθερη πτώση δεν ασθάνεται τη βαρύτητα και όταν οι επιπτώσεις αυτής είναι μη διακρίσιμες από αυτές σε κατάσταση επιτάχυνσης. Αυτό που είναι σήμερα γνωστό ως η Αρχή της Ισοδυναμίας του \textlatin{Einstein} (\textlatin{Einstein's Equivalence Principle -- EEP}), τον οδήγησε σε μια θεωρία ικανή να εξηγήσει ή έστω να περιγράψει το μεγαλύτερο μέρος της βαρυτικής φυσικής. Η βασική παρατήρηση που εξάγεται από την \textlatin{EEP} είναι ότι η βαρύτητα κατανοείται καλύτερα ως η καμπυλότητα του χωροχρόνου σε αντίθεση με τις άλλες θεμελιώδεις δυνάμεις της φύσης. Αργότερα αυτό είχε τεράστιο αντίκτυπο για το μέλλον της θεωρητικής φυσικής οδηγώντας σε μια «γεωμετροποίηση» της φυσικής και του τρόπου με τον οποίο προσεγγίζουμε τη φυσική συνολικά (π.χ. θεωρίες βαθμίδων, θεωρία χορδών κ.α.). Η Γενική Σχετικότητα, όπως περιγράφεται από τη δράση \textlatin{Einstein-Hilbert}, είναι βασισμένη στη γλώσσα της γεωμετρίας του \textlatin{Riemann} και (εκείνη την εποχή) πρωτοποριακών άρθρων αναφορικά με τη διαφορική γεωμετρία και τον τανυστικό λογισμό, που εφαρμόζονται ακόμη και σήμερα.

Αμέσως μετά της ανακάλυψης της ΓΣ, προτάθηκαν τροποποιήσεις της, αν και αρχικά ήταν οδηγημένες κυρίως από επιστημονική περιέργεια και όχι από κάποια αδυναμία της θεωρίας. Ωστόσο οι μελέτες αυτές, καθώς και ο αυτός ο τρόπος έρευνας, επιβραβεύθηκαν αργότερα από την ανάγκη για νέα χαρακτηριστικά, που η ΓΣ δεν περιλαμβάνει, όπως είχε διατυπωθεί από τον \textlatin{Einstein}. Διάφορες θεωρίες με ανάλογες επιτυχίες προτάθηκαν όπως για παράδειγμα το πρόγραμμα κβαντικής βαρύτητας βρόχου (\textlatin{loop quantum gravity -- LQG}) μαζί με άλλα που επιχειρούν να κατασκευάσουν μια κβαντική θεωρία της βαρύτητας ή άλλες όπως τα μοντέλα \textlatin{Kaluza-Klein} ή \textlatin{string theory} που στοχεύουν στην ενοποίηση της βαρύτητας με τις υπόλοιπες θεμελιώδεις δυνάμεις.

Είναι ενδιαφέρον ότι περίπου την ίδια εποχή που διαμορφώθηκε για πρώτη φορά η ΓΣ, ο \textlatin{E. Cartan} την δεκαετία του '20 ανέπτυξε έναν πολύ διαφορετικό τύπο διαφορικής γεωμετρίας, με βάση τις διαφορικές μορφές (\textlatin{differential forms}) και τις δέσμες ινών (\textlatin{fiber bundle}). Σχετικά με τη ΓΣ, εξέτασε άλλες δέσμες εκτός της εφαπτομενικής δέσμης και άλλες \textlatin{connections} εκτός της \textlatin{Levi-Civita}. Αυτό επιδιώχθηκε επίσης άκαρπα από τον \textlatin{H. Weyl} και άλλους την ίδια περίπου περίοδο. Η γενίκευση της γεωμετρίας \textlatin{Riemann} από τον \textlatin{Cartan} και οι πιο γενικές \textlatin{connections} ανακαλύφθηκαν (ή καταλληλότερα επανεξετάστηκαν) πολύ αργότερα στα έργα των \textlatin{Yang} και \textlatin{Mills} (1954). Σήμερα, οι αλληλεπιδράσεις στη φύση περιγράφονται από ένα \textlatin{gauge field} (ή \textlatin{connection}). Οι προσπάθειες τροποποίησης της ΓΣ με γενίκευση της \textlatin{connection} ή, πιο σημαντικά της υποκείμενης γεωμετρίας, είναι εμφανείς σήμερα ακόμα κι αν διαφοροποιούν τη διατύπωση της βαρυτικής θεωρίας (θεωρία μετρικής ή \textlatin{connection}).

Η γενίκευση της έννοιας του χωροχρόνου σε μια καμπύλη πολλαπλότητα $\mathcal{M}_4$ με μετρική \selectlanguage{english}$\text{g}$\selectlanguage{greek}, όπως υποτίθεται από τη Γενική σχετικότητα, επέφερε επανάσταση στον τρόπο προσέγγισης των σύγχρονων θεωριών που προσπαθούν να ενσωματώνουν τη βαρύτητα με τη συμβατική σωματιδιακή φυσική. Ίσως μία από τις πιο κρίσιμες, και ταυτόχρονα ανεπιτήδευτες, προϋποθέσεις της ΓΣ είναι η \textlatin{connection} \textlatin{Levi-Civita}, με άλλα λόγια ότι η \textlatin{connection} είναι συμβατή με την μετρική (διατηρεί τη μετρική) και συμμετρική (ελευθερία στρέψης). Η συγκεκριμένη επιλογή της \textlatin{connection} δεν είναι τυχαία, καθώς αποδείχθηκε από τον \textlatin{Levi-Civita} (και τον \textlatin{Christoffel}) ότι συνδέεται με την έννοια της παράλληλης μεταφοράς διανυσμάτων στον καμπύλο χώρο. Επιπλέον, η ιδέα της συναλλοίωτης παραγώγου (\textlatin{covariant derivative}) ενός διανύσματος κατά μήκος μιας καμπύλης γενικεύτηκε για την περίπτωση μιας γεωμετρίας \textlatin{Riemann}, την οποία ακριβώς χρειαζόταν η ΓΣ εκείνη την εποχή. Πολύ σύντομα η προϋπόθεση της \textlatin{connection} \textlatin{Levi-Civita} άρχισε να αμφισβητείται, όπως και άλλες πτυχές της ΓΣ, αρχικά στις μελέτες του \textlatin{A. Palatini}, όπου γενικεύοντας την έννοια της \textlatin{connection} σε μια ανεξάρτητη της μετρικής, μπόρεσε να αποδείξει ότι και οι δύο διατυπώσεις της ΓΣ είναι ισοδύναμες. Η κύρια διαφορά βρίσκεται στο γεγονός ότι η συνθήκη \textlatin{Levi-Civita} ανακτάται στο επίπεδο των εξισώσεων κίνησης της θεωρίας (\textlatin{on-shell}) και δεδομένου ότι ο φορμαλισμός \textlatin{Palatini} ή πρώτης τάξης της ΓΣ είναι ισοδύναμος με τη συμβατική μετρική ή δεύτερης τάξης διατύπωση το θέμα των διαφόρων διατυπώσεων τέθηκε σε αναστολή και δόθηκε ελάχιστη προσοχή μέχρι τις επόμενες μελέτες του \textlatin{Cartan} που έδειξαν ότι και οι γεωμετρίες εκτός του \textlatin{Riemann} μπορούν επίσης να υποστηρίξουν μια θεωρία βαρύτητας. 

Μετά από πρόσφατες εξελίξεις σε εκτεταμένες/τροποποιημένες θεωρίες βαρύτητας, που επικεντρώνονται κυρίως σε κοσμολογικά ζητήματα (π.χ. μοντέλα σκοτεινής ενέργειας, μοντέλα κοσμολογικού πληθωρισμού), φάνηκε ότι ο φορμαλισμός \textlatin{Palatini} αντιστοιχεί σε ένα θεμελιώδες ερώτημα σχετικά με την παραμετροποίηση των βαρυτικών βαθμών ελευθερίας. Συγκεκριμένα, το μεγαλύτερο μέρος των πληθωριστικών μοντέλων που είναι ακόμα σε συμφωνία με τις πειραματικές ενδείξεις περιέχουν ένα είδος μη-τετριμμένης ζεύξης της βαρύτητας με το βαθμωτό πεδίο \textlatin{inflaton}, γεγονός που είναι ικανό να οδηγήσει σε προβλέψεις που διαφέρουν δραματικά ανάμεσα στους δύο φορμαλισμούς (μετρικής και \textlatin{Palatini}). Ίσως ένα από τα πιο γνωστά πληθωριστικά μοντέλα με συνεχείς επιτυχίες είναι το μοντέλο \textlatin{Starobinsky} ή μοντέλο τετραγωνικής βαρύτητας ($R+R^2$), διατυπωμένο στο φορμαλισμό μετρικής. Προβλέψεις αυτού βρίσκονται εντός της επιτρεπόμενης περιοχής $1\sigma$ των παρατηρήσεων οδηγώντας σε μια πληθώρα άλλων μοντέλων που επιχειρούν να το τροποποιήσουν ή να το επεκτείνουν διατηρώντας παράλληλα ορισμένα από τα ελκυστικά χαρακτηριστικά του. Όμως, το μοντέλο \textlatin{Starobinsky} διατυπωμένο κατά το φορμαλισμό \textlatin{Palatini} περιγράφει μια τελείως διαφορετική εικόνα του σεναρίου του κοσμολογικού πληθωρισμού, γεγονός που τονίζεται ιδιαίτερα στα κύρια αποτελέσματα της διατριβής.

Με βάση τις προηγούμενες συζητήσεις φαίνεται σημαντικό να επανεξαστούν ορισμένες πτυχές του κοσμολογικού πληθωρισμού, πράγμα που γίνεται στο δεύτερο κεφάλαιο όπου τα περισσότερα από τα αποτελέσματα που χρειάζονται μετέπειτα εξάγονται με ελάχιστη μαθηματική αυστηρότητα, παραπέμποντας τον αναγνώστη στην τεράστια βιβλιογραφία για εκάστοτε θέμα αλλά και σε επόμενες ενότητες στην ίδια την διατριβή. Ξεκινώντας με μια σύντομη επισκόπηση της σύγχρονης κοσμολογίας, οδηγούμαστε αμέσως στα ερωτήματα του \textlatin{Big Bang}, που με τη σειρά τους μας οδηγούν στην κεντρική ιδέα του πληθωρισμού. Παρόλο που οι πραγματικές προβλέψεις σχετικά με την περίοδο του πληθωρισμού εξαρτώνται σε μεγάλο βαθμό από το μοντέλο, όλα όσα παρουσιάζονται στο κεφάλαιο αυτό είναι εντελώς ανεξάρτητα από το μοντέλο, θυσιάζοντας ενδεχομένως μερικά από τα συναρπαστικά αποτελέσματα που προσφέρει η θεωρία, τα οποία συζητούνται σε επόμενα κεφάλαια. Η πραγματική πρόβλεψη του πληθωρισμού, δηλαδή οι διαταραχές της πυκνότητας ενέργειας του πεδίου \textlatin{inflaton} μελετώνται και παρουσιάζονται με μεγάλη λεπτομέρεια λόγω της σημασίας τους. Αναλύοντας στη συνέχεια το απλούστερο μοντέλο ενός πληθωρισμού, ένα πραγματικό βαθμωτό πεδίο ελάχιστα συζευγμένο (\textlatin{minimal coupling}) με τη βαρύτητα με το δυναμικό αυτό-αλληλεπίδρασής του υπό την προσέγγιση αργής κύλισης (\textlatin{single-field slow-roll}), οι παράμετροι αργής κύλισης (\textlatin{slow-roll parameters}) συνδέονται με το πλάτος αυτών των διακυμάνσεων, σηματοδοτώντας έναν τρόπο ποσοτικής κατανόησης και δίνωντας την ικανότητα πρόβλεψης των παρατηρήσιμων ποσοτήτων κατά τον πληθωρισμό. Για να αντισταθμίσουμε αυτό, εξετάζουμε επίσης τη λεγόμενη προσέγγιση σταθερής κύλισης του πεδίου \textlatin{inflaton} (\textlatin{constant-roll}) και συσχετίζουμε προβλέψεις μεταξύ των δύο προσεγγίσεων για συγκεκριμένα μοντέλα σε επόμενο κεφάλαιο. Στο τέλος του κεφαλαίου παρουσιάζουμε μια σύντομη σύνοψη της περιόδου αναθέρμανσης που ακολουθεί τον πληθωρισμό, πρώτα εξετάζοντας μερικούς από τους πιθανούς μηχανισμούς και τέλος επικεντρωνόμαστε στην παραμετροποίηση της αναθέρμανσης ως προς τις πληθωριστικές παραμέτρους, διατηρώντας με αυτόν τον τρόπο μια άμεση σύνδεση μεταξύ των δύο εποχών, που μας επιτρέπει να θέσουμε πιθανώς αυστηρότερους περιορισμούς στις παραμέτρους του κάθε μοντέλου.

Στο επόμενο κεφάλαιο (Κεφάλαιο 3) αναλύουμε τον φορμαλισμό \textlatin{Palatini} ή φορμαλισμό πρώτης τάξης, που αποτελεί την κεντρική ιδέα του κύριου μέρους της διατριβής, δίνοντας ιδιαίτερη προσοχή σε αυτά που επισημαίνονται περαιτέρω στα πληθωριστικά μοντέλα. Αρχικά εφιστούμε την προσοχή σε ορισμένες πτυχές του συμβατικού φορμαλισμού μετρικής που είναι θεμελιωδώς διαφορετικές στη διατύπωση \textlatin{Palatini} της ΓΣ, όπως ο όρος \textlatin{York-Gibbons-Hawking (YGH)}. Για να καθορίσουμε την \textlatin{Palatini variation} , αρχίζουμε ορίζοντας αρχικά τους \textlatin{metric-affine} χώρους και ιδιαίτερα την έννοια της στρέψης (\textlatin{torsion}), της μη-μετρικότητας (\textlatin{nonmetricity}) και της καμπυλότητας στην $\mathcal{M}_4$, όπου οδηγούμαστε στο θέμα της \textlatin{metric-affine} \textlatin{connection}. Μετά από μια σύντομη ιστορική ανασκόπηση επανεξετάζουμε τον συμβατικό μετρικό φορμαλισμό και εξάγουμε τις γνωστές εξισώσεις πεδίου \textlatin{Einstein} της ΓΣ. Στη συνέχεια, εξάγουμε τις εξισώσεις πεδίου για την \textlatin{Lagrangian Einstein-Hilbert} υποθέτωντας τον φορμαλισμό πρώτης τάξης, δηλαδή η μετρική και η \textlatin{connection} δεν είναι εξαρτημένες μεταξύ τους εκ των προτέρων, όπου τα αρχικά αποτελέσματα του \textlatin{Palatini} (και των άλλων) αναπαράγονται καθιστώντας την ισοδυναμία της ΓΣ μεταξύ αυτών των δύο διατυπώσεων. Κλείνοντας το κεφάλαιο συζητάμε την προσπάθεια κατανόησης εάν αυτή η ισοδυναμία παραμένει στο κβαντικό επίπεδο. 

Στο τέταρτο κεφάλαιο παρουσιάζουμε τα κύρια αποτελέσματα της διατριβής χρησιμοποιώντας και συνδυάζοντας μερικές από τις ιδέες που παρουσιάστηκαν στα προηγούμενα κεφάλαια, με το επίκεντρο να είναι το πληθωριστικό μοντέλο \textlatin{Starobinsky}. Έτσι, ξεκινάμε πρώτα με μια ανασκόπηση του μοντέλου στον φορμαλισμό μετρικής, τονίζοντας πως ο βαθμωτός βαθμός ελευθερίας (γνωστός και ως \textlatin{scalaron}) που προέρχεται από τον όρο $R^2$ προκύπτει στη βαθμωτή αναπαράσταση της εν λόγω θεωρίας. Στη συνέχεια, είναι εύκολο να εξάγουμε το δυναμικό \textlatin{Starobinsky} στο σύστημα αναφοράς του \textlatin{Einstein} (\textlatin{Einstein frame}) μετά από μια επανακλιμάκωση \textlatin{Weyl} (\textlatin{Weyl rescaling}) της μετρικής και έναν επαναπροσδιορισμό του πεδίου, εφαρμόζοντας έτσι τον μηχανισμό του πληθωρισμού αργής κύλισης ενός πεδίου, όπως περιγράφεται στο δεύτερο κεφάλαιο προκειμένου να αποκτηθούν οι περίφημες προβλέψεις του μοντέλου \textlatin{Starobinsky}. Συνεχίζουμε εξετάζοντας μια σύζευξη του μοντέλου \textlatin{Starobinsky} με ένα πραγματικό βαθμωτό πεδίο και το δυναμικό του, πρώτα με ελάχιστο τρόπο (μέσω του παγκόσμιου όρου $\sqrt{-g}$) και αργότερα μέσω μιας μη-τετριμμένης σύζευξης με τον όρο \textlatin{Einstein-Hilbert} της μορφής $\xi\phi^2 R$. Είναι προφανές ότι και στις δύο περιπτώσεις η θεωρία περιέχει ουσιαστικά δύο βαθμωτούς βαθμούς ελευθερίας, το \textlatin{scalaron} $\chi$ και το αρχικό βαθμωτό πεδίο $\phi$. Εκεί παρατηρούμε ότι η εφαρμογή των μοντέλων αυτών στον πληθωρισμό περιπλέκεται αρκετά αφού και τα δύο πεδία μπορούν κατ' αρχήν να συμβάλουν στην υλοποίηση του πληθωρισμού, και ειδικά στην περίπτωση όπου οι κινητικοί όροι του κάθε πεδίου εξαρτώνται μη-τετριμμένα από το άλλο πεδίο, περιπλέκοντας την ανάλυση.

Υποθέτουμε έπειτα το μοντέλο της τετραγωνικής βαρύτητας στον φορμαλισμό \textlatin{Palatini}, όπου και αποδεικνύουμε ότι ο όρος $R^2$ δεν οδηγεί στην πραγματικότητα σε ένα δυναμικό βαθμωτό πεδίο, $\chi$. Δεδομένου ότι στον φορμαλισμό πρώτης τάξης η συνοχή και η μετρική δεν εξαρτώνται εκ των προτέρων η μία από την άλλη, μια επανακλιμάκωση της μέτρησης (\textlatin{Weyl rescaling}) αφήνει τον τανυστή \textlatin{Ricci} αμετάβλητο, καθώς σε αυτή την περίπτωση είναι καθαρά συνάρτηση της \textlatin{connection}, δηλαδή $R_{\mu\nu }(\Gamma)$. Επομένως, δεν υπάρχει τρόπος για το \textlatin{scalaron} να αποκτήσει έναν κινητικό όρο στο σύστημα αναφοράς του \textlatin{Einstein}. Τότε, οι διάφοροι συνδυασμοί των συναρτήσεων καμπυλότητας υψηλότερων τάξεων δεν είναι σε θέση να συνεισφέρουν ένα βαθμωτό βαθμό ελευθερίας, έτσι ώστε το μοντέλο να περιγράψει ενα σενάριο πληθωρισμού και το πεδίο \textlatin{inflaton} πρέπει να συμπεριληφθεί εμφανώς στη δράση με τη μορφή ενός θεμελιώδους βαθμωτού πεδίου $\phi$. Το σημείο εκκίνησής μας είναι τότε μια δράση αυτής της μορφής. Στο πλαίσιο του \textlatin{Einstein} δείχνουμε ότι το \textlatin{scalaron} $\chi$ περιλαμβάνεται σε πολλαπλασιαστικούς παράγοντες του κινητικού όρου του $\phi$ και του δυναμικού του $V(\phi)$. Στη συνέχεια, μετά από \textlatin{variation} της τελικής δράσης σε σχέση με το $\chi$, παράγουμε την εξίσωση κίνησης του $\chi$, την οποία στη συνέχεια αντικαθιστούμε στη δράση. Εκεί διαπιστώνουμε ότι η δράση αποκτά κινητικούς όρους υψηλότερης τάξης του αρχικού βαθμωτού πεδίου, $\propto(\nabla\phi)^4$, καθώς και περίπλοκες εκφράσεις για τη μη-κανονική κινητική συνάρτηση και το βαθμωτό δυναμικό. Στη συνέχεια, μπορούμε να εξάγουμε τις γενικευμένες εξισώσεις πεδίου \textlatin{Einstein} για το σύστημα μετά από \textlatin{variation} σε σχέση με τη μετρική και το βαθμωτό πεδίο και να δείξουμε ότι η εξίσωση κίνησης της \textlatin{connection} οδηγεί στη συνθήκη \textlatin{Levi-Civita} σε σχέση με την νέα μετρική \textlatin{$\bar{\text{g}}$}. Εξετάζοντας τη διατύπωση \textlatin{path integral} της ίδιας θεωρίας διαπιστώνουμε ότι η μη-δυναμική φύση του πεδίου \textlatin{scalaron} δεν περιορίζεται στην κλασική δράση, αλλά παραμένει στο κβαντικό επίπεδο ακόμα και όταν άλλα πεδία ύλης περιλαμβάνονται στην δράση, οδηγούμενοι σε τοπικούς όρους που μπορούμε με ασφάλεια να αγνοήσουμε. Κλείνοντας αυτή την ενότητα προσφέρουμε μια σύντομη συζήτηση για το θέμα των συστημάτων αναφοράς, του \textlatin{Einstein} και του \textlatin{Jordan} συγκεκριμένα, καθώς η μετάβαση από το σύστημα αναφοράς \textlatin{Jordan} στο σύστημα αναφοράς του \textlatin{Einstein} παίζει πολύ σημαντικό ρόλο για την επικείμενη ανάλυση.

Στην επόμενη ενότητα εστιάζουμε σε έναν αριθμό πληθωριστικών μοντέλων που ήδη αποκλείονται από παρατηρήσεις υποθέτοντας μια τετριμμένη σύζευξη με τον βαρυτικό τομέα $R+R^2$. Αρχικά εξετάζουμε το λεγόμενο μοντέλο \textlatin{natural inflation}, όπου δείξαμε ότι ο όρος $R^2$ έχει σημαντική συμβολή στο δυναμικό πληθωρισμού προκαλώντας μια επίπεδη περιοχή στα όρια μεγάλων τιμών του πεδίου του πληθωρισμού $\phi$. Στην πραγματικότητα, το αποτέλεσμα αυτό ισχύει γενικά για οποιαδήποτε λογική μορφή δυναμικού $V(\phi)$ που οδηγεί σε ένα οροπέδιο για πολύ γενικές συνθήκες. Στη συγκεκριμένη περίπτωση του \textlatin{natural inflation}, η εξομάλυνση του δυναμικού είναι εμφανής. Δεδομένου ότι μας ενδιαφέρει ο πληθωρισμός αργής κύλισης, παραμελούμε τη συμβολή των κινητικών όρων υψηλότερης τάξης $\propto\dot{\phi}^4$ και εφαρμόζουμε τον συμβατικό μηχανισμό του πληθωρισμού ενός πεδίου. Εκεί διαπιστώνουμε ότι όταν το μοντέλο φυσικού πληθωρισμού γενικεύεται με τον όρο \textlatin{Starobinsky} στον φορμαλισμό \textlatin{Palatini}, μπορούμε να λάβουμε αποδεκτές προβλέψεις για τα πληθωριστικά παρατηρήσιμα μεγέθη, επιτρέποντας επομένως τη δυνατότητα το μοντέλο να περιγράφει ένα αποδεκτό σενάριο κοσμολογικού πληθωρισμού σε αντίθεση με τη συνηθισμένη τους διατύπωση κατά το φορμαλισμό μετρικής όπου προβλέψεις του μοντέλου βρίσκονται εκτός της επιτρεπόμενης περιοχής.

Το απλούστερο σενάριο ενός ελεύθερου έμμαζου βαθμωτού πεδίου μοιράζεται επίσης την ίδια μοίρα με το μοντέλο \textlatin{natural inflation}, ωστόσο θεωρώντας αυτό στον φορμαλισμό \textlatin{Palatini} με τετριμμένη ζεύξη με το μοντέλο \textlatin{Starobinsky}, λαμβάνουμε τιμές των πληθωριστικών παρατηρήσιμων ποσοτήτων εντός της επιτρεπόμενης περιοχής $1\sigma$ για τιμές του όρου μάζας κοντά στα $m\sim 10^{13}\,\textlatin{GeV}$. Η κλίμακα του πληθωρισμού, που ορίζεται ως οι τιμές πεδίου του κανονικοποιημένου πεδίου \textlatin{inflaton}, είναι ελαφρώς πάνω από την κλίμακα \textlatin{Planck}.

Μετά την επιτυχία των προηγούμενων μοντέλων, μας ενδιαφέρει να αναλύσουμε το σενάριο πληθωρισμού \textlatin{Higgs} (\textlatin{Higgs inflation}) με τετριμμένη ζεύξη με τη βαρύτητα. Είναι γνωστό ότι στον φορμαλισμό μετρικής απαιτείται μια μη-τετριμμένη ζεύξη, ωστόσο στο συγκεκριμένο πλαίσιο δείχνουμε ότι η επίδραση του όρου $R^2$ επιτρέπει την επίτευξη του κοσμολογικού πληθωρισμού με την προϋπόθεση ότι ο αριθμός των $e$-\textlatin{foldings} που απαιτείται είναι μεγαλύτερος από τον συνηθισμένο, τουλάχιστον $N\sim70$ $e$-\textlatin{foldings}.

Έχοντας αναλύσει το φάσμα των ελάχιστα συζευγμένων μοντέλων, συζητούνται επίσης τα εξαιρετικά δημοφιλή μοντέλα που έχουν μια μη-τετριμμένη ζεύξη με τον όρο \textlatin{Einstein-Hilbert}. Το γενικό χαρακτηριστικό της ισοπέδωσης του βαθμωτού δυναμικού στο πλαισίο \textlatin{Einstein} παραμένει ακόμα. Εφαρμόζοντας αυτό το πρόγραμμα στο μοντέλο \textlatin{Coleman-Weinberg} και στο μοντέλο \textlatin{induced gravity} λαμβάνουμε αποδεκτές τιμές για τα παρατηρήσιμα μεγέθη κατά τον πληθωρισμό και για τα δύο μοντέλα για ένα μεγάλο μέρος του χώρου των παραμέτρων για το κάθε μοντέλο. Ιδιαίτερη σημασία έχει το σενάριο της μη-τετριμμένης ζεύξης του \textlatin{Higgs} όπου δείχνουμε ότι η σταθερά ζεύξης $\xi$ μεταξύ του \textlatin{Higgs} και της βαρύτητας μπορεί να λάβει μικρές τιμές σε σύγκριση με αυτές που αποκτούνται απουσία του όρου $R^2$. Σε αντίθεση με το προηγούμενο σενάριο όπου θεωρήσαμε τετριμμένη ζεύξη, το μοντέλο είναι ικανό να παράγει κατάλληλα πληθωριστικά παρατηρήσιμα μεγέθη για τιμές $N\in[50,60]$ $e$-\textlatin{foldings}.

Με κίνητρο την επιτυχία των προηγούμενων μοντέλων, ερευνήσαμε το σενάριο όπου ο όρος \textlatin{Starobinsky} είναι συζευγμένος μη-τετριμμένα με ένα θεμελιώδες βαθμωτό που έχει \textlatin{quartic} δυναμικό προωθώντας τη σταθερά \textlatin{Starobinsky} $\alpha$ να συμπεριλάβει λογαριθμικές διορθώσεις $\propto\log{(\phi^2/\mu^ 2)}$ του θεμελιώδους βαθμωτού πεδίου, $\alpha\mapsto\alpha(\phi)$. Η δράση που προκύπτει στο σύστημα αναφοράς του \textlatin{Einstein} έχει παρόμοια μορφή με τα μοντέλα \textlatin{Palatini}-$R^2$ που θεωρήθηκαν προηγουμένως, ωστόσο το οροπέδιο του δυναμικού παραβιάζεται λογαριθμικά για τιμές του πεδίου $\phi>\mu$. Διαπιστώνουμε ότι η πρόβλεψη του μοντέλου σχετικά με τα πληθωριστικά παρατηρήσιμα στοιχεία είναι σε καλή συμφωνία με τα πρόσφατα δεδομένα παρατήρησης, ιδίως όταν η συνάρτηση ζεύξης $\alpha(\phi)$ παίρνει μεγάλες τιμές οι τιμές των παρατηρήσιμων μεγεθών επηρεάζονται σε αντίθεση με τα προηγούμενα μοντέλα όπου μόνο το παρατηρήσιμο μέγεθος $r$ εξαρτιώταν από το $\alpha$. Αυτό υποδηλώνει ότι άλλα μοντέλα που βρίσκονται εκτός της επιτρεπόμενης περιοχής $(+)2\sigma$ μπορούν τελικά να συμφωνούν με τις παρατηρήσεις αν εισάγουμε σταθερά που εξαρτάται από το βαθμωτό πεδίο, $\alpha(\phi)$. Επιπλέον, οι τιμές του $r$ μπορεί να κυμαίνονται από μικροσκοπικές, που είναι ένα γενικό χαρακτηριστικό των μοντέλων \textlatin{Palatini}-$R^2$, έως αρκετά μεγάλες πλησιάζοντας το άνω όριο του $r$, πράγμα που σημαίνει ότι μπορούν να έρθουν σε επαφή με μελλοντικά πειράματα αναμενόμενης ακρίβειας $10^{-3}$ ή ακόμη και $10^{-4}$. Μετά το τέλος του πληθωρισμού για το μοντέλο μελετάται η διαδικασία αναθέρμανσης. Μέσω του μηχανισμού που επισημαίνεται σε προηγούμενο κεφάλαιο είναι δυνατόν να συσχετισθούν οι παράμετροι αναθέρμανσης με τις παραμέτρους του πληθωρισμού και έτσι για διαφορετικές τιμές της παραμέτρου κατάστασης αναθέρμανσης $w_\textlatin{R}$ δείξαμε ότι το εν λόγω μοντέλο είναι πράγματι ικανό να υποστηρίξει μια εποχή αναθέρμανσης για τις συγκεκριμένες τιμές των παραμέτρων του μοντέλου που λαμβάνονται κατά τη διάρκεια του πληθωρισμού, με μέγιστη θερμοκρασία αναθέρμανσης $T_\textlatin{R}\sim 10^{15}\,\textlatin{GeV}$. Υπό την υπόθεση της στιγμιαίας αναθέρμανσης, λάβαμε ένα άνω όριο για τον αριθμό των $e$-\textlatin{foldings} $N\approx 52$. Κλείνοντας αυτήν την ενότηταν εξετάστηκαν άλλες μορφές εξάρτησης της σταθεράς με το πεδίο, $\alpha(\phi)$, που διατηρούν επίσης την επιθυμητή επιπεδότητα του δυναμικού \textlatin{inflaton}. Ακόμη και σε αυτές τις περιπτώσεις μπορέσαμε να βρούμε συμφωνία με παρατηρήσεις για ένα συγκεκριμένο μέρος του χώρου παραμέτρων του μοντέλου.

Στην τελευταία ενότητα εστιάζουμε την ανάλυσή μας στην πληθωριστική φαινομενολογία του πεδίου \textlatin{Higgs} σε συνδυασμό με την τετραγωνική βαρύτητα υπό την υπόθεση της προσέγγισης σταθερής κύλισης. Παρόμοια με τα προηγούμενα μοντέλα, η προκύπτουσα δράση στο πλαισίο \textlatin{Einstein} έχει τη μορφή του γενικευμένου τύπου $k$-\textlatin{inflation}. Συγκεκριμένα. αναλύουμε και τις δύο περιπτώσεις ζεύξης, τετριμμένης και μη, του \textlatin{Higgs} με τη βαρύτητα υποθέτοντας ότι η συνθήκη σταθερής κύλισης $\ddot{\phi}\sim\beta H\dot{\phi}$ ισχύει, όπου το $\beta$ είναι μια σταθερή παράμετρος. Και στις δύο περιπτώσεις, οι προβλέψεις για τα παρατηρήσιμα μεγέθη δείχνουν σημαντική εξάρτηση από τους κινητικούς όρους υψηλότερης τάξης, σε αντίθεση με την αντίστοιχη περίπτωση όπου τα μοντέλα εξετάζονται υποθέτοντας \textlatin{slow-roll inflation}. Ειδικά για το σενάριο τετριμμένης ζεύξης διαπιστώσαμε ότι λαμβάνουμε αποδεκτές τιμές για τα παρατηρήσιμα μεγέθη για $N\in[50,60]$ $e$-\textlatin{foldings}, κάτι που έρχεται σε αντίθεση με την περίπτωση αργής κύλισης όπου είχαμε αποδείξει ότι χρειαζόμαστε $N\gtrsim70$ $e$-\textlatin{foldings} προκειμένου το $n_s$ να βρίσκεται εντός της επιτρεπόμενης περιοχής $2\sigma$.

\clearpage
\selectlanguage{english}

\thispagestyle{empty}
\addcontentsline{toc}{chapter}{\protect\numberline{}\emph{List of publications}}

%\vspace*{1.5cm}

\begin{center}
   \textbf{\emph{\LARGE{Dissemination}}}
\end{center}

\vspace{0.7cm}

\begin{center}
    \textbf{\emph{\large{List of publications}}}
\end{center}

\vspace{0.5cm}

\noindent The main results in this thesis are closely based on the following peer-reviewed publications presented in chronological order, in which the authors are listed alphabetically according to particle physics convention.

\begin{enumerate}[align=left]
    \item[\textbf{(i)}] \emph{\large{Palatini inflation in models with an $R^2$ term}} \\\hspace*{0.5cm} I. Antoniadis, A. Karam, A. Lykkas, K. Tamvakis \\ \hspace*{0.5cm} \textbf{JCAP 11 (2018) 028}
    \item[\textbf{(ii)}] \emph{\large{Rescuing Quartic and Natural Inflation in the Palatini formalism}} \\
    \hspace*{0.5cm} I. Antoniadis, A. Karam, A. Lykkas, T. Pappas, K. Tamvakis \\\hspace*{0.5cm} \textbf{JCAP 03 (2019) 005}
    \item[\textbf{(iii)}] \emph{\large{Constant-roll in the Palatini-$R^2$ models}} \\
    \hspace*{0.5cm} I. Antoniadis, A. Lykkas, K. Tamvakis \\\hspace*{0.5cm} \textbf{JCAP 04 (2020) 04}
    \item[\textbf{(iv)}] \emph{\large{Extended interactions in the Palatini-$R^2$ inflation}} \\
    \hspace*{0.5cm} A. Lykkas, K. Tamvakis \\\hspace*{0.5cm} \textbf{JCAP 08 (2021) 043},
\end{enumerate}

\vspace{0.65cm}

\fancybreak{\Huge{\ding{101}}}

\vspace{1.35cm}

\noindent During the PhD studies I have worked in parallel in the following articles.

\begin{enumerate}[align=left]
    \item[\textbf{(i)}] \emph{\large{Frame-invariant approach to higher-dimensional scalar-tensor gravity}}  \\\hspace*{0.5cm} A. Karam, A. Lykkas, K. Tamvakis \\\hspace*{0.5cm} \textbf{Phys.Rev.D 97 (2018) 12, 124036}
    \item[\textbf{(ii)}] \emph{\large{Equivalence of inflationary models between\\ the metric and Palatini formulation of scalar-tensor theories}}\\\hspace*{0.5cm} L. J{\"a}rv, A. Karam, A. Kozak, A. Lykkas, A. Racioppi, M. Saal \\\hspace*{0.5cm} \textbf{Phys.Rev.D 102 (2020) 4, 044029}
    \item[\textbf{(iii)}] \emph{\large{Palatini-Higgs inflation with nonminimal derivative coupling}} \\\hspace*{0.5cm} I. D. Gialamas, A. Karam, A. Lykkas, T. D. Pappas \\\hspace*{0.5cm} \textbf{Phys.Rev.D 102 (2020) 6, 063522}
\end{enumerate}

\clearpage

\thispagestyle{empty}
\addcontentsline{toc}{chapter}{\protect\numberline{}\emph{Acknowledgements}}

\vspace*{4.25cm}

\begin{center}
    \LARGE{\emph{\textbf{Acknowledgements}}}
\end{center}

\vspace{0.25cm}

Undeniably, in completing the journey of a doctoral dissertation numerous people have contributed, even though by boldly displaying the author's name on the front page it might suggest otherwise. Without their support, through highs and lows, I wouldn't have made it to this point.

First and foremost, I would like to thank my scientific advisor Professor Kyriakos Tamvakis. A great deal of luck played a role in starting our collaboration and I am grateful for his initial interest in me and his support throughout my years as a graduate student. His unique way of understanding physics, posing the right questions, his work ethic, passion and professionalism are some of the aspects he influenced me in my career as well as at a personal level. But most of all, I would like to thank him for his assistance and dedicated involvement in every step of the way that bolstered me through difficult times.

I would also like to take this opportunity to thank all of the faculty members of the University of Ioannina, especially the Theoretical Physics Division, for their help and support through my undergraduate and postgraduate studies. Above all, I wish to thank Professor Panagiota Kanti who ignited my interest in gravity and aspects of cosmology, that later played a critical role in my career path.

Over the years, I had the chance to collaborate with different people, many of them have inspired, influenced and contributed to different parts of the present work. I am especially thankful to Alexandros Karam and Thomas Pappas for the many fruitful discussions we shared and for always lending an ear in times of hardship.

My time in graduate school wouldn't be the same without the rest of the PhD students in the theory group, especially my ``classmates'' Lawrence Kazantzidis, Theodoros Nakas, Ilias Tavellaris and Konstantinos Violaris-Gountonis that made this journey more enjoyable through numerous discussions on and off the topic of physics. For their constant availability to discuss various matters in physics and lending their help and experience in other areas I am grateful to Athanasios Karozas, Alexandros Karam and Thomas Pappas. Special thanks go to my office-mates Ilias and Konstantinos for their sympathy, their friendship, sharing their ideas over coffee breaks, and at times helping to keep me sane.

Looking towards the past, I would like to acknowledge the huge impact of my childhood friends and those I met in Ioannina over the years. I want to take advantage of this moment to thank individually Thanos, Kris, Lazaros, Ntinos, Panos, Chris, Alexios for making my childhood years better and, among others, Christos M., Kostas T., Fotis L., Nikos A., Theofanis S., Georgios V., Theodoros M., Panos G., Ilias T., Lavrentis K., Thanos D. and Alex K. for making my college years more exciting and all the great times we shared together. Likewise, my sincere thanks go to everyone who has helped me in this journey and had infinite patience with me, even thought at times I showed the inverse of that. I couldn't have done it without you.

My highest gratitude goes to my family - my sister, parents and grandparents. I am especially thankful to my mother for her unconditional love that made every ``stage'' I stepped on feel like the whole stadium was with me, and to my little sister Nefeli (``little'' being my perspective) for always being there for me.

Finally, I am deeply grateful to my partner Tijana for her love and support. There is absolutely no way I can express in a paragraph how much our conversations and jokes, your little and large kindnesses, your passion and guidance have played a major role in shaping who I am and, more relevant, in the completion of my PhD. Thank you for being by my side through all of this and for the new heights of happiness we climb together.

\vspace*{3cm}

This research is co-financed by Greece and the European Union (European Social Fund- ESF) through the Operational Programme ``Human Resources Development, Education and Lifelong Learning'' in the context of the project ``Strengthening Human Resources Research Potential via Doctorate Research – 2nd Cycle'' (MIS-5000432), implemented by the State Scholarships Foundation (IKY).
\begin{figure}[H]
    \centering
    \includegraphics[width = 0.65 \textwidth]{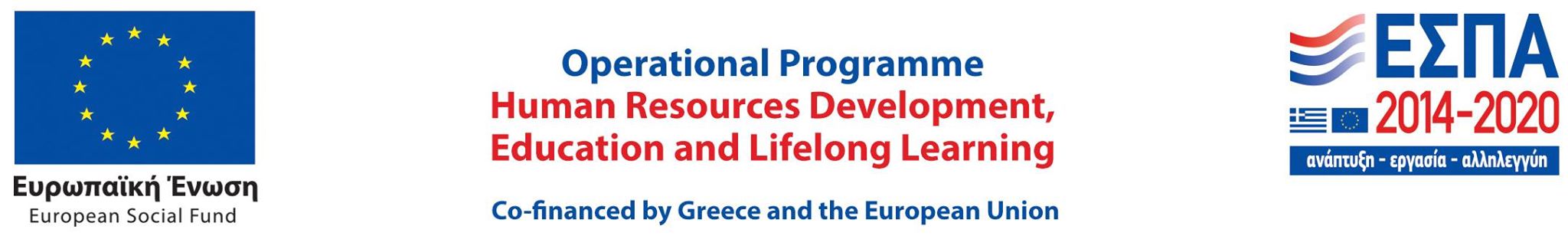}
\end{figure}
\thispagestyle{plain}

\clearpage

\begin{KeepFromToc}
  \tableofcontents
  \thispagestyle{plain}
\end{KeepFromToc}

\mainmatter
\pagestyle{ruled}
\chapterstyle{ell}
\renewcommand*{\chaptitlefont}{\normalfont\Huge\sffamily}
\setsecheadstyle{\LARGE\sffamily\bfseries}

\chapter{Introduction}

It is not an exaggeration to state that the Standard Model (SM) of particle physics~\cite{Glashow1961,Weinberg1967,Salam1968} and the General theory of Relativity (GR)~\cite{Einstein1916} constitute the pillars of our understanding of the physical world. After all, both of them are established as the most complete theories to date. 

The last piece of the puzzle of the SM came with the discovery of the Higgs boson~\cite{Higgs1964a,Higgs1964,Englert1964,Guralnik1964} at CERN's Large Hardon Collider (LHC)~\cite{Aad2012,Chatrchyan2012} that validated and solidified the predictions of the theory. Even though processes described by the SM are (or can in principle be) observed, there exist however others that SM cannot account for without nontrivial modifications to its content, leading to possibly serious ramifications. Many theoretical issues are brought up when shortcomings of the SM are discussed,\footnote{The notion of disagreement of an experiment with the SM is measured in $\sigma$, meaning that after some observation is above some $\sigma$ ($5\sigma$ is believed to be the threshold) of the SM prediction it is labeled as ``new physics''. Actually, this would be an encouraging result hinting towards a possible avenue of research, much awaited by the theoretical physics community. Currently, under serious consideration are the measured value of the anomalous dipole moment of the muon (preliminary results point at deviation of $4.2\sigma$~\cite{Abi2021}; see also ref.~\cite{Blum2013} for a review on the subject) and the $B$ meson decay (BaBar reports a $3.4\sigma$~\cite{Lees2012}; LHCb $2.1\sigma$~\cite{Aaij2015}).} some of them being the inability of the SM to provide a dark matter candidate and also to explain the dark energy of the universe via a vacuum energy density of the appropriate magnitude. Arguably the most important of them is its prediction of exactly massless neutrinos which has been observationally falsified via neutrino oscillations~\cite{Fukuda1998,AguilarArevalo2001,Boehm2001,Ahn2003,Ashie2004,Araki2005,Abe2008}. 

On the other hand, the theory of GR is validated through the years~\cite{Will2006} with its latest achievement being the observation of its predicted gravitational waves produced by a black hole merger~\cite{Abbott2017,AmaroSeoane2017}. Issues regarding GR arise primarily due to its failure to be quantised (or renormalised for that matter~\cite{Hooft1974,Goroff1985}) and, in general, quantum phenomena in curved spacetime are inadequately understood in that context.\footnote{Even though different approaches to quantising gravity have been proposed with various levels of success, the overall statement still holds at present time.} It is however entirely possible that the overall approach of quantising a theory with a geometrical interpretation is ill-defined and as such results obtained in this way should be considered approximate at best. For example, other bold claims of GR such as the black hole solutions (and the initial singularity; ``Big Bang'') were viewed as mathematical paradoxes to be snuffed out by a complete theory of gravity.\footnote{As is known a black hole has already been observed and we even managed to capture an image of it and its shadow~\cite{Akiyama2019}! There is however a justified concern regarding predictions made by the theory due to its breakdown at the point of the singularity.}

The incompatibility of the SM with GR demonstrates then that the aspiration of unifying the fundamental forces under one theoretical framework is still far away from being realised, with many possible shortcomings on the horizon. However, cosmology and particle physics are getting increasingly closer over the years, undeniably not better exemplified by developments in the theory of \emph{cosmic inflation}~\cite{Guth1981,Linde1982,Albrecht1982,Sato1981,Linde1983a}. By supplementing GR with an additional fundamental scalar degree of freedom that is able to support a quasi-de Sitter expansion during the early moments of the universe, known as inflation, we can explain the observed flatness, homogeneity and isotropy at large distances measured by precision data~\cite{Aghanim2020}. There exist however certain patches of inhomegeneity measured in the Cosmic Microwave Background (CMB)~\cite{Penzias1965} approximately at $10^{-5}$ on large energy scales. Surprisingly, these fluctuations are almost Gaussian and relatively scale-invariant and can be also explained by inflation when the classical de Sitter fluctuations are treated quantum mechanically~\cite{Hawking1982,Starobinsky1982,Guth1982}. Constraints from observations include the amplitude of these fluctuations that in turn are translated to further constraints on the proposed models and increasing accuracy of these observations can heavily restrict these models~\cite{Akrami2020,Ade2018}. In fact, a direct detection of primordial gravitational waves coming from inflation can determine the energy density of inflation via the amplitude of the tensor modes, although the growing precision of experiments can place a substantial upper bound already. Most of the proposed inflationary models are inspired in some way from developments in particle physics, perhaps demonstrated best in the Higgs inflation model~\cite{Bezrukov2008,DeSimone2009,Bezrukov2009a}, in which the Higgs boson assumes the role of the field driving inflation and serves as the simplest inflationary scenario in terms of extensions to the SM field content.

\begin{figure}[H]
    \centering
    \includegraphics[width=\textwidth]{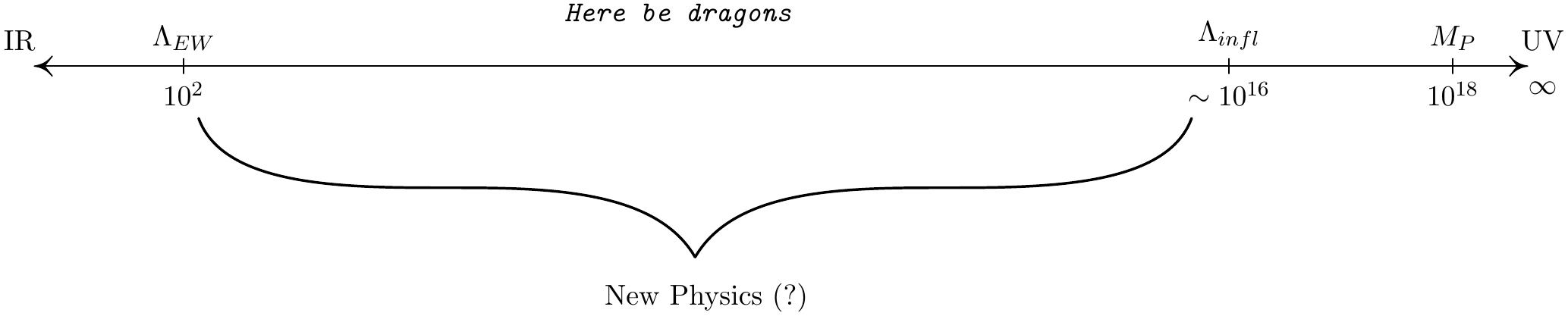}
    \caption{The difference between the EW scale, the Planck and the (supposed -- model dependent) scale of inflation. It is possible that the large chasm separating these energy scales may include New Physics phenomena or it may end up being simply a transitional era between high- and low-energy physics.}
\end{figure}

It cannot be overstated how the generalisation of the notion of spacetime to a curved manifold $\mathcal{M}$ with a metric $\text{g}$ as supposed by the theory of GR~\cite{Einstein1916} has revolutionised how one approaches modern theories that attempt to incorporate gravity with conventional particle physics. Perhaps one of the most crucial, and simultaneously unassuming, postulates of GR is the Levi-Civita connection, in other words that the connection on the manifold is \emph{metric-compatible} (preserves the metric) and \emph{symmetric} (torsion freedom). The particular choice of the connection is not all accidental since it was proven by Levi-Civita~\cite{LeviCivita1916} (and Christoffel~\cite{Christoffel1869}) that is linked to the notion of parallel transport of vectors in curved space, and additionally the idea of covariant derivative of a vector along a curve was generalised for the case of a Riemannian geometry, which is exactly what GR was in need of at the time. Similarly to other aspects of GR, it did not take long for the assumption to be challenged, done first in the works of A. Palatini~\cite{Palatini:1919}, in which by generalising the concept of a connection to a \emph{metric-affine} one he was able to show that both formulations of GR are equivalent. The main difference is that the Levi-Civita condition is recovered at the level of equations of motion of the theory (on-shell) and since the \emph{Palatini} or \emph{first-order formulation of GR} is equivalent with the conventional metric or second-order formulation the matter was put to rest with little attention paid until subsequent works of E. Cartan (see for a review~\cite{Hehl1976,Gronwald1995,Shapiro2002}) demonstrating that non-Riemannian geometries can also support a theory of gravity. Then, following recent developments in extended/modified theories of gravity, primarily focused on cosmological issues (e.g. dark energy models), it appears that the Palatini variation is translated to a fundamental question on the parametrisation of the gravitational degrees of freedom. 

\subsection*{\emph{Outline}}
\addcontentsline{toc}{subsection}{\protect\numberline{}\emph{Outline}}

In the next chapter, ch.~\ref{Ch2:Inflation}, the theory of cosmic inflation is explored and most of the physical results needed in later chapters are extracted with minimal mathematical rigor, referring the reader to the vast literature on the subject for a complete discussion. We start by analysing several aspects that make inflation appealing as a solution of the headaches induced by the initial singularity. Even though actual predictions regarding the inflationary period are highly model-dependent everything presented in the chapter is completely model-independent, potentially sacrificing some of the exciting results the theory offers which are discussed in a later chapter. The actual prediction of inflation, i.e. the energy density perturbations of the inflaton field are studied and presented in great detail due to their significance. By then analysing the simplest model of an inflation, a real scalar field minimally coupled to gravity with its self-interacting potential under the slow-roll approximation, the slow-roll parameters are connected with the amplitude of these fluctuations marking a way to quantitatively understand and make predictions about quantities during inflation that are observed today. To compensate that, we examine also the so-called constant-roll approximation and relate predictions between the two approximations for particular models in a later chapter. At the end of the chapter we present a brief overview of the reheating era that follows inflation, first by reviewing some of the mechanisms possible and finally concentrate on parametrising the reheating in terms of the inflationary parameters, maintaining in that way a direct connection between the two that is readily used to apply potentially stricter bounds on both of them. At the very least it can show if a candidate model of inflation can support a reheating phase.

In ch.~\ref{Ch3:FirstOrder} we analyse the Palatini or first-order formalism setting it up to take center stage in the main part of the thesis. First we draw attention to some aspects of the conventional metric formalism that are fundamentally different in the Palatini formulation of GR, such as the York-Gibbons-Hawking (YGH) term. In order to establish the Palatini variation we start by first defining the metric-affine spaces and particularly the notion of torsion, nonmetricity and curvature on $\mathcal{M}$, that in turn allows us to approach the subject of an affine connection. Then we derive the field equations for the Einstein-Hilbert Lagrangian now under the assumption of the first-order formalism, i.e. the metric and the connection do not have an a priori dependence on each other, in which the initial results of Palatini (and related works) are effectively reproduced via the apparent equivalence of these two formulations of GR. Closing the chapter we discuss the endeavor of understanding if that equivalence remains at the quantum level.

The ch.~\ref{Ch4:QuadGrav} is comprised primarily by the main results by using and combining some of the ideas presented in the previous chapters. We start by highlighting that modified theories of gravity such as the Starobinsky model of inflation (among others) do not have an equivalent description in the metric and Palatini formulation. Specifically, the $R^2$ term in the first-order formalism does not generate a dynamical (propagating) degree of freedom which in the usual metric formalism is identified with the inflaton field driving inflation. Therefore, in the interest of describing an inflationary phase we attempt to couple the $R+R^2$ gravitational term, either minimally or nonminimally, to a fundamental scalar field with a self-interacting potential term. There we notice that various prominent model candidates for inflation that were previously excluded or were in marginal contact with recent observations in the so-called Palatini-$R^2$ models their predictions reside within the allowed region by the Planck 2018 mission~\cite{Akrami2020}. The Higgs inflation model is singled-out due to its appeal by not extending the particle content of the SM, and predictions of the Higgs-$R^2$ model in the Palatini formalism are also analysed in the constant-roll approximation.

All of the previous chapters are supplemented by small sections, referred to usually as ``digressions'', that provide more context to each particular discussion, however these results are not applied in the main part of the thesis. Lastly, in the last chapter (ch.~\ref{Ch5:conclusions}) we summarise the main findings of the present work.

\subsection*{\emph{Notation}}
\addcontentsline{toc}{subsection}{\protect\numberline{}\emph{Notation}}

The metric signature convention used throughout the thesis is the ``mostly-positive''
\begin{equation*}
    \eta_{\mu\nu}=\text{diag}(-,+,+,+)\,.
\end{equation*}
The Einstein summation convention is also implied and we use the usual notation that Greek indices denote strictly spacetime indices, $\mu,\nu,\ldots=0,1,2,3$ and Latin indices are internal indices specifically for gravity they take up values of $i,j,\ldots=1,2,3$. Following the conventions of the community we also employ the condensed notation
\begin{equation*}
    \mathrm{d}^nx\equiv\prod_{i=1}^n\mathrm{d}x^i\,.
\end{equation*}
Throughout the thesis we also use the dot and the prime over quantities to denote derivative with respect to time and with respect to the function's argument, respectively; in other words:
\begin{equation*}
    \dot{f}(x)\equiv\frac{\mathrm{d}f(x)}{\mathrm{d}t}\,,\qquad\qquad f'(x)\equiv\frac{\mathrm{d}f(x)}{\mathrm{d}x}\,.
\end{equation*}

When we refer to the metric tensor in text we use the symbol ``$\text{g}$'' and assign $g$ strictly to its determinant that mostly shows up in the action functional in the form of $\sqrt{-g}$ with $g\equiv\text{det}(g_{\mu\nu})$. Throughout the thesis the (anti)symmetrisation of the indices is weighted by a factor of $1/n!$ where $n$ is the number of indices, e.g.
\begin{equation*}
    A{}_{(\mu}B{}_{\nu)}\equiv\frac{1}{2!}(A_\mu B_\nu+A_\nu B_\mu)\,.
\end{equation*}
Likewise, the antisymmetrisation $A_{[\mu}B_{\nu]}$ is defined with a minus on the RHS. Indices separated as shown below
\begin{equation*}
    A_{(\mu|\nu}B_{\rho)}\equiv\frac{1}{2!}(A_{\mu\nu}B_{\rho}+A_{\rho\nu}B_{\mu})\,,
\end{equation*}
are assumed to be excluded from the (anti)symmetrisation.

In chapter~\ref{Ch3:FirstOrder} we adopt the abstract notation (when convenient) instead of the explicit one in some parts of the discussion, therefore we should emphasize in the following table their equivalent form
\begin{table}[H]
    \centering
    \begin{tabular}{c c c}
    \text{Object}&\text{Abstract}&\text{Explicit}\\\hline
    \hline
     Vector field& $X$  &$X^\mu$\\
     Tensor product&$X\otimes Y$&$X^\mu Y^\nu$\\
     Covariant derivative& $\nabla X(Y)=\nabla_Y X$&$Y^\mu\nabla_\mu X^\nu$\\
     Cov. derivative tensor& $\nabla X$&$\nabla_\mu X^\nu=\partial_\mu X^\nu+{\Gamma^\nu}_{\mu\rho}X^\rho$\\
     Metric tensor&$\braket{\cdot\,,\cdot}\equiv\text{g}$&$g_{\mu\nu}$\\
     Inner product&$\braket{X,Y}\equiv\text{g}(X,Y)$&$g_{\mu\nu}X^\mu Y^\nu$\\\bottomrule
\end{tabular}
    %\caption{Caption}
    %\label{tab:my_label}
\end{table}
The Christoffel symbols - coefficients of the Levi-Civita connection - are defined by
\begin{equation*}
    {\Gamma^\rho}_{\mu\nu}=\frac{1}{2}g^{\rho\lambda}\left(\partial_\mu g_{\lambda\nu}+\partial_\nu g_{\mu\lambda}-\partial_\lambda g_{\mu\nu}\right)=:\left\{{}_\mu{}^\rho{}_\nu\right\}_\text{g}\,,
\end{equation*}
where the subscript $\text{g}$ in the last equation is occasionally neglected and is implied through context. In this work the Lie derivative is denoted by $\mathcal{L}$ and the Lagrangian density by $\mathscr{L}$, even though a possible confusion between the two is highly unlikely. We also reserve the notation $\stackrel{!}{=}$ to denote equality modulo equations of motion, however when the context allows us we opt to neglect it.

In scalar field theories it is useful to recast the dimensional quantities in the \emph{natural units} in which
\begin{equation*}
    \hbar:=1=:c\,,
\end{equation*}
allowing us then to cast the dimensions of any physical quantity in terms of mass dimensions. We may also set the reduced Planck mass to unity
\begin{equation*}
    M_P^2\equiv\frac{1}{8\pi G}:=1\,,
\end{equation*}
for ease of notation but at the expense of obscuring the dimensions of the quantities. All of the above hold unless otherwise stated.

\subsection*{\emph{Glossary of abbreviations}}
\addcontentsline{toc}{subsection}{\protect\numberline{}\emph{Abbreviations}}

\begin{multicols}{2}
\begin{center}
\begin{itemize}
    \item[\textbf{ADM}]\quad \emph{Arnowitt-Deser-Misner}
    \item[\textbf{BBN}]\quad \emph{Big Bang nucleosynthesis}
    \item[\textbf{BD}]\quad  \emph{Brans-Dicke}
    \item[\textbf{BSM}]\quad \emph{Beyond the Standard Model}
    \item[\textbf{CMB}]\quad \emph{Cosmic Microwave Background}
    \item[\textbf{CR}]\quad  \emph{Constant-roll}
    \item[\textbf{CSI}]\quad  \emph{Classical scale invariance}
    \item[\textbf{CW}]\quad \emph{Coleman-Weinberg}
    \item[\textbf{DEC}]\quad \emph{Dominant energy condition}
    \item[\textbf{DOF(s)}]\quad \emph{Degree(s) of freedom}
    \item[\textbf{EEP}]\quad \emph{Einstein equivalence principle}
    \item[\textbf{EFT}]\quad \emph{Effective field theory}
    \item[\textbf{EH}]\quad \emph{Einstein-Hilbert}
    \item[\textbf{EoM(s)}]\quad \emph{Equation(s) of motion}
    \item[\textbf{EW}]\quad \emph{Electroweak}
    \item[\textbf{FRW}]\quad \emph{Friedmann-Robertson-Walker}
    \item[\textbf{GR}]\quad \emph{General Relativity}
    \item[\textbf{GUT}]\quad \emph{Grand Unified Theory}
    \item[\textbf{HSRP(s)}]\quad \emph{Hubble slow-roll parameter(s)}
    \item[\textbf{LHC}]\quad \emph{Large Hardon Collider}
    \item[\textbf{LHS}]\quad \emph{Left-hand side}
    \item[\textbf{LQG}]\quad \emph{Loop quantum gravity}
    \item[\textbf{NEC}]\quad \emph{Null energy condition}
    \item[\textbf{($\ldots$N)NLO}]\quad \emph{($\ldots$next-to) next-to leading order}
    \item[\textbf{PSRP(s)}]\quad \emph{Potential slow-roll parameter(s)}
    \item[\textbf{QCD}]\quad \emph{Quantum Chromodynamics}
    \item[\textbf{QM}]\quad \emph{Quantum Mechanics}
    \item[\textbf{RG}]\quad \emph{Renormalisation group}
    \item[\textbf{RHS}]\quad \emph{Right-hand side}
    \item[\textbf{SEC}]\quad  \emph{Strong energy condition}
    \item[\textbf{SEP}]\quad \emph{Strong equivalence principle}
    \item[\textbf{SM}]\quad \emph{Standard Model}
    \item[\textbf{SR}]\quad \emph{Slow-roll}
    \item[\textbf{SRP(s)}]\quad \emph{Slow-roll parameter(s)}
    \item[\textbf{SSB}]\quad \emph{Spontaneous symmetry breaking}
    \item[\textbf{VEV}]\quad \emph{Vacuum expectation value}
    \item[\textbf{WEC}]\quad \emph{Weak energy condition}
    \item[\textbf{WEP}]\quad \emph{Weak equivalence principle}
    \item[\textbf{YGH}]\quad \emph{York-Gibbons-Hawking}
    \item[\textbf{YM}]\quad \emph{Yang-Mills}
\end{itemize}
\end{center}
\end{multicols}

\chapter{Inflation}
\label{Ch2:Inflation}

The idea of the Big Bang singularity is accompanied with a series of puzzles cosmological in nature, primarily regarding  the initial conditions of the universe, such as the \emph{horizon problem}, the \emph{flatness} and the \emph{magnetic monopole problem}~\cite{Guth1981,Linde1982}. In principle, most of physics deals with the detailed evolution of an initial state, within some boundaries of uncertainty. Of course, the issue of initial conditions takes center stage when one considers the initial state of the universe, however one can avoid the discussion altogether by admitting a tremendous fine-tuning of the initial conditions. The theory of \emph{Cosmic Inflation}~\cite{Guth1981,Linde1982,Albrecht1982,Sato1981,Linde1983a} was first proposed in order to address these issues of the Big Bang cosmology, and suggests a period of exponential, quasi-de Sitter expansion of space of the universe moments after its genesis, that is capable of leading the universe in that peculiar initial state. Advanced in the late 70s and early 80s, it boldly applied insights and theories from the successful particle physics frontier, yet its exact particle physics mechanism is still unknown. As such, different models of inflation are proposed that lead to various predictions, which are then linked and verified or falsified by observations. Inflation has proven to be the \emph{simplest} realisation, in terms of application and assumptions, of such an attempt to understand our early universe; however, a UV-complete theory of gravity can ultimately constitute this discussion redundant.

While inflation was constructed so that it can naturally address the issues discussed above, it includes an essential feature; the ability to seed the large-scale structure formation of the known universe~\cite{Starobinsky1979,Mukhanov1981,Hawking1982,Hawking_1983,Starobinsky1982,Guth1982} through growing quantum fluctuations of the field describing inflation, known as the \emph{inflaton}. This feature hints at the predictive power of the theory and its success in leading to an observable effect of the interplay between GR and Quantum Mechanics (QM), possibly one of the few known cases in physics today. This effect is testable experimentally in the anisotropies observed in the CMB~\cite{Aghanim2020} and, with growing precision (missions are planned for the next decades), it allows us to constrain the vast model space of the inflationary paradigm~\cite{Akrami2020,Ade2018}.

Inflation serves also in diluting the undesired relics (e.g. topological defects); as a consequence, at the end of it only the zero mode of the inflaton and tiny fluctuations of the metric remain. Therefore, it is natural to assume the possible existence of a period during which the universe \emph{thermalised}, from cold and empty right after inflation to the large energy and entropy observed at the current horizon. Throughout that period, known as \emph{reheating}, the inflaton's energy density is converted to radiation (or other massive particles) at the end of inflation through different mechanisms. In fact, in its first years of study the reheating era was thought to be largely understood via its minimal scenario, in which the inflaton field decays to other fields that it was coupled with,\footnote{The exponential expansion of the early universe proposed by inflation, would dilute the energy densities of these particles. So, these types of couplings proposed between the fields can indeed exist during inflation, even though they do not play a role during that period other than maybe inducing radiative corrections.} referred to now as the \emph{perturbative reheating} scenario~\cite{Abbott1982,Dolgov1982,Albrecht1982b}. Since then the landscape of the possible mechanism of reheating has expanded dramatically, including also nonperturbative dynamics~\cite{Traschen1990,Kofman1994,Kofman1997,Greene1997,Felder1999,Felder2001,Felder2001a,Shuhmaher2006,Dufaux2006}, implying that its underlying nature is highly complicated and uncertain at this point (see ref.~\cite{Lozanov2019} for a review).

Unfortunately, during the phase of reheating the comoving scales re-enter the horizon, making its indirect detection a challenge, unlike during inflation in which they ``freeze-out'' and leave an imprint on the CMB. Also, as would be the case in inflation and every era that precedes recombination, it is not directly detectable. Therefore, the period of reheating is hardly constrained observationally. A way to at least restrain its expansion history is to relate its comoving modes re-entering the horizon to the ones at horizon exit of inflation. Then, that kind of parametrisation can provide indirect signatures and also assist in minimizing the parameter space of the inflationary model~\cite{Dodelson2003,Liddle2003a,Dai2014,Munoz2014,Gong2015,Cook2015a}.

In this chapter, we briefly illustrate the puzzles of the Hot Big Bang, by first introducing some basic wisdom from modern cosmology, and then addressing them in the framework of inflation. The elementary implementation of an inflationary scenario into the theory is discussed so that the field dynamics are also presented, which will serve as the foundations for a large part of the thesis. Certain subtle points of slow-roll inflation and its derivatives are discussed in detail, primarily highlighting the observable quantities predicted by inflation. We direct the reader to an indicative list of reviews on inflation in refs.~\cite{Lyth1999,Weinberg2008,Linde2008,Baumann2011,Senatore2017} and references therein for further details. Then, we present the concept of reheating after inflation. After briefly reviewing some of its more intricate mechanisms, we parametrise the reheating parameters in terms of the inflationary ones and thus making possible contact with high-energy physics phenomena.

\section{A sketch of Modern Cosmology}

In order to work out the details of the inflationary era we require the introduction of some basic aspects of modern cosmology, briefly reviewed in this section. Cosmology is established under the \emph{cosmological principle}, which states that the universe viewed by two observers at two different points looks the same. Despite stated as a principle it has been observationally confirmed that universe is \emph{homogeneous} and \emph{isotropic} at large scales ($\gtrsim 100$ Mpc)~\cite{Aghanim2020}, meaning it has a translational and rotational invariance. Without loss of generality, the metric respecting these symmetries is the Friedmann-Robertson-Walker\footnote{Throughout the literature it is also referred to as the Friedmann-Lemaitre-Robertson-Walker (FLRW) or just Robertson-Walker (RW) metric. It should also be noted that we invoked our notation $c\equiv1$ in order to present the metric in such a form. In general, including the speed of light in the definition of $\mathrm{d}s^2=-c^2\mathrm{d}t^2+a^2(t)\,\mathrm{d}\Sigma^2$ means that the coordinates $x^\mu$ have the dimension of length, which in turn simplifies the dimensional analysis of the theory. More on that subject in chapter~\ref{Ch4:QuadGrav}.} (FRW) metric
\begin{equation}\label{Cond:FRW}
\mathrm{d}s^2=-\mathrm{d}t^2+a^2(t)\left(\frac{\mathrm{d}r^2}{1-k\,r^2}+r^2\left(\mathrm{d}\theta^2+\sin^2{\theta}\,\mathrm{d}\phi^2\right)\right)\,,
\end{equation}
where $a(t)$ is known as the \emph{scale factor} describing the evolution of the spatial slices $\Sigma$ with cosmic time $t$. Here, $k$ is a curvature parameter that assumes values of $k\!=\!\left\{0,+1,-1\right\}$ for these spacelike $3$-hypersurfaces that are flat (Euclidean $\mathbb{E}^3$), positively curved (spherical $\mathbb{S}^3$) and negatively curved (hyperbolic $\mathbb{H}^3$), respectively. Clearly, the metric \eqref{Cond:FRW} is invariant under a constant rescaling of the form $a\mapsto a\lambda$, $r\mapsto r/\lambda$ and $k\mapsto k\lambda^2$, utilised in setting the scale factor at present day to unity, $a_0\equiv a(t_0)\equiv 1$. For reasons that will soon become clear, it is useful to re-express the metric \eqref{Cond:FRW} in the following way
\begin{equation}\label{Cond:FRWchi}
\mathrm{d}s^2=-\mathrm{d}t^2+a^2(t)\left(\mathrm{d}\chi^2+\mathrm{d}\Omega^2\left\{\begin{matrix}
\sin^2{\chi},&k=+1\\
\chi^2,&k=0\\
\sinh^2{\chi},&k=-1
\end{matrix}\right.\ \right)\,,
\end{equation}
in terms of $\chi\equiv\int\!\mathrm{d}r/\sqrt{1-kr^2}$.

The notion of the event and particle horizon is very prominent in cosmology and very critical to the establishment of the inflationary era. In order to introduce the causal structure of spacetime we require the concept of \emph{conformal time}, defined as
\begin{equation}
\tau\equiv\int\!\frac{\mathrm{d}t}{a(t)}\,.
\end{equation}
In an isotropic space the propagation of light is then described by the line element
\begin{equation}
\mathrm{d}s^2=a^2(\tau)\left(-\mathrm{d}\tau^2+\mathrm{d}\chi^2\right)\,,
\end{equation}
which is conformally flat. Similar then to flat space, the null geodesics ($\mathrm{d}s^2=0$) of photons are given in the $\chi-\tau$ plane as
\begin{equation}
\chi(\tau)=\pm\tau+\text{const.}\,,
\end{equation}
corresponding to straight lines at $45^\circ$ angle.\footnote{Obviously if we used the proper time $t$ the light cone would be curved.} 

\begin{figure}[H]
    \centering
    \includegraphics[scale=0.8]{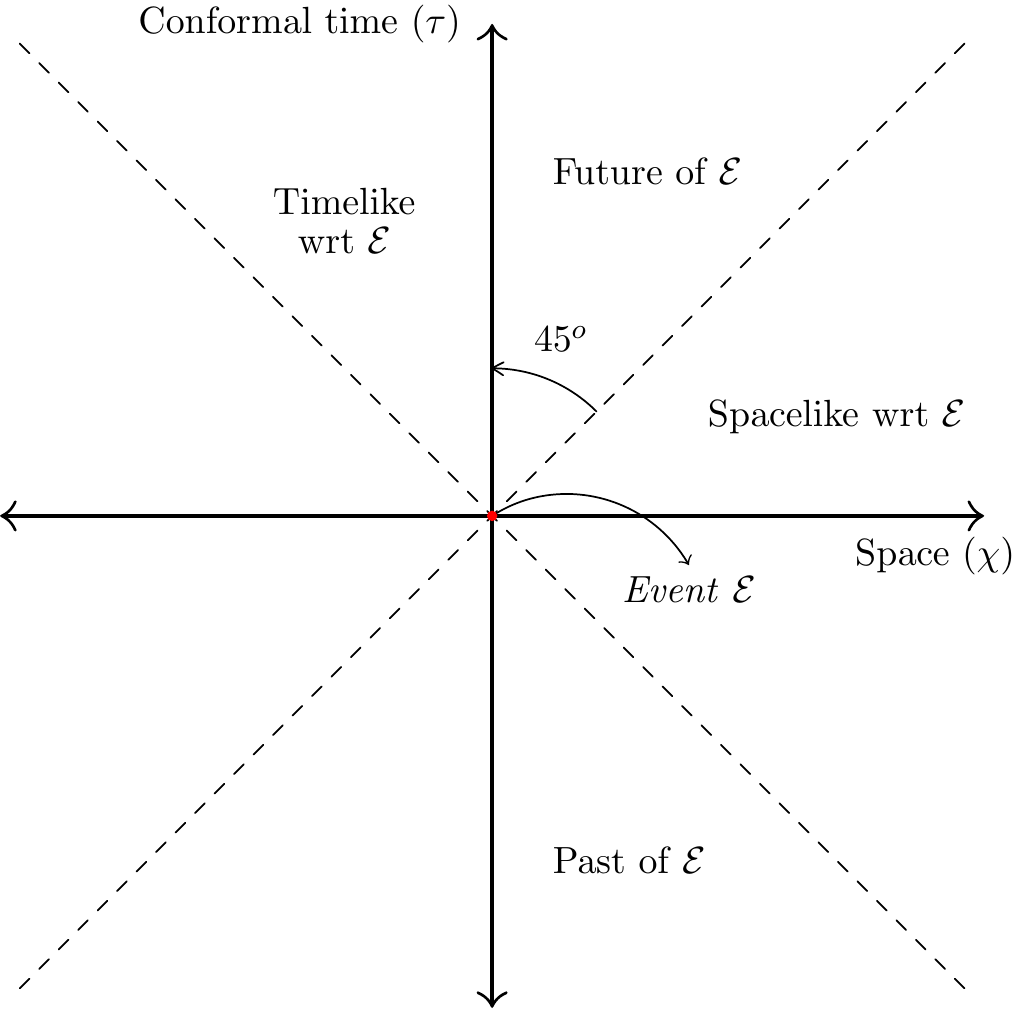}
    \caption{The light cones, future and past, are displayed by the dashed lines starting from some event $\mathcal{E}$. An event residing outside the cones is causally disconnected from $\mathcal{E}$ and travel along spacelike geodesics ($\mathrm{d}s^2<0$). Likewise, photons travel exactly on the lines of $\mathrm{d}s^2=0$, known as null geodesics and massive particles travel in timelike worldlines with $\mathrm{d}s^2>0$. Part of spacetime that lies within the interior and the light cone itself is known to be causally connected to the event $\mathcal{E}$.}
    \label{fig:Light-Cone-GEN}
\end{figure}

From a starting point in time $t_i$ until some time later $t$ light propagated a (maximum) comoving distance
\begin{equation}\label{Cond:PartHor}
\chi_p(\tau)\,\equiv\, \tau-\tau_i\,=\,\int_{t_i}^{t}\!\frac{\mathrm{d}t'}{a(t')}\,,
\end{equation}
called the (comoving) \emph{particle horizon}, with a physical size of $a(t)\chi_p$. Let us consider a single moment, say $t_\text{max}$, then there exists a maximum separation between two points after which no signal can be received between them in the future. Again, in comoving coordinates it is described as
\begin{equation}\label{Cond:EvntHor}
\chi_e(\tau)\,=\,\tau_\text{max}-\tau\,=\,\int_t^{t_\text{max}}\!\frac{\mathrm{d}t'}{a(t')}\,,
\end{equation}
and is called the \emph{event horizon}, with a physical size of $a(t)\chi_e$.

A crucial quantity often used to characterise the FRW spacetime is the rate
\begin{equation}
H(t)\equiv\frac{\dot{a}(t)}{a(t)}\,,
\end{equation}
the \emph{Hubble parameter}. The dot here denotes derivative with respect to time $t$. It is often used in characterising a specific scale in the expanding universe.

\subsection{\emph{Field equations \& fluid dynamics}}

In order to understand the evolution of the metric we require the equations that govern it, known as the Einstein field equations. Starting from the Einstein-Hilbert action, under variation with respect to the metric tensor we obtain the following famous equation
\begin{equation}
G_{\mu\nu}=M_P^{-2}\,T_{\mu\nu}\,,
\end{equation}
where $G_{\mu\nu}$ is the Einstein tensor and $T_{\mu\nu}$ denotes the energy-momentum tensor of the universe.\footnote{The cosmological constant $\Lambda$ is often included in the starting Lagrangian, but in principle can be absorbed in the energy-momentum tensor as a fluid. Pertaining to the discussion at hand the interpretation is equivalent and as such we limit ourselves to the idea of a fluid rather than a free constant of the theory. For the sake of completeness let us also include the definition of the Ricci and Einstein tensors $$R_{\mu\nu}=\partial_\rho\Gamma^\rho_{\mu\nu}-\partial_\nu\Gamma^\rho_{\mu\rho}+\Gamma^\lambda_{\mu\nu}\Gamma^\rho_{\rho\lambda}-\Gamma^\lambda_{\mu\rho}\Gamma^\rho_{\nu\lambda}\,,$$ and $$G_{\mu\nu}\equiv R_{\mu\nu}-\frac{1}{2}g_{\mu\nu}R=\left(\delta_\mu^\rho\delta^\lambda_\nu-\frac{1}{2}g_{\mu\nu}g^{\rho\lambda}\right)\left(\partial_\sigma\Gamma^\sigma_{\rho\lambda}-\partial_\lambda\Gamma^\sigma_{\rho\sigma}+\Gamma^\kappa_{\rho\lambda}\Gamma^\sigma_{\sigma\kappa}-\Gamma^\sigma_{\lambda\kappa}\Gamma^\kappa_{\sigma\rho}\right)\,.$$ } Also, $M_P$ is the reduced Planck mass and is hereafter set to unity, $M_P\equiv1$. The complete derivation of the Einstein field equations is postponed for a later chapter, and we refer the reader to sec.~\ref{Sec:MetFormRevisit} in which we delve into more details regarding the subject. Since the universe is homogeneous and isotropic the energy-momentum tensor is heavily restricted, taking the general form
\begin{equation}
T_{\mu\nu}=(\rho+p)\,u_\mu u_\nu+p\,g_{\mu\nu}\,,
\end{equation}
where $u^\mu\equiv\mathrm{d}x^\mu/\mathrm{d}\tau$ is the $4$-vector timelike velocity and, in a frame that is comoving with the perfect fluid described by $T_{\mu\nu}$, we may choose $u^\mu=\{1,0,0,0\}^\text{T}$. Here, $\rho$ is called the (rest) energy density of the system and $p$ the (principal) pressure. Conservation of the energy-momentum tensor (or through the Bianchi identity $\nabla^\mu G_{\mu\nu}=0$) it is straightforward to show\footnote{In the case of FRW symmetric metric given by $g_{00}=-1$ and $g_{ij}=a^2(t)\gamma_{ij}$, the Christoffel symbols read \begin{align*}
\Gamma^\mu_{00}&=0=\Gamma^0_{0\mu}\,,\\
\Gamma^0_{ij}&=\dot{a}a \gamma_{ij}\,,\\
\Gamma^i_{0j}&=H\delta^i_j\,,\\
\Gamma^i_{jk}&=\frac{1}{2}\gamma^{i\ell}\left(\partial_j\gamma_{k\ell}+\partial_k\gamma_{j\ell}-\partial_\ell \gamma_{jk}\right)\,,
\end{align*}
where the last one is $\Gamma^i_{jk}=0$ in the case of $\mathbb{E}^3$. More on the connection coefficients in chapter~\ref{Ch3:FirstOrder}.} that the continuity equation for the fluid reads
\begin{equation}\label{Eq:Continuity}
\nabla_\mu T^\mu_\nu=0\quad\implies\quad \frac{\mathrm{d}\rho}{\mathrm{d}t}+3H(\rho+p)=0\,,
\end{equation}
which is the first law of thermodynamics $\mathrm{d}U=-p\,\mathrm{d}V$, assuming an adiabatic expansion ($\mathrm{d}S=0$). By defining a constant state parameter
\begin{equation}
w\equiv\frac{p}{\rho}\,,
\end{equation}
we can integrate eq.~\eqref{Eq:Continuity} to obtain
\begin{equation}
\rho\propto a^{-3(1+w)}\,.
\end{equation}
All known cosmological fluids have one of three equations of state: matter, radiation and vacuum energy. Matter includes nonrelativistic particles with zero pressure, $w=0$, and free energy density decreasing as $\rho_m\propto a^{-3}$ in an expanding universe. Radiation may include actual electromagnetic radiation or relativistic particles, $w=1/3$, and their energy density falls off $\rho_r\propto a^{-4}$. Vacuum energy has negative pressure, $w=-1$, and its energy density remains constant $\rho_\Lambda\propto a^0$ during the expansion of the universe. 

Another way to derive eq.~\eqref{Eq:Continuity} is by combining the Friedmann equations reduced from the Einstein field equation
\begin{align}
H^2&=\left(\frac{\dot{a}}{a}\right)^2=\frac{\rho}{3}-\frac{k}{a^2}\,,\label{Eq:FrieEq1}\\
\dot{H}+H^2&=\frac{\ddot{a}}{a}=-\frac{1}{6}(\rho+3\,p)\,.\label{Eq:FrieEq2}	
\end{align}
In the case of a flat universe ($k=0$) we can directly solve eq.~\eqref{Eq:FrieEq1} to obtain
\begin{equation}
a(t)\propto\left\{\begin{matrix}
t^{2/(3(1+w))},&\forall\ w\neq-1\,,\\
e^{Ht},&\text{for }w=-1\,.
\end{matrix}\right.
\end{equation}

\begin{table}[H]
\centering
    \begin{tabular}{l ccc} \toprule
    {Type of fluid} & {$\rho(a)$} & {$w$} & {$a(t)$}  \\ \midrule
    {radiation}  & $a^{-4}$ & $\sfrac{1}{3}$ & $t^{\sfrac{1}{2}}$ \\
    {cold matter} & $a^{-3}$  &  $0$ &  $t^{\sfrac{2}{3}}$   \\
    {spatial curvature}  & $a^{-2}$  & $-\sfrac{1}{3}$ & $t$   \\
    {vacuum energy}  & $a^{0}$  &  $-1$ & $e^{Ht}$  \\
    {scalar field}  & $a^{-2\epsilon_H}$  & $-1+2\,\frac{\epsilon_H}{3}$ & $t^{1/\epsilon_H}$  \\ \bottomrule
\end{tabular}
    \caption{Different types of energy densities dominating a flat FRW universe and their associated state parameter values $w$ and scale factor $a(t)$ in terms of cosmic time.}
    \label{tab:my_label}
\end{table}

It is useful to discuss the contribution of various constituents to the energy density and pressure, through the fractions
\begin{equation}
\rho=\sum_i\rho_i,\qquad p=\sum_ip_i\,,
\end{equation}
where $i$ sums over all the potential contributions. Notice that the continuity equation \eqref{Eq:Continuity} holds for each constituent $\rho_i$ and $p_i$, while the Friedmann equations \eqref{Eq:FrieEq1}-\eqref{Eq:FrieEq2} only hold for the summed over $\rho$ and $p$. An important quantity, especially in astrophysics, is the present day ratio of the energy density to the critical energy density $\rho_{\text{crit},0}\equiv 3H_0^2$, defined through
\begin{equation}
\Omega_{i,0}\equiv\frac{\rho_{0,i}}{\rho_\text{crit}}\,,
\end{equation}
and similarly we may parametrise the curvature contribution by
\begin{equation}
\Omega_{k,0}\equiv-\frac{k}{\left(a_0 H_0\right)^2}\,,
\end{equation}
where the subscript ``$0$'' denotes present day values of the quantities. Normalising the scale factor at present day $a_0\equiv 1$ allows us to re-express the Friedmann equation \eqref{Eq:FrieEq1} as
\begin{equation}
\left(\frac{H}{H_0}\right)^2=\sum_i\Omega_{i,0}\,a^{-3(1+w_i)}+\Omega_{k,0}\,a^{-2}\,.
\end{equation}
Then, at present time $\sum_i\Omega_{i,0}+\Omega_{k,0}=1$. The definition of the energy fractions can be generalised to include a time dependence
\begin{equation}\label{Cond:OmegaKt}
\Omega_i(a)=\frac{\rho_i(a)}{\rho_\text{crit}(a)}\,,\qquad \Omega_k(a)=-\frac{k}{a^2H^2}\,,
\end{equation}
where $\rho_\text{crit}=3H^2$ is now time-dependent.

\section{Big Bang puzzles of initial conditions}

It should be again emphasised that the issues described in this section are not inconsistent with the standard cosmological model, but rather highlight shortcomings in its predictive power. In what follows we discuss two of these issues in detail, known as the flatness and the horizon problem, and demonstrate how the focal point of inflation can provide a natural solution to them. 

\subsection{\emph{Flatness problem}}

If we start from eq.~\eqref{Cond:OmegaKt} and assume a state parameter $w\neq-1$, we  can rewrite it as
\begin{equation}\label{Eq:OmKflat}
\frac{\partial\Omega_k}{\partial\ln{a}}=\Omega_k(1+3w)\,.
\end{equation}
This shows that values of $\Omega_k>0$ grow with time, and similarly negative values keep decreasing. Also, it seems that in the case in which $w>-1/3$ the solution of $\Omega_k=0$ is an unstable fixed point. It is then surprising that the present day observed value of $\Omega_k\sim10^{-2}$ is so close to zero. Therefore, it is expected that in earlier periods it would be even smaller, e.g. at the Big Bang nucleosynthesis (BBN) epoch it is $\Omega_k\sim10^{-16}$ and at Planck scale it would be $\Omega_k\sim10^{-61}$~\cite{Baumann2011}. One can accept $k=0$ as the precise initial state of the universe at the price of an immense fine-tuning, but a theory that dynamically explains it seems more attractive.

\subsection{\emph{Horizon problem}}

Let us rewrite the particle horizon \eqref{Cond:PartHor} as
\begin{equation}
\chi_p=\int_{a_i}^{a'}\frac{\mathrm{d}a}{a^2H}=\int_{a_i}^{a'}\!\frac{\mathrm{d}\ln{a}}{a H}\,,
\end{equation}
expressed in terms of the comoving Hubble radius $(a H)^{-1}$. Assuming a universe described by a fluid with a state parameter $w$ we obtain,
\begin{equation}
\chi_p\propto a^{2(1+3w)}\,,
\end{equation}
which implies that comoving scales entering the horizon today have not been in causal contact before that and they interact for the first time. Meaning that new regions should appear different from one another, but examining the near-homogeneity of the CMB suggests otherwise.

\subsection{\emph{Solving the problems: central idea of inflation}}

The central idea is to allow for some form of energy with a state parameter $w<-1/3$ or, in other words, a decreasing Hubble radius $(a H)^{-1}$, so that the integral of the particle horizon is dominated by early times instead of late times. Formulated in mathematical language it suggests that
\begin{equation}
\frac{\mathrm{d}}{\mathrm{d}t}\left(\frac{1}{a H}\right)<0\quad\implies\quad \ddot{a}>0\,,
\end{equation}
which implies that physical wavelengths become larger than $H^{-1}$. From eq.~\eqref{Eq:FrieEq2} it also implies that $w<-1/3$. Then, since $(aH)^{-1}$ decreases instead of increasing, the universe is driven towards flatness, and the solution of $\Omega_k=0$ becomes an attractor solution of eq.~\eqref{Eq:OmKflat}.

\begin{figure}
    \centering
    \includegraphics[scale=0.5]{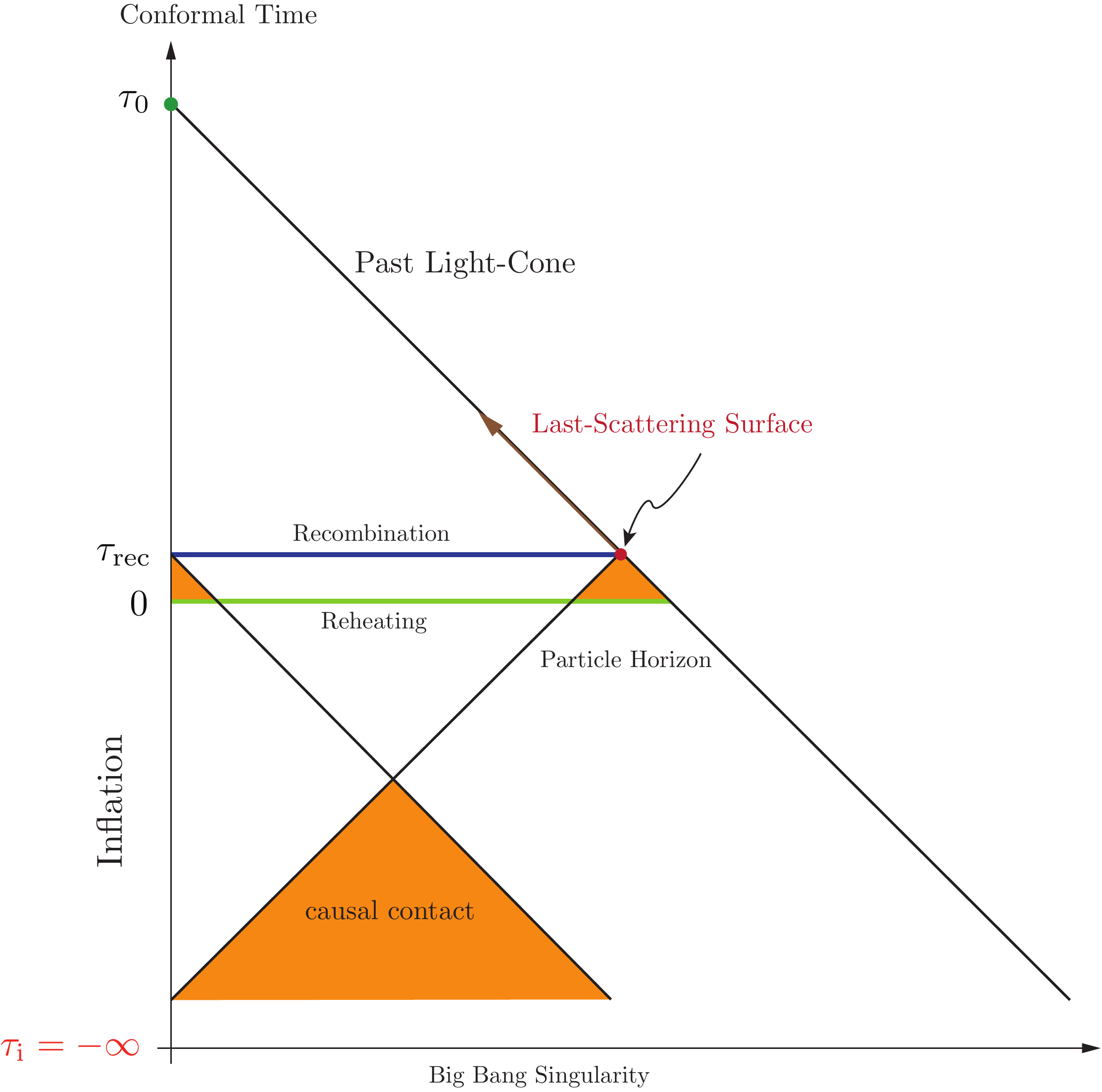}
    \caption{The cosmological histogram of inflationary scales. By extending the conformal time to negative values previously assumed casually disconnected regions of space become connected where their past light cones intersect during inflation. Credit:~\cite{Baumann2011}.}
    \label{fig:Light-Cone-SOL}
\end{figure}

Regarding the scale factor, we obtained
\begin{equation}
a\propto \tau^{2/(1+3w)}
\end{equation}
which in case $w>-1/3$ suggests that as $\tau\to0$ we are forced to the initial singularity $a\to0$. However, if we allow for a phase in which $w<-1/3$ we can extend $\tau$ to negative values and in this way making the horizon larger than $H^{-1}$ (see fig.~\ref{fig:Light-Cone-SOL}).

\subsubsection{\emph{Digression on the monopole problem}}

It is very likely that the universe underwent through a series of phase transitions during its evolution, e.g. the QCD and electroweak phase transitions and potentially others. Depending on the symmetry that is broken in the transitions topological defects form; especially in the case of string theory these defects appear in the form of magnetic monopoles. The issue of the magnetic monopoles (and the topological defects in general) is entirely model dependent and mostly tied to Grand Unified Theories (GUTs)~\cite{Pati1973,Georgi1974,Georgi1974a,Fritzsch1975,Gursey1976} that attempt to unify the three gauge interactions of the SM into one force at some unification scale $E_\text{GUT}$. These magnetic monopoles are nonrelativistic and they fall off as $\propto a^{-3}$ as opposed to the photon (or neutrino) $\propto a^{-4}$, meaning that at present times they should dominate over them, which is not the case (they are undetected). However, if inflation takes place after the phase transition the monopole density is diluted by inflation, from $a^{-3}$ to a tiny size.

\section{Inflaton field dynamics}
\label{sec:inflDynamics}

During the inflationary period we demand that the violation of the strong energy condition ($w<-1/3$; more on that in ch.~\ref{Ch4:QuadGrav}), \emph{dynamically} comes to a halt towards the end of inflation. The simplest realisation of this is achieved via a scalar field\footnote{Note that a scalar field can approximate a vacuum-like state, after all it has the same quantum number as the vacuum and can assume nonzero vacuum expectation value (VEV) and nontrivial configuration without breaking Lorentz invariance.}, dubbed the \emph{inflaton field} and usually denoted by $\phi(\mathbf{x},t)$, which is dominating the energy density of the early universe. The exact nature of the inflaton field is still highly speculative, since the physics of inflation cannot be tested in a particle accelerator due to the high energy scales.\footnote{A nontrivial assumption of the inflationary paradigm is the introduction of an additional scalar degree of freedom to the SM. The only known fundamental scalar field of the SM, that can assume the role of the inflaton, is the Higgs field, but even in that case, an extension of the interaction between the gravitational sector and the Higgs field is required (e.g. see ref.~\cite{Rubio2019}). } Meaning that the only constraints of inflation are placed on the \emph{shape} of the scalar potential $V(\phi)$ and, even then, various proposed models are able to satisfy the observational bounds within some margins of success~\cite{Akrami2020}.

Therefore, let us consider a real scalar field $\phi$ that is minimally coupled to gravity, described by the simple\footnote{We could allow for the possibility of a nonminimal interaction between the inflaton and the graviton or introduce higher-order curvature invariants that admit a propagating scalar mode. In principle, most of these models can be brought into the form of eq.~\eqref{Action:SimpleInf} via a field redefinition and/or a Weyl rescaling of the metric.} action in four dimensions
\begin{equation}\label{Action:SimpleInf}
\mathcal{S}=\int\!\mathrm{d}^4x\,\sqrt{-g}\left\{\frac{M_P^2}{2}R-\frac{1}{2}g^{\mu\nu}\partial_\mu\phi\,\partial_\nu\phi-V(\phi)\right\}\,,
\end{equation}
where $V(\phi)$ is unspecified at the moment and denotes the self-interacting potential of the inflaton. The action \eqref{Action:SimpleInf} includes two dynamical degrees of freedom, the metric tensor $g_{\mu\nu}(x)$ and the scalar field $\phi(x)$. Variation with respect to $\phi$ leads to the famous Klein-Gordon equation
\begin{equation}
\Box\phi+\frac{\mathrm{d}V(\phi)}{\mathrm{d}\phi}=0\,,
\end{equation}
where $\Box\equiv g^{\mu\nu}\nabla_\mu\nabla_\nu$ denotes the d'Alembertian operator in curved spacetime. Likewise, variation with respect to $g_{\mu\nu}$ gives rise to the Einstein field equations ($M_P^2\equiv 1$)
\begin{equation}\label{Eq:EinEoMMinInf}
G_{\mu\nu}\equiv R_{\mu\nu}-\frac{1}{2}g_{\mu\nu} R=\partial_\mu\phi\,\partial_\nu\phi-g_{\mu\nu}\left(\partial^\rho\phi\,\partial_\rho\phi+V(\phi)\right)\,,
\end{equation}
where, by definition, the RHS of eq.~\eqref{Eq:EinEoMMinInf} is identified with the scalar field energy-momentum tensor $T_{\mu\nu}(\phi)$.

Assuming a flat FRW universe, described by the metric \eqref{Cond:FRW} and a spatially homogeneous field $\phi(\mathbf{x},t)=\phi(t)$ we can rewrite the equations of motion as follows
\begin{gather}
\ddot{\phi}+3H\dot{\phi}+V'(\phi)=0\label{Eq:KG}\,,\\
3H^2=\frac{1}{2}\dot{\phi}^2+V(\phi)\label{Eq:FriedScal1}\,,\\
\dot{H}=-\frac{1}{2}\dot{\phi}^2\label{Eq:FriedScal2}\,,
\end{gather}
where, hereafter, the dot and the prime denote derivative with respect to $t$ and the function's argument, $f'(x)=\mathrm{d}f(x)/\mathrm{d}x$, respectively. Next, assuming a perfect fluid we can make use of
\begin{align}
\rho&=\frac{1}{2}\dot{\phi}^2+V(\phi)\,,\\
p&=\frac{1}{2}\dot{\phi}^2-V(\phi)\,.
\end{align}
Therefore, the equation of state for the inflaton reads
\begin{equation}\label{Cond:StateParameter-GenPhi}
w=\frac{\displaystyle{\frac{1}{2}}\dot{\phi}^2-V(\phi)}{\displaystyle{\frac{1}{2}}\dot{\phi}^2+V(\phi)}\,.
\end{equation}

\subsection{\emph{Slow-roll approximation}}

Note that eq.~\eqref{Eq:KG} is similar to the one describing a particle trajectory rolling down its potential that is also subject to a friction term, $\propto H\dot{\phi}$ (due to the expansion of the universe in this case). In a completely similar fashion, this means that the solution $\dot{\phi}\approx V'/(3H)$ is an attractor solution and the field is driven towards the minimum of the potential, for various initial conditions. The feature of the (slow-roll) attractor solution is especially appealing since it indicates that our universe will ``end up'' in the inflationary period quite generally, without fine-tuning.

\begin{figure}
    \centering
    \includegraphics[scale=1]{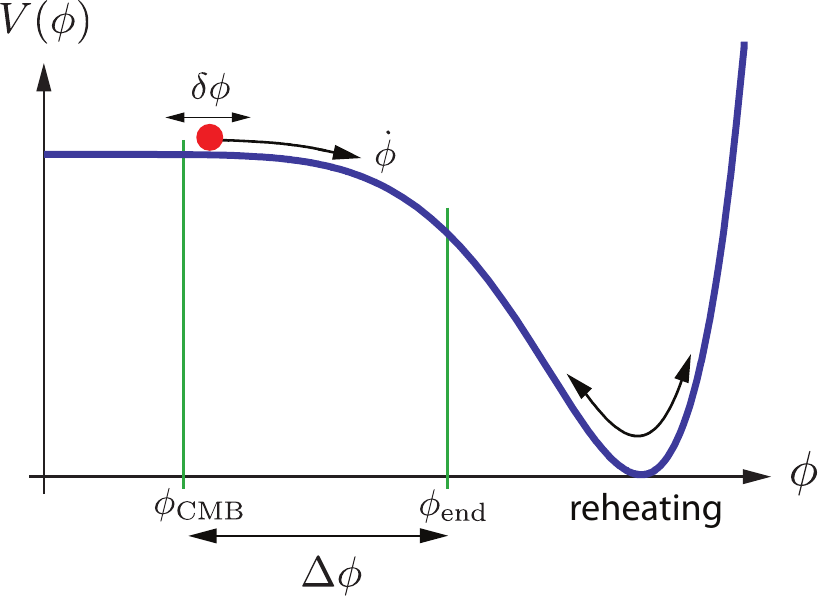}
    \caption{An exemplary inflaton potential. Inflation begins at some field value in the flat region of the potential where the potential energy dominates over the kinetic energy of the inflaton and stops approximately at some field values where it starts to contribute considerably  compared to the potential, $\dot{\phi}^2\approx V(\phi)$. Quantum fluctuations of the field $\delta\phi$ lead to different regions of space to inflate by different amounts leading in turn to observed inhomogeneities in the CMB spectrum. Credit:~\cite{Baumann2011}.}
    \label{fig:Inf Potential-GEN}
\end{figure}

Directly from eq.~\eqref{Cond:StateParameter-GenPhi} one can obtain an accelerated expansion if the scalar potential dominates over the kinetic energy. At the limit of $\dot{\phi}\to0$ (de Sitter limit) the scalar field describes a cosmological constant with negative pressure ($w=-1$) and from the continuity equation we obtain $\dot{\rho}\to0$, meaning that the inflaton has (almost) constant energy density. Then, eq.~\eqref{Eq:FrieEq1} suggests that the Hubble parameter is constant throughout inflation and the scale factor is given by an exponential function $a(t)\propto e^{Ht}$. Clearly, if we consider a constant vacuum energy describing inflation, classically the inflationary era cannot end, and even if a quantum tunnelling effect from the false to the true vacuum is considered, it would only end locally leading to a universe either too homogeneous or empty. That was the problem of the \emph{graceful exit} of the models of what is now referred to as old inflation~\cite{Guth1981}.

\subsubsection{\emph{Slow-roll parameters}}

We classify the \emph{deviation} from de Sitter space (during inflation) in terms of the so-called \emph{slow-roll parameters} (SRPs). In order to introduce them, let us rewrite eq.~\eqref{Eq:FrieEq2} in the following way:
\begin{equation}
\frac{\ddot{a}}{a}=-\frac{1}{6}(\rho+p)=H^2(1-\epsilon_H)\,,
\end{equation}
where
\begin{equation}\label{Cond:HSRP1}
\epsilon_H\equiv-\frac{\dot{H}}{H^2}
\end{equation}
and is known as the first (Hubble) slow-roll parameter (Hubble SRP or HSRP)~\cite{Liddle1994}. Now, the de Sitter limit is suggested by $\epsilon_H\to0$ and the accelerated expansion occurs when $\epsilon_H<1$. By means of eq.~\eqref{Eq:FriedScal1}, this means that if $\epsilon_H\ll 1$ we obtain
\begin{equation}
\dot{\phi}^2\ll V(\phi)\,.
\end{equation}
In order to be on the slow-roll trajectory we require also
\begin{equation}
\eta_H\equiv-\frac{\ddot{\phi}}{H\dot{\phi}}\ll 1\,,
\end{equation}
dubbed the second (Hubble) slow-roll parameter. This simplifies greatly the Klein-Gordon EoM so as to express the first SRP in terms of the potential
\begin{equation}\label{Cond:H-VSRP1}
\epsilon_H=-\frac{\dot{H}}{H^2}\approx\frac{1}{2}\left(\frac{V'}{V}\right)^2\equiv\epsilon_V\,,
\end{equation}
which is usually referred to as the potential SRP (or PSRP). Further differentiating of the Klein-Gordon EoM leads to
\begin{equation}\label{Cond:H-VSRP2}
\frac{\mathrm{d}}{\mathrm{d}t}\left(3H\dot{\phi}\approx -V'\right)\implies \eta_H+\epsilon_H\approx\frac{V''}{V}
\end{equation}
and allows us to introduce the second potential SRP as:
\begin{equation}
\eta_V\equiv\frac{\left|V''\right|}{V}\,.
\end{equation}
Equations \eqref{Cond:H-VSRP1} and \eqref{Cond:H-VSRP2} also describe the approximate\footnote{The exact relation of the potential and Hubble SRPs is given, up to first order in the SRPs, by: 
\begin{align*}
\epsilon_V&=\epsilon_H\left(\frac{3-\eta_H}{3-\epsilon_H}\right)^2\,,\\
\eta_V&=\left(\epsilon_H+\eta_H\right)\left(\frac{3-\eta_H}{3-\epsilon_H}\right)+\sqrt{2\epsilon_H}\,\frac{\eta'_H}{3-\epsilon_H}\,.
\end{align*}
} relation between the Hubble and the potential SRPs. Then, slow-roll inflation occurs when these two parameters are small
\begin{equation}
\epsilon_H\ll 1\quad \&\quad\left|\eta_H\right|\ll 1
\end{equation} 
and it ends when one of the SRPs is close to unity, namely~\cite{Liddle1994}
\begin{equation}
\epsilon_H=1\ (\epsilon_V\approx 1)\qquad \text{or/and}\qquad \eta_H=1\ (\eta_V\approx 1)\,.
\end{equation}

\subsubsection{\emph{Duration of inflation}}

The exact amount of inflation the universe went through is measured in number of $e$-foldings $N$, such that the scale factor at the start and end of inflation, $t_i$ and $t_f$ respectively, is expressed as 
\begin{equation}
\frac{a(t_f)}{a(t_i)}=e^{N}\implies \mathrm{d}N=\mathrm{d}\ln{a}\,.
\end{equation}
Then using $\mathrm{d}\ln{a}=H\mathrm{d}t$ and the approximate slow-roll expression we can write $N$ in terms of the scalar potential $V(\phi)$ as follows:
\begin{equation}
N\simeq-\int_{\phi_i}^{\phi_f}\frac{V}{V'}\,\mathrm{d}\phi=-\int_{\phi_i}^{\phi_f}\frac{\mathrm{d}\phi}{\sqrt{2\epsilon_V(\phi)}}\,,
\end{equation} 
where $\phi_i\equiv\phi(t_i)$ and $\phi_f\equiv \phi(t_f)$ are the field values at the beginning and end of inflation, respectively. Notice how $\Omega_k\propto a^{-2}$ suggests that in order to explain the observed flatness, if the universe started at GUT or Planck scale, we require about $N\sim60$ $e$-folds (see refs.~\cite{Dodelson2003,Liddle2003a}). Analogously, the horizon problem is solved if regions of the CMB were within the horizon, meaning that we require $N\gtrsim 60$ $e$-folds in that case.

The definition of the number of $e$-foldings $N$ allows us to introduce the Hubble slow-roll parameters in hierarchical order~\cite{Liddle1994}
\begin{equation}
\epsilon_{H,i+1}=\frac{1}{\left|\epsilon_{H,i}\right|}\,\frac{\mathrm{d}\left|\epsilon_{H,i}\right|}{\mathrm{d}N}\,,
\end{equation}
where $\epsilon_{H,0}=H_\text{start}/H$, with $H_\text{start}$ denoting the Hubble parameter at the start of inflation $t_i$. Again, during slow-roll inflation we expect that $\epsilon_{H,i}\!\ll\! 1$, $\forall i\!\in\!\mathbb{N}$ and they are approximately of the same order of magnitude. Inflationary observables are usually expressed in terms of leading order in the SRPs (at most NLO), however the increasing accuracy and sensitivity of the experimental missions may require higher-order terms and other types of approximations (other than the Taylor expansion discussed here).

\subsubsection{\emph{Attractor points}}

As we stated earlier, the concept of cosmological attractors is especially  desirable; after all it suggests that the inflaton scalar field eventually arrives to its preferred evolution, regardless of its initial conditions. In this case, initial data are the field value and its velocity at some point. Intuitively, we can think of the expansion of the universe as some kind of friction acting on the system, that while the field is away from the attractor solution it is negligible, but when it reaches that point it dominates and ``forces'' it into the general trajectory. Effectively, it seems that any kind of initial conditions ultimately sends the field onto the preferred trajectory from where it continues to evolve.

The actual phase space of trajectories is four-dimensional, possibly parametrised by $\left\{a,p_a,\phi,p_\phi\right\}$, where $p_a$ and $p_\phi$ are the conjugate momenta of the scale factor $a$ and inflaton field $\phi$, respectively. Then, it is not straightforward how or why the $(\dot{\phi}-\phi)\simeq\mathbb{R}^2$ space completely defines the effective phase space, in which in principle trajectories of $a-p_a$ should cross into it. Generally, a $2n$-dimensional symplectic manifold $C$ is equipped with a closed, non-degenerate $2$-form, known as the symplectic form
\begin{equation}
    \omega=\sum_{i=1}^n\mathrm{d}p_i\wedge\mathrm{d}q^i\,,
\end{equation}
where $p^i$ are local coordinates on the manifold and $q_i$ are \emph{soldered} momenta to the velocities $\mathrm{d}q^i$. It also defines the Liouville measure
\begin{equation}
    \Omega=\frac{(-1)^{\frac{n(n-1)}{2}}}{n!}\,\omega^n\,,
\end{equation}
that under Liouville's theorem of classical mechanics is conserved along the Hamiltonian flow vector $\mathcal{X}_\mathcal{H}$, meaning $\mathcal{L}_{\mathcal{X}_\mathcal{H}}\Omega=0$. Then, the space of trajectories is  given by $\Gamma=\{C/\mathcal{H}_*\}/\mathcal{X}_\mathcal{H}$, where $\mathcal{H}_*$ is the constrained Hamiltonian of the system~\cite{Gibbons1987}. In the particular case of the (canonically normalised) scalar field coupled minimally to the Einstein-Hilbert term with zero spatial curvature ($k=0$ in eq.~\eqref{Cond:FRW}), it was shown in ref.~\cite{Remmen2013} that $\Gamma$ is $2$-dimensional and in fact the measure $\mathrm{d}\dot{\phi}\wedge\mathrm{d}\phi$ is conserved under Hamiltonian flow. Therefore, in that unique case we can safely assume that other trajectories do not ``bleed'' into the $\dot{\phi}-\phi$ space and we can observe the (coordinate dependent) attractor behaviour. In what follows, we use this argument in order to draw conclusions regarding the attractor solution of the cosmological systems we study.

\subsection{\emph{Density perturbations}}

One of the most appealing feature of inflation is the prediction of the observed CMB anisotropies, when addressed under a quantum framework. It perfectly highlights the predictive power of the theory since these cosmological perturbations were only considered after the formulation of the theory before the observation of the CMB fluctuations. However, just from the apparent large-structure formation observed today it was expected that some kind of fluctuations existed at sub-Hubble scales that were then amplified (``stretched'') to large scales. Remarkably, the quantum effects in a gravitational setting (a result surprising on its own) after an exponential expansion become the source of galaxies and other structures in our universe and provide us with a connection to physics of small distances.

Deviations from the scale-invariant spectrum are designated by inhomogeneous primordial perturbations of the scalar field around its classical background, $\delta\phi(\mathbf{x},t)\equiv\phi(\mathbf{x},t)-\varphi(t)$, that in turn lead to different regions expanding by different amount. Since the inflaton field assumes the role of a local clock (measuring the amount of inflation), its spatially varying fluctuations spontaneously break time-translation (due to the uncertainty principle). Then, we can draw inspiration from particle physics and describe inflation by introducing the Goldstone boson of the broken gauge redundancy (following refs.~\cite{ArkaniHamed2004,Creminelli2006,ArkaniHamed2007,Cheung2008,Senatore2012} and a similar detailed approach in ref.~\cite{Weinberg2008a}; also see refs.~\cite{Piazza2013,Burgess2017,Senatore2017} for relevant reviews).

\vspace{0.5cm}

\subsubsection{\emph{Digression on gauge redundancy}}

A critical point to keep in mind is that the introduction of the perturbations around a classical background is not uniquely defined, and therefore depends on the coordinate frame or the choice of gauge. In other words, by defining the constant-time hypersurfaces we implicitly choose a gauge for the perturbations and, in general, we can either introduce fictitious perturbations or completely eliminate physical perturbations. It is useful then to describe them in a gauge-invariant way and include perturbations of the metric and the matter field(s).

The issue of the gauge freedom is laborious and in what follows we attempt to circumvent it as much as possible, but let us demonstrate its importance in the following simple example. Think of a quantum universe and a field $\phi(x)$ expanded around its classical background $\varphi(t)$. Thus, its vacuum state is not any more an eigenstate of its operator, say $\hat{\phi}\ket{0}\neq\phi\ket{0}$, where
\begin{equation}
\phi=\varphi(t)+\delta\phi(\mathbf{x},t)\,.
\end{equation}
Consider an infinitesimal change in the coordinates as
\begin{equation}
x^\mu \to \bar{x}^\mu=x^\mu+\xi^\mu\,.
\end{equation} 
Then, we can express the scalar field in the new coordinate frame, up to first order, as
\begin{equation}
\bar{\phi}(\bar{x}-\xi)\approx\bar{\phi}(\bar{x})-\xi^\mu\partial_\mu\bar{\phi}(\bar{x})\,.
\end{equation}
Any scalar density of weight $w$ transforms as $\bar{\phi}(\bar{x})=\mathcal{J}^w\phi(x)$, where $\mathcal{J}^w\equiv\left[\text{det}\displaystyle{\frac{\partial(\bar{x}^\nu)}{\partial x^\mu}}\right]^w$ is the Jacobian of the coordinate transformation. Making use of $\text{det}\left(\mathds{1}+A\right)\approx 1+\text{Tr}A+\ldots$ for a matrix $A$, the Jacobian becomes $\mathcal{J}^w\approx 1+w\,\partial_\mu\xi^\mu$.  Returning to the scalar field we obtain\footnote{The definition of the perturbation $\delta\phi$ is not to be confused with the usual definitions of the value variation $\delta\phi=\bar{\phi}(\bar{x})-\phi(x)$ and the form variation $\delta_0\phi(x)=\bar{\phi}(x)-\phi(x)$. }
\begin{align*}
\bar{\phi}(\bar{x}-\xi)&\approx\phi(x)+w\,\partial_\mu\xi^\mu\,\phi(x)-\xi^\mu\partial_\mu\phi(x)+\mathcal{O}(\xi^2,\xi\cdot\delta\phi)\\
&\approx\varphi(t)+\delta\phi(x)-\xi^\mu\partial_\mu\varphi(t)+w\,\partial_\mu\xi^\mu\,\varphi(t)+\mathcal{O}(\xi^2,\xi\cdot\delta\phi)\,,
\end{align*}
where we recognise the Lie derivative of a scalar density $\mathcal{L}_\xi\phi(x)=\xi^\mu\partial_\mu\phi(x)-w\,\partial_\mu\xi^\mu\phi(x)$. In the particular case of a scalar field it coincides with the directional derivative $\nabla_\xi$ and the expression is simplified further, leading to
\begin{equation}
\delta\phi(x)\to\overline{\delta\phi}(\bar{x})-\xi^0\,\dot{\varphi}(t)\,.
\end{equation}
Therefore, the fluctuations $\delta\phi$ shift under (time) diffeomorphisms, a fact that we exploit later on so that we gain a deeper understanding of the particle physics phenomenology of the inflaton.

\vfill

\subsubsection{\emph{Scalar perturbations}}

\begin{flushleft}
\underline{\emph{The inflaton in the EFT approach}}
\end{flushleft}

Following our previous discussion it is interesting to describe the dynamics of the inflaton field using the language of the Effective Field Theory (EFT) approach, meaning that the action includes all available degrees of freedom in terms of operators compatible with the symmetries. The residual gauge symmetry is the remaining time-dependent spatial diffeomorphisms $\bar{x}^i=x^i+\xi^i(\mathbf{x},t)$. The most general Lagrangian reads as~\cite{Cheung2008}
\begin{align}\label{Action:EFT-Action0}
\mathcal{S}&=\int\!\mathrm{d}^4x\,\sqrt{-g}\Bigg\{\frac{1}{2}R-c_0(t)g^{00}-c_1(t)+\frac{1}{2!}M_2(t)^4(\delta g^{00})^2+\frac{1}{3!}M_3(t)^4(\delta g_{00})^3+\\
&\qquad\qquad-\frac{1}{2}\overline{M}_1(t)^3\delta g^{00}\,\delta{K^\mu}_\mu-\frac{1}{2}\overline{M}_2(t)^2\left(\delta K{}^\mu{}_\mu\right)^2-\frac{1}{2}\overline{M}_3(t)^2\,\delta K{}^\mu{}_\nu \,\delta K{}^\nu{}_\mu+\ldots\Bigg\}\,,\nonumber
\end{align} 
where dots indicate higher-order terms in fluctuations and $c_i(t)$, $M_i(t)$ and $\overline{M}_i(t)$ are coefficients. Here we defined $\delta g^{00}\equiv g^{00}+1$ and made use of the extrinsic curvature arising from the embedding of the 3-hypersurfaces of constant time in four-dimensional spacetime
\begin{equation}
K_{\mu\nu}=h{}_\mu{}^\rho\,\nabla_\rho n_\nu,
\end{equation}
where $\nabla$ is the covariant derivative, $n_\mu$ is the timelike unit vector ($n_\mu n^\mu=-1$) normal to the $3$-hypersurface,  and $h_{\mu\nu}$ the induced metric $h_{\mu\nu}=g_{\mu\nu}+n_\mu n_\nu$ on it. Note that we denote $\delta K_{\mu\nu}=K_{\mu\nu}-a^2 H h_{\mu\nu}$ variations around a flat FRW background, then the coefficients can be easily fixed by the FRW solution (by tadpole cancellation). It is evident then that the higher-order terms (beyond the zeroth order) are model dependent. Generally, one can imagine more (or infinite) terms that contribute at first order to eq.~\eqref{Action:EFT-Action0} containing derivatives, but by integrating them by parts we obtain a combination of the terms already present.

The Friedmann equations  become
\begin{equation}
3H^2=c_0+c_1\qquad\&\qquad \dot{H}+H^2=\frac{1}{3}c_1-\frac{2}{3}c_0\,.
\end{equation}
Solving the system for the $c_i$'s and substituting them back into the action, we obtain~\cite{Cheung2008}
\begin{align}\label{Action:EFT-Action1}
\mathcal{S}&=\int\!\mathrm{d}^4x\,\sqrt{-g}\Bigg\{\frac{1}{2}R+\dot{H}g^{00}-(3H^2+\dot{H})+\frac{1}{2!}M_2(t)^4(\delta g^{00})^2+\frac{1}{3!}M_3(t)^4(\delta g_{00})^3+\\
&\qquad\qquad-\frac{1}{2}\overline{M}_1(t)^3\delta g^{00}\,\delta{K^\mu}_\mu-\frac{1}{2}\overline{M}_2(t)^2\left(\delta K{}^\mu{}_\mu\right)^2-\frac{1}{2}\overline{M}_3(t)^2\,\delta K{}^\mu{}_\nu \,\delta K{}^\nu{}_\mu+\ldots\Bigg\}\,,\nonumber
\end{align}
Notice how, in the language of the background scalar field, we can write
\begin{equation}
\int\!\mathrm{d}^4x\,\sqrt{-g}\left(-\frac{1}{2}g^{00}\dot{\varphi}(t)-V(\varphi)\right)=\int\!\mathrm{d}^4x\,\sqrt{-g}\left(\dot{H}g^{00}-3H^2-\dot{H}\right)\,,
\end{equation}
where in the last step we used the Friedmann equations. Since during inflation $H\approx\text{const.}$ and $\dot{H}\approx0$ we assume that it holds for the rest of the operators, leading to a Lagrangian that is (approximately) invariant under time translations.

At this point, it is not clear how the scalar degree of freedom is represented in the action \eqref{Action:EFT-Action1}. This is completely analogous to the case of a non-Abelian gauge group in the unitary gauge by introducing the Goldstone boson of the broken symmetry. Let us take a step back and consider the operators at the zeroth order under a time diffeomorphism $t\to\bar{t}=t+\xi^0(x)$ and $\mathbf{x}\to\bar{\mathbf{x}}$; then we obtain
\begin{equation}
\int\!\mathrm{d}^4x\,\sqrt{-g}\left( c_0 (t) g^{00}+c_1(t)\right)=\int\!\mathrm{d}^4\bar{x}\sqrt{-\bar{g}}\left\{c_1(\bar{t}-\xi^0)+c_0(\bar{t}-\xi^0)\,\frac{\partial (\bar{t}-\xi^0)}{\partial \bar{x}^\mu}\,\frac{\partial (\bar{t}-\xi^0)}{\partial \bar{x}^\nu}\,\bar{g}^{\mu\nu}\right\}\,.
\end{equation}
Next, we make the following redefinition 
\begin{equation}
-\xi^0(x(\bar{x}))\equiv \bar{\pi}(\bar{x})\,,
\end{equation}
in order to reintroduce explicitly the Goldstone boson~\cite{Senatore2017}. Then, eq.~\eqref{Action:EFT-Action1} becomes (dropping the bars for brevity)
\begin{align}\label{Action:EFT-Action2}
\mathcal{S}=\int_\mathcal{M}\!\mathrm{d}^4x&\,\sqrt{-g}\,\bigg\{\frac{1}{2}R-\left(3H^2+\dot{H}\right)(t+\pi)+\dot{H}(t+\pi)\partial_\mu(t+\pi)\,\partial_\nu(t+\pi)\,g^{\mu\nu}+\\
&\qquad\qquad+\frac{M_2(t+\pi)^4}{2!}\left[\partial_\mu(t+\pi)\,\partial_\nu(t+\pi)\,g^{\mu\nu}+1\right]^2+\nonumber\\
&\qquad\qquad+\frac{M_3(t+\pi)^4}{3!}\left[\partial_\mu(t+\pi)\,\partial_\nu(t+\pi)\,g^{\mu\nu}+1\right]^3+\ldots\bigg\}\,,\nonumber
\end{align}
where the dots here denote terms of higher-order and terms including the extrinsic curvature.

Similar to a gauge theory, the action \eqref{Action:EFT-Action2} is simplified at short distances, in which the scalar field decouples from the metric fluctuations. Above some high energy scale, terms that include derivatives are subleading to the main contribution of the kinetic term of $\pi$. Let us focus on the tadpole contribution ($M_2=0=M_3$ etc) which includes the slow-roll solution~\cite{Cheung2008}. The leading term that mixes the gravitational with the scalar degrees of freedom is
\begin{equation}
\propto \dot{H}\dot{\pi}\,\delta g^{00}\,.
\end{equation}
Now, $\delta g^{00}$ is the gravitational potential and is determined by $\pi$. Using the post-Newtonian approximation we obtain $H\partial_i \delta g^{00}\approx \dot{H}\partial_i\pi$ implying $\delta g^{00}\approx \dot{H}\dot{\pi}/H$. Then, it is straightforward to show that the mixing term is negligible in the limit of a scale $\Lambda$ above the mixing scale, which holds especially in the far UV region. Therefore, the action is further simplified as follows\footnote{Note that we have $$\frac{\partial(t+\pi)}{\partial x^\mu}\,\frac{\partial(t+\pi)}{\partial x^\nu}\,g^{\mu\nu}=g^{00}(1+\dot{\pi})^2+2g^{0i}\partial_i\pi(1+\dot{\pi})+g^{ij}\partial_i\pi\partial_j\pi\longrightarrow-1-\dot{\pi}^2-2\dot{\pi}+\frac{(\partial_i\pi)^2}{a^2}\, $$ and keep terms up to third order.}
\begin{equation}\label{Action:EFT-Last}
\mathcal{S}=\int_\mathcal{M}\left\{\frac{1}{2}R-\dot{H}\left[\dot{\pi}^2-\left(\frac{\partial_i\pi}{a}\right)^2\right]+2(M_2)^4\left[\dot{\pi}^2(1+\dot{\pi})-\dot{\pi}\left(\frac{\partial_i\pi}{a}\right)^2\right]-\frac{4}{3}(M_3)^4\dot{\pi}^3+\ldots\right\}\,.
\end{equation}

Clearly, the action presented in the decoupling limit is only useful for calculating correlation functions just after the last horizon crossing via the gauge invariant quantity, usually known as $\zeta$, that remains constant at all orders of perturbation~\cite{Salopek1990,Lyth2005}. If one is interested in non-Gaussianities of the system, encoded in the $3$-point function $\braket{\zeta(\mathbf{k}_1)\zeta(\mathbf{k}_2)\zeta(\mathbf{k}_3)}$, terms mixing with gravity and self-interactions of $\pi$ have to be considered~\cite{Cheung2008}.

\begin{flushleft}
\underline{\emph{Canonical quantisation}}
\end{flushleft}

Having nested the inflaton degree of freedom in a field-theoretic language and without leaving the gauge, we employ the Arnowitt-Deser-Misner (ADM) formalism~\cite{Arnowitt1959} to the action functional \eqref{Action:EFT-Last}. As we alluded to earlier, the ADM decomposition (or $3+1$ approach) consists of describing the spacetime $(\mathcal{M},\text{g})$ as a set of $3$-hypersurfaces $\Sigma$, constant in time, that are propagating in time, meaning $\mathcal{M}\cong\Sigma\times\mathbb{R}$.\footnote{Here we consider a case in which the manifold $\mathcal{M}$ is diffeomorphic to $\Sigma\times\mathbb{R}$, where space is denoted by $\Sigma$ and time by $t\in\mathbb{R}$. In fact, there are different ways to split spacetime by picking a diffeomorphism $f:\mathcal{M}\rightarrow\Sigma\times\mathbb{R}$, but we focus on a particular slice of $\tau=0$, where $\tau=f^*t$ is the time coordinate, and assume that is also spacelike. These foliations are allowed based in a theorem stating that in a globally hyperbolic spacetime $(\mathcal{M},\text{g})$ there exists a global time function such that each constant ``surface'' is a Cauchy surface. The subject of the ADM decomposition of GR is beyond the scope of this work and we avoid getting into detailed calculations. Although, it is completely analogous to the textbook canonical quantisation of a classical particle in a configuration space $\mathbb{R}^n$, but in this case the configuration space is the superspace defined by the Riemannian metrics on $\Sigma$, ${}^{(3)}g_{ij}$.} The line element is decomposed as
\begin{equation}
    \mathrm{d}s^2=-N^2\mathrm{d}t^2+h_{ij}\left(\mathrm{d}x^i+N^i\mathrm{d}t\right)\left(\mathrm{d}x^j+N^j\mathrm{d}t\right),
\end{equation}
where $h_{ij}$ is, once again, the induced metric, and $N$ and $N^i$ are the lapse function and shift vector, respectively. Then, using the Gauss--Codazzi relation\footnote{It is derived by relating the 3- and 4-dimensional Riemann tensors on $(\Sigma,h)$ and $(\mathcal{M}_4,\text{g})$ by means of the extrinsic curvature.} we can contract the four-dimensional Ricci scalar as:
\begin{equation}
    R={}^{(3)}\!R+\left(K_{\mu\nu}K^{\mu\nu}-K^2\right)+2\,\nabla_\nu\left(n^\nu\nabla_\mu n^\mu-n^\mu\nabla_\mu n^\nu\right),
\end{equation}
where ${}^{(3)}R(h)$ is the intrinsic curvature of the hypersurfaces. Note that, as before, the covariant derivative $\nabla_\mu$ is taken with respect to the induced metric $h$. Finally, using the ADM expression for the extrinsic curvature
\begin{equation}
    K_{ij}=\frac{1}{N}E_{ij},\qquad E_{ij}\equiv\frac{1}{2}h_{ij}-\nabla_{(i}N_{j)},
\end{equation}
and after integration by parts, the action \eqref{Action:EFT-Last} reads
\begin{align}
    \mathcal{S}=\int\!\mathrm{d}^4x\,\sqrt{h}&\Bigg\{N\cdot{}^{(3)}\!R+\frac{1}{N}(E_{ij}E^{ij}-E^2)+\nonumber\\
    &\quad+2\dot{H}(t+\pi)\left[-\frac{1}{N}(1+\dot{\pi}^2)+\frac{2}{N}(1+\dot{\pi})N^i\,\partial_i\pi-N(h^{ij}\partial_i\pi\,\partial_j\pi)-\frac{1}{N}(N^i\partial_i\pi)^2\right]+\nonumber\\
    &\quad-N\left(\dot{H}(t+\pi)+3H^2(t+\pi)\right)+\mathcal{O}(M_2,M_4,\ldots)\Bigg\}
\end{align}
For the purposes of this work we restrict ourselves in the case where $M_{2,4,\ldots}=0$, that includes the slow-roll inflation. Variation of the above action with respect to $N_i$ and $N$ leads to equations of motion, which respectively are given by:
\begin{equation}\label{Eq:ADM-EoM1}
    \nabla_i\left(\frac{E^i_j-\delta^i_j\,E}{N}\right)+\frac{2}{N}\dot{H}(t+\pi)\left[(1+\dot{\pi})\partial_i\pi-N^j\,\partial_j\pi\,\partial_i\pi\right]=0\,,
\end{equation}
and
\clearpage
\begin{align}\label{Eq:ADM-EoM2}
   {}^{(3)}\!R-\frac{1}{N^2}(E_{ij}&E^{ij}-E^2)-3H^2(t+\pi)+\dot{H}(t+\pi)+\\
   &+\frac{\dot{H}(t+\pi)}{N^2}\bigg[(1+\dot{\pi})^2-(1+\dot{\pi})N^i\partial_i\pi+N^2h^{ij}\partial_i\pi\partial_j\pi+(N^i\partial_i\pi)^2\bigg]=0\,.\nonumber
\end{align}
These are the momentum and Hamiltonian constraints, respectively defined by $\mathcal{P}_i\stackrel{!}{\approx}0$ and $\mathcal{H}\stackrel{!}{\approx}0$, where $\stackrel{!}{\approx}$ denotes equality modulo equations of motion. Then, the variables $N$ and $N_i$ are constrained and therefore they are \emph{not} physical degrees of freedom of the theory. Let us do a brief counting of the degrees of freedom; we started with $(10)$ from the symmetric metric tensor and $(1)$ from the inflaton field ($\pi$ in this case). We may also remove $(4)$ degrees of freedom from the gauge redundancy ($3$ generators from the spatial diffeomorphisms and $1$ from the time ones), and another $(4)$ from the constrained quantities $N$ and $N_i$, leaving us with just $(3)$ \emph{dynamical} degrees of freedom. Two of them represent the helicities of the graviton and the last one is the matter field. At this point (second order) we can neglect the tensor perturbations since they do not mix with the scalar perturbations.

Next we fix a gauge; we choose the $\zeta$-gauge, in which the induced metric and the time diffeomorphisms are fixed by imposing the following conditions
\begin{equation}
    h_{ij}=a^2\delta_{ij}e^{2\zeta},\qquad\&\qquad \pi=0\,.
\end{equation}
In this gauge we gain better insight into the IR behaviour of the gauge-invariant quantity $\zeta$. The eqs.~\eqref{Eq:ADM-EoM1}-\eqref{Eq:ADM-EoM2} become respectively:
\begin{equation}
    \nabla_i\left(\frac{E^i_j-\delta^i_j\,E}{N}\right)=0\,,
\end{equation}
\begin{equation}
    {}^{(3)}\!R-\frac{1}{N^2}(E_{ij}E^{ij}-E^2)-3H^2-\dot{H}+2\,\frac{\dot{H}}{N^2}=0\,.
\end{equation}
The solution for the constrained $N$ and $N_i$ yields:
\begin{equation}
    N=1+\frac{\dot{\zeta}}{H}, \qquad\&\qquad N_i=-\partial_i\left(\frac{1}{a^2}\,\frac{\dot{\zeta}}{H}+\frac{\dot{H}}{H^2}\,\frac{1}{\partial^2}\dot{\zeta}\right)\,.
\end{equation}
Substituting the above solutions back into the action leads to
\begin{equation}
    \mathcal{S}=\int\!\mathrm{d}^4x\,a^3\left(-\frac{\dot{H}}{H^2}\right)\left(\dot{\zeta}^2-\frac{1}{a^2}(\partial_i\zeta)^2\right)\,.
\end{equation}
Quantising the system is now easier, since the action resembles the one for a massless scalar field. Then, for the conjugate momentum of $\zeta$ we find
\begin{equation}
    \Pi_\zeta=\frac{\delta\mathcal{L}}{\delta\dot{\zeta}}=-2a^3\dot{\zeta}\,\frac{\dot{H}}{H^2}
\end{equation}
and impose the equal time commutation relation $[\zeta(\mathbf{x},t),\Pi_\zeta(\mathbf{x}',t)]=i\delta^{(3)}(\mathbf{x}-\mathbf{x}')$ where the classical variables have been promoted to quantum operators (note that $\hbar=1$). We are free to expand the Fourier components of $\zeta$ in terms of annihilation and creation operators 
\begin{equation}
    \bar{\zeta}(t)=\zeta_{\textbf{k}}^\text{cl}(t)a_\textbf{k}+\zeta_{-\textbf{k}}^{\text{cl},*}(t)a^\dagger_{-\textbf{k}}\,,
\end{equation}
where the usual commutation relation $[a_\mathbf{k}(t),a^\dagger_{\mathbf{k}'}(t)]=\delta^{(3)}(\mathbf{k}+\mathbf{k}')$ holds $\forall t$ and $\zeta^\text{cl}$ represents the classical solution to the equation of motion
\begin{equation}
    \frac{\delta\mathcal{L}}{\delta\zeta}=-\frac{\mathrm{d}}{\mathrm{d}t}\left(a^3\dot{\zeta}^\text{cl}\,\frac{\dot{H}}{H^2}\right)+a k^2\zeta^\text{cl}=0\,.
\end{equation}
In order to solve the equation we require two initial conditions. One condition we can impose is given by~\cite{Senatore2017}
\begin{equation}
    \zeta^\text{cl}(-k\tau\gg1)\simeq-\frac{i}{(2\epsilon)^{1/2}a^3(\tau)}\,\frac{H}{\sqrt{2k/a(\tau)}}\,e^{i k\tau}\,,
\end{equation}
where the coefficient stems from normalisation. This is true since we know that vacuum state modes inside the horizon $1/H$ should look the same as in the Minkowski space~\cite{Cheung2008}. The second order equation is then solvable if we ignore the time evolution of terms $H$ and $\dot{H}$ and calculate them at horizon exit; yielding the following solution:
\begin{equation}
    \zeta^\text{cl}_k(\tau)=\frac{H}{\sqrt{2\epsilon}\,(2k)^{3/2}}\,(1-ik\tau)e^{ik\tau}\,.
\end{equation}
Finally the power spectrum of the perturbations is given by~\cite{Cheung2008}
\begin{equation}\label{Eq:Scalar-ZetaFluct}
    \braket{\zeta_\textbf{k}(\tau)\,\zeta_{\textbf{k}'}(\tau')}=(2\pi)^3\delta^{(3)}(\textbf{k}+\textbf{k}')\,\frac{1}{4M_\text{Pl}^2\,k^3}\,\left.\left(-\frac{H^4}{\dot{H}}\right)\right|_\text{horizon exit}\,,
\end{equation}
iff $k\tau\ll1$ and $k\tau'\ll 1$. In the last step we reintroduced also the dimensionful parameters. Now, it is straightforward to recognise that the expression for the power spectrum of scalar perturbations coincides with the already familiar one.

Let us define the \emph{dimensionless} power spectrum of primordial scalar fluctuations~\cite{Lyth1999}
\begin{equation}
    \mathcal{A}_s=\frac{k^3}{2\pi^2}\,\mathcal{P}_\zeta\,,\qquad\text{where}\quad \mathcal{P}_\zeta\equiv\frac{1}{4M_\text{P}^2\,k^3}\,\left.\left(-\frac{H^4}{\dot{H}}\right)\right|_\text{horizon exit}\,,
\end{equation}
defined through eq.~\eqref{Eq:Scalar-ZetaFluct}. The deviation of the scalar power spectrum from absolute scale invariance is measured by the scalar \emph{spectral index} $n_s$ (also referred to as the \emph{spectral/primordial tilt}), defined as
\begin{equation}
    n_s-1\equiv\frac{\mathrm{d}\ln{\mathcal{A}_s}}{\mathrm{d}\ln{k}}\stackrel{\text{SR}}{\sim}-4\epsilon_H+2\eta_H\approx-6\epsilon_V+2\eta_V,
\end{equation}
where all the quantities are calculated at horizon exit $k= a H$ and $n_s=1$ represents the point of an \emph{exactly} scale-invariant spectrum. In the second equality we assumed slow-roll conditions for the scalar field allowing us to rephrase the quantities as
\begin{equation}
    \frac{\mathrm{d}\ln{\mathcal{A}_s}}{\mathrm{d}\ln{k}}=\frac{\mathrm{d}}{\mathrm{d}\ln{k}}\ln{\left(-\frac{H^4}{\dot{H}}\right)}\approx\frac{\mathrm{d}}{\mathrm{d}\ln{k}}\ln{\left(-\frac{H^4}{\dot{\phi}^2}\right)}\stackrel{\mathrm{d}\ln{k}\sim H\mathrm{d}t}{\sim}-2\,\frac{\dot{H}}{H^2}+2\,\frac{\ddot{\phi}}{H\dot{\phi}}=4\epsilon_H-2\eta_H\,,
\end{equation}
computed, for the purpose of this work, up to first order in the SRPs. 

For the sake of completeness, let us include the complete parametrisation of the power spectrum used for the analysis of the CMB anisotropies. Here, $\mathcal{P}_\zeta$ is decomposed into its scale invariant part $\mathcal{A}_s$ and parts encoding the scale dependence, as follows:
\begin{equation}
    \ln{\mathcal{P}_\zeta}=\ln{\mathcal{A}_s}+(n_s-1)\ln{\frac{k}{k_*}}+\frac{1}{2!}\,\alpha_s\ln^2{\frac{k}{k_*}}+\frac{1}{3!}\,\beta_s\ln^2{\frac{k}{k_*}}+\mathcal{O}\left(\frac{\mathrm{d}^4\mathcal{P}_\zeta}{\mathrm{d}\ln{k^4}}\right)\,,
\end{equation}
where $k_*=a_*H_*$ is a reference scale (usually $0.05\,\text{Mpc}^{-1}$) and $\alpha_s$, $\beta_s$ denote the \emph{running} and the \emph{running of the running} of the spectral index $n_s$, respectively. They are defined by
\begin{align}
    \alpha_s&\equiv\left.\frac{\mathrm{d}^2\ln{\mathcal{P}_\zeta}}{\mathrm{d}\ln{k^2}}\right|_{k=k_*}=\left.\frac{\mathrm{d}n_s}{\mathrm{d}\ln{k}}\right|_{k=k_*}\approx8\epsilon_V(2\eta_V-3\epsilon_V)-2\xi_V\,,\\
    \beta_s&\equiv\left.\frac{\mathrm{d}^3\ln{\mathcal{P}_\zeta}}{\mathrm{d}\ln{k^3}}\right|_{k=k_*}=\left.\frac{\mathrm{d}^2n_s}{\mathrm{d}\ln{k^2}}\right|_{k=k_*}\approx32\epsilon_V\left[\eta_V^2+6\epsilon_V(\epsilon_V-\eta_V)\right]+2\epsilon_V(12\xi_V-\eta_V)-2\pi_V\,,
\end{align}
where the higher-order PSRPs are given by
\begin{equation}
    \xi_V=M_P^4\,\frac{V'\,V^{(3)}}{V^2}\,,\qquad\pi_V=M_P^6\,\frac{(V')^2\,V^{(4)}}{V^3}\,.
\end{equation}
The Planck 2018 collaboration has constrained the values of the runnings $\alpha_s$ and $\beta_s$ to~\cite{Akrami2020}
\begin{equation}
    \alpha_s=0.013\pm0.010\,,\qquad \beta_s=0.022\pm0.012\,,
\end{equation}
that has helped restrict the space of the simplest models, since they predict a scale-invariant spectrum to a very high degree; by power counting, a rough estimate is $\alpha_s\sim\mathcal{O}(10^{-3})$ and $\beta_s\sim\mathcal{O}(10^{-5})$. More importantly, the data (consistent with previous missions, e.g. Planck 2015~\cite{Ade2016}) suggest a nonstandard hierarchy of the runnings, hinting to possible new dynamics to be included in the frame of inflation.

\subsubsection{\emph{Tensor perturbations}}

The FRW metric can be perturbed in the following general way~\cite{Stewart1990}:
\begin{equation}
    \mathrm{d}s^2=-(1+2\Phi)\mathrm{d}t^2+2a(t)B_i\,\mathrm{d}x^i\mathrm{d}t+a^2(t)\left[(1-2\Psi)\delta_{ij}+A_{ij}\right]\mathrm{d}x^i\mathrm{d}x^j\,.
\end{equation}
Notice that we included the potential vector perturbations even though they are not produced during inflation driven by a scalar field. We can separate these perturbations according to their transformation under rotation. Then, we conclude that $\Phi$ and $\Psi$ are scalars (helicity zero) and we can decompose $A_{ij}$ and $B_{i}$ into their scalar, vector and tensorial components. It is now possible to see that at first order, and in a rotational invariant background (like FRW) the different modes are not mixing and they evolve independently of each other. Furthermore, under a general coordinate transformation it is straightforward to show that the different helicity metric  fluctuations do not mix and they transform only by the coordinate change with the same helicity. That is why the scalar (and vector) perturbations are transforming and the tensor perturbations are gauge invariant.

Since we are interested in the tensor perturbations we can simply express the metric as
\begin{equation}
    \mathrm{d}s^2=-\mathrm{d}t^2+a^2(t)\left(\delta_{ij}+h_{ij}\right)\mathrm{d}x^i\mathrm{d}x^j\,,
\end{equation}
where $\delta_{ij}$ is the Kronecker delta and we ignore the rest of the perturbations. At first order of perturbations it is straightforward to show that only the GR part of the action contributes and the perturbed action reads
\begin{equation}
    \mathcal{S}\propto\int\!\mathrm{d}^4x\,a^3\left\{(\dot{h}_{ij})^2-\frac{1}{a^2}(\partial_kh_{ij})^2\right\}\,,
\end{equation}
where the small fluctuations $h_{ij}$ are implicitly dimensionless; an overall proportionality factor of $M_P^2/8$ has been absorbed. Next, let us decompose $h_{ij}$ into its two helicity modes as follows
\begin{equation}
    h_{ij}=\begin{pmatrix}h_+&h_\times&0\\h_\times&-h_+&0\\0&0&0\end{pmatrix}=h_+e_{+,ij}+h_\times e_{\times,ij}\,,
\end{equation}
where we assumed, without loss of generality, that the unit vector of propagation is along the $\mathbf{k}=\hat{z}$ axis. We can now expand $h_{ij}$ in its Fourier modes
\begin{equation}
    h_{ij}=\sum_{s=+,\times}\int\!\mathrm{d}^3k\,e^s_{ij}\,h^s_\mathbf{k}\,e^{i\mathbf{k}\cdot\mathbf{x}}\,,\qquad \text{with}\quad k^ie_{ij}=0,\quad e^s_{ik}\,e^{s'}_{kj}=\delta_{ij}\,\delta_{s,s'}
\end{equation}
and rewrite the action as
\begin{equation}
    \mathcal{S}=\sum_{s=+,\times}\int\!\mathrm{d}t\,\mathrm{d}^3k\ a^3\left\{\dot{h}^s_\mathbf{k}\,\dot{h}^s_{-\mathbf{k}}-\frac{k^2}{a^2}\,h^s_\mathbf{k}\,h^s_{-\mathbf{k}}\right\}\,.
\end{equation}
Then, the action is similar to the one for the scalar perturbations, but for a different normalisation of the fluctuations. Therefore, we immediately obtain the power spectrum of each of the polarisations of the tensor perturbations 
\begin{equation}
    \braket{h^s_\mathbf{k}\,h^{s'}_{\mathbf{k}'}}=(2\pi)^3\,\delta^3(\mathbf{k}+\mathbf{k}')\,\delta_{s,s'}\left.\,\frac{H^2}{M_P^2}\,\frac{1}{k^3}\right|_{k=aH}\,,
\end{equation}
where once again we reinstated the proper units. Similarly to what we did before we may write the power spectrum of the tensor perturbations as
\begin{equation}
    \mathcal{A}_t=\frac{k^3}{2\pi^2}\,\mathcal{P}_t,\qquad\text{where}\quad \mathcal{P}_t=\left.\,\frac{H^2}{M_P^2}\,\frac{1}{k^3}\right|_{k=aH}\,.
\end{equation}
Likewise, we associate the deviation from the scale invariant spectrum by its tensor tilt defined as
\begin{equation}
    n_t-1=\frac{\mathrm{d}\ln{\mathcal{A}_t}}{\mathrm{d}\ln{k}}\stackrel{SR}{\approx}-2\,\epsilon_H\,,
\end{equation}
calculated at the horizon exit. Very much like the spectral index $n_s$, we can write down the \emph{running} of $n_t$ as
\begin{equation}
    \alpha_t\approx4\epsilon_V(\eta_V-2\epsilon_V)\,.
\end{equation}
Finally, we introduce another useful (observable) quantity, the \emph{tensor-to-scalar ratio} of their respective power spectra, defined through
\begin{equation}\label{Eq:tensortoscalar}
    r\equiv\frac{\mathcal{A}_t}{\mathcal{A}_s}\approx16\,\epsilon_H\,.
\end{equation}
A measurement of $r$ is a direct measure of the energy scale of inflation since $\mathcal{A}_s$ is fixed and $\mathcal{A}_t\propto H^2$. Then large values of $r$, close to $r\geq0.01$, suggest that inflation occurs near a GUT scale ($\sim10^{16}$ GeV) and the field excursion can potentially be transPlanckian, $\Delta\phi\gtrsim M_P$.

\subsubsection{\emph{Digression on the Lyth bound}}

We can recast the tensor-to-scalar ratio $r$ in the following way
\begin{equation}
    r=\frac{8}{M_P^2}\,\frac{\dot{\phi}^2}{H^2}=\frac{8}{M_P^2}\left(\frac{\mathrm{d}\phi}{\mathrm{d}N}\right)^2\,,
\end{equation}
which connects the evolution of the inflaton with $r$. Using the slow-roll approximated results it is straightforward to show that the total field excursion from the time the observable scales exited the horizon to the end of inflation is given by
\begin{equation}
    \Delta\phi=\frac{M_P}{8\sqrt{\pi}}\sqrt{r}\left|\Delta N\right|\,.
\end{equation}
Note that at least at first order in the slow-roll approximation,\footnote{In second order in the slow-roll approximation the field excursion is also related to the scalar perturbations through~\cite{Easther2006}$$\Delta\phi^{(2)}=\frac{M_P}{2\sqrt{\pi}}\sqrt{\epsilon}\left|\Delta N\right|\left[1+(\eta-\epsilon)\Delta N\right]\,. $$} we can safely assume that $r$ is slowly varying with $N$. In Lyth's original paper the scales considered ($1<\ell\leq100$) leaving the horizon allowed for the universe to expand in that time by $\left|\Delta N\right|\approx 4$, which leads to the formula~\cite{Lyth1997}
\begin{equation}
    \Delta\phi\gtrsim M_P\sqrt{\frac{r}{4\pi}}\,,
\end{equation}
known as the \emph{Lyth bound}. The bound can be even more strict if one accounts for the entire span of inflation and obtain a more refined bound of $\Delta\phi\approx 6M_P\,r^{1/4}$, when $r\gtrsim10^{-3}$~\cite{Efstathiou2005}.

\subsection{\emph{Contact with observations}}

\begin{figure}[H]
    \centering
    \includegraphics[scale=0.8]{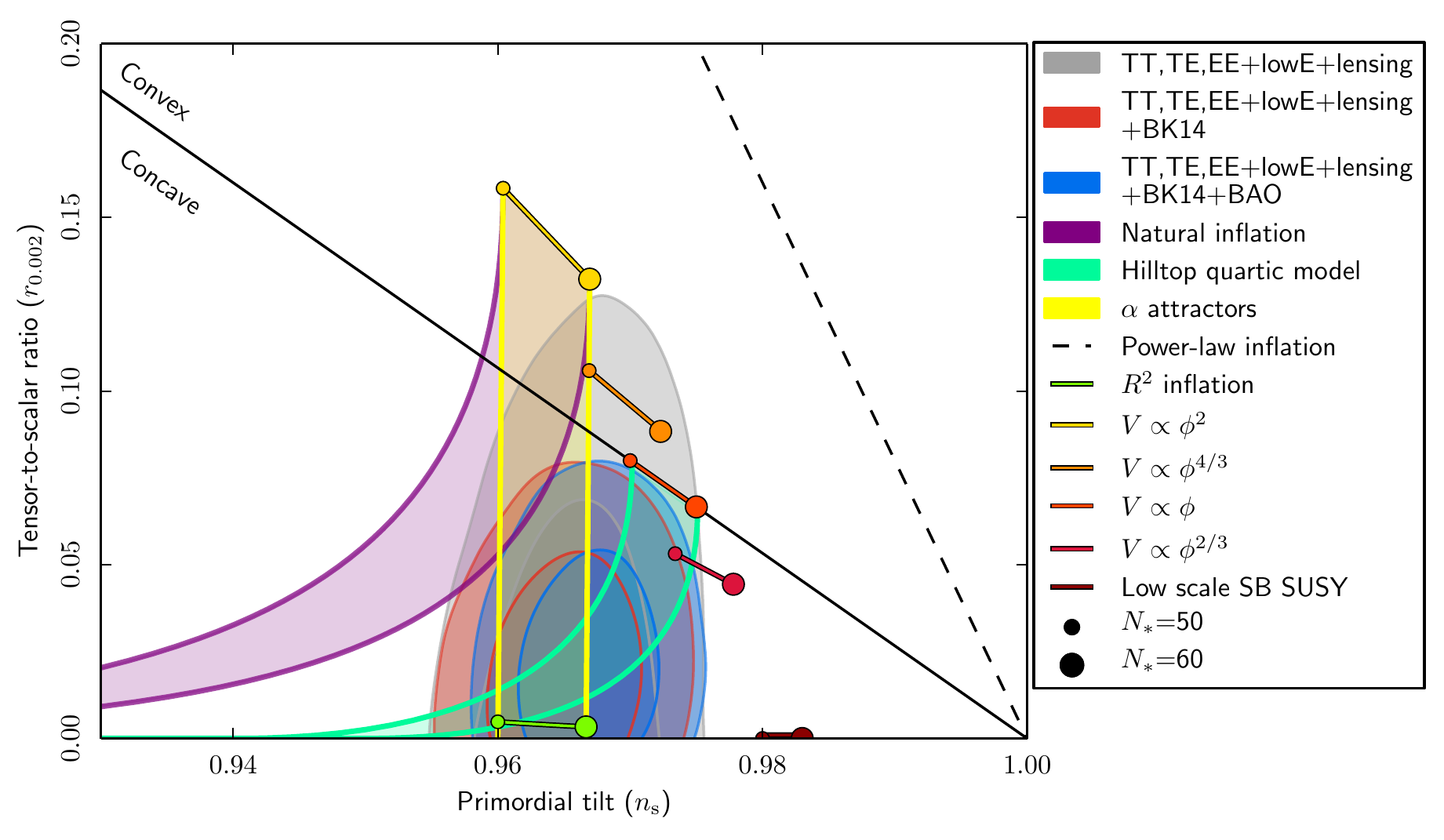}
    \caption{The results of the Planck 2018 collaboration regarding the predictions of well-known inflationary models against the latest observational data. Credit:~\cite{Akrami2020}.}
    \label{fig:Planck2018}
\end{figure}

The hypothesis of inflation is testable against observational data, albeit, disappointingly, current bounds on the plethora of models are relatively lenient. The cornerstone of our cosmological data have come through the observation of the CMB anisotropies, leading to fig.~\ref{fig:Planck2018} in which the $1\sigma$ and $2\sigma$ allowed regions of $r$ and $n_s$ are presented.

At this point our qualitative bounds on the inflationary period are given by the Planck collaboration~\cite{Akrami2020} (and similarly from BICEP2~\cite{Ade2018}) and they are
\begin{equation}
    \mathcal{A}_s\approx 2.1\times 10^{-9}\,,\qquad r\lesssim 0.056\,,\qquad n_s=\left\{\begin{matrix}(0.9607,0.9691)&\text{at }1\sigma\text{ region}\\ &\\(0.9565,0.9733)&\text{at }2\sigma\text{ region}\end{matrix}\right.
\end{equation}
Future missions with higher precision aim to place further constraints on the model space by restricting the allowed region of these quantities even more. 

It is rather incredible that a lot of the simplest single-field inflation models are already ruled out by observations or they are at marginal contact with them. There are those however, much like the \emph{Starobinsky model}~\cite{Starobinsky1980}, the \emph{Higgs inflation model}~\cite{Bezrukov2008,DeSimone2009,Bezrukov2009a} or the $\alpha$-attractor models~\cite{Kallosh2013,Kallosh2013a,Galante2015} that are persistently in good agreement with observation. These models, and others alike, usually include some kind of \emph{nonminimal coupling} of the scalar mode that assumes the role of the inflaton to the gravitational sector. So, in a way, the added degree of freedom accounts for the favourable predictions, while on the other hand one has to account for the issues concerning the nature of that coupling. Such couplings, of the form $f(\phi)R$, are expected to arise at the quantum level of a gravitational theory that includes a scalar field living in the background\footnote{See for example ref.~\cite{Parker2009} for more details. In the case of the Starobinsky model the higher-order curvature terms are also expected to appear at the quantum level of the theory~\cite{Utiyama1962,Stelle1977}. However, current results pertaining to the quantisation of gravity are always met with skepticism and at this point we should not make bold claims regarding what a quantum theory of gravity will look like.} (\emph{scalar-tensor theories} -- due to their propagating modes) or from dimensional reduction of higher-dimensional theories\footnote{For example, a simple model in which a $5$-dimensional Kaluza-Klein theory with a compactified dimension leads to a $4$-dimensional dilaton model (also referred to as the radion field). }. Another interesting possibility is the scenario that more than one scalar mode contributes to inflation, either both at the same time or one after the other. In the latter, the flat directions of the (multi-dimensional) potential allow for one field to lead inflation and it naturally stops to enter a phase in which another one contributes. For this reason, analysing the spectrum of these theories is often a daunting task, usually met with further assumptions.

There exist more parameters that can provide additional information and restrain the inflationary era. One of them is the non-Gaussianity detected in the power spectrum~\cite{Maldacena2003,Seery2007}. In fact, conventional single-field slow-roll inflation models predict a Gaussian spectrum with high accuracy, since in order for slow-roll to occur any kind of interactions of the inflaton field have to be weak. The non-Gaussianity is encoded in the three-point function $\braket{\zeta_{\mathbf{k}_1}\,\zeta_{\mathbf{k}_2}\,\zeta_{\mathbf{k}_3}}$ and its bi-spectrum can leave imprints in the angular power spectrum of the CMB fluctuations. The latest constraint on the parameter characterising the non-Gaussianity is $f_{NL}^\text{local}=-0.9\pm5.1$~\cite{Wands2002}. Even though the task of detecting them is demanding, they can have a major impact on our understanding of the physics of the early universe.

There is also a point to be made for the potentially astounding information the CMB polarisation offers. Induced by Thompson scattering, the polarisation can be decomposed into two spin-$0$ fields, known as $E$ and $B$, completely describing the linear polarisation field. It was shown~\cite{Zaldarriaga1997,Kamionkowski1997} that scalar perturbations create only $E$-modes and tensor perturbations (primordial gravitational waves) create both $E$- and $B$-modes.\footnote{Vector perturbations, that decay with the expansion of the universe, are not considered here but they do create $B$-modes as well.} Then, measuring the angular spectrum of the $B$-modes provides us with unique information about the primordial tensor modes.

\subsection{\emph{Constant-roll approximation}}
\label{subsec:CRapprox}

In principle, since most of the single-field slow-roll inflation models predict negligible non-Gaussianity (close to the order of the SRPs), one expects that any kind of detection of non-Gaussianity would exclude a lot of them, or at least heavily constrain the model space of inflation. It has been argued~\cite{Kinney2005,Namjoo2013,Martin2013}, however, that this might not be exactly the case if we allow for a ``generalisation'' of the slow-roll approximation, dubbed the \emph{constant-roll}~\cite{Motohashi2015a,Motohashi2017a,Odintsov2017,Oikonomou2017}.

It was first noticed, that in the case of the \emph{ultra} slow-roll~\cite{Martin2013}, in which the potential is almost flat ($\partial V/\partial\phi\approx0$), the scalar power spectrum is almost scale-invariant, similarly to the usual slow-roll approximation. Intuitively, we expect that since the Klein-Gordon equation yields $\ddot{\phi}/(H\dot{\phi})=-3$ the second SRP is $\eta_H\approx\mathcal{O}(1)$ and therefore the power spectrum should deviate from scale-invariance. Generalising that case we allow for the possibility of the scalar field to roll with a \emph{constant rate}, defined by~\cite{Motohashi2015a}
\begin{equation}\label{Cond:CRcondition}
\frac{\ddot{\phi}}{H\dot{\phi}}\approx \beta\,,
\end{equation}
where $\beta\in\mathbb{R}$ is a constant parameter. In the case that $\beta=0$ the standard slow-roll approximation is recovered, but if $\beta=-3$ the ultra slow-roll one is obtained. Therefore, the constant-roll encompasses both of these approximations, while also supporting a more general scenario. Clearly, not all cases of $\beta$ are allowed seeing that they do not comply with current observational data, although they lead to predictions that are in principle distinguishable from the slow-roll inflation.

Apart from accounting for possible CMB anisotropies, an attractive feature of the constant-roll scenario is that plenty of the inflaton dynamics can be studied analytically. For example, equation~\eqref{Eq:FriedScal2} can be expressed as
\begin{equation}\label{Eq:FriedScal2-CR}
    \dot{\phi}=-2M_P^2\,\frac{\mathrm{d}H}{\mathrm{d}\phi}\,,
\end{equation}
where we reintroduced the Planck scale and assumed that $\dot{\phi}\neq0$ in order to write the RHS. Direct substitution of this into eq.~\eqref{Cond:CRcondition} gives rise to the following differential equation for the Hubble parameter
\begin{equation}\label{Eq:HDiffEQ-CR}
    \frac{\mathrm{d}^2H}{\mathrm{d}\phi^2}=-\frac{\beta}{2M_P^2}\,H\,.
\end{equation}
The most general solution to this equation is given by: 
\begin{equation}
    H(\phi)=c_1\,\text{exp}\left(\sqrt{\frac{-\beta}{2}}\,\frac{\phi}{M_P}\right)+c_2\,\text{exp}\left(-\sqrt{\frac{-\beta}{2}}\,\frac{\phi}{M_P}\right)\,.
\end{equation}
Notice how $\beta\neq0$ induces linear perturbations to the standard case of slow-roll inflation (in which $\beta=0$). From eq.~\eqref{Eq:FriedScal1} we obtain the form of the inflaton potential required to support the above solution; it reads:
\begin{equation}\label{Cond:GenPot-CR}
    V(\phi)=3\,M_P^2\,H^2-2M_P^4\left(\frac{\mathrm{d}H}{\mathrm{d}\phi}\right)^2\,.
\end{equation}
It is straightforward to show that eqs.~\eqref{Cond:CRcondition}, \eqref{Eq:HDiffEQ-CR} and \eqref{Cond:GenPot-CR}  satisfy the Klein-Gordon equation eq.~\eqref{Eq:KG} trivially. 

Any solution of $H(\phi)$ may not necessarily be an attractor solution or may even evolve out of the attractor trajectory. In order to verify the stability of the solution we have to employ numerical techniques, but it might prove useful to first describe it analytically. Suppose there exists a solution to eq.~\eqref{Cond:GenPot-CR}, say $H_0(\phi)$ with a linear perturbation $\delta H(\phi)$. The linearised equation for the perturbation then becomes:
\begin{equation}
    H_0'(\phi)\,\delta H'(\phi)=\frac{3}{2M_P^2}\,H_0(\phi)\,\delta H(\phi)\,,
\end{equation}
which has the general solution
\begin{equation}
    \delta H(\phi)=\delta H(\phi_0)\,\text{exp}\left(\frac{3}{2M_P^2}\int_{\phi_0}^\phi\!\frac{H_0(\phi)}{H'_0(\phi)}\,\mathrm{d}\phi\right)\,,
\end{equation}
where $\delta H(\phi_0)$ and $\phi_0$ denote initial conditions of the cosmological system. Therefore, if the linear perturbations are decaying the attractor point is maintained and the solution is stable. To complete the analysis of the attractor behaviour one is also required to study the phase space of solutions $(\dot{\phi}-\phi)$ (usually done numerically). We discuss further the attractor behaviour of the cosmological systems in ch.~\ref{Ch4:QuadGrav}, where we consider specific inflationary models.

\section{Parametrising Reheating after Inflation}

In previous sections we demonstrated how inflation provides a solution to modern cosmological puzzles, while also making predictions of its own. One of its valuable features is the ability to dilute unwanted energy densities appearing prior to inflation due to the exponential expansion of the universe. For any particle species (relativistic or not) the matter and radiation densities decay as $\rho_m\propto a^{-3}\sim e^{-3N}\to0$ and $\rho_r\propto a^{-4}\sim e^{-4N}\to0$, respectively, for $N\gg1$ (while vacuum densities like the inflaton's remain constant). In order then to obtain the large energy and entropy observed today it is postulated that another era exists starting after the end of inflation, known as \emph{reheating}, in which the universe thermalises.

\subsection{\emph{Brief overview of reheating mechanisms}}

\subsubsection{\emph{Perturbative reheating}}

Soon after the first models capable of describing the dynamics of inflation were proposed, the first models of reheating were also suggested~\cite{Abbott1982,Dolgov1982,Albrecht1982b}, contemplating the possible transfer of the energy density of the inflaton to other fields coupled to it. The assumed interactions were purely perturbative and so the decay rates were calculated in the usual way. 

Let us introduce interactions of the inflaton with a massive scalar field $\chi$ and fermion $\psi$ in the following way:
\begin{equation}
    \mathscr{L}_\text{int}\supset-\sigma\,\phi\chi^2-h\,\phi\,\overline{\psi}\psi\,,
\end{equation}
where $\sigma$ and $h$ have to be small couplings\footnote{They are usually encountered in gauge theories with sponstaneously broken symmetries.} to avoid large radiative corrections during inflation. Then, the equation of motion for the inflaton field $\phi$ including also the effects of particle production reads~\cite{Lozanov2019}:
\begin{equation}\label{Eq:PerReh}
    \ddot{\phi}+(3H(t)+\Gamma)\dot{\phi}+m^2\phi=0\,,
\end{equation}
where $\Gamma$ is identified with the total decay rate, given by
\begin{equation}
    \Gamma=\sum_{i}\Gamma(\phi\to\chi_i\chi_i)+\sum_j\Gamma(\phi\to\overline{\psi}_j\psi_j)\,.
\end{equation}
Here we assumed that the inflaton potential is $V(\phi)=(m^2/2)\phi^2+\mathcal{O}(\phi^4)$. After the end of inflation we can safely assume $m\gg H$ and $H\sim\Gamma$ (or even if $H\gg \Gamma$ specifically at the start of reheating), thus we approximate the solution of $\phi(t)\approx\Phi(t)\cos{(mt)}$ where the amplitude is assumed to be slowly varying compared to the phase. Then, the equation of motion \eqref{Eq:PerReh} admits the following solution
\begin{equation}
    \phi(t)\approx\phi_0\,\text{exp}\left(-\frac{1}{2}(3H+\Gamma)t\right)\,\cos{(mt)}\,.
\end{equation}
The evolution of the number density and energy satisfies the Boltzman equation
\begin{equation}
    \frac{\mathrm{d}z}{\mathrm{d}t}=-\Gamma\,z\,,\qquad\text{where}\quad z=\{a^3n_\phi,\,a^3\rho_\phi\}\,,
\end{equation}
and $n_\phi=\rho_\phi/m$. 

Since the couplings are assumed to be small based on perturbativity grounds, the inflaton field loses energy at the start of reheating primarily due to the expansion of the universe. Only after the Hubble rate has reached $H\sim\Gamma$ does the particle production take effect. The decay rates are not always comparable to each other, meaning that some channels are more effective than others. If we assume the opposite, i.e. the decay products are in thermal equilibrium, we can place an upper bound on the reheating temperature, under the assumption of instantaneous reheating with relativistic degrees of freedom $g_*\approx 10^2$ that reads~\cite{Lozanov2019}
\begin{equation}
    T_\text{R}\approx \mathcal{O}(0.1)\sqrt{M_P\,\Gamma}\,.
\end{equation}
We refrain from discussing further the phenomenological aspects, since they are model dependent and the specific model of chaotic inflation assumed here as an example is already ruled out by observational data (see fig.~\ref{fig:Planck2018}).

Unfortunately, the equation of motion eq.~\eqref{Eq:PerReh} does not account for all the dynamics of the inflaton field. For example, the equation should include fluctuations that are present in such systems with dissipation~\cite{Kofman1997}. Also, even if the couplings are small enough to satisfy perturbativity, there is still the scenario that the phase space of $\chi$-particles is densely populated and therefore Bose condensation effects can enhance the decay rate~\cite{Lozanov2019}. The most crusial point however, is that the perturbative method fails (at least in the initial stage); the inflaton condensate is a coherent homogeneous field that oscillates and therefore many inflaton particles decay simultaneous and \emph{not} independent of each other. In other words, even though we are justified in describing the inflaton field classically due its large amplitude of oscillations, the decay products have to be treated quantum mechanically. Furthermore, for the perturbative reheating to end one assumes a coupling of the inflaton to fermions, which further constrains the structure of the theory allowing for reheating. In what follows we briefly describe some of the nonperturbative dynamics of reheating.

\subsubsection{\emph{Preheating \& parametric resonances}}

The study of the early stage of reheating with nonperturbative techniques is usually referred to as \emph{preheating} and falls under the paradigm of particle production in the presence of strong background fields. For illustrative purposes, let us consider once again the chaotic inflation potential. The total scalar potential reads
\begin{equation}
    V(\phi,\chi)=\frac{m^2}{2}\,\phi^2+\frac{m_\chi^2}{2}\,\chi^2+\frac{1}{2}g^2\chi^2\phi^2\,.
\end{equation}
The classical equation of motion for the $\chi$ field is given by
\begin{equation}
    \ddot{\chi}-\frac{1}{a^2}\nabla^2\chi+3H\dot{\chi}+V_\chi=0\,.
\end{equation}
Expanding it in Fourier modes
\begin{equation}
    \chi(t,\mathbf{x})=\int\!\frac{\mathrm{d}^3k}{(2\pi)^{3/2}}\left(a_\mathbf{k}\chi_\mathbf{k}(t)e^{-i\mathbf{k}\cdot\mathbf{r}}+a^\dagger_\mathbf{k}\chi^*_\mathbf{k}(t)e^{i\mathbf{k}\cdot\mathbf{r}}\right)\,,
\end{equation}
it satisfies
\begin{equation}
    \ddot{\chi}_\mathbf{k}+3H\dot{\chi}_\mathbf{k}+\left(\frac{k^2}{a^2}+m_\chi^2+g^2\phi^2(t)\right)\chi_\mathbf{k}=0\,,
\end{equation}
where $\phi(t)\approx\Phi(t)\sin{(mt)}$. This is the equation for an oscillator with a varying frequency that is further damped by the expansion of the universe $3H\dot{\chi}$. Then, we expect some modes $k$ to parametrically excite themselves, similar to the case of a parametric oscillator. The above equation can be recast into the form of a Mathieu equation by also disregarding the expansion of the universe $H\approx0$ at this point
\begin{equation}
    \frac{\mathrm{d}^2\chi_k}{\mathrm{d}z^2}+(A_k-2q\cos{2z})\chi_k=0\,,
\end{equation}
where $z\equiv mt$ is dimensionless and
\begin{equation}
    A_k\equiv\frac{k^2+m_\chi^2}{m^2}+2q\,,\qquad q=\frac{g^2\Phi^2}{4m^2}\,.
\end{equation}
The instabilities appear for certain modes $k$ that lead to exponential growth $\chi_k\propto e^{\mu_kz}$, where $\mu_k$ is called the Floquet exponent and is nonnegative. The value of the $q$ parameter leads to different types of resonances;  values of $q<1$ introduce what is known as \emph{narrow} resonances and $q>1$ induce more efficient resonance effects, \emph{broad} resonances~\cite{Kofman1997}.

Finally, if the effective frequency of oscillations of the $\chi$ field were negative it leads to tachyonic resonances, a scenario usually referred to as \emph{tachyonic preheating}. This is achieved trivially in models with negative couplings. Then, higher-order terms are added to ensure the overall stability of the system, but only contribute during inflation and are unimportant during reheating. In some cases these tachyonic instabilities prove to be even more efficient than the narrow and broad resonances.

\subsection{\emph{Parametrising reheating}}
\label{sec:ParReheating}

Undoubtedly, there is a high uncertainty concerning the theoretical part of reheating. As was illustrated there are different complicated processes one has to take into account, depending also on the type of inflation preceding reheating. Furthermore, the period of our universe from the end of inflation until the point of baryogenesis and Big Bang nucleosynthesis is relatively unknown.

Interestingly, we may parametrise the cosmic fluid during reheating by an effective constant equation of state parameter $w_\text{R}$. At the end of inflation the inflaton oscillates around its potential minimum, essentially, between the point of complete kinetic domination ($w=1$) and domination of the potential ($w=-1$). Therefore, the universe at that time is properly described by $w=0$ that increases as the inflaton decays, reaching the radiation domination era of $w=1/3$. In reality, the equation of state parameter changes very quickly to $w\sim0.25$ at the first stages of preheating (e.g. see refs.~\cite{Felder1999,Podolsky2006}) justifying the assumption of a constant $w_\text{R}$. 

\begin{figure}
    \centering
    \includegraphics[scale=1]{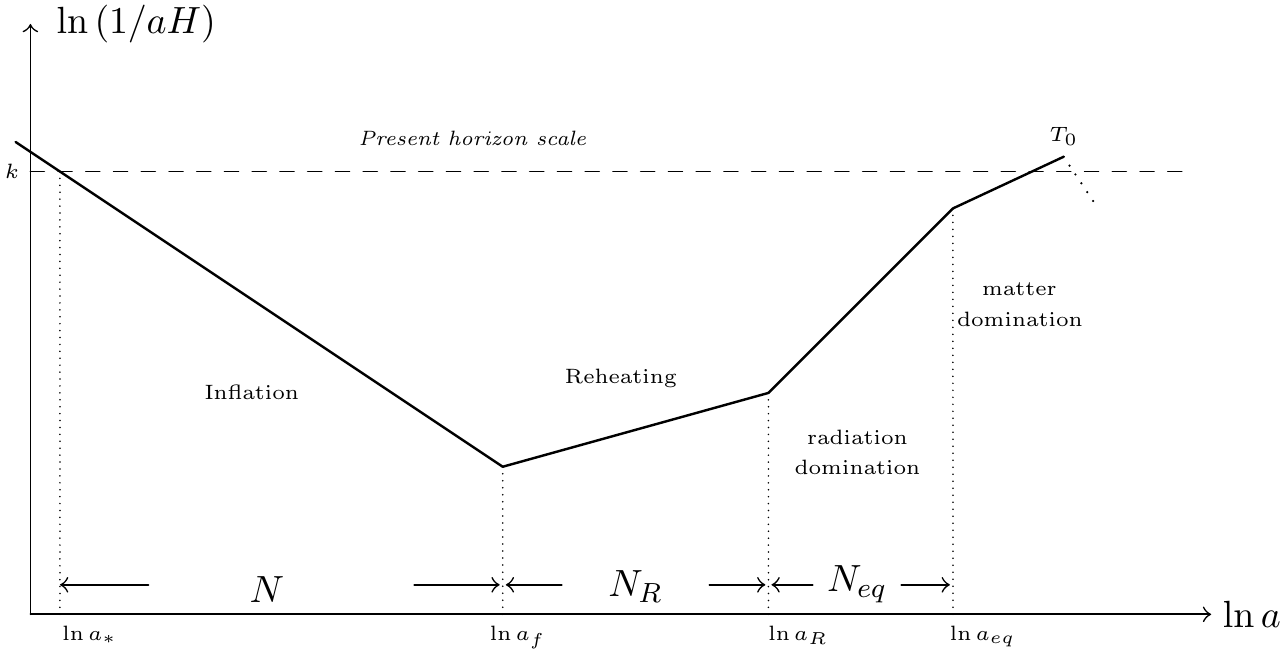}
    \caption{An example of the evolution of the comoving Hubble scale $1/aH$ displaying also the different epochs with their associated number of $e$-foldings. The phase of reheating connects the inflation and radiation era.}
    \label{fig:Reheating-Log(aH)-Horizon}
\end{figure}

Apart from the state parameter $w_\text{R}$, reheating is described also by its duration $N_\text{R}$ and its temperature $T_\text{R}$. By considering a history of the expansion of the scales from the point of horizon exit that we observe at the CMB to horizon re-entry, we can relate the reheating parameters to the inflationary ones~\cite{Dodelson2003,Liddle2003a,Dai2014,Munoz2014,Gong2015,Cook2015a}. Generally, we assume that the inflaton field $\phi$ describes inflation for $N$ $e$-foldings given a specific potential $V(\phi)$, during which the comoving Hubble horizon decreases. Then, the reheating phase begins that grows the comoving horizon for $N_\text{R}$ $e$-folds until the era of radiation domination is initiated. Finally, the radiation era proceeds for $N_\text{eq}$ $e$-folds until the point of equilibrium. It is important to note that we assume instant transition between each of these epochs and the state parameter $w$ remains constant in each of them (see fig.~\ref{fig:Reheating-Log(aH)-Horizon}).

Each period's duration is defined through its respective $e$-folds
\begin{equation}
    N=\ln{\frac{a_f}{a_*}}\,,\qquad N_\text{R}=\ln{\frac{a_\text{R}}{a_f}}\,,\qquad N_\text{eq}=\ln{\frac{a_\text{eq}}{a_\text{R}}}\,,
\end{equation}
where $a_f$ and $a_*$ are the scale factor values at the end and start of inflation, respectively. Similarly, $a_\text{R}$ and $a_\text{eq}$ are the scale factor values at the end of reheating and end of radiation, respectively. Finally, $a_0$ is the present day value of the scale factor. Considering next the relation between a pivot scale $k=a_* H_*$ at the start of inflation and the size of the present horizon $a_0H_0$, we obtain~\cite{Liddle2003a}:
\begin{equation}
    \frac{k}{a_0H_0}=\frac{a_*H_*}{a_0H_0}=\frac{a_*}{a_f}\,\frac{a_f}{a_\text{R}}\,\frac{a_\text{R}}{a_\text{eq}}\,\frac{a_\text{eq}}{a_0}\,\frac{H_\text{eq}H_*}{H_0H_\text{eq}}\,,
\end{equation}
\begin{equation}
    \therefore\ \ln{\frac{k}{a_0H_0}}=-N-N_\text{R}-N_\text{eq}+\ln{\frac{a_\text{eq}H_\text{eq}}{a_0H_0}}+\ln{\frac{H_*}{H_\text{eq}}}\,.
\end{equation}

During the period of reheating we assume $\rho\propto a^{-3(1+w)}$ and write the following
\begin{equation}
    \frac{\rho_f}{\rho_\text{R}}=\left(\frac{a_f}{a_\text{R}}\right)^{-3(1+w_\text{R})}\quad\implies\quad\ln{\frac{\rho_f}{\rho_\text{R}}}=3(1+w_\text{R})N_\text{R}\,,
\end{equation}
where $\rho_f$ is the energy density at the end of inflation at field values of $\phi=\phi_f$, approximately described by $\rho_f\approx 3V(\phi_f)/2$ with $V(\phi_f)\equiv V_f$ the potential value at field value $\phi=\phi_f$. Then, solving the equation above for the number of $e$-folds we obtain
\begin{equation}
    N_\text{R}=\frac{1}{3(1+w_\text{R})}\,\ln{\left(\frac{3}{2\rho_R}\,V_f\right)}\,.
\end{equation}
The temperature and energy density are related by\footnote{A particle species with $g$ degrees of freedom and $\mu$ chemical potential, has an equilibrium energy density\begin{equation}
    \rho=g\int\!\frac{\mathrm{d}^3p}{(2\pi)^3}\,E(\mathbf{p})\,f(\mathbf{p})\,,
\end{equation}
where $f(\mathbf{p})=1/\left[\text{exp}\left((E(\mathbf{p})-\mu)/T\right)\pm 1\right]$ is the phase space distribution in momentum space with plus (minus) for Fermi-Dirac (Bose-Einstein) statistics, and $E(\mathbf{p})=\sqrt{|\mathbf{p}|^2+m^2}$. In the case that $T\gg m$, $\mu$ the integral is exactly solvable and it leads to eq.~\eqref{Cond:EQEnergDensity}. Interestingly, for nonrelativistic species, where $T\ll m$, the energy density is exponentially vanishing $\rho\propto T^{3/2}e^{-(m-\mu)/T}$. Note also that at early times the chemical potential $\mu$ of all particles was tiny, such that we can safely neglect it at this point.}
\begin{equation}\label{Cond:EQEnergDensity}
    \rho_\text{R}=\frac{\pi^2}{30}\,g_\text{R}\,T_\text{R}^4\,,
\end{equation}
where $g_\text{R}$ denotes the relativistic degrees of freedom at the end of reheating. Substituting to the expression for $N_\text{R}$ we obtain:
\begin{equation}\label{Eq:NRfirst}
    N_\text{R}=\frac{1}{3(1+w_\text{R})}\,\ln{\left(\frac{45}{\pi^2}\,\frac{V_f}{g_\text{R}{T_\text{R}}^4}\right)}\,,
\end{equation}

From conservation of entropy between the transition of the reheating era to today, $S\propto g_*(aT)^3$, we can write the following expression
\begin{equation}
    g_\text{R}\left(a_\text{R}T_\text{R}\right)^3=g_\gamma\left(a_0T_\gamma\right)^3+g_\nu\left(a_0T_\nu\right)^3\,,
\end{equation}
where $g_\gamma$ is the photon and $g_\nu=\frac{7}{8}\cdot3\cdot2=21/4$ are three light neutrinos with 
\begin{equation}
    \left(T_\nu=\frac{4}{11}T_\gamma\right)^3
\end{equation}
and $T_\gamma\sim2.7\,\text{K}$. Therefore, the reheating temperature is expressed as
\begin{equation}
    T_\text{R}=T_\gamma\left(\frac{a_0}{a_\text{eq}}\right)e^{N_\text{eq}}\left(\frac{43}{11g_\text{R}}\right)^{1/3}\,.
\end{equation}
Next, focusing on the term of $a_0/a_\text{eq}$ we can write it down as
\begin{equation}
    \frac{a_0}{a_\text{eq}}=\frac{a_0 H_* a_*}{a_\text{eq}k}=\frac{a_0H_*}{k}\,\frac{a_*}{a_f}\,\frac{a_f}{a_\text{R}}\,\frac{a_\text{R}}{a_\text{eq}}=\frac{a_0H_*}{k}\,e^{-N}e^{-N_\text{R}}e^{-N_\text{eq}}\,,
\end{equation}
introducing a factor of unity by means of the pivot scale $k=a_*H_*$. Finally, the temperature $T_\text{R}$ reads
\begin{equation}\label{Eq:TRfirst}
    T_\text{R}=\left(\frac{T_\gamma\, a_0}{k}\right)\left(\frac{43}{11g_\text{R}}\right)^{1/3}H_*e^{-N}e^{-N_\text{R}}\,.
\end{equation}

Expectantly, an increase in the number of $e$-foldings $N_\text{R}$ suggests a decrease in the reheating temperature $T_\text{R}$, and vice versa. The bounds on the actual reheating temperature are inferred from potential impact on the lower-energy phenomenology. Meaning that temperatures below $T_\text{R}>10\,\text{MeV}$ are ruled out from BBN~\cite{Steigman2007}, but even temperatures close to the EW scale, $T_\text{R}\gtrsim100\,\text{GeV}$ may affect baryogenesis (while not strictly disallowed). On the other hand, an upper bound of $T_\text{R}\sim10^{16}\,\text{GeV}$ can be placed, in order to avoid the restoration of a potential GUT symmetry right after inflation. This was a point of intense research at early stages of the reheating paradigm, and it was also shown that in sypersymmetric theories the upper bound can be relaxed to $10^{9}\,\text{GeV}$~\cite{Moroi1993,Kawasaki1995,Gherghetta1999,Bolz2001}.

If there were stronger indications for the state parameter $w_\text{R}$, then the scenario of reheating can be better interpreted. However, we can assume an effective, constant $w_\text{R}$ in the range of $\left\{-\frac{1}{3},1\right\}$, in which $w_\text{R}=-\sfrac{1}{3}$ is the minimum value required such that inflation terminates. From direct substitution of eq.~\eqref{Eq:TRfirst} into eq.~\eqref{Eq:NRfirst} we obtain the following for $N_\text{R}$
\begin{equation}\label{Eq:NRgenw}
    N_\text{R}=\frac{4}{3(1+w_\text{R})}\left\{N+N_\text{R}+\ln{\frac{k}{a_0T_\gamma}}+\ln{\frac{V_f^{1/4}}{H_*}}+\ln{\left[\left(\frac{45}{\pi^2}\right)^{1/4}\left(\frac{11}{3}\right)^{1/3}g_\text{R}^{1/12}\right]}\right\}\,.
\end{equation}
Now, this is an algebraic equation in $N_\text{R}$ that admits a straightforward solution depending on the value of $w_\text{R}$. 

\clearpage

\begin{flushleft}
\underline{\emph{Case of $w_\text{R}\neq1/3$}}
\end{flushleft}
Let us first assume that $w_\text{R}\neq1/3$; then directly from eq.~\eqref{Eq:NRgenw} it gives rise to
\begin{equation}
    N_\text{R}=\frac{4}{1-3w_\text{R}}\left\{-N-\ln{\frac{V_f^{1/4}}{H_*}}-\ln{\frac{k}{a_0T_\gamma}}-\ln{\left[\left(\frac{45}{\pi^2}\right)^{1/4}\left(\frac{11}{3}\right)^{1/3}g_\text{R}^{1/12}\right]}\right\}\,.
\end{equation}
At the point of reheating and since we assume energy scales at least larger than the EW scales, we can liberally assume that\footnote{The proposed number of $g_\text{R}$ conservatively only includes the SM degrees of freedom. It is natural to expect beyond the Standard Model (BSM) degrees of freedom coming from the sought-after UV completeness of the theory to increase the value of $g_\text{R}$, but even if one assumes $g_\text{R}\approx\mathcal{O}(10^3)$ the number of $e$-folds $N_\text{R}$ and the reheating temperature $T_\text{R}$ are largely unaffected, attributed to the small dependence on the actual value of $g_\text{R}$.}
\begin{equation}
     g_\text{R}\gtrsim\frac{427}{4}\approx 100\,.
\end{equation}
Specifically, at a pivot scale of $k=0.05\,\text{Mpc}^{-1}$ we obtain the simplified version of this:
\begin{equation}
    N_\text{R}=\frac{4}{1-3w_\text{R}}\,\left(61.6-\ln{\frac{V_f^{1/4}}{H_*}}-N\right)\,,
\end{equation}
which after some algebra leads to the following expression for the temperature 
\begin{equation}
    T_\text{R}=\left\{\left(\frac{43}{11g_\text{R}}\right)^{1/3}\left(\frac{T_\gamma\, a_0}{k}\right)H_*e^{-N}\left(\frac{45}{\pi^2}\,\frac{V_f}{g_\text{R}}\right)^{-\frac{1}{3(1+w_\text{R})}}\right\}^{3(1+w_\text{R})/(3w_\text{R}-1)}\,.
\end{equation}

\begin{flushleft}
\underline{\emph{Case of $w_\text{R}=1/3$}}
\end{flushleft}
In the special case of $w_\text{R}=1/3$ it is straightforward to show that eq.~\eqref{Eq:NRgenw} leads to
\begin{equation}
    N=61.6-\ln{\frac{V_f^{1/4}}{H_*}}\,,
\end{equation}
following also what we assumed in the previous case. This can be seen as a constraint on the duration of inflation, through $N$, or the predicted value of $n_s$, and is a general feature of the scenario known as \emph{instant reheating} suggesting that the period of radiation starts immediately after inflation. In that case the reheating temperature $T_\text{R}$ takes its maximum allowed value depending primarily on the model of inflation.

There is a particular connection of the reheating and inflation periods (specifically the SRPs) that is apparent by using the definition of the tensor-to-scalar ratio, eq.~\eqref{Eq:tensortoscalar}, at the pivot scale $k=a_*H_*$
\begin{equation}
    r=\frac{2H_*^2}{M_P^2\,\pi^2\mathcal{A}_s}
\end{equation}
and the slow-roll approximation $r\approx 16\epsilon_V$ to obtain~\cite{Cook2015a}:
\begin{equation}
    H_*\approx M_P\,\pi\sqrt{8\,\mathcal{A}_s\,\epsilon_V(\phi_*)}\,.
\end{equation}
Finally, assuming a precise inflationary model and its potential $V(\phi)$ we can express its value $V_f$ at the end of inflation in terms of SRPs and therefore the values of $N_\text{R}$ and $T_\text{R}$ are expressed explicitly in terms of them. Further discussion on the model-dependent part of reheating is left for a later chapter (ch.~\ref{Ch4:QuadGrav}), in which we investigate if different inflationary models can support a reheating period, and if so under what conditions.

\chapter{First-order formalism}
\label{Ch3:FirstOrder}

Inevitably Einstein's astonishing insight has to be mentioned when one discusses the General theory of Relativity. It is based on his realisation that an observer that is freely falling in a frame does not experience gravity and therefore its effects are indistinguishable from those in an accelerating frame. What is today known as the Einstein Equivalence Principle (EEP), led him to a beautiful theory capable of explaining or describing most of gravitational physics, which, with the discovery of gravitational waves by LIGO~\cite{Abbott2017}, is still validated today, hundred years after its inception.

The key observation extracted from EEP is that gravity is better understood as the curvature of spacetime unlike the other external forces of nature. Later this had an immense impact on the future of theoretical physics, and, together with seminal works of others after Einstein, led to a \emph{``geometrisation''} of physics and the way we approach physics altogether (e.g. gauge theories, string theory, etc). General Relativity, described by the Einstein-Hilbert action, is nested in the language of Riemannian geometry and (at that time) pioneer works on differential geometry and $19^\text{th}$ century tensor calculus, which are applied even today (with a more convenient notation).

Immediately after its inception, modifications to GR had been proposed, albeit driven at first primarily by scientific curiosity. That avenue of exploration was later legitimised by the need for new features that GR, as formulated by Einstein, does not include. For example, alternatives to GR such as the loop quantum gravity (LQG) program~\cite{Ashtekar1986,Rovelli1995,Rovelli1998} attempt to construct a quantum theory of gravity, or others like the Kaluza-Klein or string theory models that attempt to unify gravity with other fundamental forces.

Interestingly, around the same time when GR was first formulated, E. Cartan in the 1920s developed a very different type of differential geometry, based on differential forms and fiber bundles~\cite{ASENS_1923_3_40__325_0,ASENS_1924_3_41__1_0,ASENS_1925_3_42__17_0}. Actually, most of Cartan's important contributions were developed around that time. Pertaining to GR, he considered other bundles other than the tangent bundle and other connections other than the Levi-Civita one. This was also unfruitfully pursued by H. Weyl~\cite{Weyl1921} and others around the same time. Cartan's generalisation of Riemannian geometry and, in particular, the more general connections were discovered (or more appropriately reconsidered) by physicists much later, in the works of Yang and Mills (1954)~\cite{Yang1954}. Today, interactions in nature are described by a \emph{gauge field} (or connection). Attempts to modify GR by generalising the connection, or more importantly the underlying geometry, are nowadays prominent even if they are rather divisive with respect to the metric and connection based formulations. 

There are a plethora of reasons why one should consider GR as a theory of connections (much like Yang-Mills), but historically, and most appealing, is the attempt to also canonically quantise gravity. With the development of gauge theories it was understood that they are dependent on connection $1$-forms\footnote{In fact, fiber bundles have allowed us to understand the global (or nonperturbative) aspect of these theories. For example the gauge field $A_\mu^a$ in Yang-Mills theory are connections on principal bundles and not just local $1$-forms. Also, the Yang-Mills instanton is actually the nontrivial class of the principal bundle underlying the gauge field~\cite{Schaefer1998}. In similar fashion fiber bundles govern other topologically nontrivial dynamics, for example quantum anomalies (not globally well-defined action functionals) are understood as statements of nontrivialisability of bundles.} (taking values of the same gauge group) and alongside their conjugate momenta we attempt to analyse their Hamiltonian formulation. It turns out that applying the usual Hamiltonian analysis to GR leads to many problems, for example the fact that the Hamiltonian constraints are non-polynomial in the canonical variables, as well as ordering problems of their respective operators. However, in the 1980s Ashtekar showed~\cite{Ashtekar1986,Ashtekar1987} by extending the (Palatini) action to complex values that the constraints were now closed and polynomial, with the caveat that one has to now impose reality conditions on the metric.\footnote{Later development along those lines led to LQG with many interesting and promising results (see ref.~\cite{Rovelli1998} for a review), further solidifying the search for a connection-oriented gravity.} More recently, developments regarding the Jackiw-Teitelboim (JT) gravity~\cite{Jackiw1985,Teitelboim1983,Almheiri2015} - a starting point in discussing nonperturbative quantum aspects of (lower dimensional) gravity - were made using its first-order formulation (e.g. see refs.~\cite{Saad2019,Iliesiu2019}). All of the above considerations concern quantum phenomena and owe to the fact that the first-order formulation involves momentum independent cubic interactions as opposed to the second-order (or metric) formulation that involves momentum dependent three-point and more vertices~\cite{Brandt2020}.

There is a recent effort in describing gravity in a connection-based language that can also be applied to lower energies. Typically, the two variational principles coincide (on-shell) for the Einstein-Hilbert action~\cite{Palatini:1919}, but when one assumes extensions that include nonminimal couplings with the matter sector or the introduction of higher-order curvature invariants in the theory, the two formalisms lead to wildly different results, that can potentially differentiate between the two formalisms (see ref.~\cite{Sotiriou2007} for a review). These types of modifications are especially popular in recent inflationary models and appear in a variety of cosmological studies, which are in principle classical but incorporate some quantum phenomena.

In this chapter, we begin by recalling some of the key features of the metric formulation of the Einstein-Hilbert action, which are later related and compared to its Palatini formulation. Especially, the derivation of the Einstein field equations is described in great detail for the same reason. Following that, we discuss the Palatini variation and the concept of a metric-affine connection. First we parallelise the discussion of the first- and second-order formulation of gravity with the analogous first- and second-order formulation of Electromagnetism. After that we introduce the notion of a metric-affine space and its primary components, the torsion, metricity and curvature tensors. Then, we are ready to apply the Palatini variation to the Einstein-Hilbert action and derive the complete set of field equations for the dynamical variables in the theory. The point of classical equivalence of the Einstein-Hilbert action between the two formulations is further stressed and is followed by a brief discussion on their possible equivalence at the quantum level.

\section{Metric formalism resivited}
\label{Sec:MetFormRevisit}

The most essential and revolutionary concept in GR, as opposed to Newtonian gravity, is the notion of spacetime. By incorporating the gravitational force into relativity we are led to a generalisation of the Minkowskian spacetime $(\mathbb{R}^4,\eta)$ to a four-dimensional Lorentzian manifold with a Lorentzian (nondegenerate and symmetric) metric, $(\mathcal{M},\text{g})$, defined at each point $p\in\mathcal{M}$. The idea of a \emph{curved manifold} is crucial to GR, since it can provide a local description of Euclidean space at each point $p\in\mathcal{M}$ (EEP). Additionally, another assumption is that any test-object with positive mass follows a timelike curve in $\mathcal{M}$, which in the case of free fall is a \emph{geodesic}.

The metric tensor at some point can be considered as a matrix $g_{\mu\nu}(p)$ that is dependent on the choice of coordinates. However, due to its symmetry\footnote{The symmetry stems from the definition/creation of the metric tensor through its definition of the spacetime interval equation, or the invariant line element. Since we have $[\mathrm{d}x^\mu,\mathrm{d}x^\nu]=0$ only the symmetric part of $g_{\mu\nu}$ contributes and we may as well assume that it is symmetric. In turn it ensures that the norm of vectors $\in\Gamma(T\mathcal{M})$ is nonnegative and the angle between them does not depend on the order chosen.} ($g_{[\mu\nu]}=0$) it can be diagonalised and its eigenvalues are coordinate independent. In fact, we can choose a base in which the eigenvalues are $+1$ and $-1$ and vary smoothly with $p$. The concept of distances in spacetime is encapsulated in the invariant line element
\begin{equation}
    \mathrm{d}s^2=g_{\mu\nu}\,\mathrm{d}x^\mu\otimes\mathrm{d}x^\nu\,,
\end{equation}
with the help of the metric tensor.

Together with the metric tensor, the manifold is also endowed with a connection $\nabla$ that clarifies the idea of parallel transporting data in a curved spacetime. Specifically, it is in the form of a covariant derivative that measures the change of a vector field after being parallel transported in another direction. Think of a curve $\gamma:I\rightarrow\mathcal{M}$ with two vector fields, $u^\mu$ and $v^\mu$, along that curve. The inner product of $g_{\mu\nu}u^\mu v^\nu$ changes along the curve $\gamma$
\begin{equation}
    \dot{\gamma}^\rho\,\nabla_\rho\left(g_{\mu\nu}u^\mu v^\nu\right)=\dot{\gamma}^\rho u^\mu v^\nu\,\nabla_\rho\, g_{\mu\nu}\,,
\end{equation}
unless $\nabla_\rho\, g_{\mu\nu}=0$. This is known as the \emph{metric-compatibility condition}. Then, for a Lorentzian metric $\text{g}$ there is only one connection that satisfies that condition, the \emph{Levi-Civita connection}, characterised by its connection coefficients ${\Gamma^\rho}_{\mu\nu}$, known as the Christoffel symbols and are given in terms of the metric tensor as
\begin{equation}\label{Cond:ChristoffelSymbs}
    {\Gamma^\rho}_{\mu\nu}=\frac{1}{2}g^{\rho\sigma}\left(\partial_\mu g_{\sigma\nu}+\partial_\nu g_{\mu\sigma}-\partial_\sigma g_{\mu\nu}\right)\equiv\{{}_\mu{}^\rho{}_\nu\}\,.
\end{equation}
Given now a covariant derivative one can define the curvature on the manifold
\begin{equation}
    {R_{\mu\nu\rho}}^\sigma b_\sigma\,=\,2\,\nabla_{[\mu\nu]} b_\rho\,,
\end{equation}
where $b_\mu$ is a $1$-form. The rank $(1,3)$ tensor ${R_{\mu\nu\rho}}^\sigma$ is known as the \emph{Riemann curvature tensor} and has many properties; most important are the following: ${R_{\mu\nu\rho}}^\sigma=-{R_{\nu\mu\rho}}^\sigma$, ${R_{[\mu\nu\rho]}}^\sigma=0$ and $\nabla_{[\lambda}{R_{\mu\nu]\rho}}^\sigma$ (Bianchi identity). Due to its symmetries the Riemann tensor has $n^2(n^2-1)/12$ independent components ($20$ in $n=4$ dimensions). We pay extra attention to the idea of these tensors on $\mathcal{M}$ in section~\ref{subsec:MetAffineSpace} and refer the reader there (and the vast literature on the subject e.g. refs.~\cite{Stewart1994,Carroll1997,Schutz2009,Tong2019}) for more details .

The actual dynamics of gravity are encoded in its Lagrangian that is a scalar function on $\mathcal{M}$. Besides that, we also demand that the Euler-Lagrange equations of the system are up to second order in derivatives of the metric tensor $\text{g}$.\footnote{Usually, that point becomes nontrivial when one considers generalisations of GR. It is not obvious how in GR a Lagrangian containing nondegenerate second-order derivatives $\partial^2\text{g}$ does not lead to Ostrogradski instabilities~\cite{OstrogradskyMmoiresSL}, that essentially state that if that is the case the Hamiltonian of the system has at least one linear instability. In fact, GR falls under the ``blanket'' theory of Lovelock gravity~\cite{Lovelock1971}, in which conserved second-order equations of motion can be produced in arbitrary spacetime dimensions.} From contracting the Riemann tensor we are provided with a scalar quantity, the Ricci scalar (or \emph{scalar curvature}) 
\begin{equation}
    R\,\equiv\,g^{\mu\nu}\,R_{\mu\nu}\,,
\end{equation}
where $R_{\mu\nu}=g^{\rho\sigma}R_{\mu\rho\nu\sigma}$ is known as the \emph{Ricci tensor}. Finally, we can consider the action\footnote{Suppose we ignore the reduced Planck mass for a moment. Then, in order for the action to have the right dimensions, i.e. time$\times$energy (like $\hbar$), a factor of $c^3/(16\pi G)$ has to be included~\cite{Tong2019}. Here, $c$ is the speed of light and $G$ is Newton's constant and plays the role of the coupling constant of gravitational interactions. If one is interested mainly in studying gravitational phenomena, e.g. black holes, they can safely set $c=1=G$, incidentally equating mass with length dimension. However, by fixing $\hbar=1$ allows us to rewrite the dimensional factor as\begin{equation*}
    \frac{\hbar c}{8\pi G}\equiv M_P^2\,,
\end{equation*}known as the (reduced) Planck mass. It is then better to set $c=1=\hbar$. Lastly, the fixing of $c=\hbar=G=1$ seems nonsensical, seeing that all of the dimensionful parameters in the theory are now eliminated alongside with our ability to apply dimensional analysis.}
\begin{equation}
    \mathcal{S}_{EH}[\text{g};\mathcal{V}]=\frac{M_P^2}{2}\int_{\mathcal{V}}\!\mathrm{d}\text{vol}\,R[\text{g}]\,,
\end{equation}
where $\mathcal{V}$ is a domain of $(\mathcal{M},\text{g})$ and $\mathrm{d}\text{vol}$ denotes the infinitesimal volume element generated by the metric field $\text{g}$, which in local coordinates reads $\mathrm{d}\text{vol}=\sqrt{-g}\,\mathrm{d}^4x$ (here $g$ denotes $\text{det}(g_{\mu\nu})$).

\subsection{\emph{Principle of least Action \& Field equations}}

According to the \emph{Action Principle}, also known as the Variational Principle, acceptable solutions of the physical system have to be \emph{stationary} points of the Lagrangian density. In this way, we derive partial differential equations for the tensor fields, named the Euler-Lagrange equations. These then lead to the equations of motion for the system.

Before we derive the field equations for the Einstein-Hilbert action, let us consider first a compact variation of the metric tensor
\begin{equation}
    \delta g_{\mu\nu}=\left.\frac{\mathrm{d}g_{\mu\nu}(\varepsilon)}{\mathrm{d}\varepsilon}\right|_{\varepsilon=0}\,,
\end{equation}
where $\varepsilon$ is a small parameter and
\begin{equation}
    g_{\mu\nu}(\varepsilon)=g_{\mu\nu}+\varepsilon\,\delta g_{\mu\nu}+\mathcal{O}(\varepsilon^2)\,.
\end{equation}
Note also that, in order to retain $g_{\mu\rho}\,g^{\nu\rho}=\delta_\mu^\nu$ at first order in $\varepsilon$, it follows that $g^{\mu\nu}(\varepsilon)=g^{\mu\nu}-\varepsilon\,\delta g^{\mu\nu}+\mathcal{O}(\varepsilon^2)$. Then, for the field $g_{\mu\nu}$ to be stationary in $\mathcal{S}$ for any compact variation $g_{\mu\nu}(\varepsilon)$ it suffices that
\begin{equation}
    \left.\frac{\mathrm{d}\mathcal{S}(\varepsilon)}{\mathrm{d}\varepsilon}\right|_{\varepsilon=0}\equiv\frac{\delta\mathcal{S}}{\delta g_{\mu\nu}}=0\,,
\end{equation}
where we adopted the shorthand notation $\mathcal{S}(\varepsilon)\equiv\mathcal{S}[g_{\mu\nu}(\varepsilon);\mathcal{V}]$.

Let us apply the above procedure in the case of the Einstein-Hilbert action, while also disregarding the proportionality constant $M_P^2/2$ for the time being. For the volume element we trivially obtain
\begin{equation}
    \delta\,(\mathrm{d}\text{vol})=\frac{1}{2}\,\mathrm{d}^4x\,\sqrt{-g}\,g^{\mu\nu}\delta g_{\mu\nu}=-\frac{1}{2}\mathrm{d}^4x\,\sqrt{-g}\,g_{\mu\nu}\delta g^{\mu\nu}\,.
\end{equation}
Therefore, by distributing the variation for the Ricci scalar as $\delta R=\delta g^{\mu\nu} R_{\mu\nu}+g^{\mu\nu}\delta R_{\mu\nu}$ we obtain:
\begin{equation}\label{Eq:EinHil-Variation}
    \frac{\delta\mathcal{S}_{EH}}{\delta g^{\mu\nu}}=\int_\mathcal{V}\!\mathrm{d}^4x\,\sqrt{-g}\left\{R_{\mu\nu}-\frac{1}{2}g_{\mu\nu}R\right\}\delta g^{\mu\nu}+\int_\mathcal{V}\!\mathrm{d}^4x\,\sqrt{-g}\,g^{\mu\nu}\delta R_{\mu\nu}\,.
\end{equation}

Let us ignore the second integral at this moment; we show later that it is a total derivative and does not contribute to the equations of motion of the system. If we also allow for the possibility of matter fields living on the gravitational background we can extend the total action describing the system by
\begin{equation}
    \mathcal{S}_\text{tot}=\mathcal{S}_{EH}+\mathcal{S}_m\,,
\end{equation}
where $\mathcal{S}_m[\phi,\text{g}]=\int\mathrm{d}^4x\,\sqrt{-g}\,\mathscr{L}_m[\phi,\text{g}]$ denotes the matter action over a configuration of fields denoted collectively by $\phi$. Likewise, we can write the variation of the matter action as
\begin{equation}
    \frac{\delta \mathcal{S}_m}{\delta g^{\mu\nu}}=\int_\mathcal{V}\!\mathrm{d}^4x\,\sqrt{-g}\left\{\frac{\partial\mathscr{L}_m}{\partial g^{\mu\nu}}-\frac{1}{2}g_{\mu\nu}\mathscr{L}_m\right\}\delta g^{\mu\nu}
\end{equation}
and define the energy-momentum tensor that describes the energy density and momentum distribution of the system as
\begin{equation}
    T_{\mu\nu}\equiv\frac{2}{\sqrt{-g}}\,\frac{\delta\mathcal{S}_m}{\delta g^{\mu\nu}}= \frac{\partial\mathscr{L}_m}{\partial g^{\mu\nu}}-\frac{1}{2}g_{\mu\nu}\mathscr{L}_m\,.
\end{equation}
Note that the tensor is symmetric by construction - a fundamental feature of matter fields. Finally, we are led to the Einstein field equations, which read
\begin{equation}\label{Eq:EinsteinFieldEqs}
    G_{\mu\nu}(\{\,\})\equiv R_{\mu\nu}-\frac{1}{2}g_{\mu\nu} R=T_{\mu\nu}\,,
\end{equation}
where we should point out that the Einstein tensor $G_{\mu\nu}$ is dependent entirely on the metric denoted by the short-hand notation $\{\,\}$ of the Christoffel symbols \eqref{Cond:ChristoffelSymbs}. Notice that by means of the Bianchi identity $\nabla^\mu G_{\mu\nu}=0$ the divergence of the energy-momentum tensor implicitly vanishes, $\nabla^\mu T_{\mu\nu}=0$. 

Equation~\eqref{Eq:EinsteinFieldEqs} is the central point in understanding the dynamics of different systems describing gravitational interactions. In order to do that we have to solve the system of differential equations, either by numerical methods or by imposing even more constraints on the allowed symmetry of the system at hand. Historically, by following the second way, cosmological solutions (FRW metric) were found that are capable of describing the dynamics of the cosmos and also black hole solutions (e.g. Schwarzschild metric). The case of cosmological solutions, specifically the FRW metric, plays a crucial role in understanding inflation and as such it is discussed in detail in a previous ch.~\ref{Ch2:Inflation}.

\subsubsection*{\emph{The York-Gibbons-Hawking term}}

Let us return to the second integral in the RHS of eq.~\eqref{Eq:EinHil-Variation}. We need the variation of the Ricci tensor, which after using the Palatini identity yields
\begin{equation}
    \delta R_{\mu\nu}=\nabla_\rho\left(\delta{\Gamma^\rho}_{\mu\nu}\right)-\nabla_\nu\left(\delta{\Gamma^\rho}_{\mu\rho}\right)\,.
\end{equation}
A way to see why this is the case is to think of a point $p\in\mathcal{M}$ where the Christoffel symbols vanish. The total variation of the connection coefficients reads
\begin{equation}
    \delta{\Gamma^\rho}_{\mu\nu}=\frac{1}{2}g^{\rho\sigma}\left(\nabla_\mu \delta g_{\sigma\nu}+\nabla_\nu \delta g_{\mu\sigma}-\nabla_\sigma \delta g_{\mu\nu}\right).
\end{equation}
Then, after some manipulation of the indices and using the metric-compatibility condition it is straightforward to show that
\begin{equation}
    \int_\mathcal{M}\!\mathrm{d}^4x\,\sqrt{-g}\,g^{\mu\nu}\delta R_{\mu\nu}=\int_\mathcal{M}\!\mathrm{d}^4x\,\sqrt{-g}\ \nabla_\mu\left(g^{\mu\nu}\,\delta{\Gamma^\rho}_{\rho\nu}-g^{\nu\sigma}\,\delta{\Gamma^\mu}_{\nu\sigma}\right)\equiv\int_\mathcal{M}\!\mathrm{d}^4x\,\sqrt{-g}\ \nabla_\mu V^\mu\,,
\end{equation}
where $V^\mu$ is a vector field in $\mathcal{M}$. By means of the Stokes theorem we can express the resulting integral as
\begin{equation}
    \int_\mathcal{M}\!\mathrm{d}^4x\,\sqrt{-g}\,\nabla_\mu V^\mu=\oint_\mathcal{\partial\mathcal{M}}\!\mathrm{d}\Sigma_\mu\,V^\mu=\oint_\mathcal{\partial\mathcal{M}}\!\epsilon\, V^\mu n_\mu\sqrt{|\text{det}(h)|}\,\mathrm{d}^3x'\,.
\end{equation}
We denote by $n_\mu$ the unit normal to $\partial\mathcal{M}$, and $g_{\mu\nu}=\epsilon n_\mu n_\nu+h_{\mu\nu}$ with $h_{\mu\nu}$ playing the role of the induced metric on the boundary. Here $\epsilon=n^\mu n_\mu$ is $+1$ ($-1$) if $\partial\mathcal{M}$ is timelike (spacelike). 

Next, we assume that the variation of the metric field vanishes on the boundary, meaning $\left.\delta g^{\mu\nu}\right|_{\partial\mathcal{M}}=0$, and therefore its covariant derivatives become simple derivatives. After some tedious manipulation of the indices one can show that the integral becomes
\begin{equation}
    \oint_\mathcal{\partial\mathcal{M}}\!\epsilon\, V^\mu n_\mu\sqrt{|\text{det}(h)|}\,\mathrm{d}^3x'=-\oint_{\partial\mathcal{M}}\!\epsilon\,h^{\mu\nu}\,(\partial_\rho\delta g_{\mu\nu}) n^\rho\,\sqrt{|\text{det}(h)|}\,\mathrm{d}^3x'\,.
\end{equation}
Therefore, in order to properly define the compact variation of the Einstein-Hilbert action the boundary term, known as the YGH term~\cite{York1972,Gibbons1977} has to be added, such that this contribution is exactly canceled. It is not difficult to see that the above term is the variation of $\delta(\nabla_\mu n^\mu)$, and thus the desired term reads
\begin{equation}
    \mathcal{S}_{YGH}=\oint_{\partial\mathcal{M}}\!\mathrm{d}^3x\,\sqrt{|\text{det}(h)|}\,\epsilon\,K=\oint_{\partial\mathcal{M}}\!\mathrm{d}^3x\,\sqrt{|\text{det}(h)|}\,\epsilon\,\nabla_\mu n^\mu\,,
\end{equation}
where $K$ is the trace of the intrinsic curvature.

Even though the addition of this term has incredible physical significance we are not interested in its application in this work. Furthermore, it will become apparent that in its first-order formulation the YGH term is redundant with respect to the completeness of the variational principle.

\section{Palatini variation}

In the first-order or Palatini formulation of gravity the metric and the connection do not have an a priori dependence on one another. In other words, the connection coefficients ${\Gamma^\rho}_{\mu\nu}$ are not necessarily given by the Levi-Civita condition \eqref{Cond:ChristoffelSymbs}, but instead they are \emph{dynamically} obtained at the level of the equations of motion, since now the action includes two dynamical degrees of freedom, namely the metric field and the connection. This particular idea falls under the class of \emph{metric-affine spaces}, in which unlike other generalised connections, they allow for metric-compatibility and torsion. 

\subsection{\emph{Analogy with tree-level Yang-Mills theory}}

Before discussing further the concept of metric-affine spaces, let us provide a brief comparison with the first-order formulation of a Yang-Mills (YM) theory. In its first-order formulation, the Lagrangian for a free $SU(2)$ Yang-Mills field reads
\begin{equation}\label{Eq:1YMLagr}
    \mathscr{L}_{YM}=\frac{1}{4}{F^a}_{\mu\nu}F^{a\mu\nu}-\frac{1}{2}{F^a}_{\mu\nu}\left(\partial^\mu A^{a\nu}-\partial^\nu A^{a\mu}+g f^{abc}A^{b\mu}A^{c\nu}\right)\,,
\end{equation}
where at this point the fields $A^a$ and $F^a$ are assumed to be completely independent and dynamical variables of the system. Here, $g$ is the coupling constant and $f^{abc}$ are the structure constants. Then, variation of the Lagrangian gives rise to the following equations of motion
\begin{equation}
    \partial_\mu F^{a\mu\nu}=0\,,
\end{equation}
\begin{equation}
    {F^a}_{\mu\nu}=\partial_\mu {A^a}_\nu-\partial_\nu {A^a}_\mu+g f^{abc}{A^b}_\mu {A^c}_\nu\,.
\end{equation}
Combining the two equations we are led to the same equation that describes the same classical dynamics with its usual (second-order) formulation, on-shell. Also, both formulations are invariant under local gauge transformations of the form:
\begin{equation}
    \delta {F^a}_{\mu\nu}=g f^{abc}{F^b}_{\mu\nu}\,\theta^c\,,\qquad\qquad \delta {A^a}_\mu=\partial_\mu\theta^a+g f^{abc}{A^b}_\mu\,\theta^c\,.
\end{equation}

Notice the parallels between the YM theory and GR, in which the role of $A^a$ is played by the metric tensor $\text{g}$ and the one of the field-strength $F^a$ by the connection $\Gamma$. In the case of the interacting YM theory we have to make further assumptions, e.g. that the interacting Lagrangian is explicitly independent of $\partial A^a$ such that the Noether current is unaffected. Equivalently, by considering interactions between (Palatini) gravity and matter fields we are led to assumptions regarding the underlying geometry of that theory as well.

There is a point to be made regarding the path integral quantisation of these two formulations, specifically for the case of the Yang-Mills field. The form of the Lagrangian \eqref{Eq:1YMLagr} suggests that we should expect two propagators $(AA)$ and $(FF)$, alongside a mixed propagator $(AF)$ and a vertex $\braket{AAF}$. However, if we perform a shift of
\begin{equation}
    {F^a}_{\mu\nu}\longrightarrow {F^a}_{\mu\nu}+(\partial_\mu {A^a}_\nu-\partial_\nu {A^a}_\mu+g f^{abc}{A^b}_\mu {A^c}_\nu)\,,
\end{equation}
the path integral becomes, neglecting the ghost-fixing terms,~\cite{Brandt2016,Brandt2019,Brandt2021}
\begin{equation}
    \int\![\mathcal{D}{A^a}_\mu][\mathcal{D}{F^a}_{\mu\nu}]\,\text{exp}\left(i\int\!\mathrm{d}^dx\left(\frac{1}{4}{F^a}_{\mu\nu}F^{a\mu\nu}+\mathscr{L}_{YM}^{\text{(II)}}[A^a]\right)\right)\,,
\end{equation}
where $\mathscr{L}_{YM}^{\text{(II)}}$ denotes the second-order formulation implicitly dependent solely on $A^a$. 

Then, the integral over $F^2$ decouples, recovering the known path integral for the Yang-Mills theory. In this case, we obtain two propagators of $(AA)$ and $(FF)$ alongside two vertices $\braket{AAA}$ and $\braket{AFF}$, but there is not a mixed propagator of the vector potential and the field strength. Thus, by performing a different kind of shift we can in principle exploit the fact that we can ``change'' the Feynman rules in YM, and apply it in GR. This subject is beyond the scope of this work, however some aspects of path integral quantisation will be raised in a following section.

\subsection{\emph{Metric-affine spaces}}
\label{subsec:MetAffineSpace}

A \emph{metric-affine space} is defined by $(\mathcal{M},\text{g},\nabla)$, where $\mathcal{M}$ is a differentiable manifold with a metric $\text{g}$ and a linear connection $\nabla$ (e.g. see ref.~\cite{Hehl1995}).\footnote{The notation can be potentially confusing; in fiber bundle theory $\nabla$ represents the covariant derivative while $\Gamma$ denotes the connection $1$-form. Keeping with relativistic physics notation, we regularly refer to the connection coefficients ${\Gamma^\rho}_{\mu\nu}$ as the connection, and not its map $\nabla$. We rely to context in order to resolve any further misunderstanding caused by this abuse of notation.} Formally, the affine connection is defined as the map~\cite{ChoquetBruhat1982}
\begin{equation}
    \nabla: \Gamma(T\mathcal{M})\times\Gamma(T\mathcal{M})\rightarrow\Gamma(T\mathcal{M})\,,
\end{equation}
where $\Gamma(T\mathcal{M})$ denotes the collection of vector fields on $\mathcal{M}$. The idea is to connect information about tensor fields evaluated at one point $p\in\mathcal{M}$ with their values at some other point, say $q\in\mathcal{M}$.  At any point $p\in\mathcal{M}$ we can define the connection coefficients on the tangent bundle $T_p\mathcal{M}$ in a coordinate basis ${e_\mu}$ as follows
\begin{equation}
    \nabla_{e_\mu}e_\nu={\Gamma^\rho}_{\mu\nu}e_\rho\,,
\end{equation}
where ${\Gamma^\rho}_{\mu\nu}$ are the components of the connection or \emph{connection coefficients}. To relieve some of the notation we adopt the usual notation of $\nabla_\mu\equiv\nabla_{\epsilon_\mu}$.

The connection satisfies the following conditions
\begin{align}
    \nabla_X(Y+Z)&=\nabla_XY+\nabla_XZ\,,\\
    \nabla_{(fX+hY)}Z&=f\nabla_XZ+h\nabla_YZ\,,\\
    \nabla_X(fY)&=f\nabla_XY+(\nabla_Xf)Y\,,
\end{align}
where $X,Y,Z$ are vectors on $\mathcal{M}$ and $f,h$ are smooth real-valued functions. To touch base with the usual definition of the connection in GR, we apply the covariant derivative as:
\begin{equation}
    \nabla_XY=\nabla_X(Y^\mu\epsilon_\mu)=X^\nu\epsilon_\nu\,Y^\mu\epsilon_\mu+X^\nu Y^\mu\nabla_\nu\epsilon_\mu=X^\nu\left(\epsilon_\nu Y^\mu+{\Gamma^\mu}_{\rho\nu}Y^\rho\right)\epsilon_\mu\,.
\end{equation}
Since $X^\nu$ can be factored out we denote $\nabla_XY=\nabla_\nu Y$. Finally, if we assume a coordinates basis with ${\epsilon_\mu}={\partial_\mu}$ we can write it in components:
\begin{equation}
    \nabla_\nu Y^\mu=\partial_\nu Y^\mu+{\Gamma^\mu}_{\rho\nu}Y^\rho\,.
\end{equation}
The action of the covariant derivative can be unambiguously generalised to tensor fields of general rank $(p,q)$ in the following way
\begin{align}
    \nabla_\rho\,{T^{\mu_1\mu_2...\mu_p}}_{\nu_1\nu_2...\nu_q}&=\partial_\rho\,{T^{\mu_1\mu_2...\mu_p}}_{\nu_1\nu_2...\nu_q}+\Gamma^{\mu_1}_{\rho\sigma}{T^{\sigma\mu_2...\mu_p}}_{\nu_1\nu_2...\nu_q}+\ldots+\Gamma^{\mu_p}_{\rho\sigma}{T^{\mu_1\mu_2...\mu_{p-1}\sigma}}_{\nu_1\nu_2...\nu_q}+\nonumber\\
    &\qquad-\Gamma^{\sigma}_{\rho{\nu_1}}{T^{\mu_1\mu_2...\mu_p}}_{\sigma\nu_2...\nu_q}-\ldots-\Gamma^{\sigma}_{\rho{\nu_q}}{T^{\mu_1\mu_2...\mu_p}}_{\nu_1\nu_2...\nu_{q-1}\nu_q}\,.
\end{align}
For every upstairs index we get a contribution of $\Gamma T$, while for a downstairs one we obtain $-\Gamma T$, with the appropriate index contraction.

\subsubsection*{\emph{Torsion}}
It is straightforward to show that the connection is not in fact a tensor itself. However, two other tensors can be defined from it. One of them is a rank $(1,2)$ tensor, the \emph{torsion tensor}, defined through its action on $X,Y\in\Gamma(T\mathcal{M})$ and $1$-form $\omega\in\Lambda^1(\mathcal{M})$ as~\cite{Nakahara:2003nw}
\begin{equation}
    T(\omega;X,Y)=\omega\left(\nabla_XY-\nabla_YX-[X,Y]\right)\,.
\end{equation}
In coordinate basis ${e_\mu}={\partial_\mu}$ and dual basis ${a^\mu}={\mathrm{d}x^\mu}$ we can rewrite it as:
\begin{equation}
    {T^\rho}_{\mu\nu}=a^\rho\left(\nabla_\mu\epsilon_\nu-\nabla_\nu\epsilon_\mu-[\epsilon_\mu,\epsilon_\nu]\right)={\Gamma^\rho}_{\mu\nu}-{\Gamma^\rho}_{\nu\mu}=2\,{\Gamma^\rho}_{[\mu\nu]}\,,
\end{equation}
where we used $[\partial_\mu,\partial_\nu]=0$. The torsion tensor is \emph{manifestly antisymmetric} in its lower indices and manifolds that are torsion-free, i.e. the torsion tensor vanishes, admit an additional symmetry of the connection, namely ${\Gamma^\rho}_{[\mu\nu]}=0$. However, it should be stressed that this potential symmetry relies on the condition that the commutator of the basis vectors on $T_p\mathcal{M}$ vanishes.

\begin{figure}[H]
    \centering
    \includegraphics{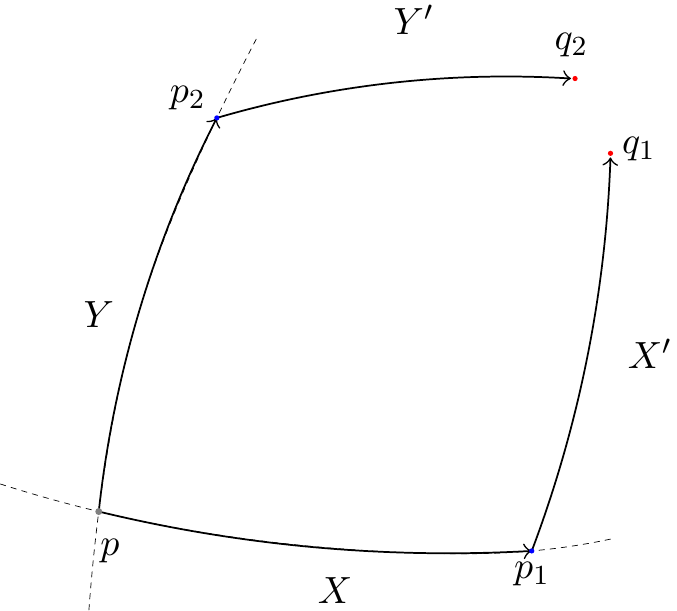}
    \caption{The infinitesimal parallelogram formed by parallel transporting vectors $X$ and $Y$ along each other, and its non-closure due to nonvanishing torsion.}
    \label{fig:torsion_parallelogram}
\end{figure}

There is a simple example that further illustrates the significance of torsion on the manifold. Suppose two vectors $X,Y\in T_p\mathcal{M}$ with coordinates $x^\mu$; then we can write $X=X^\mu\partial_\mu$ and $Y=Y^\mu\partial_\mu$. Starting from some point $p\in\mathcal{M}$ we can construct two other points very close to $p$, say $p_1$ and $p_2$, using the vectors $X,Y$; these are:
\begin{equation}
    (p_1):\ x^\mu+X^\mu\varepsilon\,,\qquad \qquad (p_2):\ x^\mu+Y^\mu\varepsilon\,,
\end{equation}
where $\varepsilon$ is an infinitesimal parameter. Next, we parallel transport the vectors with respect to each other, forming new vectors with components
\begin{equation}
    X'=\left(X^\mu-\varepsilon{\Gamma^\mu}_{\nu\rho}Y^\nu X^\rho\right)\partial_\mu\,,\qquad Y'=\left(Y^\mu-\varepsilon{\Gamma^\mu}_{\mu\rho}X^\nu Y^\rho\right)\partial_\mu\,.
\end{equation}
These new vectors form two new points, $q_1$ and $q_2$ respectively; their coordinates are
\begin{align}
    (q_1):\ &x^\mu+(X^\mu+Y^\mu)\varepsilon-\varepsilon^2{\Gamma^\mu}_{\nu\rho} Y^\nu X^\rho\,,\\
    (q_2):\ &x^\mu+(X^\mu+Y^\mu)\varepsilon-\varepsilon^2{\Gamma^\mu}_{\nu\rho} X^\nu Y^\rho\,.
\end{align}
Then we can see that the difference of the two resulting vectors is
\begin{equation}
    \propto\left({T^\rho}_{\mu\nu}\right)X^\mu Y^\nu\,.
\end{equation}
Therefore, we can think of the torsion tensor as a ``measurement'' of the (in)ability of the (infinitesimal) parallelogram to close. This point is further illustrated in fig.~\ref{fig:torsion_parallelogram}.

Another subtle point is that if we include matter on the manifold, which is the main goal of this work, the covariant derivatives do not commute; especially for a scalar field
\begin{equation}
    \nabla_{[\mu}\nabla_{\nu]}\phi\propto {T^\rho}_{\mu\nu}\nabla_\rho\,\phi\,.
\end{equation}

There has been a large amount of work on theories based on torsion. Historically, torsion was first considered to be directly sourced by spin. This was first seen in the works of Cartan and later by Kibble~\cite{Kibble1961} and Sciamma~\cite{RevModPhys.36.463}. The Einstein-Cartan-Kibble-Sciamma theory~\cite{RevModPhys.48.393} is one of the most fundamentally sound attempt to include spin-matter fields. Since then, more works have been put forth starting as modifications to what is known as the \emph{teleparallel equivalent} of GR\footnote{This is an equivalent description of GR formulated solely by the trace of the torsion tensor, $\int_\mathcal{M}\! T\,\mathrm{d}\text{vol}$.}~\cite{Maluf2013}, like the $f(T)$ gravity (see ref.~\cite{Cai2016} and references therein), in which $T$ is the trace of the torsion tensor assuming the role of the gravitational field. These works are primarily focused on cosmology and as such remain classical in nature, but provide a meaningful test of these theories against the best available cosmological data with the added possibility of falsifying them in the future. In these works (and others related also to nonmetricity) we frequently meet the torsion tensor in linear combinations, such as the \emph{contorsion tensor} defined by
\begin{equation}
    {K_{\mu\nu}}^\rho\equiv T{}_\mu{}^\rho{}_\nu-{T^\rho}_{\mu\nu}-{T_{\mu\nu}}^\rho\,,
\end{equation}
or through some of its contractions or trace. The subject of torsion is impressively deep, along with its applications, however we refrain from discussing it further, after all in the next chapter (ch.~\ref{Ch4:QuadGrav}) where we apply the same formalism to inflation, we manage to circumvent the issue of torsion altogether.

\subsubsection*{\emph{(Non)Metricity}}

Following what we have discussed in previous sections, one can impose another condition known as the \emph{metricity} or \emph{metric-compatibility condition}, requiring that the affine connection ``covariantly preserves'' the metric
\begin{equation}
    \nabla_X(\text{g})=0\quad\implies\quad \partial_\rho g_{\mu\nu}-{\Gamma^\sigma}_{\rho\mu}g_{\sigma\nu}-{\Gamma^\sigma}_{\rho\nu}g_{\mu\sigma}=0\,.
\end{equation}
That is the sole condition imposed on the affine connection in the original works of Cartan. It ensures that the inner product of two vectors is invariant under parallel transport over any curve, which is also a consequence of the EEP.

Since the connection is a tensor it must obey the Leibniz rule (linearity) and therefore it satisfies
\begin{equation}
    \nabla\left(X\otimes Y\right)=(\nabla X)\otimes Y+X\otimes(\nabla Y)\,.
\end{equation}
Then, we can also say $\nabla_\mu X^\nu=(\nabla X){}_\mu{}^\nu$ and consider $\nabla X$ as a rank $(1,1)$ tensor. Therefore we have,
\begin{equation}
    g_{\rho\nu}\nabla_\mu X^\nu=g_{\rho\nu}(\nabla X){}_\mu{}^\rho=(\nabla X)_{\mu\nu}=\nabla_\mu X_\nu\,,
\end{equation}
but also,
\begin{equation}
    \nabla_\mu X_\nu=\nabla_\mu(g_{\rho\nu}X^\rho)=g_{\rho\nu}\nabla_\mu X^\rho+(\nabla_\mu g_{\rho\nu})X^\rho\,.
\end{equation}
Clearly now for this to be consistent we must require that the \emph{nonmetricity tensor} vanishes, namely $Q_{\rho\mu\nu}\equiv\nabla_\rho g_{\mu\nu}=0$. In fact, this is another way of applying the metricity condition, since we assume that the metric is able to relate covectors and contravectors. Further than that, in a topological space we can have more than one connection but only one of them is going to be compatible with the metric if we also require ${\Gamma^\rho}_{[\mu\nu]}=0$.

\begin{figure}[H]
    \centering
    \includegraphics{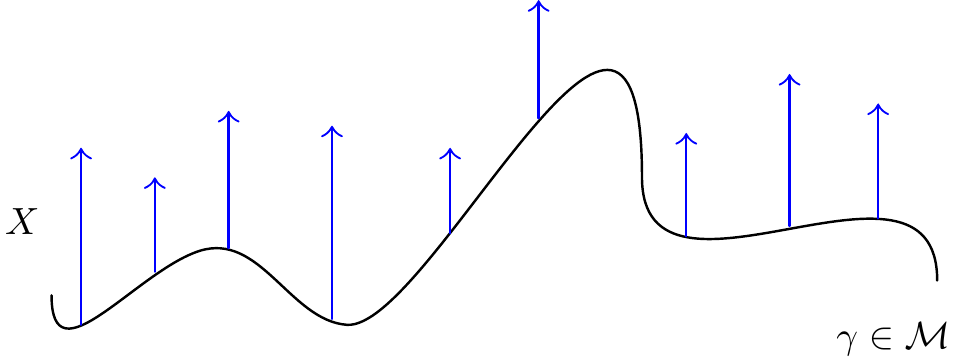}
    \caption{Variation of the length of a vector $X$ as is transported along a path, due to the nonvanishing of the nonmetricity tensor.}
    \label{fig:nonmetricity-vectors}
\end{figure}

Similar to torsion, there are (classical) theories being developed that attempt to describe gravitational interactions using only the nonmetricity tensor.\footnote{Once again, in complete analogy with the case of the torsional description of GR, there exists an equivalent description of GR utilising (contractions of) the nonmetricity tensor, $\int_\mathcal{M}\!Q\,\mathrm{d}\text{vol}$.} The studies are also focused mainly on cosmology and testing the theories against observational data~\cite{Lazkoz2019}. A general class of them is dubbed $f(Q)$ gravity theories~\cite{Mandal2020} and is described by a general function of the trace of the nonmetricity tensor $Q$, which is given by
\begin{equation}
    Q=-g^{\mu\nu}\left({L^\rho}_{\sigma\mu}{L^\sigma}_{\nu\rho}-{L^\rho}_{\sigma\rho}{L^\sigma}_{\mu\nu}\right)\,,
\end{equation}
where
\begin{equation}
    {L^\rho}_{\mu\nu}\equiv \frac{1}{2}g^{\rho\sigma}\left(Q_{\sigma\mu\nu}-Q_{\mu\sigma\nu}-Q_{\nu\sigma\mu}\right)
\end{equation}
defines the \emph{disformation}. In these theories the torsion tensor and the curvature are ``turned off'', which defines the covariant derivative on $\mathcal{M}$. This is usually done by implementing Lagrange multipliers in the theory.

\subsubsection*{\emph{Curvature: Riemann tensor and its contractions}}

For the sake of completeness let us include the other tensor we can build out of the connection. That is a rank $(1,3)$ tensor that we have already briefly covered, the \emph{curvature tensor}, known also as the Riemann tensor defined as the action on $X,Y,Z\in\Gamma(T\mathcal{M})$ and $\omega\in\Lambda^1(\mathcal{M})$
\begin{equation}
    R(\omega;X,Y,Z)=\omega\left(\nabla_X\nabla_Y Z-\nabla_Y\nabla_X Z-\nabla_{[X,Y]}Z\right)\,.
\end{equation}
In components it reads ${R^\rho}_{\sigma\mu\nu}=R[a^\rho;\epsilon_\sigma,\epsilon_\mu,\epsilon_\nu]$, and with $\epsilon_\mu=\partial_\mu$ and $a^\mu=\mathrm{d}x^\mu$ it is expressed as
\begin{equation}
    {R^\rho}_{\sigma\mu\nu}=\partial_\mu{\Gamma^\rho}_{\nu\sigma}-\partial_\nu{\Gamma^\rho}_{\mu\sigma}+{\Gamma^\lambda}_{\nu\sigma}{\Gamma^\rho}_{\mu\lambda}-{\Gamma^\lambda}_{\mu\sigma}{\Gamma^\rho}_{\nu\lambda}\,,
\end{equation}
where once again $[\partial_\mu,\partial_\nu]=0$ was used. Notice that the Riemann tensor is \emph{antisymmetric} in its last two indices ${R^\rho}_{\sigma(\mu\nu)}=0$ if the torsion tensor vanishes.

From the Riemann tensor we can build the Ricci tensor by contraction, namely
\begin{equation}
    {R^\rho}_{\mu\rho\nu}\equiv R_{\mu\nu}\,,
\end{equation}
with $R_{[\mu\nu]}=0$ again attributed to the vanishing of the torsion tensor. Next, by contracting the last two indices with the metric tensor the scalar curvature on the manifold is formed, known also as the \emph{Ricci scalar} - given by $R\equiv g^{\mu\nu}R_{\mu\nu}$ - which takes the spotlight in GR and has an essential role in the theories proposed to modify GR; discussed in detail in the next chapter.

If we allow for a general transformation of the form ${\Gamma^\rho}_{\mu\nu}\mapsto{\Gamma^\rho}_{\mu\nu}+\hat{\Gamma}{}^\rho_{\mu\nu}$ that relate two different (metric-affine) connections, the Riemann tensor transforms as
\begin{equation}
    {R^\rho}_{\sigma\mu\nu}(\Gamma)\,\longrightarrow\,{R^\rho}_{\sigma\mu\nu}(\Gamma)+2\,\hat{\Gamma}{}^\rho{}_{[\mu|\lambda}\hat{\Gamma}{}^\lambda{}_{\nu]\sigma}+2\,\nabla_{[\mu}\hat{\Gamma}{}^\rho{}_{\nu]\sigma}+{T^\lambda}_{\mu\nu}\hat{\Gamma}{}^\rho{}_{\lambda\sigma}\,.
\end{equation}
Likewise, the torsion and nonmetricity tensors become
\begin{align}
    {T^\rho}_{\mu\nu}(\Gamma)&\,\longrightarrow\,{T^\rho}_{\mu\nu}(\Gamma)+2\,\hat{\Gamma}{}^\rho{}_{[\mu\nu]}\,,\\
Q_{\rho\mu\nu}(\text{g},\Gamma)&\,\longrightarrow\, Q_{\rho\mu\nu}(\text{g},\Gamma)-2\,\hat{\Gamma}_{(\mu|\rho|\nu)}\,.
\end{align}

\subsubsection*{\emph{A menagerie of geometrical spaces and theories}}

\begin{figure}[H]
    \centering
\begin{tikzpicture}[
  ->,
  >=stealth',
  shorten >=1pt,
  auto,
  node distance=4cm,
  semithick,
  every state/.style = {drop shadow, shape=rectangle, rounded corners,
    draw, align=center,
    top color=white, bottom color=blue!30},
]
  \node[state]         (W)                     {Weyl spaces};
  \node[drop shadow,shape=rectangle,rounded corners,draw, align=center, top color=white, bottom color=purple!40]         (M) [above right of=W]  {Metric-affine spaces\\ $(\mathcal{M},\text{g},\nabla)$};
  \node[state]         (RC) [below right of=M]  {Riemann-Cartan spaces};
  \node[drop shadow,shape=rectangle,rounded corners,draw, align=center, top color=white, bottom color=purple!40]         (R) [below right of=W]  {Riemann spaces\\ $(\mathcal{M},\text{g})$};
  \node[state]          (Wz) [below right of=RC] {Weitzenb{\"o}ck spaces};
  \node[state]         (E)  [below left of=W]   {Euclidean spaces};

  \path[every node/.style={sloped,anchor=south}]
        (M) edge              node {$T=0$} (W)            
        (M) edge              node {$Q=0$} (RC)
        (W) edge                 node {$Q=0$} (R)
        (RC) edge              node {$T=0$} (R)
        (M) edge              node {} (R)
        (RC) edge[bend left]              node {$R=0$} (Wz)
        (R) edge[bend right]              node {$R=0$} (E);
\end{tikzpicture}
    \caption{A diagrammatic representation of the connection between the geometrical spaces discussed in this chapter and their relation to the general metric-affine space.}
    \label{fig:Geom_Spaces}
\end{figure}
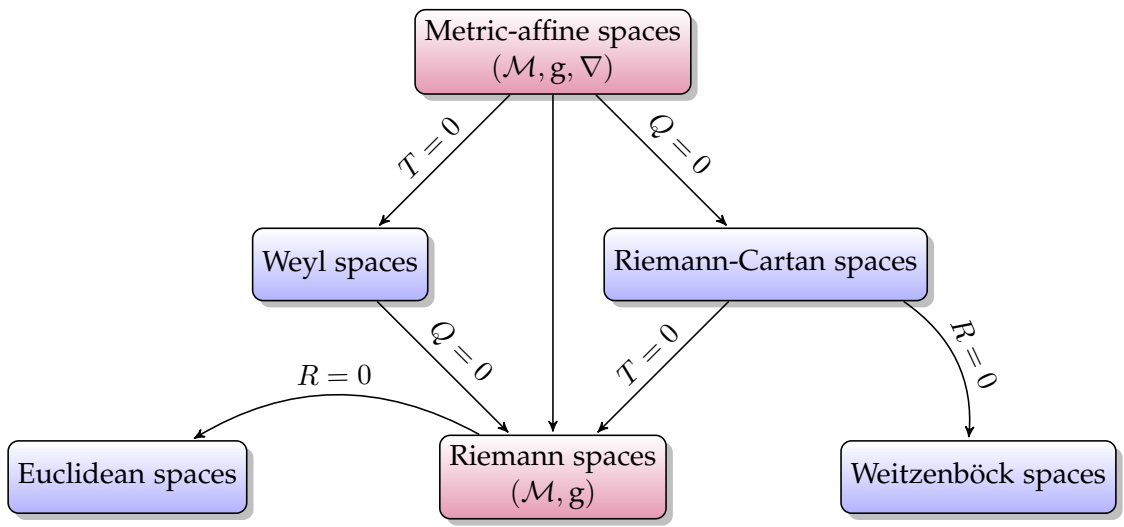

All of the above can muddle the perception of the underlying geometry of a gravitational theory. In fig.~\ref{fig:Geom_Spaces} we hope to address any misconceptions and illustrate how these spaces correspond to GR, at least diagrammatically. We start with the most general concept of a general \emph{metric-affine space} $(\mathcal{M},\text{g},\nabla)$ with a \emph{nonzero} torsion and nonmetricity tensor, $T\neq0\neq Q$. Historically, by assuming the metric-compatibility condition, $Q=0$, we are led to the (Riemann-)Cartan spaces. Even further, by switching off the torsional contribution GR is recovered, or (pseudo-)Riemannian spaces in general. On the other hand, the condition that $R=0$, alongside the condition of metric-compatibility, produces the Weitzenb{\"o}ck spaces, in which the original notion of \emph{teleparallism} was born by Einstein some years after his proposal of GR. A metric-affine space with vanishing torsion, $T=0$ is known as a Weyl space. Clearly, if we also impose that $Q=0$ we once more obtain a Riemann space. Finally, any Riemann space with zero curvature is a Euclidean or flat space.

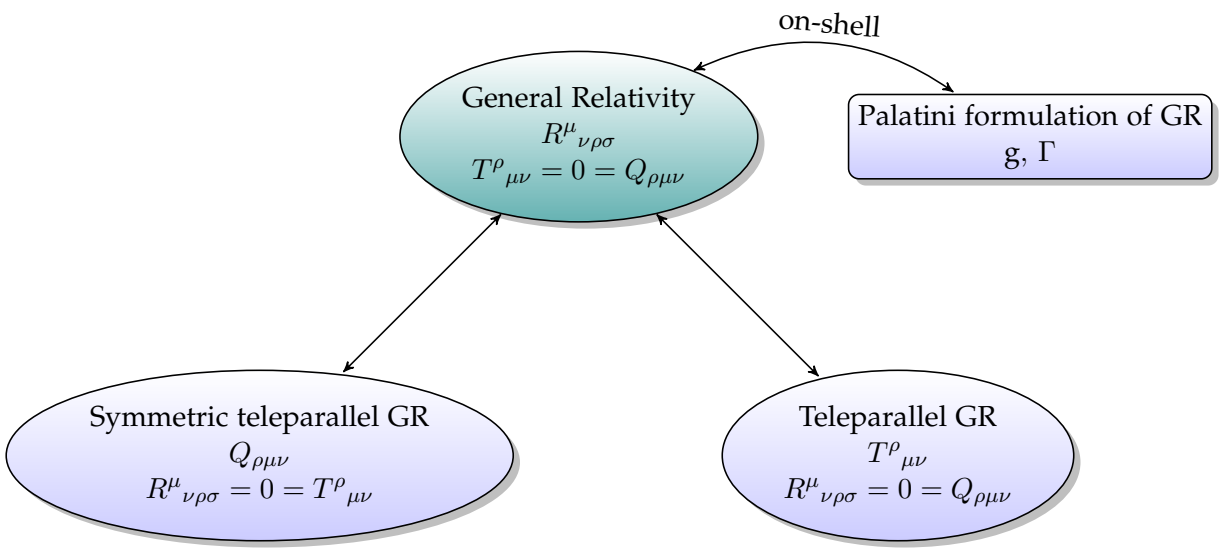
\begin{figure}[H]
    \centering
    \begin{tikzpicture}[
  <->,
  >=stealth',
  shorten >=1pt,
  auto,
  node distance=6cm,
  semithick,
  every state/.style = {drop shadow, shape=ellipse, rounded corners,
    draw, align=center,
    top color=white, bottom color=blue!20},
]

  \node[drop shadow,shape=ellipse,rounded corners,draw, align=center, top color=white, bottom color=teal!60]         (GR)                     {General Relativity \\ ${R^\mu}_{\nu\rho\sigma}$\\ ${T^\rho}_{\mu\nu}=0=Q_{\rho\mu\nu}$};
  \node[state]         (TGR) [below right of=GR]  {Teleparallel GR\\ ${T^\rho}_{\mu\nu}$\\ ${R^\mu}_{\nu\rho\sigma}=0=Q_{\rho\mu\nu}$};
  \node[state]         (STGR) [below left of=GR]  {Symmetric teleparallel GR\\ $Q_{\rho\mu\nu}$\\ ${R^\mu}_{\nu\rho\sigma}=0={T^\rho}_{\mu\nu}$};
  \node[drop shadow, shape=rectangle, rounded corners,
    draw, align=center,
    top color=white, bottom color=blue!20]  (PalGR) [right of=GR] {Palatini formulation of GR\\$\text{g},\,\Gamma$};

  \path[every node/.style={sloped,anchor=south}]
        (GR) edge              node {} (TGR)     
        (GR) edge[bend left]              node {on-shell} (PalGR)
        (GR) edge             node {} (STGR);
\end{tikzpicture}
    \caption{The equivalent representations of GR expressed solely in terms of torsion or nonmetricity. The first-order formulation of GR is not intrinsically an equivalent representation but a reparametrisation of the gravitational degrees of freedom. The above diagram holds only at tree-level.}
    \label{fig:Trio of EQ}
\end{figure}

As far as GR is concerned there exist \emph{equivalent representations} following the definitions given in fig.~\ref{fig:Geom_Spaces}. The equivalence is based on classical phenomena (currently), and holds only in vacuum, meaning no interactions, minimal or nonminimal, with other matter fields. In fig.~\ref{fig:Trio of EQ}, the equivalence between the different formulations is depicted by the arrows. Incredibly, all of them allow for a massless spin-$2$ field (graviton) respecting the equivalence principle. This topic is beyond the scope of this work and as such we refer the reader for more details to ref.~\cite{Jimenez2019} and references therein. It suffices to say that there is an extensive literature on generalisations of each of these equivalent descriptions considering modifications and extensions to them, similar to modifications of GR.

\subsection{\emph{Variational Principle \& dynamical generation of Levi-Civita connection}}

After being acquainted with the concept of a metric-affine connection, let us apply this idea to the Einstein-Hilbert action. In this case it reads as
\begin{equation}\label{Eq:EH1action}
    \mathcal{S}_{EH}[\text{g},\Gamma]=\frac{M_P^2}{2}\int\!\mathrm{d}^4x\,\sqrt{-g}\,g^{\mu\nu}\,R_{\mu\nu}(\Gamma)\,,
\end{equation}
where the scalar curvature is split into its metric-dependent part, $\sqrt{-g}\,g^{\mu\nu}$, and the one purely dependent on the connection (and its derivatives) $R_{\mu\nu}(\Gamma)$. The action contains \emph{two} (potentially) dynamical degrees of freedom, the metric and the connection, and thus we have to consider variation of the action with respect to both. 

Since the connection is not the Levi-Civita, the symmetries of the Riemann tensor do not hold in general. This means that there is actually multiple ways to construct the Ricci tensor. In fact, there are two ways to contract the indices of the Riemann tensor 
\begin{equation}
    R_{\mu\nu}={R^\rho}_{\mu\rho\nu}\qquad\text{and}\qquad \overline{R}_{\mu\nu}\equiv{R^\rho}_{\rho\mu\nu}\,.
\end{equation}
The first is the one used in conventional GR, defined as
\begin{equation}
    R_{\mu\nu}=2\,\partial_{[\rho}{\Gamma^\rho}_{\mu|\nu]}+2\,{\Gamma^\rho}_{\sigma[\rho}{\Gamma^\sigma}_{\mu|\nu]}\,,
\end{equation}
while the other one is given by
\begin{equation}
    \overline{R}_{\mu\nu}=2\,\partial_{[\mu}{\Gamma^\rho}_{\rho|\nu]}\,.
\end{equation}
Since the metric tensor is symmetric $g_{[\mu\nu]}=0$, the second Ricci tensor vanishes by contracting with the metric, i.e. $\overline{R}\equiv 2g^{(\mu\nu)}\overline{R}_{\mu\nu}=0$. Therefore, the Ricci scalar is uniquely constructed in this case as well, by $R=g^{\mu\nu}R_{\mu\nu}$. Obviously there are more contractions of the Riemann tensor, such as $R{}_\mu{}^\rho{}_{\rho\nu}$, but involve the metric tensor, and eventually lead to the same unique definition of the Ricci scalar~\cite{Sotiriou2007}.

\subsubsection*{\emph{Field equations}}

Firstly, let us consider the variation of the action \eqref{Eq:EH1action} with respect to the metric field. Since the Ricci tensor is independent of the metric we obtain:
\begin{equation}
    \delta_g\left(\sqrt{-g}\,g^{\mu\nu}R_{\mu\nu}\right)=\left\{(\delta_g\sqrt{-g})g^{\mu\nu}+\sqrt{-g}\,\delta_g (g^{\mu\nu})\right\}R_{\mu\nu}\,,
\end{equation}
where we used for brevity the shorthand notation $\delta_g\equiv\delta/\delta g^{\mu\nu}$. It is then straightforward to show that we obtain the field equations
\begin{equation}
    G_{\mu\nu}(\text{g},\Gamma)=R_{\mu\nu}(\Gamma)-\frac{1}{2}g_{\mu\nu}\,R(\text{g},\Gamma)=0\,.
\end{equation}
At this point, the Einstein tensor is dependent on the metric tensor and the connection $\Gamma$. Suppose that we include matter fields living on the background that are also explicitly independent of the connection; we can use the usual definitions to write
\begin{equation}
    G_{\mu\nu}(\text{g},\Gamma)=T_{\mu\nu}\,.
\end{equation}
Crucially, these equations \emph{resemble} the Einstein field equations, although they do not yet have the same description of gravity. The equations of motion for the connection are needed in order to derive the complete set of equations describing the physical system.

Next is the variation with respect to the connection. The only component in the action that depends on the connection is the Ricci tensor, so we have to evaluate $\delta_\Gamma R_{\mu\nu}(\Gamma)$. We drop the subscript $\Gamma$ in what follows to alleviate the notation, but we are only considering variation with respect to $\Gamma$. By definition $\delta{\Gamma^\rho}_{\mu\nu}$ is a tensor since it is a difference of the connection coefficients, and therefore we can write down its covariant derivative as
\begin{equation}
    \nabla_{\lambda}\,\delta{\Gamma^\rho}_{\mu\nu}=\partial_\lambda\,\delta{\Gamma^\rho}_{\mu\nu}+{\Gamma^\rho}_{\lambda\sigma}\,\delta{\Gamma^\sigma}_{\mu\nu}-{\Gamma^\sigma}_{\lambda\mu}\,\delta{\Gamma^\rho}_{\sigma\nu}-{\Gamma^\sigma}_{\lambda\nu}\,\delta{\Gamma^\rho}_{\mu\sigma}\,.
\end{equation}
Using this we can express the variation of the Ricci tensor as
\begin{equation}\label{Cond:DeltaRmn}
    \delta R_{\mu\nu}=\nabla_\rho\, \delta{\Gamma^\rho}_{\nu\mu}-\nabla_\nu\,\delta{\Gamma^\rho}_{\rho\mu}+{T^\lambda}_{\rho\nu}\,\delta{\Gamma^\rho}_{\lambda\mu}\,,
\end{equation}
known also as the (generalised) \emph{Palatini identity}. Notice the contribution of the torsion tensor. After some algebraic manipulation we can expand the first term in the RHS of eq.~\eqref{Cond:DeltaRmn} (analogously done for the second term) as
\begin{equation} 
    \int\!\mathrm{d}^4x\,\sqrt{-g}\,g^{\mu\nu}\,\nabla_\rho\,\delta{\Gamma^\rho}_{\nu\mu}=\int\!\mathrm{d}^4x\,\left(\sqrt{-g}\,g^{\mu\nu}\,{T^\lambda}_{\lambda\rho}-\nabla_\rho(\sqrt{-g}\,g^{\mu\nu})\right)\delta{\Gamma^\rho}_{\mu\nu}\,,
\end{equation}
up to vanishing boundary terms. We already have hints that the connection is the Levi-Civita. Returning to the complete variation of the Ricci tensor we obtain the following equation
\begin{equation}
    \nabla_\lambda\left(\sqrt{-g}\,g^{\nu\lambda}\right)\delta^\mu_\rho-\nabla_\rho\left(\sqrt{-g}\,g^{\mu\nu}\right)+\left(g^{\mu\nu}{T^\lambda}_{\lambda\rho}-g^{\nu\lambda}{T^\sigma}_{\sigma\lambda}\delta^\mu_\rho+g^{\nu\lambda}{T^\mu}_{\lambda\rho}\right)\sqrt{-g}=0\,.
\end{equation}
Next, by assuming a torsion-free manifold, in other words vanishing of the torsion tensor, and taking the trace of $\mu$ and $\rho$ in the equation we end up with
\begin{equation}
    \nabla_\rho\left(\sqrt{-g}\,g^{\mu\nu}\right)=0\,,
\end{equation}
which is the metric-compatibility condition. This is expressed in terms of components, so let us generalise the notion of metricity by assuming two vector fields $X,Y\in\mathcal{M}$ and a vector $V\in T_p\mathcal{M}$. Then a connection is said to be metric-compatible if
\begin{equation}
    V(g(X,Y))=g(\nabla_V X,Y)+g(X,\nabla_V Y)\,.
\end{equation}
The only connection that is metric-compatible as well as torsionless is the Levi-Civita connection with coefficients given by the Christoffel symbols. Its uniqueness is trivially proven if we write down the symmetry conditions for some vectors fields $X,Y,Z\in\mathcal{M}$
\begin{align}
    \nabla_X Y-\nabla_Y X&=[X,Y]\,,\nonumber\\
    \nabla_Y Z-\nabla_Z Y&=[Y,Z]\,,\\
    \nabla_Z X-\nabla_X Z&=[Z,X]\,,\nonumber
\end{align}
where brackets here denote the Lie bracket. Also the compatibility conditions
\begin{align}
    g(\nabla_X Y,Z)+g(Y,\nabla_X Z)=X g(Y,Z)\,,\nonumber\\
    g(\nabla_Y Z,X)+g(Z,\nabla_Y X)= Y g(Z,X)\,,\\
    g(\nabla_Z X, Y)+g(X,\nabla_Z Y)=Z g(X,Y)\,.\nonumber
\end{align}
Summing the first two, subtracting the last one and after using the symmetries above we obtain
\begin{equation}
    2g(\nabla_X Y,Z)=X g(Y,Z)+Yg(Z,X)-Zg(X,Y)+g(Z,[X,Y])+g(Y,[Z,X])+g(X,[Z,Y])\,,
\end{equation}
which in terms of components it leads to the Christoffel symbols given in eq.~\eqref{Cond:ChristoffelSymbs}.

Another way to obtain the Levi-Civita dynamically is to start by assuming a deviation of the general connection\footnote{In a general sense the connection coefficients can be decomposed in terms of the distortion and contorsion tensors as $${\Gamma^\rho}_{\mu\nu}=\left\{{}_\mu{}^\rho{}_\nu\right\}+{K^\rho}_{\mu\nu}+{L^\rho}_{\mu\nu}\,.$$} from the Levi-Civita one~\cite{Vitagliano2011}
\begin{equation}
    {\Gamma^\rho}_{\mu\nu}=\left\{{}_\mu{}^\rho{}_\nu\right\}+\overline{\Gamma}{}^\rho{}_{\mu\nu}
\end{equation}
and vary the action with respect to $\overline{\Gamma}$. The field equations for the metric tensor remain unchanged, however the equation for the connection becomes
\begin{equation}\label{Eq:TorsionGENEQ}
    \overline{\Gamma}{}^\nu{}^\rho{}_\rho\,\delta^\mu_\lambda+\overline{\Gamma}{}^\rho{}_{\rho\lambda}\,g^{\mu\nu}-2\,\overline{\Gamma}{}^\nu{}_\lambda{}^\mu=0\,.
\end{equation}
A (trivial) solution to the above equation is $\overline{\Gamma}=0$, meaning that $\Gamma=\{\ \}$ on-shell.

In any case, after considering the complete variation of the Einstein-Hilbert action with respect to both the dynamical variables we obtain two sets of equations, which after combining the solution of the connection constraint equation with the field equations lead to the correct partial differential equations. It is important to note that we did not need any boundary term to complete the Palatini variation, unlike the metric formulation in which the YGH term was introduced. 

In this way, the Levi-Civita connection on $\mathcal{M}$ is dynamically generated without making any assumptions at the level of the action~\cite{Palatini:1919}. In fact, if one is concerned particularly with gravitational dynamics the Levi-Civita condition is very crucial and is not at all incidental that it was initially considered as the connection. Even though the procedure of the general metric-affine connection seems attractive requiring less assumptions than its counterpart, it might as well be a mathematical framework under which we can deduce how to recover GR dynamics. In other words, the Levi-Civita condition is a sought-after outcome and if we do away with it there are serious ramifications. Some of them include the fact that the geodesic equation is not necessarily a resultant of the Euler-Lagrange equation, and the connection cannot be set locally to vanish ruining the equivalence principle~\cite{Sotiriou2007,Vitagliano2011}, which established our understanding of the gravitational interaction.

Even though the equivalence holds up to boundary terms for the Einstein-Hilbert action, the two approaches differ tremendously when one considers modified theories of gravity, specifically including higher-order curvature invariants and/or nonminimal couplings between gravity and the matter sector. Since the present work deals with inflation, the matter sector discussed here is constituted of a real scalar field and its self-interacting potential, in other words explicitly independent of the connection. However, even the simplest case is highly complicated, in which the Einstein-Hilbert term is coupled (minimally) to matter that is dependent on the connection. For example one has to define the tensor
\begin{equation}\label{Cond:GenHypermomentum}
    \Delta_{\rho}{}^{\mu\nu}\equiv-\frac{2}{\sqrt{-g}}\,\frac{\delta \mathcal{S}_m}{\delta\Gamma{}^\rho{}_{\mu\nu}}\,,
\end{equation}
where $\mathcal{S}_m$ denotes the matter action. It turns out that the form of $\Delta_\rho{}^{\mu\nu}$ is restricted at the level of the equations of motion, which in turn means that the matter Lagrangian has to be chosen such that it can satisfy these conditions~\cite{Vitagliano2011}, namely $\Delta_\mu{}^{\mu\nu}=0$. 

Suppose now that the matter action includes a (canonical) kinetic term for a scalar field and a potential term. Then we have $\nabla_\mu\phi=\partial_\mu\phi$ and thus the action is manifestly independent of the connection. Then, matter follows the geodesics predefined by a metric-compatible connection. Even more, the equation of the connection is ``demoted'' to a constraint, meaning that the connection is effectively an auxiliary field and does not carry any characteristics of the curvature. Therefore, the theory is relegated basically to a metric theory describing spacetime, just like GR. Let us clarify this point further by considering a simple example of a real scalar field $\phi(x)$ with a self-interacting potential $V(\phi)$ coupled minimally to gravity and described by the following Lagrangian
\begin{equation}
    \mathscr{L}_m=-\frac{1}{2}(\nabla\phi)^2-V(\phi)\,.
\end{equation}
Then, variation of the action with respect to $\phi$ leads to
\begin{equation}
    \frac{\delta\mathcal{S}}{\delta\phi}=\int_\mathcal{M}\!\mathrm{d}\text{vol}\left\{g^{\mu\nu}\nabla_\mu\nabla_\nu\phi-\frac{\partial V(\phi)}{\partial\phi}+\frac{1}{\sqrt{-g}}\nabla_\mu\left(\sqrt{-g}\,g^{\mu\nu}\right)\nabla_\nu\phi\right\}\delta\phi+\oint_{\partial\mathcal{M}}\!\mathrm{d}\Sigma\,n_\mu\left\{\sqrt{-g}\,g^{\mu\nu}\nabla_\nu\phi\right\}\delta\phi\,,
\end{equation}
where $n_\mu$ denotes the outward pointing unit vector normal to $\partial\mathcal{M}$ and $\mathrm{d}\Sigma$ the invariant volume on the boundary. Assuming that $\delta\phi=0$ on the boundary $\partial\mathcal{M}$, then the second integral vanishes and we are left with a deformed Klein-Gordon equation including a term that it ends up vanishing via the Levi-Civita condition on-shell. In what follows, these terms arising from the nonmetricity are assumed to be vanishing at the level of equations of motion (the connection is Levi-Civita at that point) and therefore are not included in the equations.

\subsubsection*{\emph{Projective invariance}}

As we eluded to earlier, eq.~\eqref{Eq:TorsionGENEQ} can have more general and involved solutions. It is straightforward to show that a general solution is
\begin{equation}
    {\overline{\Gamma}{}^{\rho}}_{\mu\nu}=\delta^\rho_\nu\, V_\mu\,,
\end{equation}
for an arbitrary vector field $V_\mu$. This is known in the literature as a \emph{projective} transformation
\begin{equation}
    \left\{{}_\mu{}^\rho{}_\nu\right\}\ \longrightarrow\ \left\{{}_\mu{}^\rho{}_\nu\right\}+\delta^\rho_\nu \,V_\mu\,,
\end{equation}
transforming the Ricci tensor as
\begin{equation}
    R_{\mu\nu}\ \longrightarrow\ R_{\mu\nu}-2\,\partial_{[\mu}V_{\nu]}\,.
\end{equation} 
Since the metric is symmetric the scalar curvature is invariant under the transformation. Moreover, one can show by using the definition of the torsion tensor that
\begin{equation}
    {T^\nu}_{\mu\nu}=3V_{\mu}\,,
\end{equation}
which should vanish on-shell to define the Levi-Civita connection. Then one can enforce this condition by employing different techniques, such as the implementation of Lagrange multipliers. However, in a way similar to the case of eq.~\eqref{Cond:GenHypermomentum}, by demanding the trace of the torsion tensor to vanish\footnote{Notice that only the trace of the tensor vanishes and not necessarily the torsion itself.} the type of connection allowed in the theory is restricted. The fact that the Einstein-Hilbert action is invariant under projective transformations but the matter action is not necessarily can lead to inconsistencies at the level of the equations of motion of the theory.

\subsection{\emph{Beyond tree-level equivalence}}

As noted in earlier sections and throughout the literature, the equivalence between the two approaches holds at the \emph{classical level}, as far as the Einstein-Hilbert action is concerned. It is possible then to think of $\text{g}$ and $\Gamma$ as independent quantities that in general propagate and are dynamical. At tree level we showed that the connection plays the role of an auxiliary field and as such is fixed by its constraint equation and does not propagate, however this does not hold necessarily at the quantum level. If the first-order formalism is to be considered something more than an academic game, the quantum nature of the connection has to be understood through its quantum effects, if any. In what follows, by recalling conventional approaches to analysing quantum (field) interactions, we attempt to outline the influence of the connection at the quantum level.

\subsubsection*{\emph{Path integral formulation}}

Directly from the action given in eq.~\eqref{Eq:EH1action}, there is an interaction between the dynamical quantities $\text{g}$ and $\Gamma$ that result in a vertex of $\braket{\text{g}\Gamma\Gamma}$ and a very complicated nondiagonal matrix propagator of $(\text{g}\text{g})$, $(\Gamma\Gamma)$ and $(\text{g}\Gamma)$, reminiscent of the first-order formulation of the Yang-Mills theory. It seems then that the familiar problem one runs into when attempting to quantise GR, namely the infinite series of momentum-dependent vertices, has been traded away for a new headache of involved mixed propagators. However, it was shown~\cite{Brandt2015} that by utilising a shift in the variables the contribution of the connection at one-loop is vanishing. In what follows, we sketch the results of refs.~\cite{Brandt2015,Brandt2016,Brandt2017,Brandt2021} in order to gain a better understanding of the contribution of $\Gamma$ at the (one-loop) quantum level.

Let us start by rephrasing the dynamical variables as~\cite{Brandt2015}
\begin{equation}
    \phi^{\mu\nu}\equiv\sqrt{-g}\,g^{\mu\nu}\,,\qquad {G^\rho}_{\mu\nu}\equiv {\Gamma^\rho}_{\mu\nu}-\delta{}^\rho{}_{(\mu}{\Gamma^\lambda}_{\nu)\lambda}\,.
\end{equation}
Then, the $d$-dimensional Lagrangian reads
\begin{equation}
    \mathscr{L}_{EH}=-{G^\rho}_{\mu\nu}\,\partial_\rho \phi^{\mu\nu}+\frac{1}{2}M{}^{\mu\nu}{}_\lambda{}^{\rho\sigma}{}_\tau {G^\lambda}_{\mu\nu}{G^\tau}_{\rho\sigma}\,,
\end{equation}
where 
\begin{equation}
    M{}^{\mu\nu}{}_\lambda{}^{\rho\sigma}{}_\tau(\phi)\equiv2\left(\frac{1}{d-1}\delta{}^{(\rho|}{}_\tau\delta{}^{(\mu}{}_\lambda\phi{}^{\nu)|\sigma)}-\delta{}^{(\rho|}{}_\lambda\delta{}^{(\mu}{}_\tau\phi{}^{\nu)|\sigma)}\right)\,.
\end{equation}
After performing a shift of the form
\begin{equation}
    {G^\lambda}_{\mu\nu}\longrightarrow{G^\lambda}_{\mu\nu}+(M^{-1}){}^\lambda{}_{\mu\nu}{}^\tau{}_{\rho\sigma}\partial_\tau \phi^{\rho\sigma}
\end{equation}
we may find that the path integral formulation of the action results in the following generating functional~\cite{Brandt2016}
\begin{equation}
    \mathcal{Z}=\int\![\mathcal{D}\phi(x)][\mathcal{D}G(x)]\Delta_{FP}(\phi)\, \text{exp}\left\{i\int\!\mathrm{d}^dx\left(\frac{1}{2}{G^\rho}_{\mu\nu}M{}^{\mu\nu}{}_\lambda{}^{\rho\sigma}{}_\tau{G^\tau}_{\rho\sigma}+\frac{1}{2}\partial_\lambda \phi^{\mu\nu} M{}^\lambda{}_{\mu\nu}{}^\tau{}_{\rho\sigma}\partial_\tau \phi^{\rho\sigma}+\mathscr{L}_\text{gf}\right)\right\}\,.
\end{equation}
Here $[\mathcal{D}\phi]$ and $[\mathcal{D}G]$ denote integration over all possible paths $\phi$ and $G$ respectively, and $\Delta_{FP}$ is the Faddeev-Popov determinant associated with the gauge fixing term $\mathscr{L}_\text{gf}$~\cite{Faddeev1967}. Notice that we chose to ignore an overall normalisation factor $1/N$. Let us assume that we can expand $\phi^{\mu\nu}(x)$ around a flat background of the form:
\begin{equation}
    \phi^{\mu\nu}(x)=\eta^{\mu\nu}+h^{\mu\nu}(x)\,.
\end{equation}
Since the matrix $M{}^{\mu\nu}{}_\lambda{}^{\rho\sigma}{}_\tau$ is linear in $\phi$ we can expand it as $M(\phi)=M(\eta)+M(h)$ meaning that contributing diagrams to the Green's function have the field ${G^\rho}_{\mu\nu}$ in closed loops with its momentum-independent propagator~\cite{Brandt2016}. In turn, this means that  the integral associated with loop contributions of ${G^\rho}_{\mu\nu}$ has the form
\begin{equation}
    \int\!\mathrm{d}^dk\,P(k^\mu)\,,
\end{equation}
where $P(k^\mu)$ is a polynomial in $k^\mu$. Assuming dimensional regularisation these integrals vanish and therefore the only contribution with $h^{\mu\nu}$ on external legs comes from the second term, which is the same as if we considered the usual metric formulation of the Einstein-Hilbert action.

Quantum effects in gravity is a sensitive and heavily debatable subject and as such it is unsure if the above analysis describes exactly the quantum nature of the field $\Gamma$ and its interaction with other fields. It is however an indication of its effect, at least at one-loop level and in complete vacuum, meaning no interaction with matter fields. Results are also presented using the background field expansion method, however in principle one can perform the same analysis using the Heat Kernel method~\cite{Gilkey1994}, which is more direct but more involved depending on the form of the action.

\subsubsection*{\emph{Hamiltonian analysis}}

Since the discovery of the ADM variables the idea to canonically quantise GR seemed feasible, at least at initial stages. The program, arguably, has since failed but the ADM decomposition is used still, proving its capability and potential. However, attempts to canonically quantise the Palatini action (Einstein-Hilbert action assuming the first-order formalism) has left the scientific community puzzled and is mainly attributed to the confusion around how to handle the arising second-class constraints. Different methods seem to focus on eliminating time-independent fields via equations of motion, although in the meantime a generator of the gauge transformations is lost~\cite{McKeon2010,Kiriushcheva2010}.

For transformations of $x^\mu\rightarrow x^\mu+\xi^\mu$ one can show that the metric transforms as
\begin{equation}
    \delta g^{\mu\nu}=2\,\nabla^{(\mu}\xi^{\nu)}\,,
\end{equation}
which is generally covariant since $\xi^\mu$ is a true vector. Similarly, for the Christoffel symbols we can directly show that
\begin{equation}
    \delta {\Gamma^\rho}_{\mu\nu}=(\partial_\lambda\xi^\rho){\Gamma^\lambda}_{\mu\nu}-\xi^\lambda\,\partial_\lambda{\Gamma^\rho}_{\mu\nu}-2\,{\Gamma^\rho}_{(\mu|\lambda}\partial_{\nu)}\xi^\lambda-\partial_\mu\partial_\nu\xi^\rho\,.
\end{equation}
Since there are second derivatives of the parameter $\propto\partial^2\xi$ in the transformation it suggests that the generators must have the same order of derivatives, meaning that tertiary constraints should exist. In fact, this point breaks the analogy between the first-order formulation of Electromagnetism and GR, since in the first case the variation of the field strength is zero and as such there is no increase in order of the gauge parameter, as pointed out in refs.~\cite{Kiriushcheva2011,Kiriushcheva2010,Kiriushcheva2012}. The subject of canonical quantisation of the Palatini action is still open with active research developing and using different methods. It seems however that the dynamical degrees of freedom in the theory are the same as in GR~\cite{Dadhich2012}, although the exact approach is still questionable.

\chapter{Quadratic gravity coupled to matter}
\label{Ch4:QuadGrav}

In the previous chapters specific ideas of inflation and the Palatini formalism were highlighted serving as the background to the main part of the thesis, presented in this chapter. The following sections include results from the merger of these two concepts.

When the first-order formalism was discussed in ch.~\ref{Ch3:FirstOrder}, the larger part of the discussion was devoted to its equivalence with the conventional metric formalism and as such the subject of more complicated actions was avoided, even though it was claimed at that point that these action functionals can generally lead to different predictions. This fact now takes the spotlight in the following section by comparing the famous Starobinsky model (or $R+R^2$) of inflation~\cite{Starobinsky1980} within its two formulations explicitly highlighting their inequivalence. It will become obvious that in the Palatini formalism the Starobinsky model does \emph{not} include an additional scalar mode as in its metric counterpart~\cite{Antoniadis2018,Enckell2019,Edery2019} and therefore it is incapable of describing inflation in that formulation. It is then necessary to couple the $R+R^2$ term with a matter sector that manifestly includes the inflaton field and its self-interacting potential. In doing so we noticed that different models of inflation that were initially disfavoured by observations, such as the quadratic model (see fig.~\ref{fig:Planck2018}), now in their first-order formulation (coupled to the Starobinsky term) are capable of leading to an adequate inflationary era. However, it should be noted that the same form of the action in the metric formalism can possibly provide us with inflation that is also within the observational bounds, however the analysis of these theories is considerably harder than their Palatini formulation, since it involves a higher-dimensional field space including fields that are able to contribute in ``driving'' inflation. In later sections we examine some prominent inflationary models that are also motivated by lower-energy particle physics and entertain various cases of minimal or nonminimal coupling with the gravitational sector. Primarily, our investigation is focused under which conditions, namely which region of the model parameter space, the observational bounds set by the Planck collaboration are satisfied suggesting that the model at hand is capable of providing a successful inflation in the Palatini formalism.

\section{Metric \& Palatini formulation of the Starobinsky model}

Most of our investigation revolves around the Starobinsky model, and as such it is necessary to cover some of its features in this section. In what follows, we provide some details that are identical in both formulations and later the analysis is divided in two sections discussing various effects in each of them. Let us first begin with the action describing the model, which reads as~\cite{Starobinsky1980}
\begin{equation}\label{Action:StarOrig}
    \mathcal{S}=\int\!\mathrm{d}^4x\,\sqrt{-g}\left\{\frac{M_P^2}{2}\,R+\frac{M_P^2}{12m^2}\,R^2\right\}\,,
\end{equation}
where $R$ is the Ricci scalar and $R^2\equiv R\cdot R=(g^{\mu\nu}R_{\mu\nu})(g^{\rho\sigma} R_{\rho\sigma})$. The parameter $m$ has mass-dimensions and can be identified with the inflaton mass (in the metric formulation). It will prove more convenient, especially when we examine the first-order formulation of the model, to define a new dimensionless parameter, referred to also as the \emph{Starobinsky parameter} (or constant), as
\begin{equation}
    \alpha\equiv\frac{M_P^2}{6m^2}\,.
\end{equation}
Notice how in the IR limit, in which $R\ll m^2$, the model is reduced to GR with a small term that can in principle be identified with the cosmological constant (after an enormous fine-tuning). However, when $R\sim m^2$ the second term can have important contribution.

In its original proposal the model predicted (and since then readily supports) an inflationary de Sitter expansion of the early universe. It was motivated by the idea that gravitational quantum corrections should play a role in the stages of the early universe, where curvature was assumed to be strong (strong gravity limit) and therefore higher-order curvature invariants should be included in the total action.\footnote{Since it was shown that the Einstein-Hilbert action was nonrenormalisable~\cite{Hooft1974,Goroff1985}, higher-derivative (of the metric) theories became alluring. After all, the Einstein gravity produces a graviton propagator that is nonrenormalisable $\propto k^{-2}$ at large $k^2$ and a Ricci scalar squared results in a renormalisable propagator $\propto k^{-4}$ at large $k^2$. However, it seems that even if we obtain a gravitational theory that is (even perturbatively) renormalisable, other fundamental properties of a quantum field theory might be sacrificed~\cite{Stelle1977,Stelle1978,Antoniadis1986}, such as unitarity of the theory - linked to the ability of understanding the theory in a probabilistic way - or relativistic invariance, as e.g. in Ho{\v r}ava gravity~\cite{Horava2009} in which Lorentz invariance emerges as an approximate symmetry at low energies and is violated at high energies. The issue of renormalisability is brought forward for its significance; in the present work we do not attempt to study features of what would be a quantum gravity, but mainly draw inspiration from relevant works.} At one-loop order the quantum corrected action involves operators of the form~\cite{Stelle1977}
\begin{equation}
    \propto R^2,\quad R^{\mu\nu}R_{\mu\nu},\quad R^{\mu\nu\rho\sigma}R_{\mu\nu\rho\sigma}\,,
\end{equation}
however, the last two include ghost fields\footnote{The notion of a ghost field was encountered in ch.~\ref{Ch3:FirstOrder} where the Faddeev-Popov ghosts~\cite{Faddeev1967} were briefly mentioned. These are usually referred to as ``good'' ghosts since they are included in a gauge theory to keep its gauge invariance, in contrast to the ``bad'' ghosts inducing unphysical states in a theory.} that have negative Dirac norm or energies unbounded from below, violating either unitarity or causality of the theory. As far as inflation is concerned, in this thesis we include only the $R^2$ term in the action and model predictions of the inflationary observables are calculated from the $R+R^2$ action, also referred to as \emph{quadratic gravity}.

%\subsubsection*{Scalar representation}

At this point it is not obvious how the model described by the action \eqref{Action:StarOrig} leads to the usual spin-$2$ graviton and an additional scalar mode. Let us then introduce an auxiliary scalar field $\chi$ and rephrase the original Lagrangian as follows
\begin{equation}\label{Action:StarScalRep}
    \mathcal{S}=\int\!\mathrm{d}^4x\,\sqrt{-g}\left\{\frac{1}{2}\left(M_P^2+2\alpha\chi^2\right)R-\frac{\alpha}{2}\chi^4\right\}\,.
\end{equation}
Then, variation of the action with respect to $\chi^2$ leads to its constraint equation $\chi^2=R$ and substitution of that back into the action reduces to the original action in eq.~\eqref{Action:StarOrig}. One can in general introduce the field as $\bar{\chi}\equiv\chi^2$, however we found that it is more convenient to apply tools of dimensional analysis in the action as presented in eq.~\eqref{Action:StarScalRep}.

This is known as the \emph{scalar representation} of the action \eqref{Action:StarOrig} and can be generalised for any function $f(R)$ of the scalar curvature in the following way.
\begin{equation}
    \int\!\mathrm{d}\text{vol}\,f(R)\ \longrightarrow\ \int\!\mathrm{d}\text{vol}\left\{f'(\chi)(R-\chi)+f(\chi)\right\}\equiv\int\!\mathrm{d}\text{vol}\left\{\Omega^2(\chi) R-V(\chi)\right\}\,,
\end{equation}
where it is assumed that $f''(R)\neq0$, $\forall R$ and in the last equality we used the definition of $\Omega^2(\chi)\equiv f'(\chi)$ and $V(\chi)\equiv f(\chi)-\chi f'(\chi)$. In fact, the Starobinsky model can be thought of as a specific case of the general class of $f(R)$ theories with $f(R)\propto R+R^2$. Another example one can immediately think of is an expansion of $f(R)$ around the mass parameter $m^2$, then one obtains the following theory containing higher powers of the scalar curvature $R$
\begin{equation}
    f(R)= \frac{M_P^2}{2}\,R+M_P^2\sum_{n=2}^\infty a_n\,m^2\left(\frac{R}{m^2}\right)^n\,,
\end{equation}
where $a_n$ denote the coefficients in the expansion. Clearly, there exist more intricate cases of $f(R)$ theories, however they should be accompanied by studies of possible pathologies they might be contained in each case. It is not as trivial as it first seems to generalise GR in the context of $f(R)$ extended theories and there is obviously the issue of motivation behind each case. For the rest of this work we confine ourselves only in the Starobinsky model and avoid discussing further different cases of $f(R)$ theories.

\subsection{\emph{Metric formalism}}

\begin{figure}
    \centering
    \includegraphics[scale=0.43]{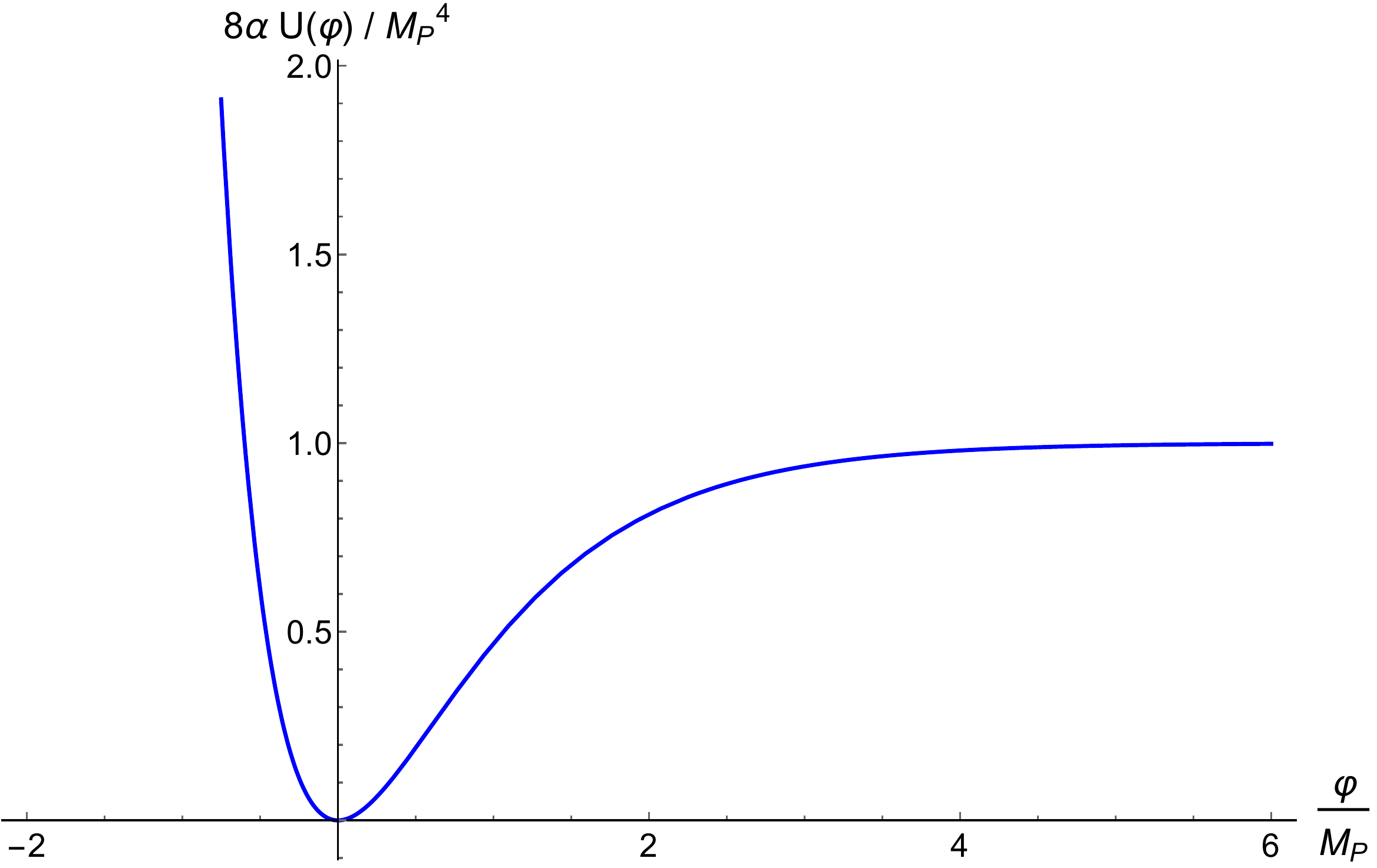}
    \caption{The form of the famous Starobinsky potential in the Einstein frame given by eq.~\eqref{Cond:OriginalStaroPot}. The axes have been rescaled in order to properly describe the qualitative behaviour of the potential without necessarily assuming any values for the model parameters. At the start of inflation the field assumes values in the flat region of the potential and eventually runs to smaller values ending up at the minimum of the potential.}
    \label{fig:OriginalStaroPot}
\end{figure}

Let us redirect our focus on the scalar representation of the Starobinsky model, given in eq.~\eqref{Action:StarScalRep}, and consider a Weyl rescaling of the metric as follows
\begin{equation}
    \bar{g}_{\mu\nu}(x)=\Omega^2(\chi)\, g_{\mu\nu}(x)\,,
\end{equation}
where now the conformal factor is defined as
\begin{equation}\label{Cond:OmegaMetStar}
    \Omega^2(\chi)\equiv\frac{M_P^2+2\alpha\chi}{M_P^2}\,.
\end{equation}
It is straightforward to show that the Christoffel symbols, given in eq.~\eqref{Cond:ChristoffelSymbs}, transform as
\begin{equation}
    \overline{\left\{{}_\mu{}^\rho{}_\nu\right\}}=\left\{{}_\mu{}^\rho{}_\nu\right\}+\Omega^{-1}\bigg(2\delta{}^\rho{}_{(\mu}\nabla_{\nu)}\Omega-g_{\mu\nu}\nabla^\rho\Omega\bigg)\,.
\end{equation}
After a lengthy calculation one can show that the Ricci scalar also transforms as
\begin{equation}
    R=\Omega^{2}\,\overline{R}+6\,\frac{\Omega^2}{\sqrt{-\overline{g}}}\,\overline{\nabla}_\mu\left(\sqrt{-\overline{g}}\,\overline{\nabla}^\mu\ln{\Omega}\right)-6\,\overline{g}^{\mu\nu}\,\overline{\nabla}_\mu\Omega\,\overline{\nabla}_\nu\Omega\,,
\end{equation}
where $\overline{\nabla}$ denotes the covariant derivative with respect to the metric $\bar{\text{g}}$. Finally, using all of the above for the $4$-dimensional Starobinsky model we can rephrase it as
\begin{equation}
    \mathcal{S}[\chi,\overline{\text{g}};\alpha]=\int\!\mathrm{d}^4x\,\sqrt{-\overline{g}}\left\{\frac{M_P^2}{2}\overline{R}-3M_P^2\,\frac{\overline{\nabla}^\mu\Omega(\chi)\overline{\nabla}_\mu\Omega(\chi)}{\Omega^2(\chi)}-\frac{V(\chi)}{\Omega^4(\chi)}\right\}+\int\!\mathrm{d}^4x\,\overline{\nabla}_\mu\left(\sqrt{-\overline{g}}\,\overline{\nabla}^\mu\ln{\Omega}\right)\,,
\end{equation}
where the last integral contributes a surface term\footnote{Notice that the metricity condition, namely $\nabla_\rho g_{\mu\nu}=0$, has already been used at this point.} that we ignore hereafter, and we reuse the definition of $V(\chi)=\alpha\chi^2/2$. Let us substitute the form of the function $\Omega(\chi)$, given in eq.~\eqref{Cond:OmegaMetStar}; the action functional becomes:
\begin{equation}\label{Action:MetricStarNonCan}
    \mathcal{S}=\int\!\mathrm{d}^4x\,\sqrt{-\overline{g}}\left\{\frac{M_P^2}{2}\,\overline{R}-\frac{3M_P^2\alpha^2}{\left(M_P^2+2\alpha\chi\right)^2}\,\overline{\nabla}^\mu\chi\,\overline{\nabla}_\mu\chi-\frac{\alpha M_P^4\chi^2}{2\left(M_P^2+2\alpha\chi\right)^2}\right\}\,.
\end{equation}
This is reminiscent of a scalar-tensor theory with a scalar field minimally coupled to gravity and a noncanonical kinetic term. In order to obtain a scalar field with a canonical kinetic term we apply a field redefinition of the form\footnote{Note that in eq.~\eqref{Cond:StarRedef} we assumed the form of the integration constants in terms of the reduced Planck mass, such that the canonical field has the correct dimensions.}
\begin{equation}
    -\frac{3M_P^2\alpha^2}{\left(M_P^2+2\alpha\chi\right)^2}\,\overline{\nabla}^\mu\chi\,\overline{\nabla}_\mu\chi\longmapsto-\frac{1}{2}(\overline{\nabla}\varphi)^2
\end{equation}
\begin{equation}\label{Cond:StarRedef}
    \therefore\frac{\varphi}{M_P}=\pm\frac{\sqrt{3}}{2}\,\ln{\frac{M_P^2+2\alpha\chi}{M_P^2}}\,,\qquad \forall\chi>-\frac{M_P^2}{2\alpha}\,.
\end{equation}
The action then becomes
\begin{equation}\label{Action:StarCanScal}
    \mathcal{S}=\int\!\mathrm{d}^4x\,\sqrt{-\overline{g}}\left\{\frac{M_P^2}{2}\,\overline{R}-\frac{1}{2}\left(\partial\varphi\right)^2-U(\varphi)\right\}\,,
\end{equation}
where the potential term reads as~\cite{Whitt1984,Maeda1988,Barrow1988}
\begin{equation}\label{Cond:OriginalStaroPot}
    U(\varphi)\equiv\frac{M_P^4}{8\alpha}\left(1-e^{-\frac{\varphi}{M_P}\frac{2}{\sqrt{3}}}\right)^2\,.
\end{equation}
In order for the above scalar potential to behave asymptotically as the one expressed in terms of the original $\chi$ field we are forced to pick the positive sign in eq.~\eqref{Cond:StarRedef}. Furthermore, for large values of the canonical field $\varphi$ it is trivial now to see that the potential is dominated by a vacuum energy
\begin{equation}
    U(\varphi\rightarrow\infty)=\frac{M_P^2}{8\alpha}\,.
\end{equation}
Notice that the potential has a global minimum at its origin, $\varphi=0$, that is a stable $U''(\varphi)>0$, and the field starting from large field values around the flat region of the potential is led to the origin naturally, as displayed in fig.~\ref{fig:OriginalStaroPot}.

In the action \eqref{Action:StarCanScal} the additional scalar mode coming from the $R^2$ term is perfectly manifested as a real scalar field $\varphi$ with a potential term $U(\varphi)$. As shown in figure~\ref{fig:OriginalStaroPot} the potential is asymptotically flat for large (positive) values of the field, which as was discussed in ch.~\ref{Ch2:Inflation} it is crucial for the slow-roll inflation.

The equations of motion for the system described by the Lagrangian in eq.~\eqref{Action:StarCanScal} are the Einstein field equations for the metric $\bar{\text{g}}$ and the Klein-Gordon equation for the scalar field; namely:
\begin{align}
    G_{\mu\nu}(\,\overline{\{\ \}}\,)\equiv \overline{R}_{\mu\nu}-\frac{1}{2}\overline{g}_{\mu\nu}\overline{R}=\overline{\nabla}_\mu\varphi\,\overline{\nabla}_\nu\varphi-\overline{g}_{\mu\nu}\left(\overline{\nabla}^\rho\varphi\,\overline{\nabla}_\rho\varphi+V(\varphi)\right)
\end{align}
and
\begin{equation}
    \overline{g}^{\mu\nu}\,\overline{\nabla}_\mu\overline{\nabla}_\nu\varphi+\frac{\mathrm{d}V(\varphi)}{\mathrm{d}\varphi}=0\,.
\end{equation}
Naturally then, one can readily apply the single-field inflation mechanism as is described in ch.~\ref{Ch2:Inflation}.

\subsubsection{\emph{Application to slow-roll inflation}}

For the sake of completeness let us also include some brief remarks on the usual $R^2$ inflation model, as was first formulated in the metric formalism. Starting from the action \eqref{Action:StarCanScal} we can effortlessly apply the mechanism of the single-field slow-roll inflation that was described in sec.~\ref{sec:inflDynamics}. First let us consider the number of $e$-folds; the integral is exactly solvable and results in
\begin{equation}
    N(\varphi)=\frac{1}{4M_P}\left(3M_P e^{\frac{\varphi}{M_P}\frac{2}{\sqrt{3}}}-\sqrt{6}\,\varphi\right)+C_0\sim\frac{3}{4}\,e^{\frac{\varphi}{M_P}\frac{2}{\sqrt{3}}}\,,
\end{equation}
where $C_0$ is an integration constant. Since the value of the integral is dominated by the large values of the inflaton field, with $\varphi_i\gg\varphi_f$ (recall that $N=N(\varphi_i)-N(\varphi_f)$), we simply keep terms that are dominating at large field values $\varphi\to\infty$. It is straightforward to show that in the same field limit the spectral tilt and the tensor-to-scalar ratio are given by
\begin{equation}
    n_s\approx1-\frac{2}{N}\qquad\quad\&\qquad\quad r\approx\frac{12}{N^2}\,, \qquad \text{as }\ \varphi\to\infty\,.
\end{equation}

These are the celebrated results of the Starobinsky model that is in persistent contact with observations~\cite{Akrami2020}. Other than the model being motivated by possible gravitational quantum corrections (albeit ad hoc from a theoretical point of view) and its ability to predict preferable values for the $n_s$ in a wide range of $N\in[50,60]$, the main advantage comes from its suppression of the tensor-to-scalar ratio, $r\sim10^{-3}$. It is expected however that future experiments are able to probe the region of $r\sim10^{-3}$ (and maybe even $10^{-4}$)~\cite{Matsumura2016,Kogut2011,Sutin2018}, which will in principle be able to falsify some of the models lying in that region, the Starobinsky model (and some of its generalisations) being one of them. Additionally, an upper bound can be placed on the dimensionless parameter $\alpha$ from the power spectrum of scalar perturbations
\begin{equation}
    \mathcal{A}_s\stackrel{\text{SR}}{\approx}\left.\frac{U(\varphi)}{24\pi^2M_P^4\,\epsilon_V(\varphi)}\right|_{\varphi=\varphi_i}\approx\mathcal{O}\left(\frac{10^{-3}}{\alpha}\right)\,.
\end{equation}
In order for the model to admit the observed amplitude of scalar perturbations the constant takes approximate values of $\alpha\sim\mathcal{O}(10^6)$, meaning that the mass parameter defined through $\alpha$ is $m\approx 10^{15}\,\text{GeV}$.

A subtle point to be made regarding inflation is the scale of inflation, meaning at which energy scales are the predictions of the models sensible (cutoff scale) placing also constraints on the associated inflaton field excursion. Clearly, the discussion pertains to each inflationary model separately and was not addressed in an earlier chapter for that reason. To properly obtain the cutoff scale of each model, the models are understood as quantum field theories interacting with gravity in a weak field expansion of the form $g_{\mu\nu}=\eta_{\mu\nu}+h_{\mu\nu}/M_P$. Then, depending on model parameters (but not necessarily), the validity of the inflationary predictions is checked by calculating the quantum corrections and their contribution. For example, it was shown that the cutoff energy scale of the $R^2$ inflation model is $\Lambda=M_P$~\cite{Hertzberg2010}. It should be noted that the nature of the issue is more complicated and in fact there are often disagreements regarding the results obtained in the way highlighted above, however it is a point to keep in mind regarding inflation and its (range of) validity.

\subsubsection{\emph{Coupling with matter}}

\begin{flushleft}
\underline{\emph{Minimal coupling}}
\end{flushleft}

A natural extension to the Starobinsky model is to attempt a coupling of it with a matter Lagrangian. Presently, we are interested only in the minimal coupling of quadratic gravity with a scalar field, that could also be fundamental. The action in the Jordan frame that describes such a system reads as
\begin{equation}\label{Action:MetStarMinScal}
    S=\int\!\mathrm{d}^4x\,\sqrt{-g}\left\{\frac{M_P^2}{2}R+\frac{\alpha}{2}R^2-\frac{1}{2}(\partial\phi)^2-V(\phi)\right\}\,.
\end{equation}
Notice that it is straightforward to generalise the above action in order to describe multiple scalar fields coupled minimally with gravity, where $\phi$ would then collectively denote all of the scalar fields. Following the same steps as we did to rephrase the pure $R^2$ model in eq.~\eqref{Action:MetricStarNonCan}, we can express the above action functional as
\begin{equation}
    \mathcal{S}=\int\!\mathrm{d}^4x\,\sqrt{-\overline{g}}\left\{\frac{M_P^2}{2}\,\overline{R}-\frac{3M_P^2\alpha^2}{(M_P^2+2\alpha\chi)^2}\,\overline{g}^{\mu\nu}\,\overline{\nabla}_\mu\chi\,\overline{\nabla}_\nu\chi-\frac{M_P^2}{2(M_P^2+2\alpha\chi)}\,\overline{g}^{\mu\nu}\,\overline{\nabla}_\mu\phi\overline{\nabla}_\nu\phi-U(\phi,\chi)\right\}\,,
\end{equation}
where the potential term is given by
\begin{equation}
    U(\phi,\chi)\equiv\frac{V(\phi)+V(\chi)}{\Omega^4(\chi)}=M_P^4\,\frac{\frac{\alpha}{2}\chi^2+V(\phi)}{(M_P^2+2\alpha\chi)^2}\,.
\end{equation}

Let us canonically normalise the kinetic terms of the scalar fields. Clearly, since the kinetic function of the $\phi$ field mixes with the field $\chi$, the kinetic term of $\phi$ has to be noncanonical (or generally one of the scalar fields would have a noncanonical kinetic term). Assuming a field redefinition of the form of eq.~\eqref{Cond:StarRedef} we can express the total action as
\begin{equation}\label{Action:MetNonminStarMULTIFIELD}
    \mathcal{S}=\int\!\mathrm{d}^4x\,\sqrt{-\overline{g}}\left\{\frac{M_P^2}{2}\,\overline{R}-\frac{1}{2}\mathscr{G}^{IJ}(\Phi)\,\overline{g}^{\mu\nu}\,\overline{\nabla}_\mu\Phi^I\,\overline{\nabla}_\nu\Phi^J-U(\Phi)\right\}\,,
\end{equation}
where $\Phi=\{\varphi,\phi\}$ denotes collectively all the scalar fields, the metric $\mathscr{G}^{IJ}(\Phi)$ is the metric on the field-space manifold indicating that the space spanned by these two fields is not flat anymore. Here the indices $I,J$ run over the different scalar fields $\{\varphi,\phi\}$. The functions are then defined as
\begin{equation}
    \mathscr{G}_{IJ}=\begin{pmatrix}1&0\\0&\text{exp}\left(-\frac{\varphi}{M_P}\frac{2}{\sqrt{3}}\right)
    \end{pmatrix}\,,\qquad U(\varphi,\phi)=\frac{M_P^4}{8\alpha}\left(1-e^{-\frac{\varphi}{M_P}\frac{2}{\sqrt{3}}}\right)^2+e^{-\frac{\varphi}{M_P}\frac{4}{\sqrt{3}}}V(\phi)\,.
\end{equation}
Ultimately the form of the two-dimensional potential $U(\Phi)$ depends on the potential $V(\phi)$. For the purposes of inflation the analysis is more complicated since it is multidimensional in the field space, however it is possible in principle that inflation happens in a flat direction of the potential $U$ (in $\phi$ in this specific case), and therefore only one of the scalar fields drives inflation (being $\chi\mapsto\varphi$ here). Interestingly enough, there is a possibility that depending on the potential both of the fields can contribute consecutively, however it cannot be realised in this particular example due to the form of the potential term induced by the $\chi$ field.

The equations of motion derived from the variation of the action \eqref{Action:MetNonminStarMULTIFIELD} with respect to the $\Phi^K$ fields and the metric $\overline{\text{g}}$ are given by
\begin{equation}\label{Eq:MetNonminStarFields}
    \mathscr{G}_{KJ}\,\overline{g}^{\mu\nu}\,\overline{\nabla}_\mu\overline{\nabla}_\nu\Phi^J+\gamma_{IJK}\,\overline{g}^{\mu\nu}\,\overline{\nabla}_\mu\Phi^I\overline{\nabla}_\nu\Phi^J-\frac{\partial U(\Phi)}{\partial \Phi^K}=0\,,
\end{equation}
\begin{equation}\label{Eq:MetNonminStarFieldsMetric}
    M_P^2\,G_{\mu\nu}=\mathscr{G}_{IJ}\,\overline{\nabla}_\mu\Phi^I\overline{\nabla}_\nu\Phi^J-\overline{g}_{\mu\nu}\left(\frac{1}{2}\mathscr{G}_{IJ}\,\overline{g}^{\rho\sigma}\,\overline{\nabla}_\rho\Phi^I\overline{\nabla}_\sigma\Phi^J-U(\Phi)\right)\,,
\end{equation}
where we also defined the connection coefficients on the field space manifold $\mathcal{M}_\Phi$ as
\begin{equation}
    \gamma_{IJK}\equiv\frac{1}{2}\left(\frac{\partial\mathscr{G}_{IK}}{\partial\Phi^J}+\frac{\partial\mathscr{G}_{KJ}}{\partial\Phi^I}-\frac{\partial\mathscr{G}_{IJ}}{\partial\Phi^K}\right)\,.
\end{equation}
These are the equations governing the evolution of the collection of scalar fields $\Phi$ in a curved spacetime.

\begin{flushleft}
\underline{\emph{Nonminimal coupling}}
\end{flushleft}

The model of eq.~\eqref{Action:MetStarMinScal} can be further generalised by assuming a nonminimal coupling between the scalar field $\phi$ and the Einstein-Hilbert term; it reads as
\begin{equation}\label{Action:StarScalar}
    S=\int\!\mathrm{d}^4x\,\sqrt{-g}\left\{\frac{1}{2}(M_P^2+\xi\phi^2)R+\frac{\alpha}{2}R^2-\frac{1}{2}(\partial\phi)^2-V(\phi)\right\}\,,
\end{equation}
where $\xi$ is a dimensionless coupling constant. A coupling of the form of $\propto\phi^2R$ is expected to arise due to quantum corrections coming from a scalar field in the gravitational background and, other than that, it is generally considered in modified theories of gravity and models describing inflation. In the case that $\phi$ is identified as the dilaton field coming from the compactification of higher dimensions a nonminimal coupling with gravity is also expected to appear.

Regardless, after performing a Weyl rescaling of the metric as
\begin{equation}
    \tilde{g}_{\mu\nu}(x)=\Lambda^2(\chi,\phi)\,g_{\mu\nu}(x)\,,\qquad\text{where}\quad \Lambda^2\equiv\frac{M_P^2+\xi\phi^2+2\alpha\chi}{M_P^2}\,,
\end{equation}
we can express the action in the Einstein frame as follows:
\begin{equation}\label{Action:MetStarfinal}
    \mathcal{S}=\int\!\mathrm{d}^4x\,\sqrt{-\tilde{g}}\left\{\frac{M_P^2}{2}\,\tilde{R}-\frac{1}{2}\mathscr{Z}^{IJ}\,\tilde{g}^{\mu\nu}\tilde{\nabla}_\mu\Psi^I\tilde{\nabla}_\nu\Psi^J-\mathcal{V}(\Psi)\right\}\,,
\end{equation}
where we define as $\Psi=\{\chi,\phi\}$ and the metric on the curved field-space manifold is given by
\begin{equation}
    \mathscr{Z}^{IJ}\equiv\begin{pmatrix}\displaystyle{\frac{6M_P^2\,\alpha^2}{(M_P^2+\xi\phi^2+2\alpha\chi)^2}}&\displaystyle{\frac{3M_P^2\,\alpha\,\xi}{2(M_P^2+\xi\phi^2+2\alpha\chi)^2}}\\&\\\displaystyle{\frac{3M_P^2\,\alpha\,\xi}{2(M_P^2+\xi\phi^2+2\alpha\chi)^2}}&\displaystyle{M_P^2\,\frac{6\xi^2+M_P^2+\xi\phi^2+2\alpha\chi}{(M_P^2+\xi\phi^2+2\alpha\chi)^2}}\end{pmatrix}\,,\qquad I,J,\ldots=\{\chi,\phi\}\,,
\end{equation}
and the scalar potential reads
\begin{equation}
    \mathcal{V}(\Psi)\equiv\frac{V(\chi)+V(\phi)}{\Lambda^4(\chi,\phi)}=M_P^4\,\frac{\frac{\alpha}{2}\chi^2+V(\phi)}{(M_P^2+\xi\phi^2+2\alpha\chi)^2}\,.
\end{equation}
One immediate consequence of the nonminimal interaction of the $\phi$ field is the appearance of a mixed kinetic term of the form $\propto(\tilde{\nabla}\chi)(\tilde{\nabla}\phi)$, evident also from the fact that $\left.\mathscr{Z}_{\chi\phi}\right|_{\xi\to0}\to0$ meaning that the terms decouple in that limit. The equations of motion derived in this case have the same form as the ones derived for the action \eqref{Action:MetNonminStarMULTIFIELD}, presented in eqs.~\eqref{Eq:MetNonminStarFields}-\eqref{Eq:MetNonminStarFieldsMetric}, with the metric $\mathscr{Z}_{IJ}$ now describing the field space $\mathcal{M}_\Phi$.

Clearly, due to the involved kinetic terms the analysis of the model is almost impossible without any assumptions that might decoupled them. One may think to introduce a new field in a linear combination of the old ones, $\chi$ and $\phi$. A trivial redefinition of that nature would be the following
\begin{equation}
    (\sigma(x))^2\equiv 6\left[M_P^2+\xi(\phi(x))^2+2\alpha\chi(x)\right]\,,
\end{equation}
where the proportionality factor is included to canonically normalise (up to an overall factor) the kinetic terms of the two fields. In turn, the action becomes
\begin{equation}
    \mathcal{S}=\int\!\mathrm{d}^4x,\sqrt{-\tilde{g}}\left\{\frac{M_P^2}{2}\,\tilde{R}-\frac{1}{2}\left(\frac{6\,M_P^2}{\sigma^2}\right)\delta^{IJ}\,\tilde{g}^{\mu\nu}\tilde{\nabla}_\mu\tilde{\Psi}^I\tilde{\nabla}_\nu\tilde{\Psi}^J-\tilde{\mathcal{V}}(\tilde{\Psi})\right\}\,,
\end{equation}
where $\delta_{IJ}$ denotes the Kronecker delta and $\tilde{\Psi}=\{\sigma,\phi\}$. It is straightforward to show that the potential now reads
\begin{equation}
    \tilde{\mathcal{V}}=\frac{36\,M_P^4}{\sigma^4}\,V(\phi)+\frac{1}{8\alpha}\left(\frac{\sigma^2}{6}-M_P^2-\xi\phi^2\right)^2\,.
\end{equation}

There exist other field redefinitions one might try, e.g. extending the definition of the scalar degree of freedom as in refs.~\cite{Maeda1988,Maeda1989,He2018}, that ultimately result in two scalar fields that have noncanonical kinetic terms but do not mix.

As a closing note, the coupling of the $R+R^2$ gravity to a matter sector, be that minimal or nonminimal, gives rise to a theory with some type of mixing between the matter fields and the scalar degree of freedom (\emph{scalaron field}). Since for most inflationary models the matter sector contains a scalar field and its potential, the predictions regarding inflation are best approached in what is known as a \emph{multifield} framework. In the next section considering the same class of models described by an action functional as in eq.~\eqref{Action:StarScalar} under the Palatini formalism we show that the scalaron field is nondynamical and can in fact be integrated out of the theory. Therefore the only dynamical scalar field capable of assuming the role of the inflaton is the original one, denoted by $\phi$ in the initial action.

\subsubsection*{\emph{Digression on the Weyl vs. conformal transformations}}

The Weyl rescaling of the metric is often confused or referred to interchangeably in the literature with the conformal transformation of the metric. Actually, the Weyl transformation, or as is more formally known the scale transformations, is not a coordinate transformation at all and is not a symmetry that is respected by the laws of physics as we know them, e.g. the SM.\footnote{An exception to that can be models that are specifically build to be (classically) scale invariant. We discuss these models in further detail in a later section, sec.~\ref{subsec:CWmodel}.} Schematically, a Weyl rescaling reads 
\begin{equation}
    x\mapsto x,\qquad g_{\mu\nu}(x)\mapsto\Omega(x)\,g_{\mu\nu}(x)\,,
\end{equation}
which changes the physical distances at each point $p\in\mathcal{M}$ by a factor of $\mathrm{d}s'^2=\Omega(x)\mathrm{d}s^2$ that may depend on the place, but it does not depend on the direction of the line we measure on.

Contrarily, conformal transformations are a subset of coordinate transformations, that, as will become obvious by their definition, include isometries as a subset. The conformal symmetry is an extension of the Poincar{\'e} group that also includes five additional degrees of freedom for the four conformal transformations and one for the dilation; in $\mathbb{R}^{N,1}$ the conformal group is $\cong\!SO(N+1,2)$, while on the other hand, the group associated with the Weyl transformations is infinite-dimensional~\cite{di1996conformal}. The transformation is defined as follows
\begin{equation}
    x\rightarrow x',\qquad g'_{\mu\nu}(x')=\frac{\partial x^\rho}{\partial x'^\mu}\,\frac{\partial x^\sigma}{\partial x'^\nu}\,g_{\rho\sigma}(x)=\Lambda(x)\,g_{\mu\nu}(x)\,.
\end{equation}
Therefore, the conformal transformation is a diffeomorphism that also scales the metric by a conformal factor $\Lambda(x)$. In fact, since the metric is invariant up to that scaling factor, the angles are preserved and we can write that $\mathrm{d}s'^2=\mathrm{d}s^2$ since we have simply relabeled the points. See ref.~\cite{Farnsworth2017} for a more detailed discussion on the role of Weyl and conformal transformations (particularly invariance) in physics today.

\subsection{\emph{Palatini formalism}}
\label{subsec:PalatiniForm}

Let us consider the Starobinsky cosmological model in the Palatini formalism, where the connection and the metric have no a priori dependence on one another. The action resembles the original one but with key differences that are further highlighted below
\begin{equation}
    \mathcal{S}[\text{g},\Gamma]=\int\!\mathrm{d}^4x\,\sqrt{-g}\left\{\frac{M_P^2}{2}g^{\mu\nu}R_{\mu\nu}(\Gamma)+\frac{\alpha}{4}\,\left(g^{\mu\nu}R_{\mu\nu}(\Gamma)\right)^2\right\}\,.
\end{equation}
The action is explicitly dependent on the two dynamical variables, namely the metric tensor $\text{g}$ and the connection $\Gamma$. Note that the Ricci tensor $R_{\mu\nu}$ is expressed purely in terms of the connection coefficients and their derivatives. Therefore, under a Weyl transformation of the form
\begin{equation}
    g_{\mu\nu}(x)\longmapsto \Omega^{-2}(x)\,g_{\mu\nu}(x)\,,
\end{equation}
the two terms appearing in the action transform as
\begin{align}
    \propto\sqrt{-g}\,g^{\mu\nu}R_{\mu\nu}(\Gamma)\quad&\longmapsto\quad\Omega^{-2}\sqrt{-g}\,g^{\mu\nu}R_{\mu\nu}(\Gamma)\,,\\
    \propto\sqrt{-g}\,g^{\mu\nu}g^{\rho\sigma}R_{\mu\nu}(\Gamma)R_{\rho\sigma}(\Gamma)\quad&\longmapsto\quad\sqrt{-g}\,g^{\mu\nu}g^{\rho\sigma}R_{\mu\nu}(\Gamma)R_{\rho\sigma}(\Gamma)\,.
\end{align}
It may be worth noting that the $R^2$ term is manifestly Weyl invariant in the first-order formalism.\footnote{A straightforward generalisation of the statement to $D$ dimensions suggests that the term $\sqrt{-g}\,R^{D/2}$ is Weyl invariant in that context.} Before we continue with the Weyl transformation, let us first consider the action in its scalar representation by introducing the auxiliary $\chi$ field, similarly to what we was done in the metric formulation of the theory. Then without loss of generality we can write\footnote{Notice that the auxiliary field is redefined in order to have the appropriate dimensions, meaning that at the level of equations of motion it satisfies $\chi^2=R$, unlike the previous section where $[\chi]_\text{m}=2$, ($\chi=R$).}
\begin{equation}
    \mathcal{S}=\int\!\mathrm{d}^4x\,\sqrt{-g}\left\{\frac{1}{2}(M_P^2+\alpha\chi^2)g^{\mu\nu}R_{\mu\nu}(\Gamma)-\frac{\alpha}{4}\chi^4\right\}\,,
\end{equation}
which after a Weyl rescaling of the metric we can absorb the factor of $R$ and obtain the action in the Einstein frame, reading
\begin{equation}
    \mathcal{S}=\int\!\mathrm{d}^4x\,\sqrt{-\overline{g}}\left\{\frac{M_P^2}{2}\,\overline{g}^{\mu\nu}R_{\mu\nu}(\Gamma)-V(\chi)\right\}\,,
\end{equation}
where 
\begin{equation}
    \overline{g}_{\mu\nu}(x)=\Omega^2(\chi)\,g_{\mu\nu}(x)\equiv\frac{M_P^2+\alpha\chi^2}{M_P^2}\,g_{\mu\nu}(x)\,,
\end{equation}
and the potential term is obtained as
\begin{equation}
    V(\chi)=\frac{\alpha\,M_P^4}{4}\,\frac{\chi^4}{(M_P^2+\alpha\chi^2)^2}\,.
\end{equation}
For large values of the auxiliary field the potential term also tends to a constant value of $\propto M_P^4/\alpha$. 

The crucial difference between the metric and the Palatini formulations is that in the latter one there is no additional (dynamical) scalar mode present in the theory~\cite{Antoniadis2018,Enckell2019}, which is effectively described by the Einstein-Hilbert term with a potential term in the Einstein frame. Thus, it contains the usual spin-$2$ graviton together with a potential term that can play the role of a cosmological constant (after fine-tuning). This is based on the equivalent description of Einstein-Hilbert action in the two formulations, which was discussed in the previous chapter, ch.~\ref{Ch3:FirstOrder}.

Obviously then the Starobinsky model in its first-order formulation is incapable of describing inflation and therefore it has to be coupled with another scalar field that would play the role of the inflaton field. With this in mind, the conventional idea of single-field inflation seems attractive and some of the inflationary models, that in their metric counterpart have already been ruled out by observational data, may in principle be ``rescued''. In order to do that let us include an exemplary scalar field, say $\phi$, with a most general nonminimal coupling with the Einstein-Hilbert term in the form of $f(\phi)R$ in the action functional:
\begin{equation}
    \mathcal{S}=\int\!\mathrm{d}^4x\,\sqrt{-g}\left\{\frac{1}{2}f(\phi)R+\frac{\alpha}{4}R^2-\frac{1}{2}(\partial\phi)^2-V(\phi)\right\}\,,
\end{equation}
where for ease of notation we reintroduced the Ricci scalar with the implicit dependence on the connection through the Ricci tensor, and $(\partial\phi)^2\equiv g^{\mu\nu}\nabla_\mu\phi\nabla_\nu\phi$. Expressed in its scalar representation it reads
\begin{equation}
    \mathcal{S}=\int\!\mathrm{d}^4x\,\sqrt{-g}\left\{\frac{1}{2}(f(\phi)+\alpha\chi^2)R-\frac{1}{2}(\partial\phi)^2-V(\phi)-\frac{\alpha}{4}\chi^4\right\}\,.
\end{equation}

Considering a Weyl rescaling of the metric as follows
\begin{equation}
    \overline{g}_{\mu\nu}(x)=\Omega^2(\phi,\chi)\,g_{\mu\nu}(x)\equiv\left(\frac{f(\phi)+\alpha\chi^2}{M_P^2}\right)g_{\mu\nu}(x)\,,
\end{equation}
we obtain the action in the Einstein frame as follows
\begin{equation}
    \mathcal{S}=\int\!\mathrm{d}^4x\,\sqrt{-\overline{g}}\left\{\frac{M_P^2}{2}\overline{g}^{\mu\nu}\,R_{\mu\nu}(\Gamma)-\frac{1}{2}\Omega^{-2}(\phi,\chi)\overline{g}^{\mu\nu}\,\overline{\nabla}_\mu\phi\overline{\nabla}_\nu\phi-\overline{V}(\phi,\chi)\right\}\,,
\end{equation}
where the potential is simply
\begin{equation}
    \overline{V}(\phi,\chi)\equiv\frac{V(\phi)+\frac{\alpha}{4}\chi^4}{\Omega^4(\phi,\chi)}\,.
\end{equation}

As expected the $\phi$ field is still the unique propagating scalar degree of freedom in the theory. Since the $\chi$ field is auxiliary it can be integrated out via its equation of motion, which it is straightforward to show that it is simply a constraint.\footnote{The process of ``integrating out'' the auxiliary field is rather misleading at this point, since it is usually associated with the path integral formulation of a theory. The effect of an auxiliary field in a quantum or classical theory is the same as a result of their nonpropagating nature and they have been used throughout physics to help simplify the calculations. Some of the most notable examples are the complex scalar field $F$ and the real scalar field $D$ appearing in the $F$- and $D$-terms, respectively, in supersymmetric theories that are used to close the supersymmetric algebra (e.g. see ref.~\cite{Gates1983}). Others include those used in string theory in order to substitute the Nambu-Goto with the Polyakov Lagrangian~\cite{Zwiebach2006}.} Computing the variation of the action functional with respect to $\chi$ we obtain the equation of motion~\cite{Antoniadis2018}
\begin{equation}
    \frac{\delta\mathcal{S}}{\delta\chi}=0\ \implies\ \alpha\,M_P^2\,\frac{\chi^3\left(\alpha(\overline{\nabla}\phi)^2-M_P^2f(\phi)\right)+\chi\left(f(\phi)(\overline{\nabla}\phi)^2+4M_P^2V(\phi)\right)}{(f(\phi)+\alpha\chi^2)^3}=0\,,
\end{equation}
which if $\chi\neq0$ and $\chi\neq-\sqrt{f(\phi)/\alpha}$  holds $\forall\phi$, it leads to the following constraint
\begin{equation}
    \chi^2=\frac{\displaystyle{\frac{4\,V(\phi)}{f(\phi)}+\frac{(\overline{\nabla}\phi)^2}{M_P^2}}}{\displaystyle{1-\alpha\,\frac{(\overline{\nabla}\phi)^2}{M_P^2\,f(\phi)}}}\,.
\end{equation}
In the case of auxiliary fields appearing in the Lagrangian in a bilinear form we can express them in terms of the other fields coupled to them, in this case the $\phi$ field. A direct substitution of the algebraic relation of $\chi$ in terms of $\phi$ in the action functional gives rise to the following~\cite{Enckell2019}
\begin{equation}\label{Action:PalStarfinal}
    \mathcal{S}=\int\!\mathrm{d}^4x\,\sqrt{-\overline{g}}\left\{\frac{M_P^2}{2}\,\overline{g}^{\mu\nu}\,R_{\mu\nu}(\Gamma)-\frac{1}{2}\,M_P^2 K(\phi)(\overline{\nabla}\phi)^2+\frac{1}{4}\,\alpha\frac{K(\phi)}{f(\phi)}(\overline{\nabla}\phi)^4-M_P^4\,\frac{K(\phi)}{f(\phi)}V(\phi)\right\}\,,
\end{equation}
where we used the following definition of the noncanonical kinetic function
\begin{equation}
    K(\phi)\equiv\frac{f(\phi)}{(f(\phi))^2+4\alpha V(\phi)}\,.
\end{equation}

Let us assume that the original potential is given by - or at least approximated by at large field values - a polynomial function $V(\phi)\propto\phi^n$ with $n\in\mathbb{N}^*$, and the nonminimal coupling function has a scale-invariant form $f(\phi)\propto\phi^2$. Then, due to the auxiliary field the rescaled potential in the large field limit tends to
\begin{equation}
    U(\phi)\equiv M_P^4\,\frac{K(\phi)}{f(\phi)}V(\phi)\stackrel{\phi\rightarrow\infty}{\approx}\frac{M_P^4}{\phi^{4-n}+4\alpha}=\left\{\begin{matrix}\displaystyle{\frac{M_P^4}{4\alpha}}\,,&\forall n\geq 4\,,\\ & \\0\,,&\forall n<4\,.\end{matrix}\right.
\end{equation}
Even though the $R^2$ term ultimately does not lead to a dynamical degree of freedom it can contribute nontrivially in the inflationary potential by inducing a flat region at large values of the inflaton field~\cite{Antoniadis2018}. In other words, it can help flatten a quite general class of inflationary potentials $V(\phi)$, thus allowing for the possibility of them supporting an inflationary epoch for some range of the parameter $\alpha$ (depending on the rest of the model parameters as well).

Comparing the form of the final actions between the two formulations, namely equations \eqref{Action:PalStarfinal} and \eqref{Action:MetStarfinal}, it is evident that we have effectively traded the two-dimensional field space with a one-dimensional field space that includes higher-order kinetic terms, $\propto(\bar{\nabla}\phi)^4$. These nonstandard terms are not unusual, in fact if we think of $\phi$ as some moduli field (like the dilaton) in string theory, the $\alpha'$ corrections predict a series of higher-derivative terms in the effective action. 

In the case that eq.~\eqref{Action:PalStarfinal} describes an inflationary model the terms quadratic in kinetic energy are highly suppressed by the potential and, during that period, they are presumably negligible. However, there is a specific type of inflation, known as $k$-inflation (``$k$'' for kinetic)~\cite{ArmendarizPicon1999,ArmendarizPicon2001}, in which the model includes higher-derivative kinetic terms that drive inflation without the need of a potential term. Since then, generalisations of the theory were considered in which a potential term was also included, bringing the theory schematically similar to the action derived in eq.~\eqref{Action:PalStarfinal}, i.e. $\mathscr{L}\sim A(\phi)(\nabla\phi)^2+B(\phi)(\nabla\phi)^4+V(\phi)$.

Let us relabel the function coefficients in the action eq.~\eqref{Action:PalStarfinal} as follows
\begin{equation}
    \mathcal{S}=\int\!\mathrm{d}^4x\,\sqrt{-\overline{g}}\left\{\frac{M_P^2}{2}\,\overline{g}^{\mu\nu}\,R_{\mu\nu}(\Gamma)+k_0(\phi)X(\phi)+k_2(\phi)(X(\phi))^2-U(\phi)\right\}\,,
\end{equation}
where $X\equiv\frac{1}{2}(\overline{\nabla}\phi)^2$. Then, variation of the action with respect to the connection $\Gamma$, the metric $\overline{\text{g}}$ and the field $\phi(x)$ leads to their respective equations of motion which read\footnote{Note also that if we were to  first vary the action ${\mathcal{S}}[\overline{\text{g}},\phi,\chi]$ with respect to $\overline{\text{g}}_{ \mu\nu}$ and substitute the solution for the auxiliary $\chi$ in the resulting Einstein equation, we would obtain the same result as in eq.~\eqref{Eq:PalatiniGenFieldEQs}, meaning that the variations with respect to $\overline{g}_{ \mu\nu}$ and $\chi$ can be interchanged. Schematically this reads as follows
$$\mathcal{S}[\overline{\text{g}},\chi,\phi]\ \longrightarrow\ \delta_\chi\mathcal{S}[\overline{\text{g}},\chi,\phi]\stackrel{!}{=}0\ \curvearrowright\ \mathcal{S}[\overline{\text{g}},\chi,\phi]\ \longrightarrow\ \delta_{\bar{\text{g}}}\mathcal{S}[\overline{\text{g}},\phi]\stackrel{!}{=}0\ \implies\ \text{Eq.}~\eqref{Eq:PalatiniGenFieldEQs}$$
is equivalent to
$$\mathcal{S}[\overline{\text{g}},\chi,\phi]\ \longrightarrow\ \left\{\begin{matrix}\delta_\chi\mathcal{S}[\overline{\text{g}},\chi,\phi]\stackrel{!}{=}0 & \curvearrowright\ \delta_{\bar{\text{g}}}\mathcal{S}[\overline{\text{g}},\chi,\phi]\stackrel{!}{=}0 \\ &\\ \wedge &  \\\delta_{\bar{\text{g}}}\mathcal{S}[\overline{\text{g}},\chi,\phi]\stackrel{!}{=}0 & \end{matrix}\right.\ \implies\ \text{Eq.}~\eqref{Eq:PalatiniGenFieldEQs}\,,$$
where the curved arrow denotes substitution of the element in the LHS into the ones in RHS.}~\cite{Lykkas2021}
\begin{align}
    &{\Gamma^\rho}_{\mu\nu}=\overline{\left\{{}_\mu{}^\rho{}_\nu\right\}}\equiv\frac{1}{2}\overline{g}^{\rho\sigma}\left(\partial_\mu\overline{g}_{\sigma\nu}+\partial_\nu\overline{g}_{\mu\sigma}-\partial_\sigma\overline{g}_{\mu\nu}\right)\,,\\
    &\overline{G}_{\mu\nu}(\overline{\text{g}},\Gamma)\equiv R_{\mu\nu}-\frac{1}{2}\overline{g}_{\mu\nu}\overline{R}=-(k_0+2k_2 X)\overline{\nabla}_\mu\phi\overline{\nabla}_\nu\phi+\overline{g}_{\mu\nu}\left(k_0 X+k_2 X^2-U\right)\,,\label{Eq:PalatiniGenFieldEQs}\\
    &\left(k_0+2k_2X\right)\overline{g}^{\mu\nu}\overline{\nabla}_\mu\overline{\nabla}_\nu\phi+2k_2\overline{g}^{\mu\nu}\partial_\mu X\partial_\nu\phi+k_0' X+3k_2'X^2+U'+(k_0+2k_2X)\overline{\nabla}_\mu\left(\sqrt{-\overline{g}}\,\overline{g}^{\mu\nu}\right)\overline{\nabla}_\nu\phi=0\,,
\end{align}
where the Ricci scalar is defined as $\overline{R}=\overline{g}^{\mu\nu}R_{\mu\nu}(\Gamma)\stackrel{!}{=}\overline{g}^{\mu\nu}R_{\mu\nu}(\overline{\text{g}},\partial\overline{\text{g}})$ and the very last term in the last equation vanishes identically due to the Levi-Civita condition.

\subsubsection{\emph{Scalaron (non)propagation at the $1\ell$ quantum level}}

Admittedly, since a full definition of quantum gravity does not exist at the moment, the claim that we obtain results at the quantum level is misleading. Specifically, if we consider a path integral for the usual Einstein gravity there are several problems connected with, but not limited to, the integration over metrics, $[\mathcal{D}\text{g}]$. Instead, results pertaining to gravity are obtained in a semi-classical way around the theory's saddle points and in this approach, even though not exhaustive, it has led in the past to many developments (e.g. see refs.~\cite{Gibbons1977,Witten1998}).

In the path integral approach to quantisation usually one considers a generating functional of the form of
\begin{equation}
    \mathcal{Z}=\int\![\mathcal{D}\Phi]\,e^{\frac{i}{\hbar}\mathcal{S}[\Phi]}\,,
\end{equation}
where $\Phi(x)$ denotes collectively the set of all classical fields present in the theory described by the action $\mathcal{S}[\Phi]$. The measure of the path integral $[\mathcal{D}\Phi]$ denotes integration over all possible configurations (or ``paths'') of $\Phi$
\begin{equation}
    \int\![\mathcal{D}\Phi]=\prod_n\int\!\frac{\mathrm{d}\Phi^n}{\sqrt{2\pi i}}\,,
\end{equation}
where $n$ is a DeWitt index denoting the different species of $\Phi^n(x)$ that are in principle dependent on $x$ and therefore implying that the product runs over points in spacetime.

In the case of the pure $R+R^2$ model (without matter fields) we showed that in the Palatini formalism it culminates to Einstein gravity with a potential term. In principle, quantum corrections of the field $\chi$ can generate a kinetic term effectively making it dynamical. Therefore, a starting point is the path integral
\begin{equation}
    \mathcal{Z}=\int\![\mathcal{D}\text{g}]\,[\mathcal{D}\Gamma]\,\Delta_{FP}\ \text{exp}\left\{i\int\!\mathrm{d}^4x\,\sqrt{-g}\left(\frac{1}{2}R+\frac{\alpha}{4}R^2+\mathscr{L}_\text{gf}+\mathscr{L}_{FP}\right)\right\}\,,
\end{equation}
where we set $\hbar\!=\!1\!=\!M_P$ in order to alleviate some of the notation and $\Delta_{FP}$ denotes the Faddeev-Popov determinant. In what follows we are not concerned with the subtleties of the gauge fixing $\mathscr{L}_\text{gf}$ and the ghost Lagrangian $\mathscr{L}_{FP}$ and thus are ignored at this point. Then, we can introduce a factor of unity in the form of a Gaussian path integral 
\begin{equation}
    \int[\mathcal{D}\chi]\,\text{exp}\left\{\pm i\,\frac{\alpha}{4}\int\!\mathrm{d}^4x\sqrt{-g}\left(\chi-R\right)^2\right\}=\left(\text{det}(\sqrt{-g})\right)^{\mp1/2}\,,
\end{equation}
integrating over a field $\chi(x)$. Thus, the total path integral becomes
\begin{equation}
    \mathcal{Z}=\int[\mathcal{D}\text{g}]\,[\mathcal{D}\Gamma]\,[\mathcal{D}\chi]\,\left(\text{det}(\sqrt{-g})\right)^{-1/2}\text{exp}\left\{i\int\!\mathrm{d}^4x\,\sqrt{-g}\left(\frac{1}{2}(1+\alpha\chi)R-\frac{\alpha}{4}\chi^2\right)\right\}\,.
\end{equation}
A factor of $\alpha^2/4$ (or $(\alpha/2)^{D/2}$ in $D$ dimensions) is absorbed in the redefinition of the measurement of the path integral, which has to be reparametrisation invariant and as such contains a normalisation of its own~\cite{Falls2019}. Under a Weyl rescaling of the form $\bar{g}_{\mu\nu}(x)=(1+\alpha\chi)g_{\mu\nu}(x)$ we may rewrite it as
\begin{equation}
    \mathcal{Z}=\int[\mathcal{D}\overline{\text{g}}]\,[\mathcal{D}\Gamma]\,[\mathcal{D}\chi]\,\left(\text{det}(\sqrt{-\overline{g}})\right)^{-1/2}\left(\text{det}(1+\alpha\chi)\right)^{-1}\text{exp}\left\{i\int\!\mathrm{d}^4x\,\sqrt{-\overline{g}}\left(\frac{1}{2}\overline{g}^{\mu\nu}R_{\mu\nu}(\Gamma)-\frac{\alpha}{4}\frac{\chi^2}{(1+\alpha\chi)^2}\right)\right\}\,.
\end{equation}
Next, by allowing for a redefinition of $\chi$ in terms of 
\begin{equation}
    \overline{\chi}\equiv\frac{\chi}{1+\alpha\chi}\,,
\end{equation}
and since $\left(\text{det}(1+\alpha\chi)\right)^{-1}=\mathcal{J}(\overline{\chi};\chi)$ is exactly the Jacobian of the transformation $\overline{\chi}\rightarrow\chi$, the path integral over the auxiliary field becomes
\begin{equation}
    \int\![\mathcal{D}\overline{\chi}]\,\text{exp}\left\{-i\,\frac{\alpha}{4}\int\!\mathrm{d}^4x\,\sqrt{-\overline{g}}\,\overline{\chi}^2\right\}=\left(\text{det}(\sqrt{-\overline{g}})\right)^{1/2}\,.
\end{equation}
Finally, we obtain
\begin{equation}
    \mathcal{Z}=\int[\mathcal{D}\overline{\text{g}}]\,[\mathcal{D}\Gamma]\,\text{exp}\left\{\frac{i}{2}\int\!\mathrm{d}^4x\,\sqrt{-\overline{g}}\,\overline{g}^{\mu\nu}\,R_{\mu\nu}(\Gamma)\right\}\,.
\end{equation}

Therefore, the pure Starobinsky model in its first-order formulation is equivalent at the quantum level to Einstein gravity in the Palatini formalism, which in turn is equivalent, at least at the classical level, to its conventional metric description. Actually, it is already known that at tree-level any Palatini $f(R)$ theory can be understood as a metric theory with an Einstein-Hilbert term and a potential (or constant) term. The above statement can be readily generalised to any $f(R)$ beyond the $R+R^2$, however the specifics of the (in)equivalence between the two formulations for the Einstein-Hilbert term is not considered here and we refer the reader to ch.~\ref{Ch3:FirstOrder} and references therein for a limited discussion on the subject.

A compelling idea would be to include matter fields in the theory in terms of a scalar field $\phi(x)$ coupled nonminimally to gravity through a term $\xi\phi^2R$, similar to our attempt at tree-level. In this way, a kinetic term for the $\chi$ field can be generated at the quantum level through its interaction with the scalar field $\phi$. Then, the initial path integral reads as
\begin{equation}
    \mathcal{Z}=\int\![\mathcal{D}\text{g}]\,[\mathcal{D}\Gamma]\,[\mathcal{D}\phi]\,\text{exp}\left\{i\int\!\mathrm{d}^4x\,\sqrt{-g}\left(\frac{1}{2}(1+\xi\phi^2)R+\frac{\alpha}{4}R^2-\frac{1}{2}(\nabla\phi)^2-V(\phi)\right)\right\}\,,
\end{equation}
which, after following the procedure of introducing the Gaussian path integral over $\chi$, becomes
\begin{align}
    \mathcal{Z}=\int\![\mathcal{D}\text{g}]\,[\mathcal{D}\Gamma]&\,[\mathcal{D}\phi]\,[\mathcal{D}\chi]\,\left(\text{det}(\sqrt{-g})\right)^{-1/2}\times\nonumber\\
    &\times\text{exp}\left\{i\int\!\mathrm{d}^4x\,\sqrt{-g}\left(\frac{1}{2}(1+\xi\phi^2+\alpha\chi)R-\frac{1}{2}(\nabla\phi)^2-V(\phi)-\frac{\alpha}{4}\chi^2\right)\right\}\,.
\end{align}
Thus, after a metric rescaling $\tilde{g}_{\mu\nu}=(1+\xi\phi^2+\alpha\chi)g_{\mu\nu}$ we obtain
\begin{align}
    \mathcal{Z}=\int\![\mathcal{D}\tilde{\text{g}}]\,[\mathcal{D}\Gamma]&\,[\mathcal{D}\phi]\,[\mathcal{D}\chi]\,\left(\text{det}(\sqrt{-\tilde{g}})\right)^{-1/2}\text{det}\left(\frac{1}{1+\xi\phi^2+\alpha\chi}\right)\times\nonumber\\
    &\times\text{exp}\left\{i\int\!\mathrm{d}^4x\,\sqrt{-\tilde{g}}\left(\frac{1}{2}\tilde{R}-\frac{1}{2(1+\alpha\chi+\xi \phi^2)}(\tilde{\nabla} \phi)^2-\frac{V(\phi)+\frac{\alpha}{4}\chi^2}{(1+\alpha\chi+\xi \phi^2)^2}\right)\right\}\,.
\end{align}
Collectively, we can rewrite it as
\begin{equation}
    \mathcal{Z}=\int\![\mathcal{D}\tilde{\text{g}}]\,[\mathcal{D}\Gamma]\,[\mathcal{D}\phi]\,[\mathcal{D}\chi]\,\left(\text{det}(\sqrt{-\tilde{g}})\right)^{-1/2}\text{exp}\left\{i\,\mathcal{S}_\text{eff}+\frac{i}{2}\int\!\mathrm{d}^4x\,\sqrt{-\tilde{g}}\,\tilde{g}^{\mu\nu}R_{\mu\nu}(\Gamma)\right\}\,,
\end{equation}
where we defined the effective action as
\begin{equation}
    \mathcal{S}_\text{eff}\,[\chi,\phi,\tilde{\text{g}}]\equiv i\, \text{Tr}\log{(1+\alpha\chi+\xi\phi^2)}+\int\!\mathrm{d}^4x\,\sqrt{-\tilde{g}}\left(-\frac{1}{2(1+\alpha\chi+\xi \phi^2)}(\tilde{\nabla} \phi)^2-\frac{V(\phi)+\frac{\alpha}{4}\chi^2}{(1+\alpha\chi+\xi \phi^2)^2}\right)\,,
\end{equation}
where we applied the identity $\text{det}(A)=\text{exp}(\log{(A)})$ for a general matrix $A$.

Next we employ the semiclassical approximation, known also as the saddle point expansion\footnote{The WKB (Wentzel–Kramers–Brillouin) method  used often in QM, can be thought of as a semiclassical approximation and sometimes is used while referring to the saddle point expansion.}, in which we expand the action around its classical solution $\delta_\chi\mathcal{S}=0\implies\chi=\chi_\text{c}$ as\footnote{Notice that the generating functional can now be approximated by
$$\mathcal{Z}=\int\![\mathcal{D}\bar{\text{g}}]\,[\mathcal{D}\Gamma]\,[\mathcal{D}\phi]\left(\text{det}(\sqrt{-\bar{g}})\right)^{1/2}\left(\text{det}\left(\frac{\delta^2\mathcal{S}_\text{eff}}{\delta\chi\delta\chi}\right)\right)^{-1/2} \,\text{exp}\left(\frac{i}{2}\int\!\mathrm{d}^4x\,\sqrt{-\bar{g}}\,\bar{g}^{\mu\nu}R_{\mu\nu}(\Gamma)+i\mathcal{S}_\text{eff}(\chi_\text{c})\right)\,.$$}
\begin{equation}
    \mathcal{S}_\text{eff}[\chi]\approx\mathcal{S}_\text{eff}[\chi_c]+\int\!\mathrm{d}^4x'\left.\frac{\delta\mathcal{S}_\text{eff}}{\delta\chi'}\right|_{\chi=\chi_c}\delta\chi'+\frac{1}{2!}\int\!\mathrm{d}^4x'\int\!\mathrm{d}^4x''\,\left.\frac{\delta^2\mathcal{S}_\text{eff}}{\delta\chi'\delta\chi''}\right|_{\chi=\chi_c}\delta\chi'\delta\chi''+\ldots
\end{equation}
where we denote $\chi(x')\equiv\chi'$, $\chi(x'')=\chi''$ , etc, and
\begin{equation}
    \chi(x)=\chi_c(x)+\delta\chi(x)\,.
\end{equation}
The path integral measure under such a shift trivially becomes $[\mathcal{D}\chi]\rightarrow[\mathcal{D}\delta\chi]$. Clearly, the second term in the expansion vanishes by definition, $\delta^{(1)}_\chi\mathcal{S}(\chi_\text{c})=0$, and the first contributing term involves the secondary functional variation $\delta^{(2)}_\chi\mathcal{S}_\text{eff}$ which reads:
\begin{align}
    \frac{1}{2}\int\!\mathrm{d}^4x'\int\!\mathrm{d}^4x''\frac{\delta^2\mathcal{S}_\text{eff}[\chi_c(x)]}{\delta\chi(x')\delta\chi(x'')}\delta\chi'\delta\chi''=&-\frac{i}{2}\text{Tr}\left[\left(\alpha\,\frac{1+\xi\phi^2-\alpha(\tilde{\nabla}\phi)^2}{(1+\xi\phi^2)^2+4\alpha V}\right)^2\right]\,\delta\chi^2+\\
    &\quad-\frac{1}{2}\int\!\mathrm{d}^4x'\,\delta\chi'\,\bar{D}(x')\,\delta\chi'\,,\nonumber
\end{align}
where we made the definition of $\bar{D}(x)\equiv\sqrt{\tilde{g}}\,D(x)$ with
\begin{equation}
    D(x)\equiv\frac{\alpha}{2}\,\left[(1+\xi\phi^2)^2+4\alpha V\right]^{-3}\,\sum_{n=0}^4\!c_{2n}(\phi)(\tilde{\nabla}\phi)^{2n}\,.
\end{equation}
Here the coefficients $c_{2n}(\phi)$ are defined as
\begin{align}\label{coef}
    c_0(\phi)&=1+4\xi\phi^2+6\xi^2\phi^4+4\xi^3\phi^6+\xi^4\phi^8\,,\nonumber\\
    c_2(\phi)&=-4\alpha(1+3\xi\phi^2+3\xi^2\phi^4+\xi^3\phi^6)\,,\nonumber\\
    c_4(\phi)&=6\alpha^2(1+2\xi\phi^2+\xi^2\phi^4)\,,\\
    c_6(\phi)&=-4\alpha^2(1+\xi\phi^2)\,,\nonumber\\
    c_8(\phi)&=\alpha^4.\nonumber
\end{align}
Then, $\bar{D}(x)$ can be seen as a Strum-Liouville opeartor with, in principle, eigenvalues and eigenvectors given by
\begin{equation}
    \bar{D}(x)\delta\chi_i(x)=\lambda_i\,\delta\chi_i(x),\qquad \lambda_i<\lambda_{i+1}\,\forall i\in\mathbb{N}^*\,.
\end{equation}
Provided we can solve the eigenvalue problem, we can then reduce the calculation to simple Gaussian integrals, and thus obtain the following path integral
\begin{align}
    \mathcal{Z}&=\int\![\mathcal{D}\tilde{\text{g}}]\,[\mathcal{D}\Gamma]\,[\mathcal{D}\phi]\left[\text{det}(\hat{D}(x))\right]^{1/2}\left[\text{det}\left(\frac{4\alpha V+(1+\xi\phi^2)^2}{1+\xi\phi^2-\alpha(\bar{\nabla}\phi)^2}\right)\right]^{-1}\\
    &\quad\text{exp}\left\{i\int\!\mathrm{d}^4x\,\sqrt{-\tilde{g}}\left[\frac{1}{2}\,\bar{g}^{\mu\nu}R_{\mu\nu}(\Gamma)+\frac{1}{(1+\xi\phi^2)^2+4\alpha V}\left[-\frac{1}{2}(1+\xi\phi^2)(\tilde{\nabla}\phi)^2+\frac{\alpha}{4}\,(\tilde{\nabla}\phi)^4-V(\phi)\right]\right]\right\}\nonumber
\end{align}
where now $\hat{D}(x)$ is defined as
\begin{equation}
    \hat{D}(x)=\bar{D}(x)-\frac{i}{\sqrt{-\tilde{g}}}\delta^{(4)}(x)\,\text{Tr}\left[\left(\alpha\,\frac{1+\xi\phi^2-\alpha(\tilde{\nabla}\phi)^2}{(1+\xi\phi^2)^2+4\alpha V}\right)^2\right]\,.
\end{equation}

Crucially, all of terms generated are local, in other words they are dependent on specific spacetime points $x$ and, most notably, on gradients of the field $\phi(x)$, thus we can ignore them and ultimately the effective action seemingly does not obtain any corrections due to $\chi$. Therefore, we claim that the action obtained at the classical level is robust to quantum corrections of the $\chi$ field, which at $1\ell$ remains nondynamical.

\subsubsection*{\emph{Digression on pure $R^2$ gravity}}

Let us pay closer attention to the Weyl-invariant $R^2$ term in the Palatini formalism (see also ref.~\cite{Ghilencea2020}). In fact any term built from the metric tensor schematically as $\propto\sqrt{-\text{det}(\text{g})}\,R^2$ is invariant under a Weyl transformation of the metric and can be generalised in $D$ spacetime dimensions to $\propto\sqrt{-\text{det}(\text{g})}\,R^{(D/2)}$. The action functional in the usual four dimensions is expressed as
\begin{equation}
    \mathcal{S}[\text{g},\Gamma]=\frac{\alpha}{2}\int\!\mathrm{d}^4x\,\sqrt{-g}\,R^2\,,
\end{equation}
where $\alpha$ is now some dimensionless constant. Similarly to what was done before we can obtain the action in its scalar representation by introducing an auxiliary field as follows
\begin{equation}
    \mathcal{S}=\int\!\mathrm{d}^4x\,\sqrt{-g}\left\{\chi^2g^{\mu\nu}R_{\mu\nu}(\Gamma)-\frac{\chi^4}{\alpha}\right\}\,.
\end{equation}
It might be worth noting that the above action has the same form in both the metric and the Palatini formalism. After a Weyl rescaling of the metric as $\hat{g}_{\mu\nu}=(2\chi^2/M_P^2)g_{\mu\nu}$ we obtain the action in the Einstein frame as
\begin{equation}
    \mathcal{S}=\int\!\mathrm{d}^4x\,\sqrt{-\hat{g}}\left\{\frac{M_P^2}{2}\,\hat{g}^{\mu\nu}R_{\mu\nu}(\Gamma)-\frac{M_P^4}{4\alpha}\right\}\,.
\end{equation}
Note that the constant term in the action is also obtained in the large field limit of the pure Starobinsky model, $R+R^2$, which is also equivalent to the strong gravity limit of $R+R^2\sim R^2$. In any case, the constant term can be identified with the cosmological constant after a substantial fine-tuning of the $\alpha$ parameter.

The equations of motion of the theory as was initially formulated in the Jordan frame\footnote{In the Einstein frame they are simply the Einstein field equations following our discussion in ch.~\ref{Ch3:FirstOrder}.} are 
\begin{equation}
    R R_{\mu\nu}-\frac{1}{4}g_{\mu\nu}R^2=0\,,
\end{equation}
and for the connection
\begin{equation}
    \alpha\left(g_{\mu\nu}\nabla^\rho-\delta^\rho_\nu\nabla_\mu\right)R+\mathcal{O}(\nabla^\rho g_{\mu\nu})+\mathcal{O}({T^\rho}_{\mu\nu})=0\,,
\end{equation}
where the terms proportional to the nonmetricity and torsion tensors are grouped. A trivial solution for the connection coefficients is given by ${\Gamma^\rho}_{\mu\nu}=\hat{\{{}_\mu{}^\rho{}_\nu\}}$, namely the Christoffel symbols for the metric $\hat{\text{g}}$ with $\chi^2=\alpha R/2$. Note that in the equations of motion for the metric a factor of $R$ is kept since the Weyl transformation $\text{g}\rightarrow\hat{\text{g}}$ becomes singular if $R=0$. This point is crucial when one discusses the expansion of the metric around a background, usually taken to be Minkowski. Then, similar to the metric case~\cite{Englert1975,Higgs1959,Kounnas2015,AlvarezGaume2016}, an expansion is made around a de Sitter background (also can be taken to be anti-de Sitter) in order to show that indeed the theory includes a spin-$2$ massless graviton and no other propagating degrees of freedom. The analysis trails closely to the metric one and we refrain from discussing further details. Interactions of the pure $R^2$ action and matter in the Palatini formalism are discussed in further detail in ref.~\cite{Ghilencea2020}.

\subsubsection*{\emph{Digression on all-inclusive quadratic gravity \& amalgamations of the Riemann tensor}}

The most general Lagrangian including curvature terms quadratic in contractions of the Riemann tensor in the metric formulation is given by
\begin{equation}
    \mathcal{S}^{(2)}=\int\!\mathrm{d}^4x\,\sqrt{-g}\left\{\alpha\,R^2+4\beta\,R_{\mu\nu}R^{\mu\nu}+\gamma\,R_{\mu\nu\rho\sigma}R^{\mu\nu\rho\sigma}\right\}\,.
\end{equation}
In the special case where $\alpha=-\beta=\gamma$ then the combination of these curvature terms is known as the Gauss-Bonnet term and is a topological invariant of the theory, specifically known as the Euler characteristic of the manifold. That means that it vanishes at the level of equations of motion, however, in general $D\neq4$ dimensions it is dynamical and generally contributes. Obviously, it might be the case that the Gauss-Bonnet term is coupled nonminimally with another field, say for example in a term schematically reading as $\propto f(\phi)\mathcal{E}_{\text{GB}}$.

In contrast, in the Palatini formalism the Gauss-Bonnet term is not necessarily a topological invariant and should in principle include dynamical degrees of freedom that can also be ghosts. In fact, the most general action quadratic in curvature terms contains additional contractions of the Riemann tensor due to the loss of the symmetries as a result of the Levi-Civita connection. As we discussed in ch.~\ref{Ch3:FirstOrder} it is possible to contract the Riemann tensor in different ways as follows
\begin{equation}
    R_{\mu\nu}={R^\rho}_{\mu\rho\nu}\,,\qquad \overline{R}_{\mu\nu}={R^\rho}_{\rho\mu\nu}\,,\qquad \hat{R}{}^\mu{}_\nu=g^{\rho\sigma}{R^\mu}_{\sigma\nu\rho}\,.
\end{equation}
Then the action containing all the possible contractions reads as~\cite{Borunda2008}
\begin{align}
    \mathcal{S}^{(2)}=\int\!\mathrm{d}^4x\,\sqrt{-g}&\Big\{\alpha R^2+\beta_1 R_{\mu\nu}R^{\mu\nu}+\beta_2 R_{\mu\nu}R^{\nu\mu}+\beta_3 \hat{R}_{\mu\nu}\overline{R}^{\mu\nu}+\beta_4 R_{\mu\nu}\overline{R}^{\mu\nu}+\beta_5\overline{R}_{\mu\nu}\overline{R}^{\mu\nu}+\nonumber\\
    &\quad+\beta_6R_{\mu\nu}\hat{R}^{\nu\mu}+\beta_7 R_{\mu\nu}\hat{R}^{\mu\nu}+\beta_8 \hat{R}_{\mu\nu}\hat{R}^{\mu\nu}+\beta_9\hat{R}_{\mu\nu}\hat{R}^{\nu\mu}+\\
    &\qquad+\gamma_1R_{\mu\nu\rho\sigma}R^{\mu\nu\rho\sigma}+\gamma_2R_{\mu\nu\rho\sigma}R^{\nu\mu\rho\sigma}+\gamma_3R_{\mu\nu\rho\sigma}R^{\mu\rho\nu\sigma}+\nonumber\\
    &\qquad\quad+\gamma_4R_{\mu\nu\rho\sigma}R^{\nu\rho\mu\sigma}+\gamma_5R_{\mu\nu\rho\sigma}R^{\rho\nu\mu\sigma}+\gamma_6 R_{\mu\nu\rho\sigma}R^{\rho\sigma\mu\nu}\Big\}\,.\nonumber
\end{align}
Notice that there are no terms involving $\overline{R}^{(\mu\nu)}$ since it is antisymmetric and it vanishes identically. We can further group the terms illustrating the possible symmetries; for a specific set of constants $\beta_i$ and $\gamma_i$ we can rewrite the action as
\begin{align}
    \mathscr{L}^{(2)}\supset \alpha R^2+&\beta'_1R_{\mu\nu}R^{(\mu\nu)}+\beta'_2R_{\mu\nu}\hat{R}^{(\mu\nu)}+\beta'_3\hat{R}_{\mu\nu}\hat{R}^{(\mu\nu)}+\left(\beta'_4R_{\mu\nu}+\beta'_5\overline{R}_{\mu\nu}+\beta'_6\hat{R}_{\mu\nu}\right)\overline{R}^{\mu\nu}+\\
    &\quad+\gamma'_1R_{\mu\nu\rho\sigma}R^{\mu\nu\rho\sigma}+\gamma'_2R_{\mu\nu\rho\sigma}R^{\mu[\nu\rho\sigma]}+\gamma'_3R_{\mu\nu\rho\sigma}R^{[\{\mu\nu\}\{\rho\sigma\}]}\,,
\end{align}
where we abused the notation slightly by defining $[\{ab\}\{cd\}]=abcd-cdab$ to mean antisymmetrisation of the set of indices appearing in the brackets. It is now straightforward to show that the Levi-Civita condition trivially reproduces the action in the metric formalism by using the Riemann tensor symmetries, for specific values of the constants $\beta_i$ and $\gamma_i$. Regardless, if the connection is metric-affine, calculations involving the complete quadratic action are complicated and beyond the scope of this work.

As an aside, it is trivial to realise that the conventional conformal gravity in the metric formulation (known also as Weyl gravity), comprised from the squared of the Weyl tensor, is not actually invariant under Weyl transformations in its first-order formulation. The Lagrangian in general $D$ spacetime dimensions reads
\begin{equation}
    \mathscr{L}=C_{\mu\nu\rho\sigma}C^{\mu\nu\rho\sigma}\equiv C^2=R_{\mu\nu\rho\sigma}R^{\mu\nu\rho\sigma}-\frac{4}{D-2}R_{\mu\nu}R^{\mu\nu}+\frac{2}{(D-1)(D-2)}R^2\,,
\end{equation}
where the Weyl tensor is defined as
\begin{equation}
    C_{\mu\nu\rho\sigma}\equiv R_{\mu\nu\rho\sigma}-\frac{4}{D-2}\,g_{\mu\lambda}g_{\nu\kappa}g{}^{[\lambda}{}_{[\rho} R{}_{\sigma]}{}^{\kappa]}+\frac{2}{(D-1)(D-2)}R\,g{}_{\mu[\rho}g{}_{\sigma]\nu}\,.
\end{equation}
The tensor has many important properties; one of them is that in the case that it vanishes the metric is locally conformally flat. Especially, in dimensions $D=2$ the tensor vanishes identically meaning that any $2$-dimensional (smooth) Riemannian manifold is conformally flat, and in $D=3$ dimensions the condition of a vanishing Cotton tensor - built from contractions of the metric and the Weyl tensor - is necessary and sufficient for the metric to be conformally flat. In the general case of $D\geq4$ the condition of a vanishing Weyl tensor is simply sufficient.

Evidently, the term $C^2(\Gamma)$ is not invariant under a rescaling of the metric since it manifestly includes terms of the Riemann tensor. It was shown in ref.~\cite{Alvarez2017} that in the Palatini formalism a generalisation of the action $\int C^2\mathrm{d}\text{vol}$ that still respects the Weyl invariance and in which the Weyl tensor has the same symmetries as in the metric formulation does not exist.

\subsection{\emph{Frames of action}}

In what was presented until this very point we avoided discussing the issue of the Jordan and Einstein frame and in many cases their equivalence is implicit. In fact, this is largely the stance we adopt in this work, meaning that as far as classical (or semiclassical for that matter) theories are concerned the two frames are mathematically equivalent and observationally indistinguishable.

Let us begin by formally defining what these frames are. In the Jordan frame, denoted hereafter by $\mathscr{J}$, gravity is nonminimally coupled with matter field(s) in the form of $f(\Phi)R(\text{g})$, in which if we restrict ourselves to scalar fields the function of the matter field(s) $\Phi$ is quadratic in them. Lagrangians that include these kinds of couplings were first considered in the $60$'s, e.g. in the bosonic string theory the action given in the formally known as string frame\footnote{For the purposes of this work the string and Jordan frame are effectively the same, or at least refer to the same dynamics, namely the nonminimal coupling of the scalar field to the Einsten-Hilbert term.}~\cite{Polchinski2007}
\begin{equation}
    \mathcal{S}=\int\!\mathrm{d}^{n}x\,\sqrt{-g}\,\frac{e^{-2\phi}}{2\kappa_n^2}\left\{R-4g^{\mu\nu}\nabla_\mu\phi\nabla_\nu\phi-\frac{1}{12}H_{\mu\nu\rho}H^{\mu\nu\rho}-\frac{n-26}{3\,\ell_s^2}\right\}\,.
\end{equation}
Also significant impact had the Brans-Dicke (BD) theory given as~\cite{Brans1961}
\begin{equation}
\mathcal{S}=\frac{1}{16\pi}\int\!\mathrm{d}^4x\,\sqrt{-g}\left\{\phi R-\frac{\omega}{\phi}g^{\mu\nu}\nabla_\mu\phi\nabla_\nu\phi\right\}+\int\!\mathrm{d}^4x\,\sqrt{-g}\,\mathscr{L}_m\,,
\end{equation}
where $\mathscr{L}_m$ denotes the matter Lagrangian that is coupled universally with gravity via the $\sqrt{-g}$ term. The BD theory, and other descendant modern scalar-tensor theories, are inspired by the original paper of Jordan~\cite{Jordan1955}. Many of these theories, like the ones considered here, are motivated by results obtained at the quantum level, in which a coupling of the scalar field to the Einstein-Hilbert term arises due to quantum corrections of the field (at the one-loop level). In this way the Jordan frame is intrinsically connected with high energy physics, or at least its origins were, and as such any questions regarding its nature are actually better posed in the context of our lackluster understanding of high energy physics phenomena, specifically the inexact methods used in obtaining these effective actions.

The Einstein frame, denoted hereafter by $\mathscr{E}$, is defined in a straightforward way as the frame in which the gravitational part is simply the Einstein-Hilbert term and matter fields are coupled to gravity minimally (through the universal term $\sqrt{-g}$). Another point is that the kinetic terms of the matter fields are canonical, however here we extend slightly the definition of $\mathscr{E}$ to include noncanonical kinetic terms, only for the sake of not introducing more terminology to refer to an action functional in an ``intermediate frame''. For example the bosonic string theory can be rewritten in the Einstein frame as
\begin{equation}
    \mathcal{S}=\frac{1}{2\kappa_n^2}\int\!\mathrm{d}^nx\,\sqrt{-\bar{g}}\left\{\bar{R}-\frac{4}{n-2}\bar{g}^{\mu\nu}\bar{\nabla}_\mu\bar{\phi}\bar{\nabla}_\nu\bar{\phi}-\frac{e^{-8\bar{\phi}/(n-2)}}{12}H_{\mu\nu\rho}H^{\mu\nu\rho}-e^{4\bar{\phi}/(n-2)}\frac{n-26}{3\ell_s^2}\right\}\,,
\end{equation}
after a Weyl rescaling of the metric as
\begin{equation}
    \bar{g}_{\mu\nu}(x)=e^{-2\phi(x)}g_{\mu\nu}(x)\,.
\end{equation}
In fact, the existence of the Weyl rescaling is at the heart of the transformation from the Jordan to the Einstein frame, used for the first time in ref.~\cite{Wagoner1970}. Moreover, usually the Einstein frame is discussed in context of passing from 
\begin{equation}
    \{\text{g},\Phi\}\in\mathscr{J}\longmapsto\{\bar{\text{g}},\bar{\Phi}\}\in\mathscr{E}\,
\end{equation} 
and rarely referred to as a standalone frame.\footnote{Interestingly, the transition $\mathscr{E}\rightarrow\mathscr{J}$ is never considered.
%which has further solidified the position that $\mathscr{E}$ is an artificial frame in the eyes of many in the scientific community.
} Note that the matter Lagrangian is rescaled by the transformation and thus conventional wisdom from $\mathscr{J}$ is not transferred trivially to $\mathscr{E}$ but rather obtains a spacetime dependence.

While the two frames are mathematically equivalent that does not necessarily imply \emph{physical} equivalence and, in general, there are three standpoints regarding to which frame is the ``correct'' one,\footnote{Unfortunately the stances are not often discussed in the literature, apart from specific papers discussing the (non)equivalence of the two frames, and as such it is usually masked or quietly implied. More than that many authors tend to cast the issue under a philosophical light, in many ways relegating it to something reminiscent of discussions on interpretations of quantum mechanics.} in terms of truly describing physics~\cite{Faraoni1999}; (i) $\mathscr{J}$ is physical and $\mathscr{E}$ is not, (ii) $\mathscr{E}$ is physical while $\mathscr{J}$ is not, (iii) both frames give an equivalent description of physics. The question can be extended to if other (infinite in principle) frames that are conformally related to the initial Jordan frame, via a metric rescaling of $\text{g}\mapsto\hat{\text{g}}$, are also physical. Usually the first viewpoint originates from particle physics intuition, since in $\mathscr{J}$ the interactions between matter fields and, more generally, nongravitational physics is well understood, as opposed to the gravitational action which in this case is complicated. In contrast, the second viewpoint is adopted mainly by cosmologists, since in the $\mathscr{E}$ frame gravity is described by the well-known Einstein-Hilbert term and hence GR is inferred, with the ``sacrifice'' that the matter sector has now an involved expression or is subject to field redefinitions in order to obtain a canonical kinetic term (with a nonpolynomial potential). These comments mostly concern how close is one frame to conventional wisdom and as such cannot be taken into full consideration if we are to ultimately label a conformal frame as ``physical''.

It is usually cited that an issue with $\mathscr{E}$ is that the Weak Equivalence Principle (WEP) is violated, unless matter is conformally coupled with gravity such that its stress-energy tensor is covariantly conserved in $\mathscr{E}$. However, the WEP can very well be violated in nature, and is in fact an avenue of on-going research together with possible detectable violations of EEP and the Strong Equivalence Principle (SEP), for example in a quantum system with a gravitational potential the WEP is respected locally and only for specific forms of the potential. Another point is that in $\mathscr{J}$ the Weak Energy Condition (WEC) is possibly violated, while in $\mathscr{E}$ the energy density is positive definite~\cite{Magnano1994,Faraoni1999}. However, the inconsistency is not measurable observationally since there does not exist a physical observable that for timelike vectors $u^\mu$ has a predicted value of $T_{\mu\nu}u^\mu u^\nu$ that is conformally invariant~\cite{Flanagan2004}. Moreover, violation of the energy conditions is not uncommon and historically some of them have been abandoned, like the Trace Enegy Condition (TEC), while the Strong Energy Condition (SEC) being on the fence as of now. Accepted wisdom suggests that at least the Null Energy Condition (NEC) should be satisfied, however there are indications that it too can be violated by quantum corrections~\cite{Visser2000}.

It seems that most of the arguments arise from quantum effects violating commonplace classical intuition in, at least, one of the frames and possibly in others conformally related to them. After all, the quantum behaviour of most of these systems is not completely understood and therefore this approach can be misleading. A point of caution would be the interpretation of scales from frame to frame (e.g. see ref.~\cite{Duff2002}), since they differ between them, but local or nongravitational physics remain the same.\footnote{For example, in $\mathscr{J}$ and $\mathscr{E}$ one can read off the Planck mass as $M_\mathscr{J}\equiv M\Omega$ and $M_\mathscr{E}=M$ respectively, under a rescaling $\text{g}\mapsto\overline{\text{g}}= \Omega^2 \text{g}$. Then distances measured in Planck units are invariant; schematically
$$M_\mathscr{J}^2\,\mathrm{d}s^2=M_\mathscr{J}^2\,g_{\mu\nu}\mathrm{d}x^\mu\mathrm{d}x^\nu=M_\mathscr{E}^2\,\overline{g}_{\mu\nu}\mathrm{d}x^\mu\mathrm{d}x^\nu=M_\mathscr{E}^2\,\mathrm{d}\overline{s}^2\,.$$} Thus, since the discrepancies between $\mathscr{J}$ and $\mathscr{E}$ cannot be measured observationally\footnote{Or they cannot possibly falsify one of the two. For example let us consider one frame which is related to the initial $\mathscr{J}$ by $\mathrm{d}s^2\longrightarrow\mathrm{d}\overline{s}^2=-\mathrm{d}\tau^2+\mathrm{d}\Sigma$, which is conformally flat. In this frame, say $\mathscr{M}$, since the universe is static the photons do not redshift due to the vanishing Hubble flow. Naively we might say that it is unphysical, however since the electron mass varies in time in $\mathscr{M}$ due to the conformal transformation, we obtain $\overline{m}(\tau)=m/(1+z)$, where $1+z\equiv a^{-1}(\tau)$. Then, the energy level in $\mathscr{M}$ is $\overline{E}_n=E_n/(1+z)$, where $E_n$ is the energy level in $\mathscr{J}$, and therefore in a level transition $n\rightarrow n'$ the frequency of photons is $\overline{E}_{nn'}=E_{nn'}/(1+z)$. Exactly what is predicted by Hubble's law in $\mathscr{J}$. See ref.~\cite{Domenech2016} for a more details regarding the subject.} (at the moment) it hints at a more conservative point of view, somewhere in between mathematical and physical equivalence, starting from the energy scale of classical physics to possibly high-energy physics. An alternative avenue of addressing the issue is to reformulate any theory in terms of quantities that transform covariantly~\cite{Burns2016,Karamitsos2018} or are invariant under conformal transformations~\cite{Catena2007,Jaerv2015,Jaerv2015a,Kuusk2016a,Kuusk2016,Karam2018}, however either the theories cannot be conventionally interpreted (and/)or they are assumed ad hoc with uncertain origins.

\section{Minimally coupled matter fields}

In this and following sections we shift our focus to understanding the inflationary predictions and the high energy behaviour of specific (inflationary) models coupled minimally (and in next section nonminimally) to gravity including higher-order curvature invariants under the assumption of the first-order formalism. Specifically, we attempt to merge the Starobinsky model of inflation, which as was illustrated in previous sections does not provide a dynamical degree of freedom in the Palatini formulation, with other prominent inflationary models. 

\subsection{\emph{Natural inflation}}

One interesting model capable of describing inflation is the one dubbed the \emph{natural inflation} model. When it was first introduced~\cite{Freese1990,Adams1993} it had the attractive feature that the potential has a flat enough plateau capable of generating an ample amount of (slow-roll) inflation. Since then, the feature of a flat potential is not so hard to come by, especially when the identification of the proposed inflationary model with low-energy physical models is lost. In fact, following our previous discussion (in sec.~\ref{subsec:PalatiniForm}), in the Palatini--$R^2$ models the potential in the Einstein frame is able to generate a flat region that in principle can provide a successful inflation.

In natural inflation the inflaton is an axionic field in the sense that its potential is shift-symmetric protecting it from quantum corrections, which in general can ruin the flat slope of any potential, thus elevating the form of the potential to ``natural''. Then, during the early universe an explicit breaking of the shift symmetry results to slow-roll expansion realising the inflaton as a pseudo-Nambu-Goldstone boson. 

In order to highlight exactly that, let us consider a complex field $\Phi$ with two degrees of freedom in the following representation
\begin{equation}
    \Phi=-\vartheta\,e^{i\phi}\,,
\end{equation}
where $\vartheta$ and $\phi$ are real fields. Notice that the complex field $\Phi$ is exactly invariant under the shift transformation of $\phi\mapsto\phi+2\pi$. There are many ways to introduce such a field (even without considering string theory) into the theory, one of them originally proposed a coupling to the SM as a second Higgs doublet known as the Peccei-Quinn-Weinberg-Wilczek (PQWW) axion with the primary objective to solve the strong CP problem~\cite{Peccei1977,Peccei1977a,Wilczek1978,Weinberg1978}, however it has since been excluded by experiment but other forms of axions still survive. A kinetic term for $\Phi$ can be expressed as
\begin{equation}
    \left|\partial\Phi\right|^2=(\partial\vartheta)^2+\vartheta^2(\partial\phi)^2\,.
\end{equation}
Assuming a large VEV for $\vartheta\approx v$ we can canonically normalise $\phi$ as $\bar{\phi}=\phi/v$ and decouple the radial from the angular component. It was proposed that since $\vartheta\approx\text{const.}$ it can be identified with Dark Energy and $\phi$ with Dark Matter (e.g. see refs.~\cite{Duffy2009,Brandenberger2021} and references therein). 

Next, let us generalise this idea to a charged scalar field $\Phi$ under a continuous global $U(1)$ symmetry. The most general renormalisable potential reads
\begin{equation}
    V(\Phi)=-m_\Phi^2|\Phi|^2+\lambda_\Phi|\Phi|^4\,,
\end{equation}
where it is assumed that $\lambda_\Phi>0$. Then a spontaneous symmetry breaking can occur when $\vartheta\rightarrow\lambda_\Phi m_\Phi^2$ and the potential can be written as~\cite{Freese1990}
\begin{equation}
    V(\phi)=\Lambda^4\left(1\pm \cos{\frac{\phi}{v}}\right)\,,
\end{equation}
where $\Lambda$ would be the scale of nonperturbative physics at which the shift symmetry is broken. It is straightforward to see that the potential is periodic, therefore it has at least one maximum and one minimum in an interval of $\phi/v\in[-\pi,\pi]$, and still respects the residual shift symmetry $\phi\mapsto\phi+2n\pi v$ meaning that only a subgroup of the original minima survive.

In what follows, we consider only the positive root of the potential, which has been shown to drive inflation in terms of appropriate scales $\Lambda\!\approx\!M_\text{GUT}$ and $v\!\approx\!M_P$, however its predictions lie on the unfavourable region of observations (see fig.~\ref{fig:Planck2018}). 

It is interesting then to see how predictions of the natural inflation model change in the framework of the Palatini-$R^2$; the total Lagrangian at the scale of inflation reads
\begin{equation}
    \mathscr{L}=\frac{M_P^2}{2}R+\frac{\alpha}{4}R^2-\frac{1}{2}(\partial\phi)^2-V(\phi)\,,
\end{equation}
where $M_P$ is the reduced Planck mass and the scalar potential is of the form
\begin{equation}
    V(\phi)=\Lambda^4\left(1+\cos{\frac{\phi}{f}}\right)\,,
\end{equation}
where $f$ is some scale that the global shift symmetry of the inflaton was spontaneously broken and $\Lambda$ is the soft explicit symmetry breaking scale giving the boson its mass. Then after we express the total action functional in its scalar representation by introducing an auxiliary field $\chi$ to assume the role of the $R^2$ term, as was done in previous section, and performing a Weyl rescaling 
\begin{equation}
    \overline{g}_{\mu\nu}(x)=\left(1+\frac{\alpha\chi^2}{M_P^2}\right)g_{\mu\nu}(x)\,,
\end{equation}
we obtain the action in the Einstein frame~\cite{Antoniadis2019}:
\begin{equation}\label{Action:PalatiniNatural}
    \mathcal{S}=\int\!\mathrm{d}^4x\,\sqrt{-\overline{g}}\left\{\frac{M_P^2}{2}\,\overline{g}^{\mu\nu}\,R_{\mu\nu}(\Gamma)-\frac{1}{2}\frac{(\overline{\nabla}\phi)^2}{\left(1+\displaystyle{\frac{4\alpha}{M_P^4}}V(\phi)\right)}-\frac{V(\phi)}{1+\displaystyle{\frac{4\alpha}{M_P^4}}V(\phi)}+\mathcal{O}\left((\overline{\nabla}\phi)^4\right)\right\}\,,
\end{equation}
where higher-order kinetic terms $\mathcal{O}\left((\overline{\nabla}\phi)^4\right)$ are neglected considering slow-roll inflation. 

The equations of motion derived from the above action are a special case of the ones given in eq.~\eqref{Eq:PalatiniGenFieldEQs}. The generalised Einstein field equations read as
\begin{equation}
    M_P^2\left(\overline{R}_{\mu\nu}-\frac{1}{2}\overline{g}_{\mu\nu}\overline{R}\right)=\frac{1}{\left(1+4\tilde{\alpha}V(\phi)\right)}\left\{\overline{\nabla}_\mu\phi\overline{\nabla}_\nu\phi-\overline{g}_{\mu\nu}\left(\frac{1}{2}(\overline{\nabla}\phi)^2+V(\phi)\right)\right\}\,,
\end{equation}
while the generalised Klein-Gordon equation, after assuming a spatially homogeneous field $\phi(x)=\phi(\mathbf{x})$ in a flat FRW background with a metric $\mathrm{d}s^2=-\mathrm{d}t^2+a^2(t)\mathrm{d}x^2$, becomes
\begin{equation}
    3M_P^2H^2=\frac{1}{\left(1+4\tilde{\alpha}V(\phi)\right)}\left(\frac{1}{2}\dot{\phi}^2+V(\phi)\right)\,,
\end{equation}
where we defined $\tilde{\alpha}\equiv \alpha/M_P^4$.

\begin{figure}
    \centering
    \includegraphics[scale=0.5]{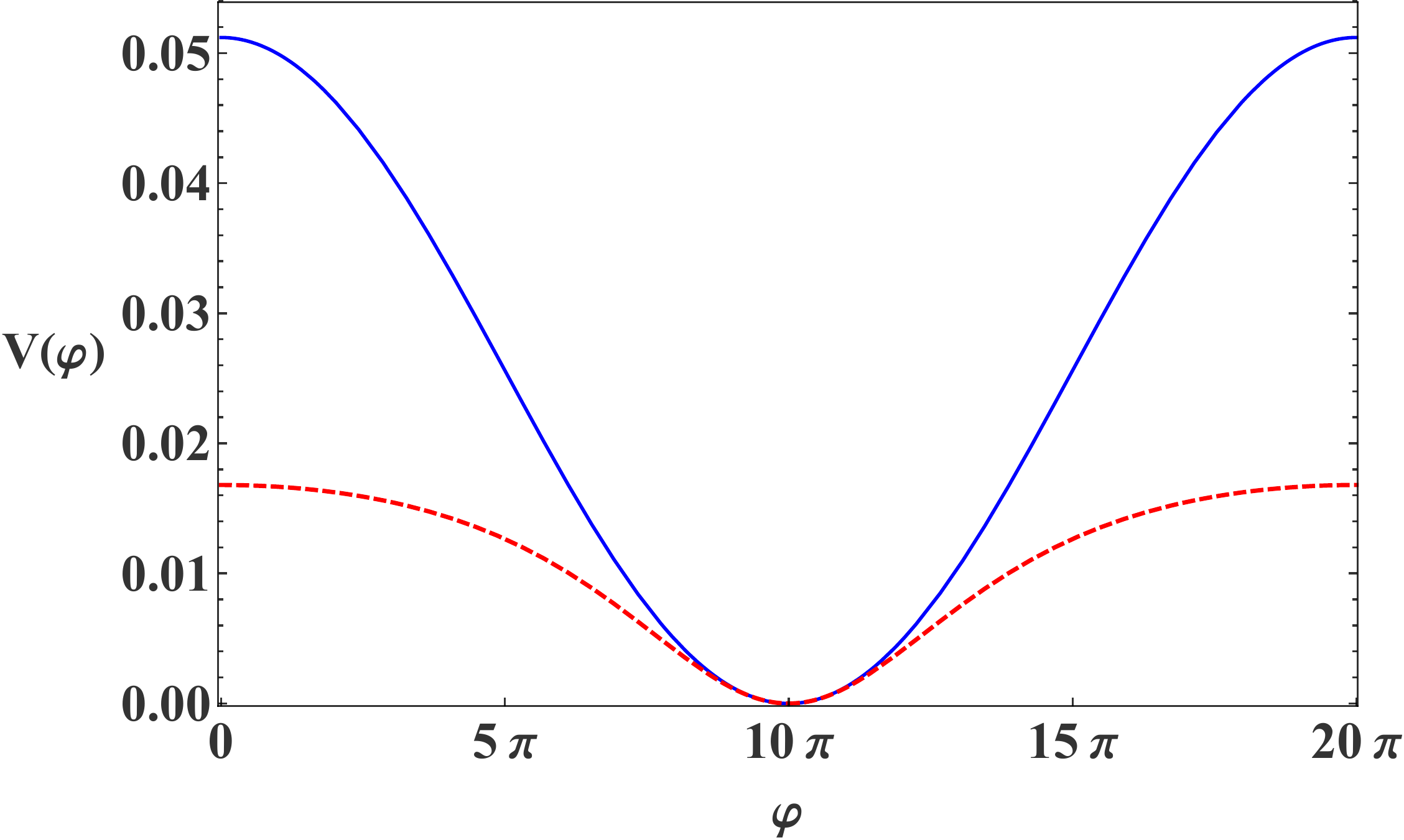}
    \caption{Plot of the original potential in the Jordan frame (blue solid line) against the scalar potential in the Einstein frame after it is flattened by the $R^2$ term (red dashed line). The chosen values of the parameters are $\Lambda=0.4$, $\alpha=10$, $f=10$ and $M_P=1$.}
    \label{fig:NaturalInf-PotU}
\end{figure}

Next, we can canonically normalise the scalar field by
\begin{align}
    \varphi=\int_0^\phi\!\frac{\mathrm{d}\phi'}{\sqrt{1+\displaystyle{\frac{4\alpha}{M_P^4}}V(\phi')}}&=\int_0^\phi\!\frac{\mathrm{d}\phi'}{\sqrt{1+\displaystyle{\frac{4\alpha\Lambda^4}{M_P^4}}\left(1+\cos{\frac{\phi'}{f}}\right)}}\nonumber\\
    &=\int_0^\phi\!\frac{\mathrm{d}\phi'}{\sqrt{1+\displaystyle{\frac{8\alpha\Lambda^4}{M_P^4}}\left(1-\sin^2{\frac{\phi'}{2f}}\right)}}\nonumber\\
    &=\frac{2fM_P^2}{\sqrt{M_P^4+8\alpha\Lambda^4}}\int_0^\frac{\phi}{2f}\!\frac{\mathrm{d}(\phi'/2f)}{\sqrt{1-\displaystyle{\frac{8\alpha\Lambda^4}{M_P^4+8\alpha\Lambda^4}\,\sin^2{\frac{\phi'}{2f}}}}}\,.
\end{align}
Then the integral can be represented using the incomplete elliptic integral of the first kind $\mathcal{F}$ defined as
\begin{equation}\label{Cond:EllipticIntegral1st}
    \mathcal{F}(\vartheta\,|\,k^2)=\int_0^\vartheta\frac{\mathrm{d}x}{\sqrt{1-k^2\sin^2{x}}}\,.
\end{equation}
Then we obtain
\begin{equation}
    \varphi=\frac{2fM_P^2}{\sqrt{8\alpha\Lambda^4+M_P^4}}\,\mathcal{F}(\frac{\phi}{2f}\,|\,\frac{1}{1+\displaystyle{\frac{M_P^4}{8\alpha\Lambda^4}}})\,.
\end{equation}
Now, the inflaton potential in terms of the canonically normalised field $\varphi$ reads
\begin{equation}
    U(\varphi)\equiv\frac{V(\varphi)}{1+\frac{4\alpha}{M_P^4}V(\varphi)}=\frac{\text{cn}^2(\frac{\varphi}{2fM_P^2}\sqrt{M_P^4+8\alpha\Lambda^4}\,|\,\frac{8\alpha\Lambda^4}{8\alpha\Lambda^4+M_P^4})}{1+\frac{8\alpha\Lambda^4}{M_P^4}\,\text{cn}^2(\frac{\varphi}{2fM_P^2}\sqrt{M_P^4+8\alpha\Lambda^4}\,|\,\frac{8\alpha\Lambda^4}{8\alpha\Lambda^4+M_P^4})}\,,
\end{equation}
where $\text{cn}(\varphi\,|\,k^2)$ is the Jacobi elliptic function. In fig.~\ref{fig:NaturalInf-PotU} we show that the potential in the Einstein frame is flattened compared to the one in the Jordan frame. The matter Lagrangian can be expressed in terms of the field $\varphi$ reading
\begin{equation}
    \mathscr{L}\supset-\frac{1}{2}(\overline{\nabla}\varphi)^2-U(\varphi)\,,
\end{equation}
meaning that we can employ the usual first-order expressions for the inflationary observables $n_s$ and $r$ in terms of the first and second slow-roll parameters $\epsilon_V$ and $\eta_V$, with respect to the canonically normalised field $\varphi$.

Since the expressions are rather involved for both the normalised field $\varphi$ as well as the initial field $\phi$, the calculations are done numerically and the results are presented below. Importantly, the power spectrum of scalar perturbations\footnote{We can safely use the convectional expressions for the inflationary observables obtained in the metric formalism, since the action contains the EH term and a nontrivial matter sector. However, see ref.~\cite{Tamanini2011} for a detailed analysis of the inflationary spectrum strictly in the Palatini formalism.} reads
\begin{equation}
    \mathcal{A}_s=\left.\frac{f^2\Lambda^4}{6\pi^2M_P^2}\,\frac{\text{cn}^2(\frac{\varphi}{2fM_P^2}\sqrt{M_P^4+8\alpha\Lambda^4}\,|\,\frac{8\alpha\Lambda^4}{8\alpha\Lambda^4+M_P^4})}{\text{sn}^2(\frac{\varphi}{2fM_P^2}\sqrt{M_P^4+8\alpha\Lambda^4}\,|\,\frac{8\alpha\Lambda^4}{8\alpha\Lambda^4+M_P^4})}\right|_{\varphi=\varphi_i}\sim2\times10^{-9}\,,
\end{equation}
where $\text{sn}(\varphi\,|\,k^2)$ is again a Jacobi elliptic function. It is important to note that in the Einstein frame the Levi-Civita condition is a solution to the equation of motion for $\Gamma$, thus it is possible to employ the conventional methods used in the metric single-field inflation.

\begin{figure}
    \centering
    \includegraphics[scale=0.5]{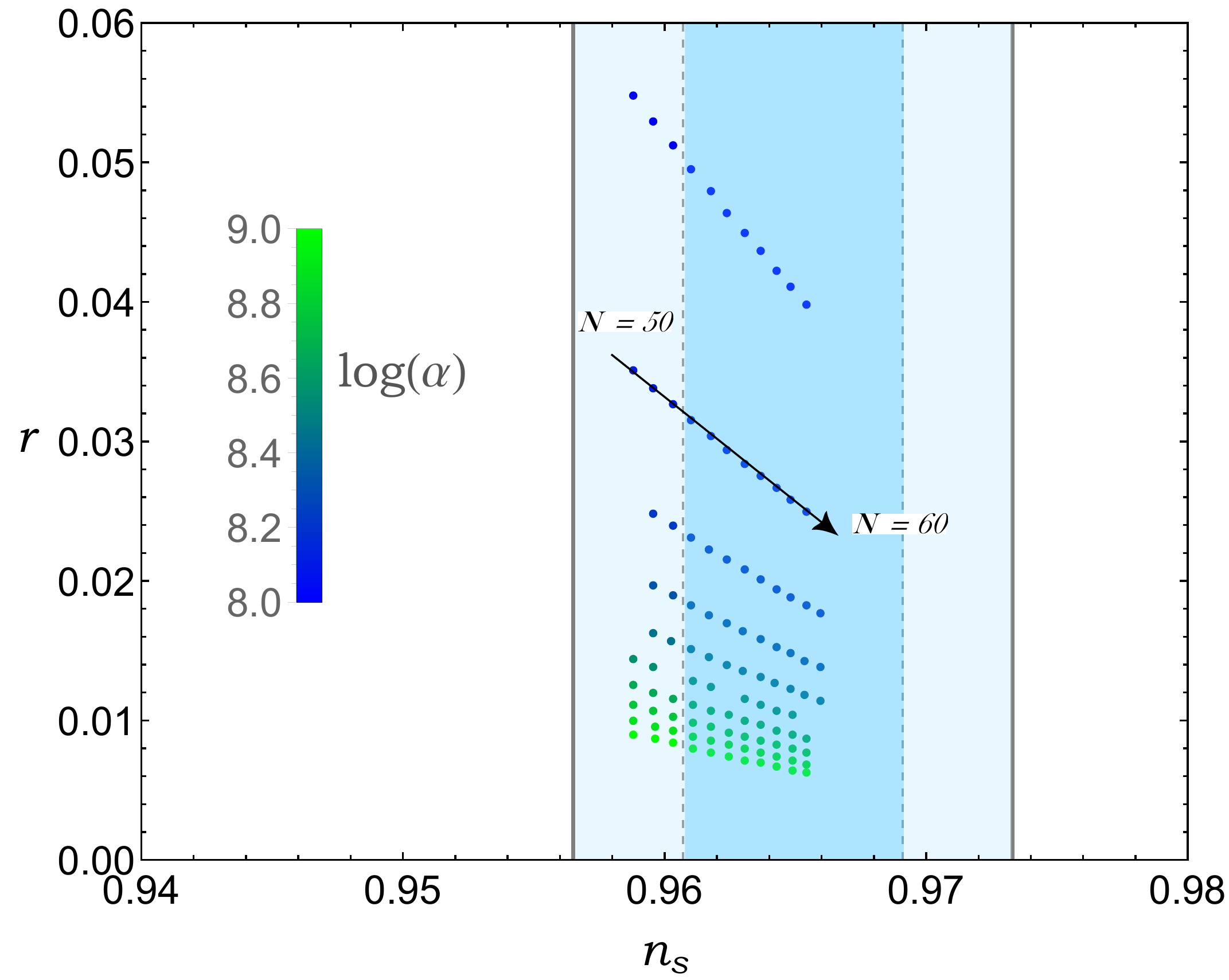}
    \caption{A plot of the $r-n_s$ region for various values of $\alpha\in[10^8,10^9]$ and $\Lambda=0.008$, $f=10$ and $M_P=1$. The light and dark blue region denote the $2\sigma$ and $1\sigma$ allowed region from the Planck2018 collaboration. The values of $\alpha$ are specifically chosen such that the plot is within the bounds of $r\lesssim0.056$. Each dot is a solution for a specific value of $N$, which from left to right is increasing up to a maximum value of $N=60$ and each ``line'' is formed for a specific value of $\alpha$, as displayed in the colour coding of the legend.}
    \label{fig:NaturalInf-RNs}
\end{figure}

In fig.~\ref{fig:NaturalInf-RNs} we plot the values of the predicted tensor-to-scalar ratio against the spectral index $n_s$, after numerically solving the equation of $\epsilon_V(\varphi_i)\equiv1$ in order to obtain the field value at the end of inflation $\varphi=\varphi_i$ and using that to compute $\varphi_f$ for various number of $e$-foldings $N\in\left[50,60\right]$. The values of the free parameters are $\Lambda=8\times10^{-3}$, $f=10$ and $\alpha\in[10^8,10^9]$ with $M_P$ fixed to unity. Values of $f\lesssim7$ (in natural units) are excluded by the recent Planck collaboration, fact which we abide by in this study. Note that for values of $\alpha\lesssim10^{7}$ (for the specific values of the other parameters) are also excluded since the tensor-to-scalar ratio escapes the preferred region. By increasing the value of $\alpha$ we obtain lower values of $r$, however $n_s$ remains unchanged. This is in fact a general feature of the Palatini-$R^2$ models, which is also discussed in ref.~\cite{Enckell2019} in a model independent way. All of the parameter values presented here lead to the correct value for the scalar power spectrum $\mathcal{A}_s\sim10^{-9}$.

In the next figure, fig.~\ref{fig:NaturalInf-RNsf}, we obtain a similar graph to the one in fig.~\ref{fig:NaturalInf-RNs}, by varying the parameter $f$ and keep $\alpha$ constant instead. The values of the parameters are $\Lambda=8\times10^{-3}$, $\alpha=10^9$ and $f\in[5,10]$, while once again fixing $M_P=1$. The plot is reminiscent of the one given by the Planck collaboration for the natural inflation, however in the case of Palatini-$R^2$ values of $f\gtrsim6$ are allowed leading to acceptable values of $r$. This is in contrast with the conventional metric case in which the model lies outside even the $2\sigma$ region of observations.

\begin{figure}
    \centering
    \includegraphics[scale=0.5]{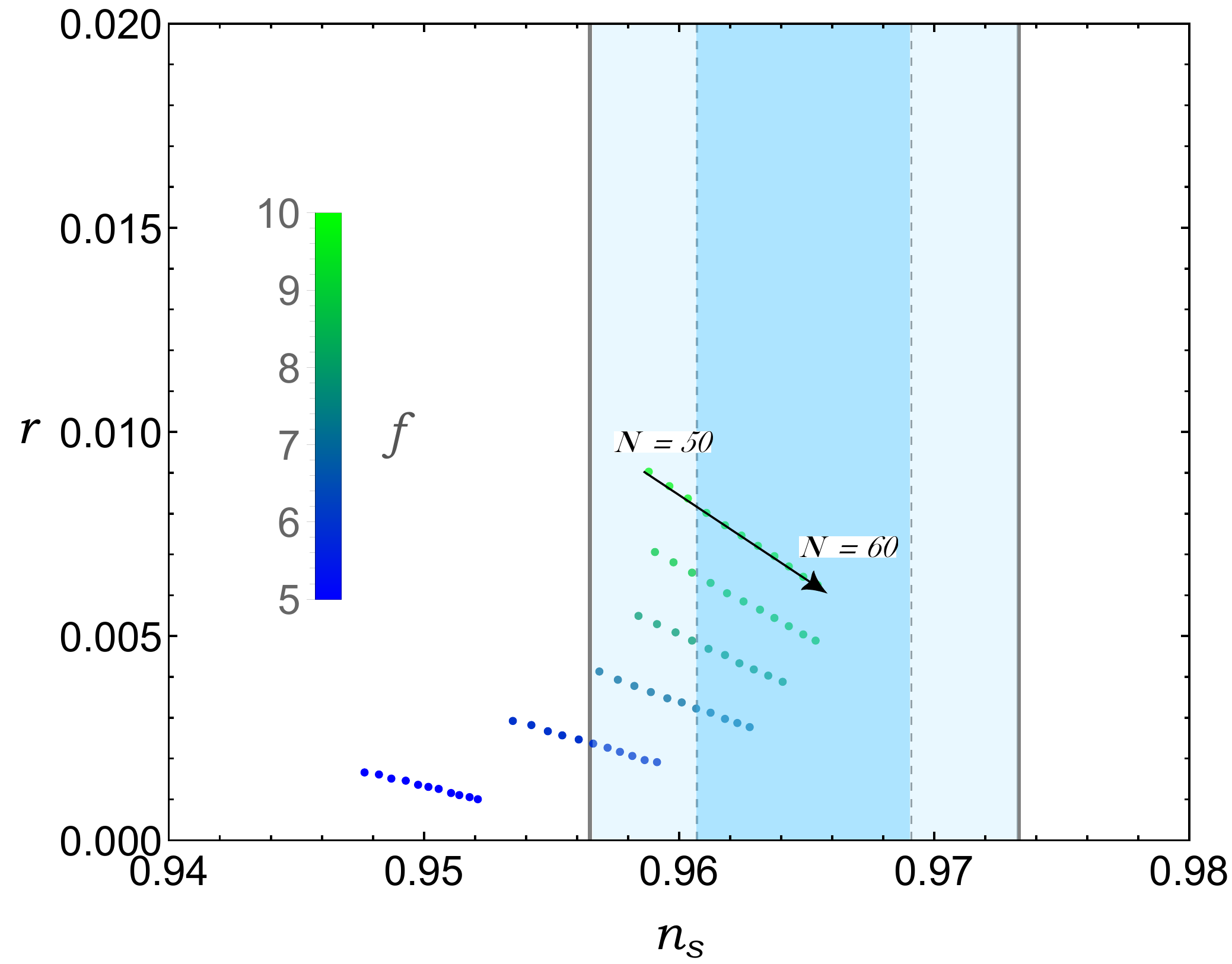}
    \caption{A plot of the $r-n_s$ region for various values of $f\in[5,10]$ and $\Lambda=0.008$, $\alpha=10^9$ and $M_P=1$. The light and dark blue region denote the $2\sigma$ and $1\sigma$ allowed region from the Planck2018 collaboration. Each dot is a solution for a specific value of $N$, which from left to right is increasing up to a maximum value of $N=60$ and each ``line'' is formed for a specific value of $f$ based on the colour coding of the legend.}
    \label{fig:NaturalInf-RNsf}
\end{figure}

\begin{figure}
    \centering
    \includegraphics[scale=0.25]{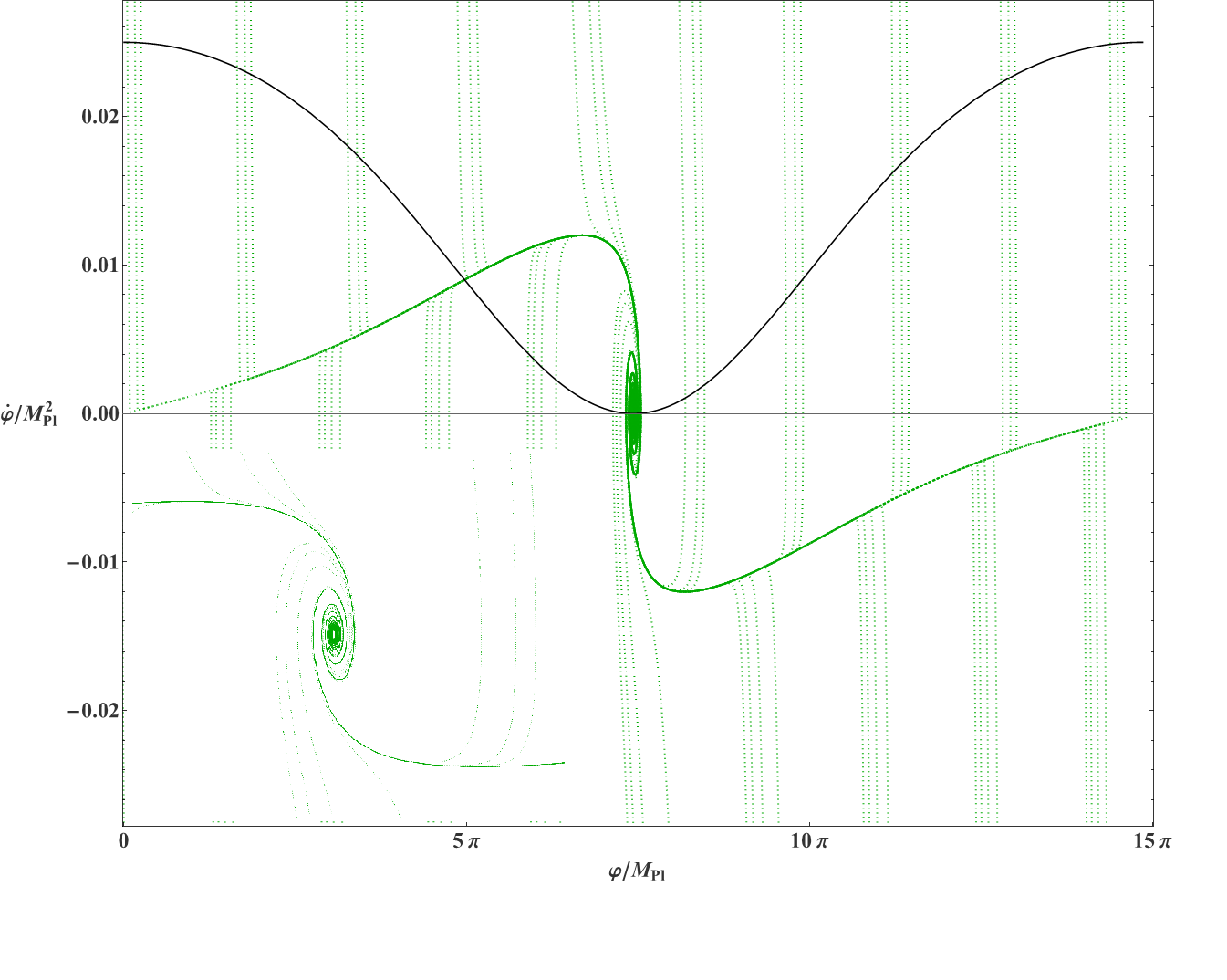}
    \caption{A plot of the phase-space trajectories $\dot{\phi}-\phi$ highlighting the attractive behaviour of the scalar potential $U(\phi)$ for a specific set of the parameters $\alpha=10^9$, $f=10$, $\Lambda=0.008$ and $M_P=1$. In the bottom left there is superimposed a magnified plot of the attractor point of the potential. The solid black curve corresponds to the potential of the canonically normalised field $\varphi$.}
    \label{fig:NaturalInf-attractor}
\end{figure}

Finally, we present a study of the phase-space flow diagram in terms of the numerical solutions of the generalised Klein-Gordon equation. For different initial conditions of the inflaton field and its velocity, the trajectories of $\dot{\phi}-\phi$ settle on the slow-roll trajectory and are led to the minimum of the potential. Then, from fig.~\ref{fig:NaturalInf-attractor} it is clear that the potential has indeed an attractor behaviour.

\subsection{\emph{Massive scalar}}

One of the simplest models realising inflation is the one described by a free massive scalar field with a potential
\begin{equation}
    V(\phi)=\frac{m^2}{2}\phi^2\,.
\end{equation}
The metric formulation of this model coupled minimally to gravity has been excluded by observations (see fig.~\ref{fig:Planck2018}), so let us consider here its Palatini counterpart but with a minimal coupling to an extended gravitational sector of $R+R^2$~\cite{Antoniadis2018}. The total final action is of the same form as the one for the previous model of natural inflation, given in eq.~\eqref{Action:PalatiniNatural}; only the scalar potential has a different form. Then the canonically normalised scalar field $\varphi$ in this case reads
\begin{equation}
    \varphi=\int\!\frac{\mathrm{d}\phi}{\sqrt{1+\displaystyle{\frac{2\alpha m^2}{M_P^4}}\phi^2}}=\frac{M_P^2}{m\sqrt{2\alpha}}\,\sinh^{-1}(m\sqrt{2\alpha}\,\phi/M_P^2)\,,
\end{equation}
which in terms of the original field can be also inverted as
\begin{equation}
    \phi=\frac{M_P^2}{m\sqrt{2\alpha}}\,\sinh(m\sqrt{2\alpha}\,\varphi/M_P^2)\,,
\end{equation}
where $m>0$ and $\alpha>0$ is assumed throughout. Next, substituting the above relation of $\phi=f(\varphi)$ in the expression of the scalar potential we obtain~\cite{Antoniadis2018}
\begin{equation}
    U(\varphi)=\frac{M_P^2}{4\alpha}\,\tanh^2{(m\sqrt{2\alpha}\,\varphi/M_P^2)}\,.
\end{equation}
This is a well-known potential, usually referred to under the general class of the T-models, in which the potential term of the canonically normalised field is given by $V(\varphi)\propto\bar{\alpha}\tanh^{2n}{(\varphi/\sqrt{6\bar{\alpha}})}$, where now $\bar{\alpha}$ is another constant not to confused with the one coming from the $R+\alpha R^2$ term. These models gathered considerable attention after it was shown that they arise in supergravity models~\cite{Kallosh2013b,Lahanas2015,Carrasco2015,Carrasco2015a}. In cosmology they can be realised more generally in the context of a scalar field with a pole in the kinetic term. In fact, these models fall under the same universality class, known as the \emph{attractor models}\footnote{In the present case of $V(\varphi)\propto\bar{\alpha}\tanh^{2n}{(\varphi/\sqrt{6\bar{\alpha}})}$ they would be the $\alpha$-attractor models.}, because their predictions are insensitive to the features of the potential $V(\phi)$ for small values of the parameter $\bar{\alpha}$ and are largely dependent on the order and residue of the pole~\cite{Kallosh2013b}.

The slow-roll parameters in terms of the canonically normalised field $\varphi$ are found to be~\cite{Antoniadis2018}
\begin{align}
    \epsilon_V&=\frac{m^2}{M_P^2}\,\frac{16\alpha}{\sinh^2{(2m\sqrt{2\alpha}\,\varphi/M_P^2)}}\,,\\
    \eta_V&=16\alpha\,\frac{m^2}{M_P^2}\,\left(\frac{2-\cosh^2{(2m\sqrt{2\alpha}\,\varphi/M_P^2)}}{\sinh^2{(2m\sqrt{2\alpha}\,\varphi/M_P^2)}}\right)\,.
\end{align}
Then the end of inflation can be analytically obtained by solving $\epsilon_V(\varphi=\varphi_f)\equiv1$, yielding
\begin{equation}
    \varphi_f\approx\frac{M_P^2}{2m\sqrt{2\alpha}}\,\sinh^{-1}{(2m\sqrt{2\alpha}/M_P^2)}\,.
\end{equation}
The scalar power spectrum is similarly given by 
\begin{equation}
    \mathcal{A}_s=\frac{M_P^4}{1536\alpha^2\pi^2 m^2}\,\sinh^2{(2m\sqrt{2\alpha}\,\varphi/M_P^2)}\tanh^2{(2m\sqrt{2\alpha}\,\varphi/M_P^2)}\,,
\end{equation}
which, once again, evaluated at horizon exit $\varphi=\varphi_i$ should yield its observed value of $\mathcal{A}_s\approx2.1\times10^{-9}$. Finally, the integral for the number of $e$-folds in terms of the canonical field is 
\begin{equation}
    N=\frac{M_P^2}{16\alpha m^2}\,\int_{2m\varphi_i\sqrt{2\alpha}/M_P}^{2m\varphi_f\sqrt{2\alpha}/M_P}\!\frac{\mathrm{d}x}{x}\,\sqrt{\sinh{x}}\,.
\end{equation}

\begin{figure}
    \centering
    \includegraphics[scale=0.5]{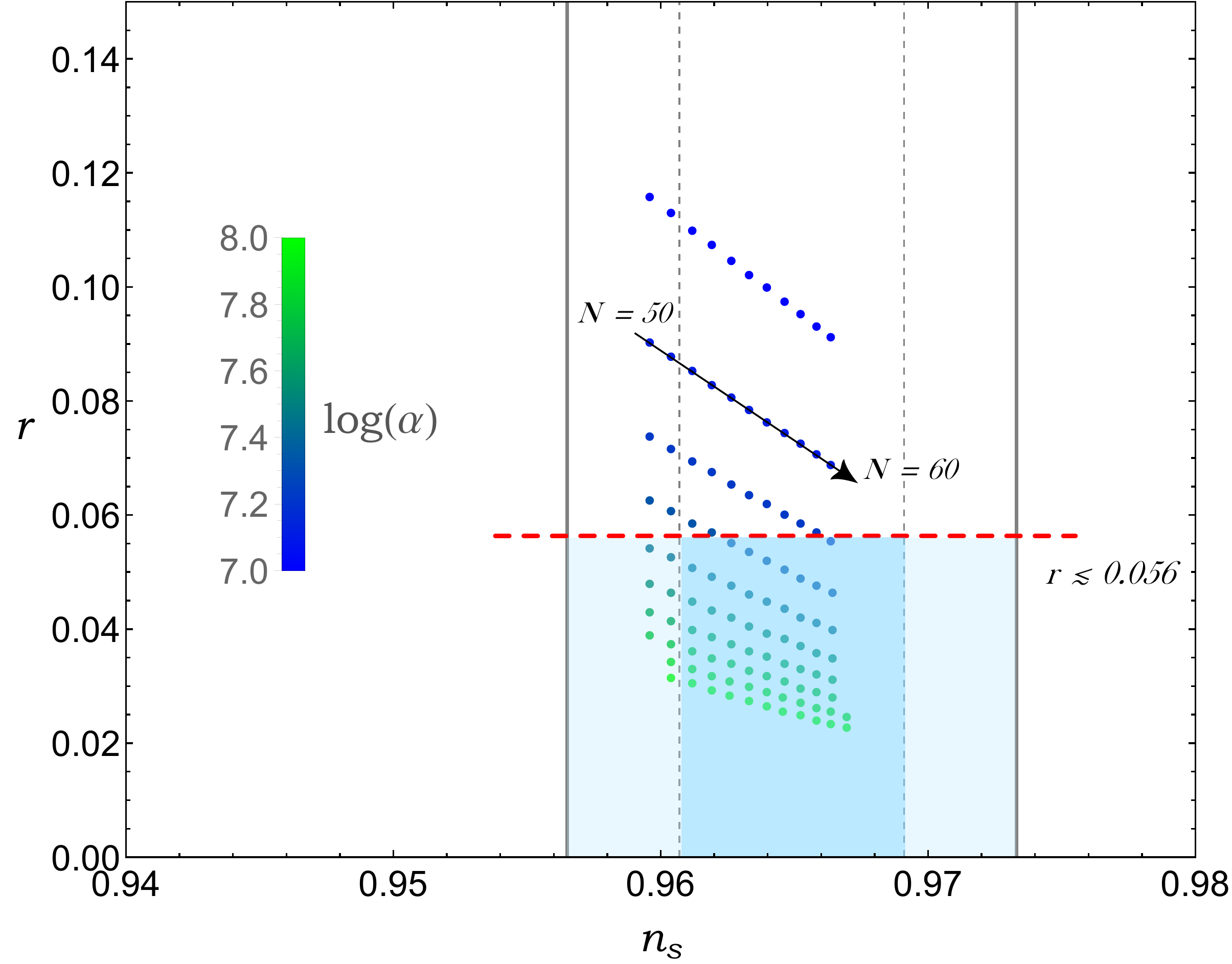}
    \caption{A plot of the $r-n_s$ region for various values of $\alpha\in[10^7,10^8]$, $m=10^{-5}$ (which is $10^{13}\,\text{GeV}$) and $M_P=1$. The light and dark blue region denote the $2\sigma$ and $1\sigma$ allowed region from the Planck2018 collaboration, while the red dashed line denotes the bound on $r\lesssim0.056$. Each dot is a solution for a specific value of $N$, which from left to right is increasing up to a maximum value of $N=60$ and each ``line'' is formed for a specific value of $\alpha$ based on the colour coding of the legend.}
    \label{fig:Massive-RNs}
\end{figure}

Unfortunately the expressions become complicated to sort out analytically and we resort to numerical methods to solve the system. In fig.~\ref{fig:Massive-RNs} the values of the tensor-to-scalar ratio $r$ are plotted against the spectral index $n_s$ for various values of the parameter $\alpha$ in a range of $N\in[50,60]$ $e$-foldings. We observe that for increasing values of $\alpha$ the tensor-to-scalar ratio is decreasing while $n_s$ remains largely unaffected. In fact $\alpha$ can assume values larger than what is presented in the figure suppressing further the predicted value of $r$.  It is also evident that as $\alpha$ increases the predictions of the observables asymptote to those produced by the simple (without the $R^2$ term) quadratic model, presented also in fig.~\ref{fig:Planck2018}. The value of the mass parameter in the figure is $m=10^{-5}$ ($10^{-13}\,\text{GeV}$) in terms of the fixed Planck mass $M_P=1$, and altogether the parameters manage to produce the appropriate value of the power spectrum $\mathcal{A}_s$. The scale of inflation is preferably described by the field values of the inflaton field during inflation, which in principle varies with the different values of the model parameters and the number of $e$-foldings. However, a general statement of their values includes
\begin{equation}
    \varphi_i\approx 10 M_P,\,\qquad\quad\&\quad\qquad\varphi_f\approx M_P\,.
\end{equation}
In terms of the initial field $\phi$ the field values of $\phi_f$ and $\phi_i$ are very similar to the ones produced for the canonical field $\varphi$. Therefore, inflation in this case happens above the Planck scale, also referred to as transPlanckian.

\subsubsection*{\emph{Digression on attractor models}}

Let us illustate the above point by considering a toy model with a Lagrangian reading 
\begin{equation}
    \mathscr{L}=\frac{M_P^2}{2}\,R-\frac{\omega(\Phi)}{2}(\nabla\Phi)^2-V(\Phi)\,,
\end{equation}
where $\Phi$ is a scalar field with a general potential $V(\Phi)$. The following discussion is universal in the metric and Palatini formulations so we ignore the subject at the moment. The noncanonical kinetic function has a general form of
\begin{equation}
    \omega(\Phi)\equiv\frac{\alpha_p}{|\phi-\phi_0|^p}\,,
\end{equation}
with $\alpha_p$ a positive constant in order to avoid ghost instabilities. Near the pole of the inflaton field $\Phi=\Phi_0$ the potential can be expanded as $V(\Phi)\approx V(\Phi_0)+\Phi V'(\Phi_0)$ suggesting that the potential should be at least well-behaved near the pole.

Let us define the canonically normalised field through
\begin{equation}
    \frac{\mathrm{d}\Psi}{\mathrm{d}\Phi}=\frac{1}{\sqrt{\omega(\Phi)}}\,,
\end{equation}
and by expressing the slow-roll parameters in terms of it we obtain
\begin{equation}
    \epsilon=\frac{M_P^2}{2}\left(\frac{V'(\Psi)}{V(\Psi)}\right)^2=\frac{M_P^2}{2}\frac{1}{\omega(\Phi)}\left(\frac{V'(\Phi)}{V(\Phi)}\right)^2\,,
\end{equation}
and
\begin{equation}
    \eta=M_P^2\,\frac{V''(\Psi)}{V(\Psi)}=M_P^2\,\frac{1}{\omega(\Phi)}\,\frac{V''(\Phi)}{V(\Phi)}\left(1-\frac{1}{2}\frac{\omega'(\Phi)}{\omega(\Phi)}\,\frac{V'(\Phi)}{V''(\Phi)}\right)\,.
\end{equation}
The number of $e$-foldings can also be expressed in terms of the original field as
\begin{equation}
    N(\Phi)=\int_{\Psi_i}^{\Psi_f}\!\frac{1}{M_P}\,\frac{\mathrm{d}\Psi}{\sqrt{2\epsilon}}=\frac{1}{M_P^2}\int_{\Phi_i}^{\Phi_f}\!\omega(\Phi)\,\frac{V(\Phi)}{V'(\Phi)}\,\mathrm{d}\Phi\,.
\end{equation}
It is straightforward then to show that the observable quantities read
\begin{equation}
    n_s=1-6\epsilon-2\eta=\frac{M_P^2}{\omega(\Phi)}\left(2\,\frac{V''(\Phi)}{V(\Phi)}-3\left(\frac{V'(\Phi)}{V(\Phi)}\right)^2-\frac{\omega'(\Phi)}{\omega(\Phi)}\,\frac{V'(\Phi)}{V(\Phi)}\right)\,,
\end{equation}
and
\begin{equation}
    r=16\epsilon=8\,\frac{M_P^2}{\omega(\Phi)}\left(\frac{V'(\Phi)}{V(\Phi)}\right)^2\,.
\end{equation}

If we expand around the inevitable point of the pole at $\Phi=\Phi_0$ we obtain
\begin{align}
    \epsilon_0&\approx\frac{M_P^2}{2\alpha_p}\,|\Phi-\Phi_0|^p\left(\frac{V'_0}{V_0}\right)^2\left(1+2(\Phi-\Phi_0)\left(\frac{V''_0}{V'_0}-\frac{V'_0}{V_0}\right)\right)\,,\\
    \eta_0&\approx\frac{M_P^2}{\alpha_p}\,|\Phi_0-\Phi|^p\left\{\frac{p}{2}\,\frac{\text{sgn}(\Phi-\Phi_0)}{|\Phi-\Phi_0|}\,\frac{V'_0}{V_0}+\frac{V''_0}{V_0}+\frac{p}{2}\left[\frac{V''_0}{V_0}-\left(\frac{V'_0}{V_0}\right)^2\right]+\ldots\right\}\,,
\end{align}
where $\text{sgn}(x)$ is the sign function. It is straightforward to show that the tensor-to-scalar ratio and the spectral index are given by
\begin{equation}
    n_s-1\approx\frac{M_P^2}{\alpha_p}\left|\Phi-\Phi_0\right|^p\left\{p\,\frac{\text{sgn}(\Phi-\Phi_0)}{|\Phi-\Phi_0|}\,\frac{V'_0}{V_0}+(p+2)\frac{V''_0}{V_0}-(p+3)\left(\frac{V'_0}{V_0}\right)^2+\ldots\right\}\,,\qquad \forall\,p\in\mathbb{Z}^+
\end{equation}
and
\begin{equation}
    r\approx\frac{8M_P^2}{\alpha_p}\,|\Phi-\Phi_0|^p\left(\frac{V'_0}{V_0}\right)^2\left(1+2(\Phi-\Phi_0)\left(\frac{V''_0}{V'_0}-\frac{V'_0}{V_0}\right)+\ldots\right)\,,\qquad \forall\,p\in\mathbb{Z}^+\,.
\end{equation}

In order to proceed we have to identify two cases of the order of the pole $p=1$ and $p\neq1$. Then, the number of $e$-foldings can be approximated as
\begin{align}
    N&\approx\frac{\alpha_p}{M_P^2}\,\frac{V_0}{V_0'}\,\ln{|\Phi-\Phi_0|}-\{\Phi\rightarrow\Phi_f\}\,,\qquad (p=1)\,.\\
    N&\approx\frac{\alpha_p}{M_P^2}\,\frac{V_0}{V'_0}\,\frac{|\Phi-\Phi_0|^{1-p}}{p-1}-\underbrace{\left\{\Phi\rightarrow\Phi_f\right\}}_{\equiv\mathcal{C}}\,\qquad (p\neq1)\,,
\end{align}
where the last term in both equations is simply the first term with $\Phi=\Phi_f$.

Interestingly, in the case of $p\neq1$, which is also the more common, the leading term of the spectral index becomes
\begin{equation}
    n_s-1\approx\text{sgn}(\Phi-\Phi_0)\,\frac{p}{p-1}\,\frac{1}{N+\mathcal{C}}\,,\qquad (p\neq1)\,.
\end{equation}
Notice that the spectral index is completely independent of the inflaton potential and solely determined by the order of the pole. Let us return to the case of T-models and take a look at a limiting case of large $\bar{\alpha}$. Then, the spectral index and the tensor-to-scalar ratio are~\cite{Kallosh2013b}
\begin{equation}
    n_s\approx1-\frac{2}{N}-\frac{n-1}{8n}r\,,\qquad \qquad r\approx\frac{24\,n\,\bar{\alpha}}{N(3\bar{\alpha}+2nN)}\stackrel{N\gg 1}{\approx}\frac{12}{N^2}\,\bar{\alpha}\,.
\end{equation}
Then, only $r$ of the two has a multiplicative dependence on $\bar{\alpha}$ that can lead to suppressed values of $r<10^{-3}$. This is completely model \emph{independent} and hinges only on the fact that the kinetic term of the inflaton field has a pole at some field value $\Phi_0$ of order $p$.

\subsection{\emph{Higgs field}}
\label{subsec:MinHiggs}

In an attempt to connect low energy particle physics with high energy phenomena it is tempting to identify the inflaton field with the sole observed scalar field, the Higgs boson. This is not only a scenario appealing to theoretical physics, but also very \emph{predictive} in the sense that we can in principle ``match'' observational data from cosmology and particle physics. The total action in question then reads
\begin{equation}
    \mathcal{S}=\int\!\mathrm{d}^4x\,\sqrt{-g}\left\{\frac{M_P^2}{2}R+\frac{\alpha}{4}R^2+\mathscr{L}_\text{SM}\right\}\,,
\end{equation}
where $\mathscr{L}_\text{SM}$ denotes SM Lagrangian. This is the most straightforward way to construct such a model of inflation, with the exception of $\alpha=0$ being the true minimal construction.

\subsubsection*{\emph{Revisiting the SM Higgs}}

Assuming that the Higgs field plays the role of the inflaton we can safely ignore the interactions with other fields and in this case obtain
\begin{equation}
    \mathscr{L}_\text{SM}\supset-(D^\mu H)^\dagger (D_\mu H)-V(|H|)\,,
\end{equation}
where $H$ is the Higgs field and $V(|H|)$ its self-interacting potential given by
\begin{equation}
    V(|H|)=-\mu^2 H^\dagger H+\lambda (H^\dagger H)^2\,,
\end{equation}
where $\mu^2>0$ is the Higgs mass term and $\lambda>0$ is its self-coupling. For the SM to maintain the $SU(2)_L\otimes U(1)_Y$ invariance the covariant derivative is introduced as
\begin{equation}
    D_\mu=\nabla_\mu-i g_2\,t^a W^a_\mu-i g_Y Y B_\mu\,,
\end{equation}
where $g_2$ and $g_Y$ are the $SU(2)$ and $U(1)_Y$ gauge coupling constants and $W^a$ and $B$ are the gauge fields corresponding to their generators, i.e. $SU(2)_L\rightarrow\{W^1_\mu,W^2_\mu,W^3_\mu\}$ and $U(1)_Y\rightarrow\{B_\mu\}$. Notice that the notion of the flat spacetime partial derivative is preemptively generalised to curved spacetime by $\partial_\mu\rightarrow\nabla_\mu$. 

Let us introduce the Higgs doublet in the spinor representation 
\begin{equation}
    H\equiv\begin{pmatrix}\phi^+\\\phi^0\end{pmatrix}\,.
\end{equation}
It is straightforward to show that due to the symmetry of the potential $V(|H|)$ there exist an infinite number of states satisfying $H^\dagger H=v^2/2$, where
\begin{equation}
    v^2=\frac{\mu^2}{\lambda}\,
\end{equation}
is the minimum of the potential, also identified with the EW scale and its observed value is $v=246\,\text{GeV}$. Then we can choose in complete generality its VEV to be
\begin{equation}
    \braket{H}=\frac{1}{\sqrt{2}}\begin{pmatrix}0\\v\end{pmatrix}\,.
\end{equation}
In this specific choice the component $\phi^0$ is identified with the neutral component. This is associated with the famous breaking of the symmetry $SU(2)_L\otimes U(1)_L\rightarrow U(1)_{em}$. It is easy to see that the vacuum is invariant under $U(1)_{em}$; a sketch of that is as follows
\begin{align}
    e^{i\alpha Q}\braket{H}&\approx(1+i\alpha Q)\braket{H}\\
    &=\braket{H}+i\alpha\left(T_3+\frac{1}{2}Y\right)\braket{H}\\
    &=\braket{H}+i\frac{\alpha}{2}\left[\begin{pmatrix}1&0\\0&-1\end{pmatrix}+\begin{pmatrix}1&0\\0&1\end{pmatrix}\right]\begin{pmatrix}0\\\displaystyle{\frac{v}{\sqrt{2}}}\end{pmatrix}\\
    &=\braket{H}\,,
\end{align}
where for the electric charge we used the Gell-Mann-Nishijima relation $Q=T_3+\frac{1}{2}Y$, with $T_3$ being a generator of the $SU(2)$ group. Therefore, the photon remains massless. The rest of the gauge fields that correspond to the broken generators $T_1$ and $T_2$ acquire a mass; this mechanism is known as the \emph{Higgs mechanism}.

\myastr

Leaving the low energy physics and returning to the inflationary regime we obtain the Higgs sector with its self-interactions only. Considering the Higgs in the unitary gauge
\begin{equation}
    H\equiv\frac{1}{\sqrt{2}}\begin{pmatrix}0\\h\end{pmatrix}\,.
\end{equation}
the total action becomes
\begin{equation}
    \mathcal{S}=\int\!\mathrm{d}^4x\,\sqrt{-g}\left\{\frac{M_P^2}{2}R+\frac{\alpha}{4}R^2-\frac{1}{2}(\partial h)^2-V(h)\right\}\,,
\end{equation}
with the potential term given by
\begin{equation}
    V(h)=\frac{\lambda}{4}(h^2-v^2)^2\,.
\end{equation}
Then we can employ the mechanism highlighted in previous sections, namely express the action in its scalar representation and transform it from the Jordan to the Einstein frame. The resulting action has the same form as the one given in eq.~\eqref{Action:PalatiniNatural}; restated here in this case as~\cite{Antoniadis2019}
\begin{equation}
    \mathcal{S}=\int\!\mathrm{d}^4x\,\sqrt{-\overline{g}}\left\{\frac{M_P^2}{2}\,\overline{g}^{\mu\nu}\,R_{\mu\nu}(\Gamma)-\frac{1}{2}\frac{(\overline{\nabla} h)^2}{\left(1+\displaystyle{\frac{4\alpha}{M_P^4}}V(h)\right)}-\frac{V(h)}{1+\displaystyle{\frac{4\alpha}{M_P^4}}V(h)}+\mathcal{O}\left((\overline{\nabla} h)^4\right)\right\}\,.
\end{equation}
In what follows we assume that it is safe to approximate the Higgs potential as
\begin{equation}
    V(h)\approx\frac{\lambda}{4}h^4\,,\qquad\quad h\gg v\,,
\end{equation}
since we expect the field to assume values far away from its VEV during inflation. It is however unfortunate that the Higgs self-coupling is open to interpretation at these energy scales. Meaning that the running of the coupling following the conventional SM renormalisation group equations suggests that it \emph{decreases} with energy to values of $\lambda\sim10^{-13}$ (see ref.~\cite{Rubio2019}). All of that depends on the exact interplay between the Higgs mass and the top quark coupling and, as expected, is very sensitive to BSM degrees of freedom. Unfortunately, the self-coupling can even assume \emph{negative} values close to the Planck scale, $\mu_\text{c}\sim M_P$, leading to the issue known as the \emph{metastability} of the Higgs vacuum~\cite{Casas1996,Isidori2001,Ellis2009,EliasMiro2012,Degrassi2012,Buttazzo2013}. In order to avoid the issue we assume tiny positive values of the coupling $0<\left.\lambda\right|_{\Lambda=M_P}\ll1$ close to the Planck scale, and if that it is not the case we assume that other degrees of freedom can stabilise the potential.

\begin{figure}
    \centering
    \includegraphics[scale=0.5]{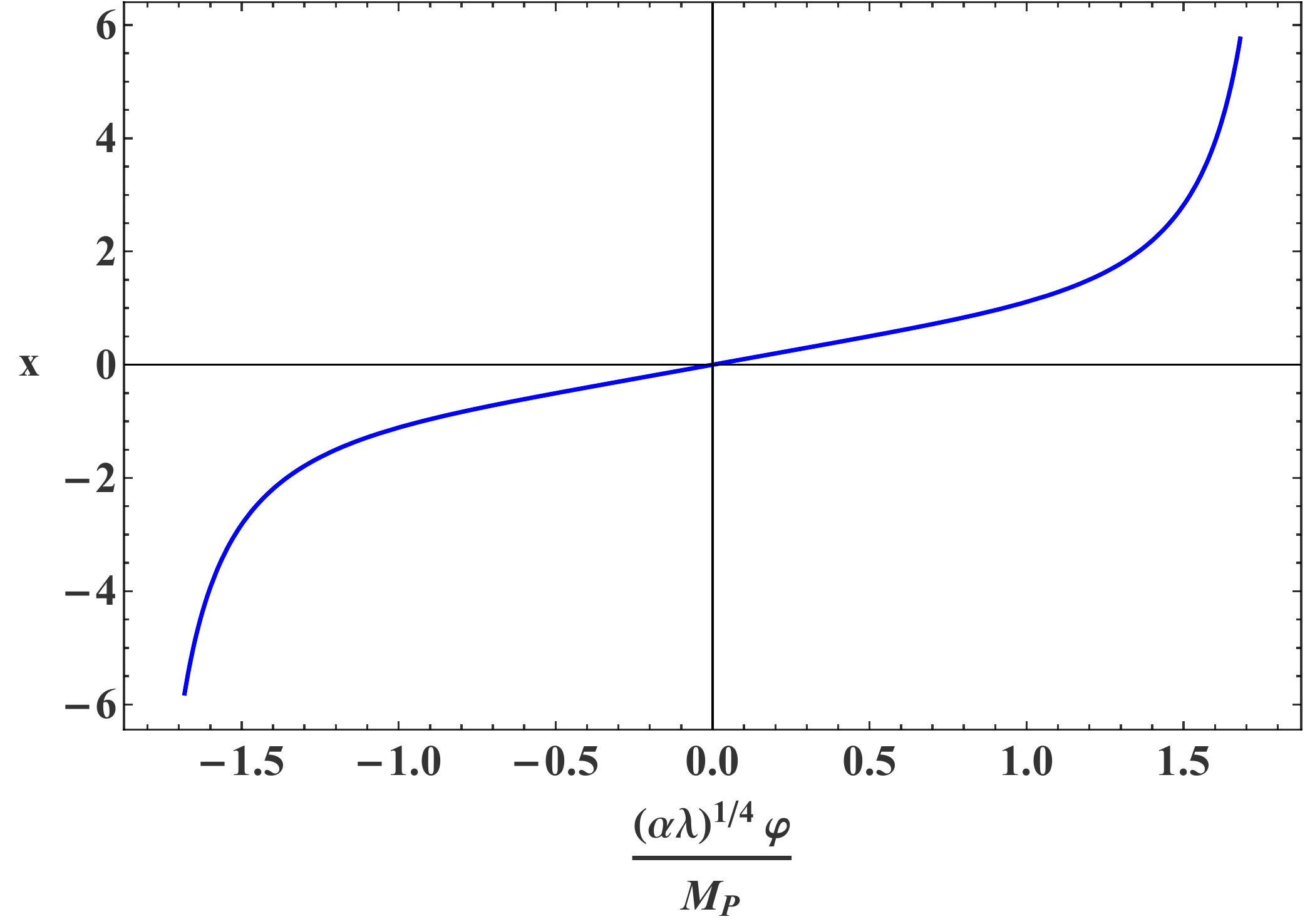}
    \caption{A plot of relation between the field $x=f(\varphi)$.}
    \label{fig:MinimalHiggs-Fields}
\end{figure}

Regarding the prediction of inflationary observables the analysis follows closely those presented for the two previous models. At the level of equations of motion the connection satisfies the Levi-Civita condition and assuming that a field redefinition exists such that the kinetic term can be canonical the conventional machinery of slow-roll inflation can be applied directly. In order to do that let us begin by canonically redefining the inflaton as follows
\begin{equation}
    \varphi=\int\!\frac{\mathrm{d}h}{\sqrt{1+\displaystyle{\frac{4\alpha}{M_P^4}}V(h)}}=\frac{M_P}{(\alpha\lambda)^{1/4}}\int\!\frac{\mathrm{d}x}{\sqrt{1+x^4}}=\frac{M_P}{(\alpha\lambda)^{1/4}}\left(\frac{4}{\sqrt{\pi}}\left(\Gamma(5/4)\right)^2-\frac{1}{2}\mathcal{F}(y,1/\sqrt{2})\right)\,,
\end{equation}
where $\mathcal{F}$ is the elliptic integral of the first kind defined previously in eq.~\eqref{Cond:EllipticIntegral1st} and we also made the following definitions
\begin{equation}
    x\equiv\frac{\alpha \lambda }{M_P}\,h\,,\qquad\quad \cos{y}\equiv\frac{x^2-1}{x^2+1}\,.
\end{equation}
We can in fact invert the expression of $\varphi(h)$ to obtain $h(\varphi)$ in terms of one of Jacobi's elliptic functions, similarly to what was done right after eq.~\eqref{Cond:EllipticIntegral1st}. In fig.~\ref{fig:MinimalHiggs-Fields} we present a plot of the field $x$ in terms of $\bar{\varphi}=(\alpha\lambda)^{1/4}\varphi/M_P$, where it is noticeable that the expression saturates at some value of $\bar{\varphi}_0$ that is connected with the value of $\frac{4}{\sqrt{\pi}}\left(\Gamma(5/4)\right)^2\approx1.85407$. This can be also verified analytically by assuming a large field expansion of $\mathcal{F}$ in terms of $x$.

The total scalar potential in the Einstein frame is given in terms of $x(\varphi)$ by
\begin{equation}
    U(x)=\frac{M_P^4}{4\alpha}\,\frac{x^4(\varphi)}{1+x^4(\varphi)}\,.
\end{equation}
Using this expression we can directly compute the slow-roll parameters as follows
\begin{equation}
    \epsilon_V=\frac{8\sqrt{\alpha\lambda}}{x^2(1+x^4)}\,,\qquad\qquad\eta_V=\frac{12(1-x^4)\sqrt{\alpha\lambda}}{x^2(1+x^4)}\,,
\end{equation}
and the number of $e$-foldings is given by
\begin{equation}
    N=-\frac{1}{M_P}\int_{\varphi_i}^{\varphi_f}\!\frac{\mathrm{d}\varphi}{\sqrt{2\epsilon_V(\varphi)}}=\frac{1}{M_P}\int_{x_i}^{x_f}\!\frac{\mathrm{d}x}{\sqrt{2\epsilon_V(x)}\,\sqrt{1+x^4}}=\frac{1}{8\sqrt{\alpha\lambda}}\,(x_i^2-x_f^2)\,,
\end{equation}
where $x_i$, $\varphi_i$ and $x_f$, $\varphi_f$ are the field values at the start and end of inflation of the $x$ and $\varphi$ fields respectively. We can obtain the field value at the end of inflation via~\cite{Antoniadis2019}
\begin{equation}
    \epsilon_V(x_f)\simeq 1\quad\implies\quad x_f^2\approx8\sqrt{\alpha\lambda}\,,\quad \text{iff }
    \sqrt{\alpha\lambda}<10^{-2}\,.
\end{equation}
Therefore, we may obtain an approximate expression for $N$ in terms of $x_i$, $(N+1)=x_i^2/(8\sqrt{\alpha\lambda})$. The power spectrum of scalar perturbations becomes
\begin{equation}
    \mathcal{A}_s=\frac{x_i^6}{768\pi^2\alpha\sqrt{\alpha\lambda}}\approx\frac{2\lambda}{3\pi^2}(N+1)^3\,.
\end{equation}
This in turn leads to values of the self-coupling $\lambda\sim10^{-13}$ in order to satisfy the observed value of $\mathcal{A}_s\approx10^{-9}$.

\begin{figure}
    \centering
    \includegraphics[scale=0.5]{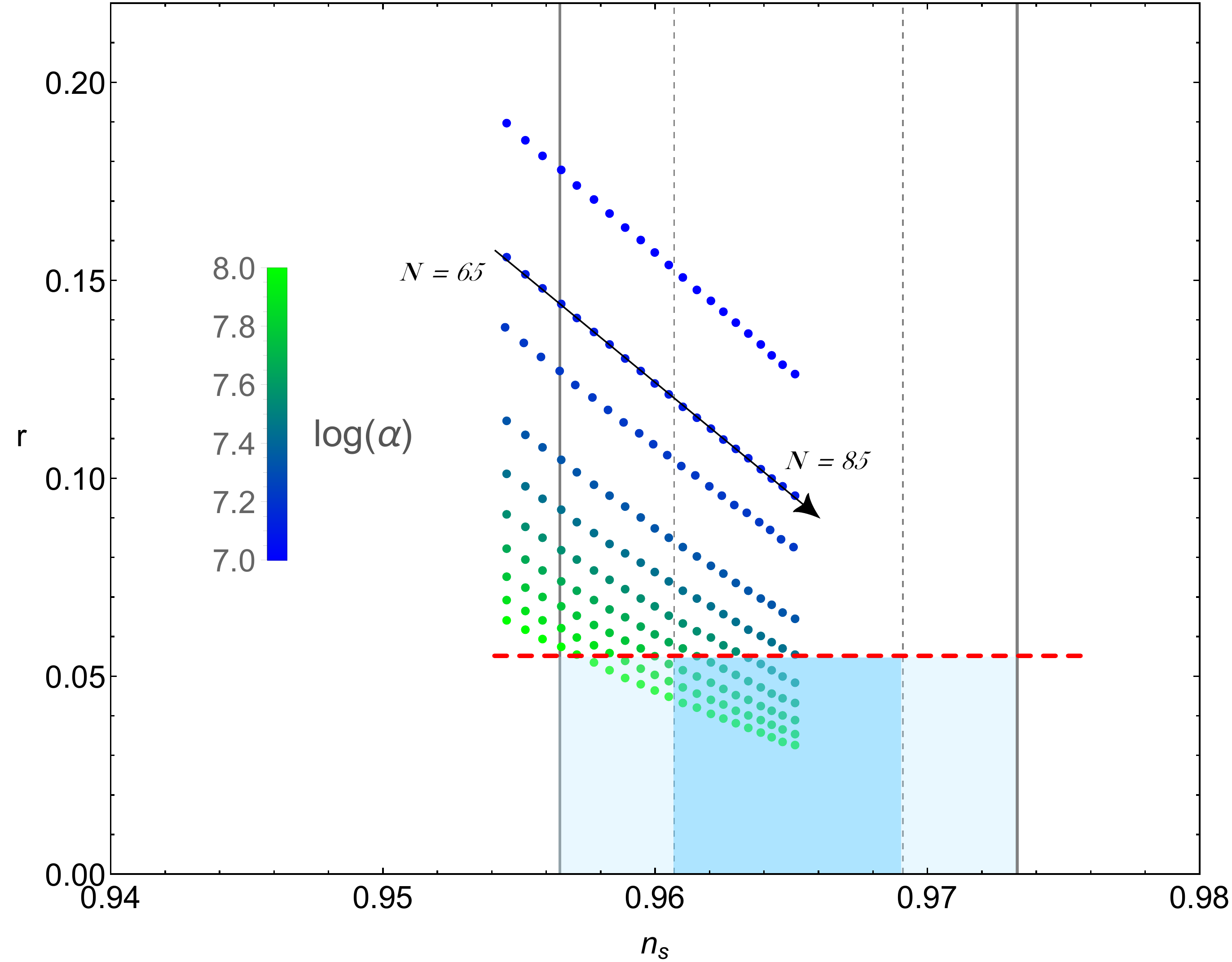}
    \caption{A plot of the $r-n_s$ region for various values of $\alpha\in[10^7,10^8]$, $\lambda=10^{-13}$ and $M_P=1$. The light and dark blue region denote the $2\sigma$ and $1\sigma$ allowed region from the Planck2018 collaboration, while the red dashed line denotes the bound on $r\lesssim0.056$. Each dot is a solution for a specific value of $N$, which from left to right is increasing from a starting value of $N=65$ to $N=85$ $e$-foldings, and each ``line'' is formed for a specific value of $\alpha$ based on the colour coding of the legend.}
    \label{fig:MinimalHiggs-RNs}
\end{figure}

Let us consider the other observable quantities; starting from the spectral index we may substitute the value of $x_i$ in terms of $N$ to obtain:
\begin{equation}
    n_s=1-6\epsilon_V(x_i)+2\eta_V(x_i)=1-\frac{24\sqrt{\alpha\lambda}}{x_i^2}\approx\frac{N-2}{N+1}\stackrel{N\gg 1}{\approx}1-\frac{3}{N}+\mathcal{O}(1/N^2)\,.
\end{equation}
Notice that it is manifestly independent of the parameter $\alpha$ and its approximate expression suggests a larger than usual amount of $e$-foldings is required in order to satisfy the bounds of $n_s$. The tensor-to-scalar ratio reads
\begin{equation}
    r=16\epsilon_V(x_i)=\frac{128\sqrt{\alpha\lambda}}{x_i^2(1+x_i^4)}\approx\frac{16}{(N+1)(1+8\alpha\lambda(N+1)^2)}\stackrel{N\gg1}{\approx}\frac{2}{\alpha\lambda}\left(\frac{1}{N^3}-\frac{3}{N^4}+\mathcal{O}(1/N^5)\right)\,.
\end{equation}
Then, in contrast to $n_s$, large values of the parameter $\alpha$ can in principle suppress $r$. It is possible to derive a lower bound on the product of $\alpha\lambda$ by demanding that 
\begin{equation}
    r\lesssim0.056\quad\implies\quad \alpha\lambda\gtrsim\mathcal{O}(10^{-5})\,, \quad\text{for }N\approx 75\ e\text{-folds}\,,
\end{equation}
where we allowed for larger values of $N$ in order to satisfy the bounds on $n_s$. Then, in order for the discussion above to make sense the product of these parameters is bounded as follows
\begin{equation}
    \mathcal{O}(10^{-5})\lesssim\alpha\lambda<10^{-4}\,,
\end{equation}
where the last inequality stems from the condition that $x_f$ is a solution to $\epsilon_V(x_f)=1$. Since the power spectrum demands values of the self-coupling close to $\lambda\sim10^{-13}$ we can derive a bound on $\alpha$ reading
\begin{equation}
    10^{8}\lesssim\alpha<10^9\,,
\end{equation}
where once again the upper bound is limited to the approximation used here and is not a physical limitation of the model. In fact, in fig.~\ref{fig:MinimalHiggs-RNs} it is clear that arbitrarily large values of $\alpha$ are also acceptable, with the ``disadvantage'' that $r$ is highly suppressed.

\begin{figure}
    \centering
    \includegraphics[scale=0.45]{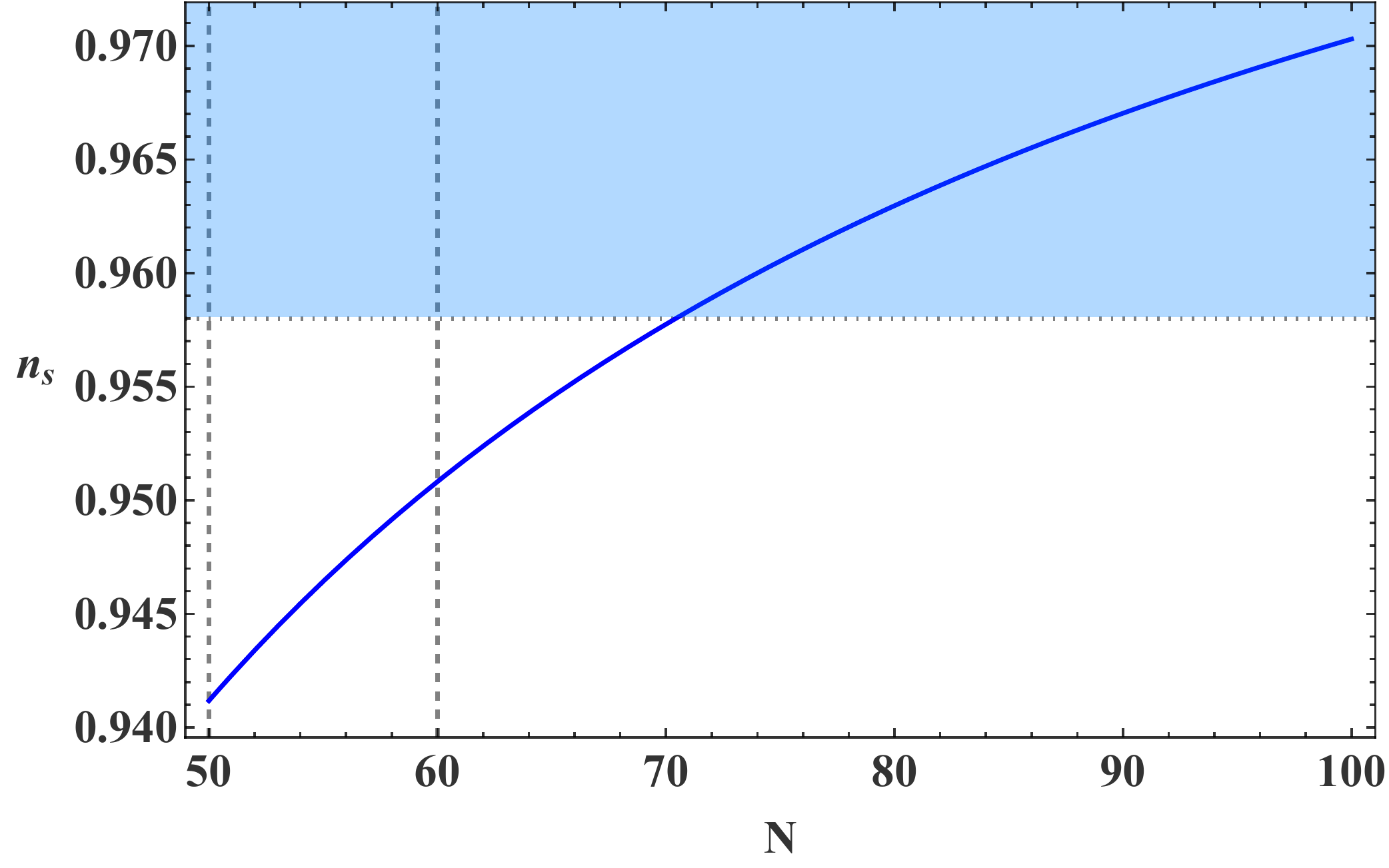}
    \caption{A plot of the relation between $n_s=f(N)$. The blue shaded region represents the $2\sigma$ allowed region of $n_s$ and the dashed lines the conventional range of $e$-foldings $N\in[50,60]$.}
    \label{fig:MinimalHiggs-NsN}
\end{figure}

In fig.~\ref{fig:MinimalHiggs-NsN} we present in a more comprehensive way the fact that in order for the predictions to reside in the allowed region of observations for $n_s$ (displayed by the light blue shaded region) a large amount of $N\gtrsim70$ $e$-foldings is required. This is made blatantly clear in the next figure, fig.~\ref{fig:MinimalHiggs-RNs}, in which we plot the numerical results for the predicted values of $n_s$ and $r$ for various values of $\alpha\in[10^7,10^8]$ and a fixed value of $\lambda=10^{-13}$ and $M_P=1$. As expected, a large number of $N\in(70,80)$ is required which is not necessarily out of line with high-scale inflation and suggests a period of slower expansion compared to the standard radiation domination.

The initial and final values of the inflaton field, for a characteristic set of the parameters $\lambda=10^{-13}$, $\alpha=10^8$ and $N\approx75$, are approximately given by
\begin{equation}
    \varphi_f\approx 3M_P\,,\qquad\&\qquad\varphi_i\approx 20M_P\,,
\end{equation}
which also confirm that the field excursion is inside the well-defined range of $[-\bar{\varphi_0},\bar{\varphi}_0]$, after all we have $\bar{\varphi}_i\approx 1.1\,M_P$. In terms of the initial field $h$ we also obtain similar values of $h_f\approx3\,M_P$ and $h_i\approx 25\,M_P$.

\begin{figure}
    \centering
    \includegraphics[scale=0.5]{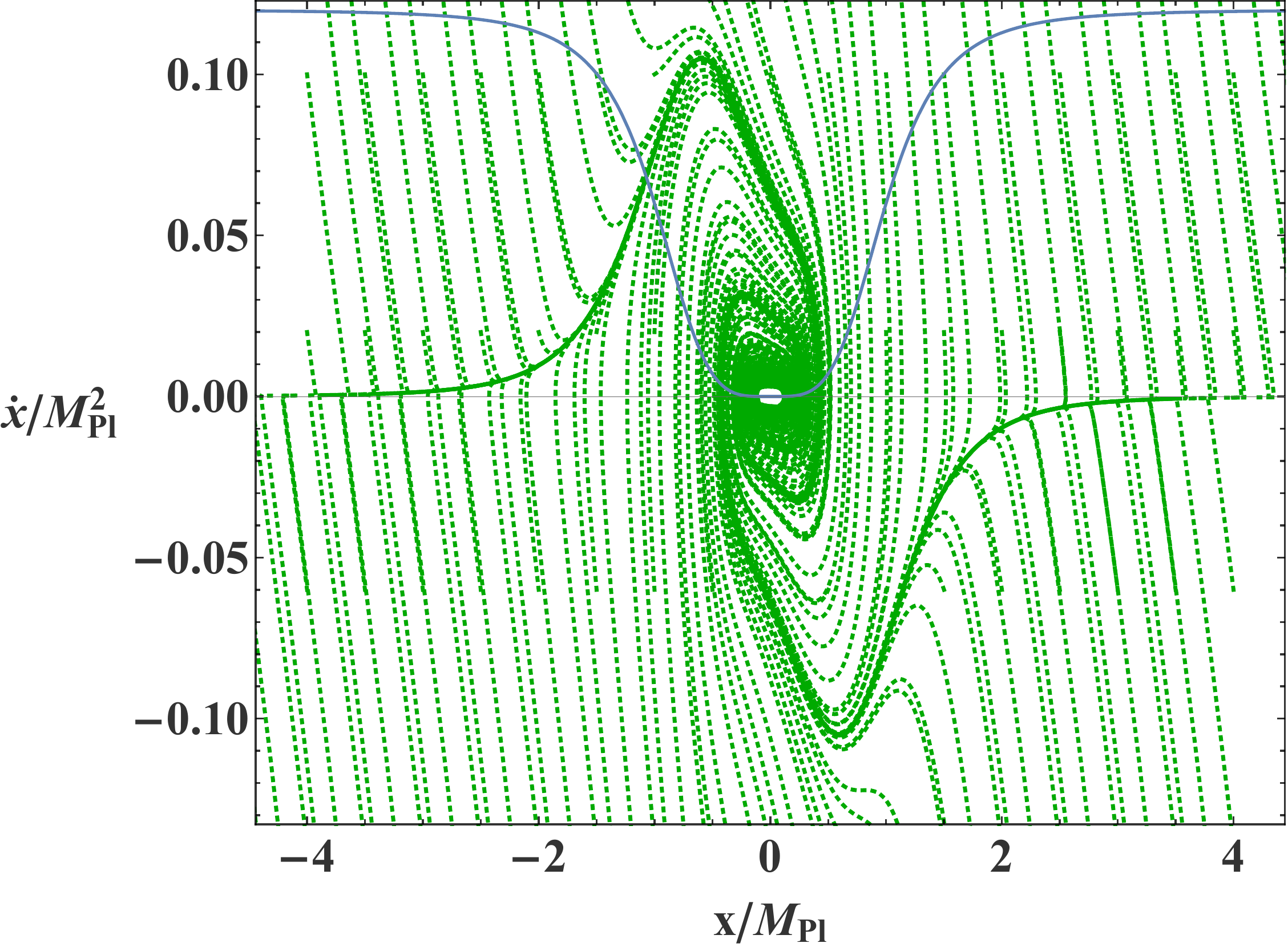}
    \caption{A plot of the phase-space trajectories $\dot{x}-x$ highlighting the attractive behaviour of the scalar potential $U(x(\varphi))$ for a specific set of the parameters $\alpha=10^8$, $\lambda=10^{-13}$ and a fixed value of $M_P=1$. The solid blue curve corresponds to the potential of the inflaton field $x$, namely $U(x)$.}
    \label{fig:MinimalHiggs-attractor}
\end{figure}

In fig.~\ref{fig:MinimalHiggs-attractor} we solve numerically the generalised Klein-Gordon equation for the field $x\equiv (\alpha\lambda)h/M_P$ for various initial conditions of the inflaton field and a specific set of parameters $\alpha=10^8$ and $\lambda=10^{-13}$. The trajectories $\dot{x}-x$ are then presented, which regardless of the initial conditions converge to minimum of the potential. Therefore, it is clear that it exhibits an attractive behaviour. A similar figure can be produced for the canonically normalised field $\varphi$.

\subsubsection{\emph{A general class of monomial potential terms}}

The last two inflationary models, namely the free massive scalar field and the Higgs inflation model, fall under the general class of monomial potentials, of which an example would be
\begin{equation}
    V(\phi)=\frac{\lambda}{4}\,\frac{\phi^{2n}}{M_P^{2(n-2)}}\,,
\end{equation}
where $\phi$ plays the role of the inflaton field, $n\in\mathbb{Z}^+$ and $\lambda$ is the self-coupling constant. Then, the discussion follows similarly to what was done previously, meaning that the canonically normalised field can be approximated by~\cite{Antoniadis2019}
\begin{equation}
    \varphi=\int\!\frac{\mathrm{d}\phi}{\sqrt{1+\displaystyle{\frac{\alpha\lambda}{M_P^{2n}}}\phi^{2n}}}\sim c_0-c_1/x^{n-1}\,,
\end{equation}
where $c_0$ and $c_1$ are integration constants and the field $x\equiv (\alpha\lambda)^{1/2n}\phi/M_P$ is the generalised case of the previously defined $x$. Then the Einstein-frame total scalar potential becomes
\begin{equation}\label{Cond:MonomialPot-EF}
    U(x)=\frac{M_P^4}{4\alpha}\,\frac{x^{2n}}{1+x^{2n}}\,,
\end{equation}
and follows a similar behaviour to $\varphi$, reaching a plateau in the large field limit. It is straightforward to show that the slow-roll parameters are
\begin{equation}
    \epsilon_V=\frac{2n^2(\alpha\lambda)^{1/n}}{x^2(1+x^{2n})}\,\qquad\qquad\eta_V=2n(\alpha\lambda)^{1/n}\,\frac{2n-1-(n+1)x^{2n}}{x^2(1+x^{2n})}\,.
\end{equation}
Then the field value at the end of inflation is obtained by demanding $\epsilon_V(x=x_f)=1$ implying that $x_f^2\approx2n^2(\alpha\lambda)^{1/n}$ for small $(\alpha\lambda)^{1/n}\ll1$. Next, the number of $e$-foldings are given by
\begin{equation}
    N=\frac{1}{4n(\alpha\lambda)^{1/n}}\left(x_i^2-x_f^2\right)\,,
\end{equation}
which can be expressed in terms of $x_i$ only, yielding $x_i^2=4n(\alpha\lambda)^{1/n}(N+n/2)$. Thus, from all of the above we are able to approximate the value of the spectral index in terms of the number of $e$-foldings and the power $n$ of the monomial; reading
\begin{equation}
    n_s(x_i)=\frac{2N-n-2}{2N+n}\,\stackrel{N\gg1}{\sim}\,1-\frac{n+1}{N}\,.
\end{equation}

\begin{figure}
    \centering
    \includegraphics[scale=0.4]{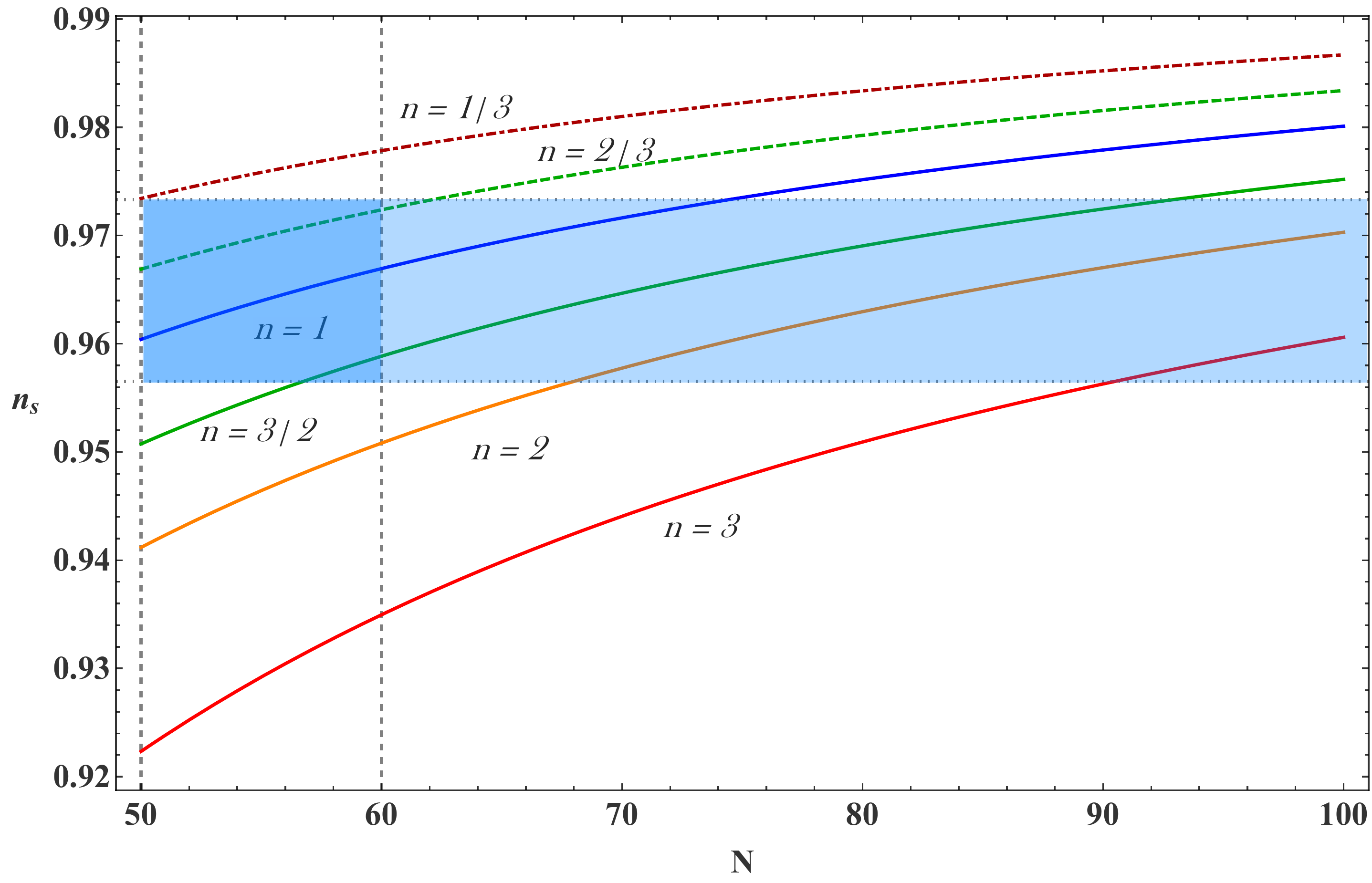}
    \caption{A plot of the relation between $n_s=f(N)$. The blue shaded region represents the $2\sigma$ allowed region of $n_s$ and the dashed vertical lines the conventional range of $e$-foldings $N\in[50,60]$. Each curve represents different values of the power of the monomial potential $n\in\left\{\sfrac{1}{3},\sfrac{2}{3},1,\sfrac{3}{2},2,3\right\}$}
    \label{fig:Monomials-NsN}
\end{figure}

Therefore the best fit of $n_s$ corresponds to values of $n$ close to unity, meaning the quadratic potential, and values of $n>1$ that require larger amount of inflation. Especially values of $n\geq3$ are excluded since they tend to unusually large values of $N$. However, there are rational values of $1<n=q/p<2$ that can provide satisfactory values for $n_s$. A plot of $n_s$ in terms of the number of $e$-foldings $N$ for different values of $n$ is presented in fig.~\ref{fig:Monomials-NsN}, in which that exact relation can be better understood visually by its ``limiting'' cases of $n\approx\sfrac{1}{3}$ and $n\approx2$.

A similar discussion follows for the tensor-to-scalar $r$ which can be rephrased as
\begin{equation}
    r(x_i;n,N)=16\epsilon_V\approx\frac{2^{3-2n}n^{1-n}}{\alpha\lambda}\frac{1}{\left(N+\displaystyle{\frac{n}{2}}\right)^{n+1}}\stackrel{N\gg1}{\sim}\frac{\mathcal{O}(1)}{\alpha\lambda N^{n+1}}\,,\qquad\forall\,n\in\mathbb{Z}^*\,,
\end{equation}
where $\lambda$ is in general the self-coupling of the $\phi^{2n}$ term; not to be confused with the Higgs self-coupling discussed previously. Then, the power spectrum of scalar perturbations leads to 
\begin{equation}
    \mathcal{A}_s=\frac{4^{n-2}n^{n-1}}{3\pi^2}\,\lambda\,\left(N+\frac{n}{2}\right)^{n+1}\,.
\end{equation}
Therefore, we may invert $\lambda$ in terms of $\mathcal{A}_s$ and substitute into the expression of $r$ amounting to
\begin{equation}
    r=\frac{1}{6\pi^2\mathcal{A}_s\alpha}\qquad\implies\qquad\alpha=\frac{1}{6\pi^2r\mathcal{A}_s}\gtrsim10^8\quad\forall\,n\,.
\end{equation}
Importantly, the value of the constant of the $R^2$ term does not depend on the form of the potential $(\lambda,n)$ and is simply chosen such that the predicted observable quantities are within the allowed region of observations. This result is in agreement with previous sections where the cases of $n=1$ and $n=2$ were discussed. 

The argument can be generalised for the values of $\lambda$ for each power $n$. Starting from the requirement that $r\lesssim0.056$ we obtain
\begin{equation}
    \alpha\lambda N^{n+1}\gtrsim\mathcal{O}(10)\quad\implies\quad\alpha\lambda\gtrsim\mathcal{O}(10^{-2n-1})\,.
\end{equation}
Therefore, by using the result that $\alpha$ assumes the same values for each $n$ it is straightforward to show that $\lambda$ must satisfy
\begin{equation}
    \lambda\gtrapprox\mathcal{O}(10^{-2n-9})\quad\forall n\,.
\end{equation}
This is also in agreement with results of previous sections, where for example in the Higgs field model (approximately a quartic monomial potential) we obtain $\lambda\approx10^{-13}$ and for the free massive scalar field (quadratic potential) we obtain $\lambda=2m^2\approx2\times10^{-10}$. 

Obviously the approximations cannot describe completely the dynamics of each model, but they do however paint an intuitive picture of the predicted results of a general class of these models, that is, for slow-roll inflation to take place the constant $\alpha$ has to assume large values and the self-coupling constant $\lambda$ of the potential $V(\phi)\propto\lambda\phi^{2n}$ takes up tiny values. Clearly, there are intricate points in each particular model, such as in the case of $n=2$ in which the amount of $e$-foldings required is larger than usual.

\section{Nonminimally coupled matter fields}

In this section we allow for the possibility of extended interactions of the inflaton field $\phi$ with the gravitational sector $R+\alpha R^2$ through a scale-invariant coupling of the form $\propto\xi\phi^2R$. As was stated in a previous chapter these types of couplings arise at the quantum level due to quantum corrections of $\phi$ in the presence of a curved gravitational background and as such they play a vital role to the renormalisability of the scalar field theory in a curved background. General cases of the nonminimal coupling $f(\phi)R$ can also be considered and they are especially interesting in studies pertaining to cosmology and inflation. The total Lagrangian relative to the inflationary epoch is given by
\begin{equation}
    \mathscr{L}\supset \frac{1}{2}(M^2+\xi\phi^2)R+\alpha R^2-\frac{1}{2}(\partial\phi)^2-V(\phi)\,,
\end{equation}
where $\alpha$ and $\xi$ are dimensionless constants and $M$ is the bare Planck mass later to be identified with the Planck scale. An analysis of the above action was presented in a completely model independent way in a previous section, sec.~\ref{subsec:PalatiniForm}, and therefore some of the calculations are retraced here.

\subsection{\emph{Coleman-Weinberg model}}
\label{subsec:CWmodel}

During inflation the higher-order curvature invariants tend to dominate over the linear Einstein-Hilbert term suggesting that gravity at high energies might be \emph{scale-invariant}. This means that there are no mass scales present in the theory at that energy scale and the theory is said to be (classically) scale-invariant. Let us first introduce the concept of scale-invariance and then present an application to the particular inflationary program described in the previous section.

\subsubsection*{\emph{Classical Scale Invariance}}

In the EFT approach the Lagrangian is written as an expansion in operators~\cite{Wilson1974,Wilson1975}
\begin{equation}
    \mathscr{L}=\sum_{i}c_i\,\frac{O_i}{\Lambda_i^{4-d_i}}=\mathscr{L}_{d\leq4}+\frac{\mathscr{L}_5}{\Lambda}+\frac{\mathscr{L}_6}{\Lambda^2}+\ldots\,,
\end{equation}
where $\Lambda$ is some high-energy scale that in principle suppresses the higher-dimensional operators $O_i$. In $D$ spacetime dimensions, operators of dimension $d$ belong to one of three groups: the \emph{relevant} operators with $d<D$, the \emph{marginal} operators with $d=D$ and the \emph{irrelevant} operators with $d>D$. The SM has only one relevant operator (all the others are renormalisable) the Higgs mass parameter. If we assume that the cut-off scale $\Lambda$ is very large compared to the EW scale  then the absence of operators $d>5$ can be explained, however a fine-tuning of the Higgs mass term is needed in order to explain its size, leading to the issues of \emph{naturalness}.\footnote{More than that EFTs are naturally nonrenormalisable theories since in order to absorb loop divergences with arbitrarily many insertions of $\mathscr{L}_{d\leq5}$, operators of arbitrarily high dimension are needed.}

This infinite set of higher-dimensional operators is not generated if the theory is (classically) scale-invariant, and can be thought of as a special case of an EFT when $\Lambda\to\infty$. In other words, in $D=4$ dimensions only operators with mass-dimensions of $d=4$ are expected to appear at least at the classical level of the theory. Operators of $d\neq D$ then break \emph{explicitly} the scale invariance by setting some unique scale and thus are not allowed, meaning that for example in the SM the Higgs mass term is not considered fundamental and a mechanism must be introduced in order to \emph{dynamically} generate that term. 

Before we move on to describing how one can generate a scale dynamically in a theory without any scales, let us first begin by considering a simple toy model of a free scalar field
\begin{equation}
    \mathcal{S}[\phi]=\int\!\mathrm{d}^Dx\left(-\frac{1}{2}\partial^\mu\phi\,\partial_\mu\phi-\frac{m^2}{2}\phi^2-\frac{g}{4!}\phi^4\right)\,,
\end{equation}
under a scale transformation of
\begin{equation}
    x\to x'=\lambda\,x\, \qquad\&\qquad\phi(x)\to\phi'(x)=\lambda^{-\Delta}\phi(\lambda^{-1}x)\,,
\end{equation}
where $\Delta$ here is the \emph{scaling dimension} (also known as canonical dimension) of the scalar field, which in the case of a free scalar is obtained to be $\Delta=(D-2)/2$, where $D$ are the spacetime dimensions. The total action functional $\mathcal{S}[\phi]$ transforms to $\mathcal{S}'[\phi']$ where the kinetic term becomes
\begin{equation}
    \mathscr{L}\supset-\frac{1}{2}\left(\frac{\partial x'^\mu}{\partial x^\nu}\,\frac{\partial\phi'(x)}{\partial x'^\mu}\right)^2=-\frac{1}{2}\left(\lambda^{-(1+\Delta)}\,\frac{\partial\phi(\lambda^{-1}x)}{\partial x'^\mu}\right)^2=-\frac{\lambda^{-2(1+\Delta)}}{2}\left(\frac{\partial\phi(\lambda^{-1}x)}{\partial(\lambda^{-1}x^{\mu})}\right)^2\,.
\end{equation}
Then it is straightforward to show after a redefinition of $y^\mu\equiv \lambda^{-1}x^\mu$ that the transformed action reads
\begin{align}
    \mathcal{S'}[\phi']=\int\!\mathrm{d}^Dy\left\{-\lambda^{D-2(1+\Delta)}\frac{1}{2}\left(\frac{\partial\phi(y)}{\partial y^\mu}\right)^2-\lambda^{D-2\Delta}\,\frac{m^2}{2}\phi^2(y)-\lambda^{D-4\Delta}\,\frac{g}{4!}\phi^4(y)\right\}\,,
\end{align}
which after substituting $\Delta=(D-2)/2$ we obtain
\begin{equation}
    \mathcal{S'}[\phi']=\int\!\mathrm{d}^Dy\left\{-\frac{1}{2}\left(\frac{\partial\phi(y)}{\partial y^\mu}\right)^2-\lambda^{2}\,\frac{m^2}{2}\phi^2(y)-\lambda^{4-D}\,\frac{g}{4!}\phi^4(y)\right\}\,.
\end{equation}
This means that in $D=4$ dimensions the only term that violates the (would-be) scale-invariance of $\mathcal{S}[\phi]\to\mathcal{S}'[\phi']$ is the mass term, and as such we demand that these terms vanish. For a general monomial term we obtain 
\begin{equation}
    \propto\phi^n(x)\to\lambda^{D-n\Delta}\phi(y)=\lambda^{\frac{(2-n)D}{2}+n}\phi(y)\stackrel{D=4}{=}\lambda^{4-n}\phi(y)\,.
\end{equation}
As expected only terms of $\propto\phi^4$ are scale-invariant in $D=4$ dimensions.

Next, we can calculate the (non)conserved Noether current which after some manipulations of the transformations $\delta x^\mu=-\lambda x^\mu$, $\delta \phi(x)=\lambda(\Delta+x^\mu\partial_\mu)\phi(x)$ we obtain
\begin{equation}
    \delta\mathscr{L}=\partial_\mu(x^\mu\mathscr{L})+m^2\phi^2\,\quad\implies\qquad \partial_\mu J^\mu=m^2\phi^2\,.
\end{equation}
In general we can construct it as usual through the energy-momentum tensor
\begin{equation}
    T^{\mu\nu}=2\frac{\delta}{\delta g_{\mu\nu}}\int\!\mathrm{d}^4x\,\mathscr{L}\,\quad \implies\quad J^\mu=T^{\mu\nu}x_{\nu}\,.
\end{equation}
Then, conservation of the scale current means that the energy-momentum tensor has to be traceless (similar to theories that are conformally invariant)
\begin{equation}
    \partial_\mu J^\mu={T^\mu}_\mu\,.
\end{equation}

Even if the theory is scale-invariant at the classical level, quantum corrections can break the symmetry logarithmically via the RG running of the couplings, which in that case the classical scale invariance (CSI) is referred to as anomalous\footnote{Such a breaking is termed ``natural'' in contrast to the one associated with power-law divergences~\cite{Hooft1980}. This is similar to GUTs, in which for example by embedding the electroweak $SU(2)\otimes U(1)$ to an $SU(5)$ GUT the radiative corrections to lighter masses often involve the larger scales. This is the motivation behind the softly broken supersymmetry, which protects this effect. Taking this a step further we can see how EFTs, that usually have power-law radiative corrections of the high energy scale, cannot describe a scale-invariant theory below that mass scale.} (similar to conformal anomaly). However, there are theories\footnote{Also known as the fixed points of the corresponding RG flow.} that are exactly (quantum) scale-invariant (e.g. see refs.~\cite{Shaposhnikov2009,MarquesTavares2014,Abel2014,Ferreira2018}), meaning that the $\beta$ functions of the couplings vanish at all orders of perturbation theory; e.g. $\mathcal{N}=4$ super-Yang-Mills theory and in principle Conformal Field Theories (CFTs)~~\cite{di1996conformal}. It is however desirable that the scale invariance is softly broken~\cite{Bardeen:1995kv,Foot2008,Foot2010} at some point since we already know that it is at best an approximate symmetry -- after all physical phenomena are different at different scales. Then one can imagine for example the SM as a theory embedded in a UV-complete theory with scale symmetry restored at the UV limit of that theory (see refs.~\cite{Allison2014,Kannike2014,Einhorn2015,Kannike2015,Ferreira2016,Kannike2017,Ferreira2017}).

\subsubsection*{\emph{The Coleman-Weinberg mechanism}}

The concept of generating a scale (nonzero VEV) through radiative corrections in a theory without any energy scales is based on the pioneer work of S. Coleman and E. Weinberg~\cite{Coleman1973}, based also on an earlier work of G. Jona-Lasinio~\cite{JonaLasinio1964}. In this section we briefly highlight the \emph{Coleman-Weinberg} (CW) \emph{mechanism} with the main goal to extract the $1$-loop corrected potential, after which will be considered as an inflationary model.

Let us first consider a (classically) massless complex scalar field charged under a local $U(1)$ symmetry described by the following Lagrangian
\begin{equation}
    \mathscr{L}=(D_\mu\Phi)^\dagger(D^\mu\Phi)+\frac{1}{4}F^{\mu\nu}F_{\mu\nu}-\frac{\lambda}{4!}\,|\Phi|^4\,,
\end{equation}
where $\Phi$ is the complex scalar field $\Phi=(\phi_1+i\phi_2)/\sqrt{2}$. The covariant derivative is $D_\mu=\partial_\mu-igA_\mu$ with $g$ denoting the gauge coupling and $F_{\mu\nu}$ the field strength. Since the Lagrangian is gauge invariant the effective potential depends only on $\phi^2={\phi_1}^2+{\phi_2}^2$. In what follows we calculate the one-loop corrected potential via cut-off regularisation of the integrals at some cut-off energy scale $\Lambda$. It seems counter-intuitive to introduce explicitly a scale $\Lambda$ in the theory since it manifestly breaks the scale-invariance, however the same results are obtained by using other schemes such the $\overline{\text{MS}}$ subtraction scheme and the dimensional regularisation. 

It will prove beneficial if we expand the Lagrangian in terms of its field components and obtain
\begin{equation}\label{Action:ColemanWeinbergMechanism-Extended}
    \mathscr{L}=\frac{1}{2}(\partial\phi_1)^2+\frac{1}{2}(\partial\phi_2)^2-\frac{g}{\sqrt{2}}A_\mu\,(\phi_1\overset{\longleftrightarrow}{\partial^\mu}\phi_2)-\frac{1}{4}(F_{\mu\nu})^2+2g^2A_{\mu}^2\phi^2-\frac{\lambda}{4!}\phi^4\,.
\end{equation}
Next, we employ an expansion around the scalar field's classical solution defined here are as
\begin{equation}
    \phi(x)=\phi_c(x)+\hat{\phi}(x)\,,
\end{equation}
where $\hat{\phi}$ is assumed to be a perturbation. The action is expanded as:
\begin{equation}
    \mathcal{S}_\text{eff}[\phi]\approx\mathcal{S}_\text{eff}[\phi_c]+\int\!\mathrm{d}^4x'\left.\frac{\delta\mathcal{S}_\text{eff}}{\delta\phi'}\right|_{\phi=\phi_c}\delta\phi'+\frac{1}{2!}\int\!\mathrm{d}^4x'\int\!\mathrm{d}^4x''\,\left.\frac{\delta^2\mathcal{S}_\text{eff}}{\delta\phi'\delta\phi''}\right|_{\phi=\phi_c}\delta\phi'\delta\phi''+\ldots
\end{equation}
The equation of motion, or $\delta_\phi\mathcal{S}=0$, is simply the familiar Klein-Gordon one
\begin{equation}
    \partial_{\mu}\partial^\mu \phi=V'(\phi)\,.
\end{equation}
Therefore, trivially the second variation of the action gives rise to the following integral
\begin{equation}
    \delta^{(2)}_\phi\mathcal{S}=\int\!\mathrm{d}^4x\,\hat{\phi}^T\left(\partial^2-V''(\phi)\right)\hat{\phi}\,.
\end{equation}
Then $\hat{\phi}$ can be integrated out and the final path integral reads as follows:
\begin{equation}
    \mathcal{Z}\propto\int\!\left(\text{det}D(x)\right)^{-1/2}\,e^{i\mathcal{S}[\phi_c]}\,,
\end{equation}
where
\begin{equation}
    D(x)\equiv\partial_\mu\partial^\mu-V''(\phi)\,.
\end{equation}
Finally we can include in the effective action the correction coming from the determinant and obtain the following expression
\begin{equation}
    \Gamma[\phi]=S[\phi]+\frac{i\hbar}{2}\text{Tr}\log{D(x)}\,.
\end{equation}
Now, the functional trace is a trace over the space which the operator $D$ acts, i.e.
\begin{equation}
    \left(D\,\hat{\phi}\right)_x=\sum_yD_{xy}\hat{\phi}_y =\left(\partial^2_x-V''\right)\hat{\phi}
\end{equation}
and matrix elements $(\partial^2-V'')_{xy}=\delta^{(4)}(x-y)(\partial^2_x-V'')$. Here we assumed that the background field $\phi$ is constant and not the same as the fluctuation $\hat{\phi}$. We can represent the trace-log as a position integral
\begin{align}
    \text{Tr}\log{(\partial^2-V'')}&=\int\!\mathrm{d}^4x\bra{x}\log(\partial^2-V'')\ket{x}\nonumber\\
    &=\iiint\!\mathrm{d}^4x\,\frac{\mathrm{d}^4k}{(2\pi)^4}\,\frac{\mathrm{d}^4k'}{(2\pi)^4}\,\braket{x|k'}\bra{k'}\log(k^2-V'')\ket{k}\braket{k|x}\nonumber\\
    &=\iint\mathrm{d}^4x\,\frac{\mathrm{d}^4k}{(2\pi)^4}\log(k^2-V'')\,,
\end{align}
where we used $\mathds{1}=\int\!\mathrm{d}^4k\,\ket{k}\bra{k}$ and $||\braket{x|k}||^2=1$. Since the logarithm of a dimensionful quantity is ill-defined we have to regulate it by choosing the additive constant (regulating tadpole diagrams). Then all the terms are added to the effective potential which reads
\begin{equation}\label{1lpot}
    U(\phi)=V(\phi)-\frac{i\hbar}{2}\int\!\frac{\mathrm{d}^4k}{(2\pi)^4}\,\log{\left(\frac{k^2-V''}{k^2}\right)}+\mathcal{O}(\hbar^2)
\end{equation}
Finally we can calculate the integral up to a cut-off scale $k^2=\Lambda^2$ after a Wick rotation $k\to k_E$ to limit the integration contour $\{k_E|\,k_E^2\leq\Lambda^2\}$ and obtain the following effective potential
\begin{equation}
    U(\phi)=V(\phi)+\frac{\Lambda^2}{32\pi^2}\,V''(\phi)-\frac{(V''(\phi))^2}{64\pi^2}\ln{\left(\frac{\sqrt{e}\Lambda^2}{V''(\phi)}\right)}\,.
\end{equation}

There is still a contribution left coming from the path integral over the gauge field $A^\mu$, which in the Landau gauge $\partial_\mu A^\mu=ik_\mu A^\mu=0$ is equivalent to
\begin{equation}
    \int\![\mathcal{D}A_\mu]\,\text{exp}\left\{\frac{i}{\hbar}\int\!\mathrm{d}^4x\left(-\frac{1}{4}(F_{\mu\nu})^2+g^2\phi^2A_\mu^2\right)\right\}=\left(\text{det}(\partial^2+2g^2\phi^2)\right)^{-3/2}\,.
\end{equation}
After following the same process described earlier it brings the one-loop corrected potential to its final form of
\begin{equation}
    V_{1\ell}(\phi)=\frac{1}{4!}\lambda\phi^2+\frac{1}{2}Z_m\phi^2+\frac{1}{4!}Z_\lambda\phi^4+\frac{\Lambda^2\phi^2}{64\pi^2}(\lambda+6g^2)+\frac{\lambda^2\phi^4}{256\pi^2}\left(\ln{\frac{\lambda\phi^2}{2\Lambda^2}}-\frac{1}{2}\right)+\frac{3\lambda^2\phi^4}{64\pi^2}\left(\ln{\frac{g^2\phi^2}{\Lambda^2}}-\frac{1}{2}\right)\,,
\end{equation}
where we included the counterterms $Z_m$ and $Z_\lambda$ required to absorb the divergences from the loop diagrams.\footnote{Note that this result is the same as if we blindly used the general formula for the one-loop effective potential
\begin{equation*}
    \Delta V_\text{eff}(\phi)=\sum_\text{dof}\frac{(-1)^f}{2^8\pi^2}\,(V'')^2\left(\ln{\frac{V''(\phi)}{2\Lambda^2}}-\frac{3}{2}\right)\,,
\end{equation*}
where the sum is over all the degrees of freedom and $(-1)^f$ is $-1$ for fermions and $+1$ for bosons.} We may impose the following renormalisation conditions 
\begin{equation}
    \frac{\mathrm{d}^2V_{1\ell}(\phi)}{\mathrm{d}\phi^2}=0\,,\qquad\&\qquad\left.\frac{\mathrm{d}^4V_{1\ell}(\phi)}{\mathrm{d}\phi}\right|_{\Lambda=\mu}=\lambda\,,
\end{equation}
where the first one imposes the massless condition and the second one defines the quartic coupling $\lambda$ at some renormalisation scale $\mu$. It is then straightforward to show after some algebra that the effective potential reads
\begin{equation}
    V_{1\ell}(\phi)=\frac{\lambda}{4!}\phi^4+\left(\frac{\lambda^2}{256\pi^2}+\frac{3g^4}{64\pi^2}\right)\phi^4\left(\ln{\frac{\phi^2}{\mu^2}}-\frac{25}{6}\right)\,.
\end{equation}

The interactions of the gauge bosons and the scalar field dynamically break the symmetry at perturbative couplings, and since $\lambda\sim g^4$ we can drop the quadratic term $\lambda^2$ and express the potential as
\begin{equation}
    V_{1\ell}(\phi)=\frac{\lambda}{4!}\phi^4+\frac{3g^4}{64\pi^2}\phi^4\left(\ln{\frac{\phi^2}{\mu^2}}-\frac{25}{6}\right)\,.
\end{equation}
Next, if we choose the renormalisation scale at the VEV of $\phi$, say $\mu=\braket{\phi}$, there is a nontrivial minimum of the effective potential at
\begin{equation}
    \lambda=\frac{33}{8\pi^2}g^4\,.
\end{equation}
Notice that in this case we can balance the quartic coupling $\lambda$ against the gauge coupling $g$. After substituting all of the above we obtain the final form of the effective potential 
\begin{equation}
    V_{1\ell}(\phi)=\frac{3g^4}{64\pi^2}\phi^4\left(\ln{\frac{\phi^2}{\braket{\phi}^2}}-\frac{1}{2}\right)\,.
\end{equation}

This is the CW mechanism in a nutshell, in which the scale $\braket{\phi}$ is generated dynamically and exchanged for the dimensionless coupling $\lambda$, in a process formally known as \emph{dimensional transmutation}. The masses of the particles in this particular theory are calculated directly -- recall that eq.~\eqref{Action:ColemanWeinbergMechanism-Extended} -- as
\begin{equation}
    m^2_A=g^2\braket{\phi}^2\,,\qquad\&\qquad  m_\phi^2=\frac{3g^4}{8\pi^2}\,\braket{\phi}^2=\frac{3g^2}{8\pi^2}\,m_A^2\,,
\end{equation}
for the vector boson and the scalar boson respectively. The $\beta$ functions of the couplings have the following form~\cite{Coleman1973}
\begin{align} 
    \beta_g&=\frac{g^3}{48\pi^2}\,,\\
    \beta_\lambda&=\frac{1}{4\pi^2}\left(\frac{5}{6}\lambda^2-3\lambda g^2+9g^4\right)\,.
\end{align}
Notice that the $\beta$ function of $\lambda$ is positive suggesting that the value of $\lambda$ decreases with energy. There exists then one energy scale, say $\mu_c$, where the value of the coupling is $\lambda(\mu_c)=33 g^4(\mu_c)/(8\pi)$, denoting exactly the energy scale of symmetry breaking. Therefore, we can understand the relation between couplings as a general result triggered by their RG running and not some sort of fine-tuning.

\subsubsection*{\emph{Return to slow-roll inflation}}

Motivated by the possible scale-invariance at high energy scales we are interested in studying a theory of a classically (quasi-)scale-invariant spectrum described by a Lagrangian such as~\cite{Antoniadis2018}
\begin{equation}\label{Action:CW-InitialLagr}
    \mathscr{L}=\frac{1}{2}\xi\phi^2g^{\mu\nu}R_{\mu\nu}(\Gamma)+\frac{\alpha}{4}R^2(g,\Gamma)-\frac{1}{2}(\nabla\phi)^2-\frac{\lambda}{4}\phi^4-\Lambda^4\,,
\end{equation}
where $\Lambda$ plays the role of a cosmological constant and is the only dimensionful parameter in the theory. We show later that during inflation close to the Planck scale $\Lambda\ll M_P$, meaning that scale-invariance is only softly broken. The coupling $\xi$ quantifies the nonminimal coupling of the scalar field $\phi(x)$ with gravity. In the absence of the $R^2$ term it was shown~\cite{Racioppi2017} that for $\xi\gtrsim 0.1$ the above action leads to linear inflation~\cite{Kannike2016}.

To this Lagrangian we should add in principle the Lagrangian $\mathscr{L}(\Phi,\psi,A_\mu)$ that contains the possible scale-invariant interactions of $\phi$ with all matter fields. As discussed previously these interactions at the quantum level will generate radiative corrections, which calculated in the (flat space) Jordan frame we group them in a general potential $V_{1\ell}(\phi)$. Assuming that the field content is such that the CW mechanism can be implemented the one-loop effective potential reads~\cite{Kannike2016,Racioppi2017,Antoniadis2018}
\begin{equation}
    V_{1\ell}(\phi)=\Lambda^4\left[1+\frac{\phi^4}{\braket{\phi}^4}\left(2\ln{\frac{\phi^2}{\braket{\phi}^2}}-1\right)\right]\,.
\end{equation}

An important note is that the Lagrangian as presented in eq.~\eqref{Action:CW-InitialLagr} lacks the scale of gravity, in other words the Planck mass is explicitly absent. We can circumvent that by demanding that the nonzero VEV of the scalar field obeys
\begin{equation}
    \braket{\phi}^2=\frac{M_P^2}{\xi}\,.
\end{equation}
Then the constant term $\Lambda$ ensures that the potential vanishes at the minimum in order to avoid issues related to the possibility of eternal inflation. Alternatively, it is possible to replace that condition by assuming the vanishing of the overall scalar potential $\bar{V}\propto V_{1\ell}(\phi)+\alpha\chi^4$ with the effective potential then given by $V_{1\ell}=M^4\phi^4(2\ln\frac{\phi^2}{\braket{\phi}^4}-1)$, where the coefficient $M$ depends on the characteristics of the matter field content. Otherwise, the potential can be brought to its final form given by~\cite{Racioppi2017,Antoniadis2018}
\begin{equation}
    V_{1\ell}(\phi)=\Lambda^4\left[1+\frac{\xi^2\phi^4}{M_P^4}\left(2\ln{\frac{\xi\phi^2}{M_P^2}}-1\right)\right]
\end{equation}

Next, we can follow the same steps as in the previous section of minimally coupled fields, i.e. express the total action in its scalar representation
\begin{equation}
    \mathcal{S}[\text{g},\Gamma,\phi,\chi]=\int\!\mathrm{d}^4x\,\sqrt{-g}\left\{\frac{1}{2}(\xi\phi^2+\alpha\chi^2)g^{\mu\nu}R_{\mu\nu}(\Gamma)-\frac{1}{2}(\nabla\phi)^2-\frac{\alpha}{4}\chi^4-V_{1\ell}(\phi)\right\}\,,
\end{equation}
perform a Weyl rescaling of the metric
\begin{equation}
    \overline{g}_{\mu\nu}(x)=\frac{\xi\phi^2+\alpha\chi^2}{M_P^2}\,g_{\mu\nu}(x)\,,
\end{equation}
to obtain the action in the Einstein frame
\begin{equation}
    \mathcal{S}[\overline{\text{g}},\Gamma,\phi,\chi]=\int\!\mathrm{d}^4x\,\sqrt{-\overline{g}}\left\{\frac{M_P^2}{2}\,\overline{g}^{\mu\nu}R_{\mu\nu}(\Gamma)-\frac{1}{2}(\overline{\nabla}\phi)^2-\overline{V}(\phi,\chi)\right\}\,,
\end{equation}
where we defined
\begin{equation}
    \overline{V}=M_P^2\,\frac{V_{1\ell}+\displaystyle{\frac{\alpha\chi^4}{4}}}{\xi\phi^2+\alpha\chi^4}\,.
\end{equation}

Variation of the action with respect to the connection results to the usual Levi-Civita condition
\begin{equation}
    \frac{\delta\mathcal{S}}{\delta\Gamma}=0\,\quad\implies\quad {\Gamma^\rho}_{\mu\nu}\stackrel{!}{=}\overline{\{{}_\mu{}^\rho{}_\nu\}}\,.
\end{equation}
The equation of motion for the auxiliary field $\chi$ is~\cite{Antoniadis2018}
\begin{equation}
\chi^2=\frac{4V_{1\ell}(\phi)+\displaystyle{\frac{\xi\phi^2(\overline{\nabla}\phi)^2}{M_P^2}}}{\xi\phi^2-\alpha\displaystyle{\frac{(\overline{\nabla}\phi)^2}{M_P^2}}}\,,  
\end{equation}
which after its substitution to the starting action functional we obtain
\begin{equation}
    \mathcal{S}[\overline{\text{g}},\phi]=\int\!\mathrm{d}^4x\,\sqrt{-\overline{g}}\left\{\frac{M_P^2}{2}\overline{R}-\frac{1}{2}\left(\frac{\xi\phi^2M_P^2}{\xi^2\phi^4+4\alpha V_{1\ell}(\phi)}\right)(\overline{\nabla}\phi)^2-\left(\frac{M_P^4}{\xi^2\phi^4+4\alpha V_{1\ell}(\phi)}\right)V_{1\ell}(\phi)+\mathcal{O}((\overline{\nabla}\phi)^4)\right\}\,.
\end{equation}

We can introduce the canonically normalised scalar field $\varphi$ through
\begin{equation}
    \varphi=\int\!\mathrm{d}\phi\,\sqrt{\frac{\xi\phi^2}{\xi^2\phi^4+4\alpha V_{1\ell}(\phi)}}\stackrel{\Lambda\ll M_P}{\approx}\frac{M_P}{2\sqrt{\xi}}\,\ln{\frac{\xi\phi^2}{M_P^2}}\,,
\end{equation}
which then brings the Lagrangian to its canonical form of
\begin{equation}
    \mathscr{L}\supset-\frac{1}{2}(\overline{\nabla}\varphi)-V(\varphi)\,,
\end{equation}
where the scalar potential is now given by~\cite{Antoniadis2018}
\begin{equation}
    V(\varphi)=\frac{M_P^4\,\Lambda^4\left(4\sqrt{\xi}\displaystyle{\frac{\varphi}{M_P}}-1+\text{exp}\left(-4\sqrt{\xi}\displaystyle{\frac{\varphi}{M_P}}\right)\right)}{M_P^4+4\alpha\Lambda^4\left(4\sqrt{\xi}\displaystyle{\frac{\varphi}{M_P}}-1+\text{exp}\left(-4\sqrt{\xi}\displaystyle{\frac{\varphi}{M_P}}\right)\right)}\,,
\end{equation}
which for large values of the normalised field $\varphi$ tends to $U(\varphi\to\infty)\to M_P^4/4\alpha$, while giving rise to a minimum at field values of $\varphi=0$. The asymptotic flatness of the inflationary potential is not at all accidental, it is generated due to the $R^2$ term similar to the case of minimally coupled models studied in the previous section. It is straightforward to show that the potential includes other inflationary models as limiting cases, e.g. as $\sqrt{\xi}\ll1$ (or equivalently $\braket{\phi}\ll1$) the potential is approximated by
\begin{equation}
    V(\varphi)\stackrel{\sqrt{\xi}\ll1}{\approx}8\xi\,\frac{\Lambda^4}{M_P^2}\,\varphi^2+\mathcal{O}(\varphi^3)\,,
\end{equation}
which is a quadratic monomial and at first order expansion it is independent of $\alpha$.

Since the action functional is given in terms of a canonically normalised scalar $\varphi$ with its self-interacting potential $V(\varphi)$ coupled minimally to the Einstein-Hilbert term we can obtain the canonical Friedmann equations by assuming a time-dependent homogeneous field $\varphi(x^\mu)=\varphi(t)$ and an FRW metric $\mathrm{d}s^2=-\mathrm{d}t^2+a^2(t)\mathrm{d}x^2$
\begin{equation}
    3M_P^2H^2=\frac{1}{2}\dot{\varphi}^2+V(\varphi)\,.
\end{equation}
Then, we can safely employ the slow-roll approximation and obtain the first order slow-roll parameters as~\cite{Antoniadis2018}
\begin{align}
    \epsilon_V&=\frac{M_P^2}{2\left(\varphi-\displaystyle{\frac{M_P}{4\sqrt{\xi}}}\right)^2\left[1+16\alpha\sqrt{\xi}\displaystyle{\frac{\Lambda^4}{M_P^5}}\left(\varphi-\displaystyle{\frac{M_P}{4\sqrt{\xi}}}\right)\right]^2}\,,\\
    \eta_V&=-\frac{32\alpha\sqrt{\xi}\,\displaystyle{\frac{\Lambda^4}{M_P^4}}}{\left(\varphi-\displaystyle{\frac{M_P}{4\sqrt{\xi}}}\right)^2\left[1+16\alpha\sqrt{\xi}\displaystyle{\frac{\Lambda^4}{M_P^4}}\left(\varphi-\displaystyle{\frac{M_P}{4\sqrt{\xi}}}\right)\right]^2}\,.
\end{align}
The field value at the end of inflation can be determined via the condition of its termination $\epsilon_V(\varphi_f)\simeq1$, leading to
\begin{align}
    \varphi_f&=\frac{M_P^4}{32\Lambda^4\alpha\sqrt{\xi}}\left(\sqrt{1+\frac{\sqrt{2}M_P^4}{16\alpha\Lambda^4\sqrt{\xi}}}-1\right)+4M_P^4\sqrt{\xi}\\
    &\approx4M_P^4\sqrt{\xi}+\frac{M_P}{2\sqrt{2}}-2\alpha\sqrt{\xi}\,\frac{\Lambda^4}{M_P^3}\,,\nonumber
\end{align}
which after inserting it to the definition of the number of $e$-foldings it gives rise to the following expression
\begin{equation}
    N\approx\ln\frac{\left(\frac{\varphi_i}{M_P}-4\sqrt{\xi}\right)}{\left(\frac{\varphi_f}{M_P}-4\sqrt{\xi}\right)}+16\alpha\sqrt{\xi}\,\frac{\Lambda^4}{M_P^5}(\varphi_i-\varphi_f)\,.
\end{equation}

\begin{figure}
    \centering
    \includegraphics[scale=0.35]{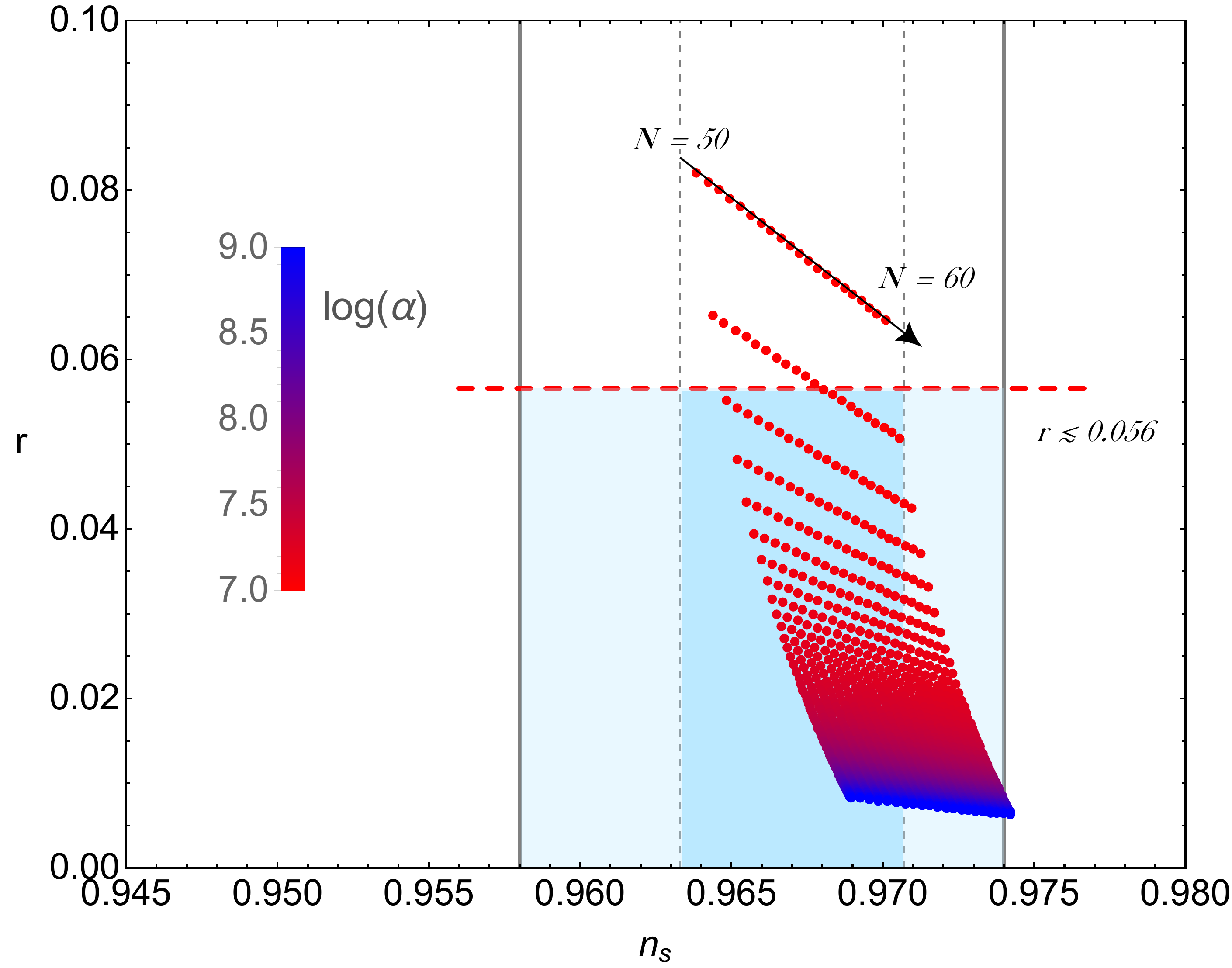}
    \caption{A plot of the $r-n_s$ region for various values of $\alpha\in[10^7,10^9]$, $\xi=10^{-3}$, $\Lambda=10^{-2}$ and $M_P=1$. The light and dark blue region denote the $2\sigma$ and $1\sigma$ allowed region from the Planck2018 collaboration, while the red dashed line denotes the bound on $r\lesssim0.056$. Each dot is a solution for a specific value of $N$, which from left to right is increasing from a starting value of $N=50$ to $N=60$ $e$-foldings, and each ``line'' is formed for a specific value of $\alpha$ based on the colour coding of the legend.}
    \label{fig:CW-RNsa}
\end{figure}

In fig.~\ref{fig:CW-RNsa} we present the numerical analysis of the model and its predictions regarding the inflationary observables $r$ and $n_s$ for various values of the parameter $\alpha$. Specifically, we considered different cases of $\alpha\in[10^7,10^9]$ with fixed values of the nonminimal coupling $\xi=10^{-3}$ and the scale $\Lambda=10^{-2}$, satisfying the initial assumption of $\Lambda\ll M_P=1$. This particular parameter space is capable of reproducing the correct values of the power spectrum $\mathcal{A}_s\propto10^{-9}$. It is noticeable that for smaller values of $\alpha<10^7$ the predictions lie outside the allowed region with the limiting case of $\alpha=0$ leading to values of $r\sim10^{-1}$. The effect of the parameter $\alpha$ is effectively to reduce the tensor-to-scalar ratio and in turn lead the model back into the $2\sigma$ (if not the $1\sigma$) of observations. This behaviour was first highlighted in ref.~\cite{Racioppi2017,Antoniadis2018} and subsequent studies included different features~\cite{Bostan2020,Racioppi2020} such as the reheating phase of the present model~\cite{Gialamas2020a}.

If we assume varying values of $\Lambda$ the observables quickly saturate at values close to $r\approx10^{-1}$ and $n_s\sim0.965$ when $\Lambda\sim10^{-3}$. In contrast, increasing values of $\xi$ tend to decrease the tensor-to-scalar ratio but increase the scalar spectral index to the point where it is outside the $2\sigma$ region at $\xi\sim0.02$, with larger values of $\xi\gtrsim0.02$ being excluded. In the limit of $\sqrt{\xi}\ll1$ the values of the observables tend to the those predicted by a quadratic monomial. All of these are illustrated in the next figure, fig.~\ref{fig:CW-RNsxi}, where the same plot is reproduced but now for varying values of $\xi\in[10^{-4},10^{-2}]$ and fixed values of $\alpha=10^8$, $\Lambda=10^{-2}$. 

\begin{figure}
    \centering
    \includegraphics[scale=0.35]{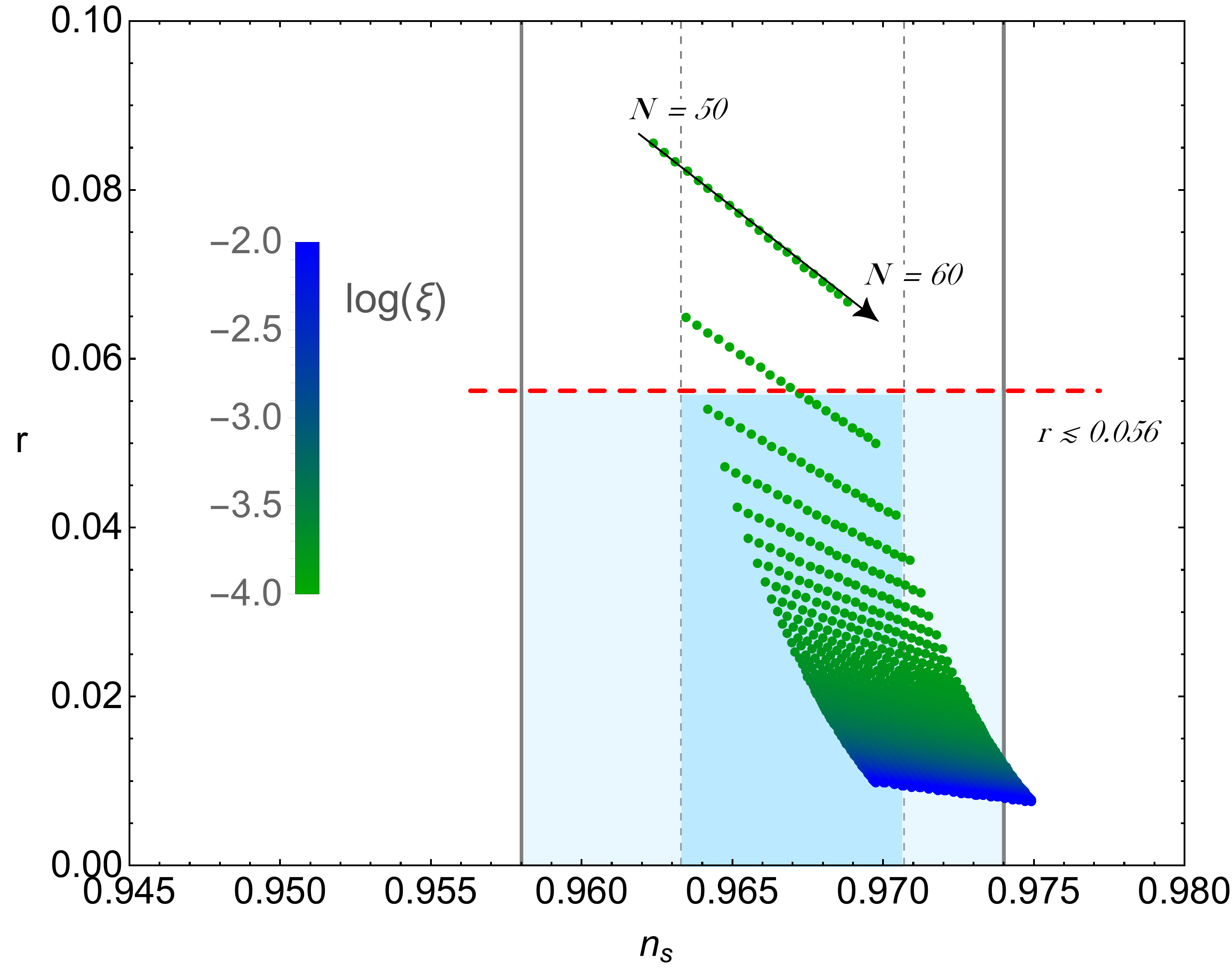}
    \caption{A plot of the predicted values of $r-n_s$ for varying values of $\xi\in[10^{-4},10^{-2}]$ and fixed values of $\alpha=10^8$, $\Lambda=10^{-2}$ and $M_P=1$. The values of the power spectrum are not entirely respected and the plot it mostly to illustrate the behaviour of the observables with respect to varying values of $\xi$. In fact, increasing values of $\xi$ suggest decreasing values of $\Lambda$, and vice versa, in order to obtain the observed value $\mathcal{A}_s$. The figure follows the ``notation'' of the previous figure, fig.~\ref{fig:CW-RNsa}.}
    \label{fig:CW-RNsxi}
\end{figure}

\subsection{\emph{Induced gravity model}}

Another way of spontaneous symmetry breaking (SSB) that can also dynamically generate the Planck scale can be achieved by assuming a scalar field with a Higgs-like potential
\begin{equation}\label{Cond:InducedGrav-Pot}
    V(\phi)=\frac{\lambda}{4}\left(\phi^2-v^2\right)^2\,,
\end{equation}
where $v=\braket{\phi}$ is the VEV of the field. In this case the breaking occurs due to a condensate of the Ginzburg-Landau type. Then, a nonminimal coupling of the scalar $\phi$ to the Einstein-Hilbert term through
\begin{equation}
    \mathcal{S}=\int\!\mathrm{d}^4x\,\sqrt{-g}\left\{\frac{1}{2}\xi\phi^2R+\frac{\alpha}{4}R^2-\frac{1}{2}(\nabla\phi)^2-\frac{\lambda}{4}(\phi^2-v^2)^2\right\}\,,
\end{equation}
is able to generate dynamically the Planck scale when the field acquires its VEV 
\begin{equation}
    \phi\ \longrightarrow\ \braket{\phi}=\frac{M_P}{\sqrt{\xi}}\,.
\end{equation}

These models originate from early attempts to reconcile the dynamics of SSB and gravity~\cite{Zeldovich1967,Zee1979}. They were celebrated for their ability to ``induce gravity'' through the VEV of scalar field, however since their initial formulation many more have been proposed and what was once their main attraction has now been largely forgotten. Nowadays, most of them are considered as inflationary models~\cite{Accetta1985,Kao1990} and as such, in this section, we study the behaviour of this particular model when coupled also with an $R^2$ term under the Palatini formalism.

Following what we did in the previous section we obtain a similar effective action in the Einstein frame~\cite{Antoniadis2018}
\begin{equation}
    \mathcal{S}[\overline{\text{g}},\phi]=\int\!\mathrm{d}^4x\,\sqrt{-\overline{g}}\left\{\frac{M_P^2}{2}\overline{R}-\frac{1}{2}\left(\frac{\xi\phi^2M_P^2}{\xi^2\phi^4+4\alpha V(\phi)}\right)(\overline{\nabla}\phi)^2-\left(\frac{M_P^4}{\xi^2\phi^4+4\alpha V(\phi)}\right)V(\phi)+\mathcal{O}((\overline{\nabla}\phi)^4)\right\}\,,
\end{equation}
where the potential $V(\phi)$ is given in eq.~\eqref{Cond:InducedGrav-Pot}. The action is produced after the substitution of the solution to the constraint equation $\delta_\chi\mathcal{S}=0$ and the Levi-Civita condition with respect to the metric $\overline{\text{g}}$.

We may introduce the canonical scalar via the redefinition
\begin{equation}
    \varphi=\frac{M_P^2}{2}\sqrt{\frac{\xi}{\xi^2+\alpha\lambda}}\,\ln\left(\phi^2(\alpha\lambda+\xi^2)-\alpha\lambda v^2+\sqrt{\alpha\lambda+\xi^2}\sqrt{\xi^2\phi^4+\alpha\lambda(\phi^2-v^2)^2}\right)\,,
\end{equation}
thus the inflaton potential expressed in terms of $\varphi$ becomes~\cite{Antoniadis2018}
\begin{equation}
    \bar{V}(\varphi)\equiv\frac{M_P^4}{\xi^2\phi^4+4\alpha V(\phi)}\,V(\phi)=\frac{\lambda M_P^4}{4(\xi^2+\alpha\lambda)}\left(\frac{\text{exp}\left(\frac{4\varphi\sqrt{\alpha\lambda+\xi^2}}{M_P\sqrt{\xi}}\right)-2\,\text{exp}\left(\frac{2\varphi\sqrt{\alpha\lambda+\xi^2}}{M_P\sqrt{\xi}}\right)-\alpha\lambda M_P^4}{\alpha\lambda M_P^4+\text{exp}\left(\frac{4\varphi\sqrt{\alpha\lambda+\xi^2}}{M_P\sqrt{\xi}}\right)}\right)^2\,,
\end{equation}
where we identified the VEV of the scalar field with $v^2=M_P^2/\xi$.

Then the matter Lagrangian is given in terms of scalar field with a canonical kinetic term and a self-interacting scalar potential, schematically of the form of $\mathscr{L}\supset-\frac{1}{2}(\overline{\nabla}\varphi)^2-\bar{V}(\varphi)$. Applying the usual slow-roll approximation $\dot{\varphi}^2\ll V(\varphi)$ the first-order slow-roll parameters have the following form~\cite{Antoniadis2018}
\begin{equation}
        \epsilon_V(\varphi)=\frac{32M_P^4(\alpha\lambda+\xi^2)\text{exp}\left(\frac{4\varphi\sqrt{\alpha\lambda+\xi^2}}{M_P\sqrt{\xi}}\right)}{\xi\left(\text{exp}\left(\frac{4\varphi\sqrt{\alpha\lambda+\xi^2}}{M_P\sqrt{\xi}}\right)+\alpha\lambda M_P^4\right)^2}\left(\frac{\xi\text{exp}\left(\frac{4\varphi\sqrt{\alpha\lambda+\xi^2}}{M_P\sqrt{\xi}}\right)+2\alpha\lambda M_P^2\text{exp}\left(\frac{2\varphi\sqrt{\alpha\lambda+\xi^2}}{M_P\sqrt{\xi}}\right)-\alpha\lambda\xi M_P^4}{\text{exp}\left(\frac{4\varphi\sqrt{\alpha\lambda+\xi^2}}{M_P\sqrt{\xi}}\right)-2\xi M_P^2\text{exp}\left(\frac{2\varphi\sqrt{\alpha\lambda+\xi^2}}{M_P\sqrt{\xi}}\right)-\alpha\lambda M_P^4}\right)^2\,,
\end{equation}
and
\begin{align}
    \eta_V(\varphi) &=\frac{16 M_P^2 \left(\alpha  \lambda +\xi ^2\right)}{\xi } \left(\frac{6 \alpha ^2 \lambda ^2 M_P^6}{\left(e^{\frac{4 \varphi  \sqrt{\alpha  \lambda +\xi ^2}}{\sqrt{\xi } M_P}}+\alpha  \lambda  M_P^4\right)^2}+\frac{2 M_P^4 \left(\alpha  \lambda +\xi ^2\right) \left(2 \xi  e^{\frac{2 \varphi  \sqrt{\alpha  \lambda +\xi ^2}}{\sqrt{\xi } M_P}}+\alpha  \lambda  M_P^2\right)}{\left(-2 \xi  M_P^2 e^{\frac{2 \varphi  \sqrt{\alpha  \lambda +\xi ^2}}{\sqrt{\xi } M_P}}+e^{\frac{4 \varphi  \sqrt{\alpha  \lambda +\xi ^2}}{\sqrt{\xi } M_P}}-\alpha  \lambda  M_P^4\right)^2}       \right. \nonumber \\
& \qquad \qquad \qquad \qquad\qquad \left. -\frac{6 \alpha  \lambda  M_P^2}{e^{\frac{4 \varphi  \sqrt{\alpha  \lambda +\xi ^2}}{\sqrt{\xi } M_P}}+\alpha  \lambda  M_P^4}+\frac{2 M_P^2 \left(\alpha  \lambda +\xi ^2\right)-\xi  e^{\frac{2 \varphi  \sqrt{\alpha  \lambda +\xi ^2}}{\sqrt{\xi } M_P}}}{-2 \xi  M_P^2 e^{\frac{2 \varphi  \sqrt{\alpha  \lambda +\xi ^2}}{\sqrt{\xi } M_P}}+e^{\frac{4 \varphi  \sqrt{\alpha  \lambda +\xi ^2}}{\sqrt{\xi } M_P}}-\alpha  \lambda  M_P^4}\right)\,.
\end{align}

\begin{figure}
    \centering
    \includegraphics[scale=0.4]{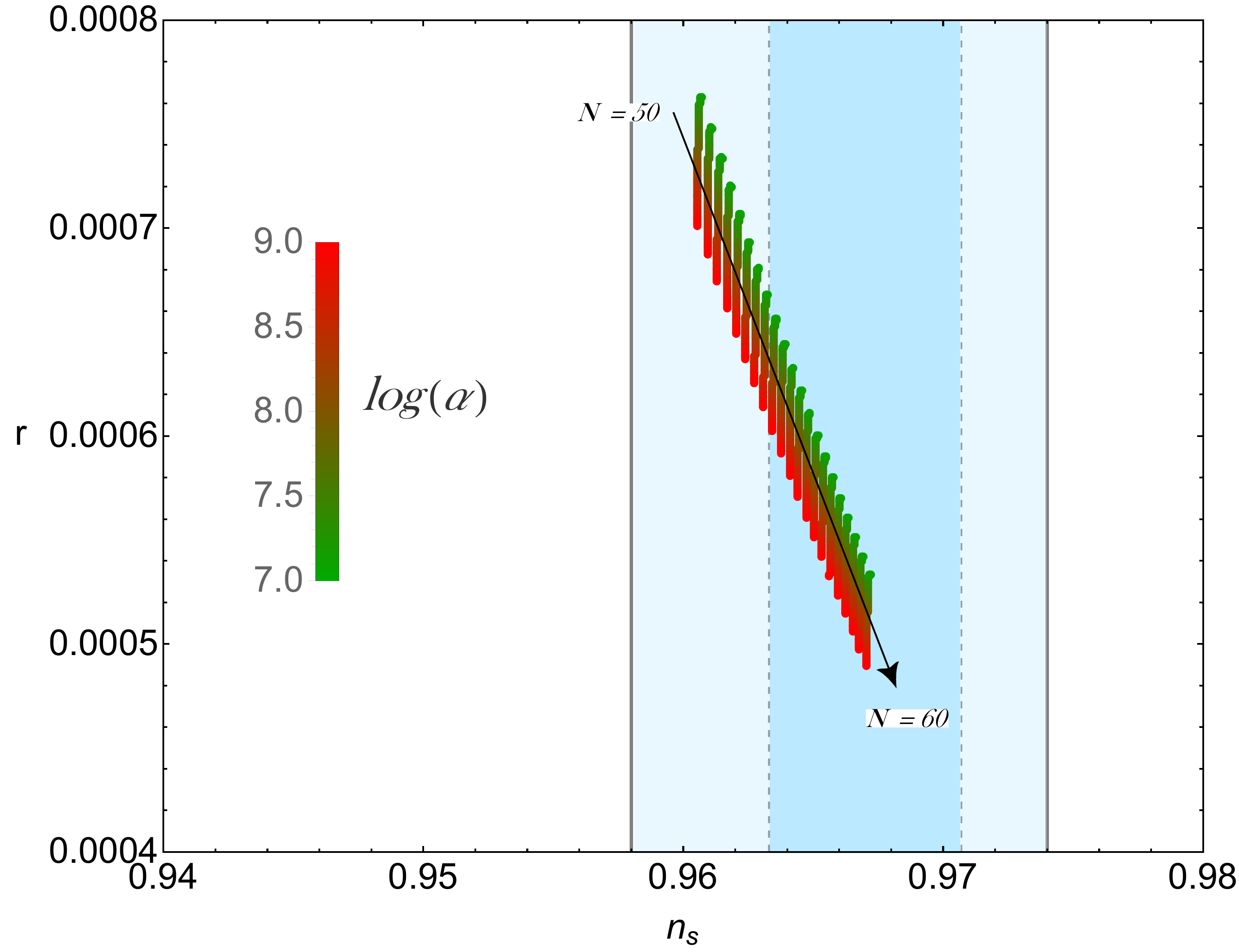}
    \caption{A plot of the $r-n_s$ region for various values of $\alpha\in[10^7,10^9]$, $\xi=1$, $\lambda=10^{-10}$,$v=M_P/\sqrt{\xi}=M_P$ and $M_P=1$. The light and dark blue region denote the $2\sigma$ and $1\sigma$ allowed region from the Planck2018 collaboration. Each dot is a solution for a specific value of $N$, which from left to right is increasing from a starting value of $N=50$ to $N=60$ $e$-foldings, and each ``line'' is formed for a specific value of $\alpha$ based on the colour coding of the legend.}
    \label{fig:Induced-RNsa}
\end{figure}

\begin{figure}
    \centering
    \includegraphics[scale=0.5]{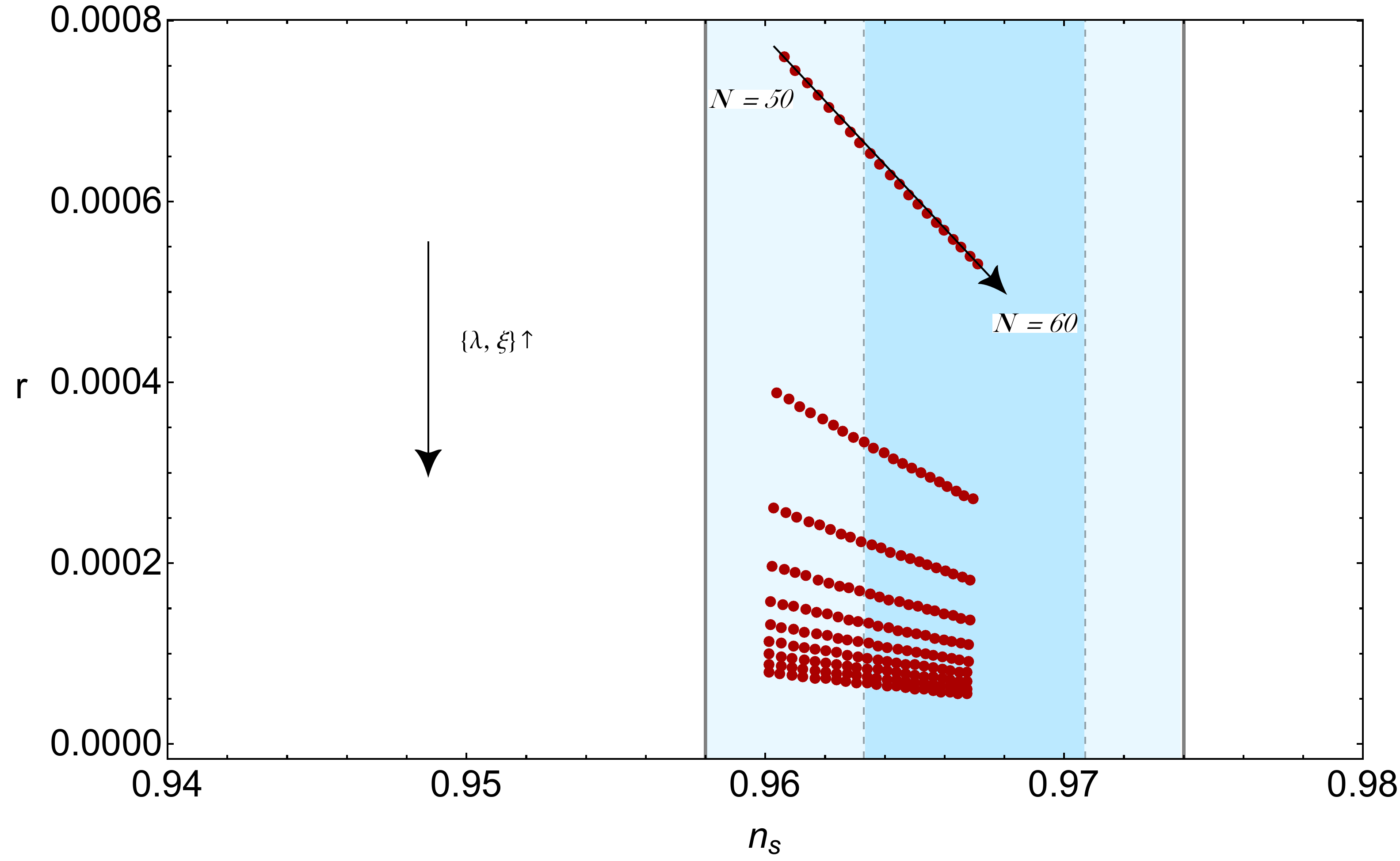}
    \caption{A plot of the $r-n_s$ region for various values of $\xi\in[1,10]$, $\lambda\in[10^{-10},10^{-9}]$ and $v=M_P/\sqrt{\xi}$, while keeping fixed the values of $\alpha=10^8$ and $M_P=1$. The light and dark blue region denote the $2\sigma$ and $1\sigma$ allowed region from the Planck2018 collaboration. Each dot is a solution for a specific value of $N$, which from left to right is increasing from a starting value of $N=50$ to $N=60$ $e$-foldings.}
    \label{fig:Induced-RNsxi}
\end{figure}

Due to the complexity of the multi-dimensional parameter space we resort to numerical methods to study the model. The condition for the inflation to end at some field value $\epsilon_V(\varphi_f)=1$ is solved numerically for a specific set of the parameters $\{\alpha,\lambda,\xi\}$. Then, by requiring inflation to last at least $N\in\{50,60\}$ $e$-foldings we obtain the field value at the start of inflation, $\varphi_i$. In fig.~\ref{fig:Induced-RNsa} we present the results for the inflationary observables $r-n_s$. The dashed and solid lines (or equivalently the shaded dark blue and light blue regions) represent the $1\sigma$ and $2\sigma$ range of $n_s$ set by the Planck 2018 collaboration. The bound on the power spectrum, $\mathcal{A}_s\approx10^{-9}$, is also satisfied approximately by parameters chosen in that figure and is largely independent of (large) values of $\alpha$. We notice that for increasing values of the $R^2$ parameter $\alpha$ the $n_s$ remains the same at each $e$-fold $N$ while $r$ is suppressed further to values $r\sim10^{-4}$. Further investigation reveals that ratio of the parameters has to be of the order of $\lambda/\xi\sim10^{-10}$ in order that the power spectrum $\mathcal{A}_s\approx10^{-9}$. This leads us to produce another figure, fig.~\ref{fig:Induced-RNsxi}, in which we plot the values of $r-n_s$ for $\alpha=10^{8}$ but varying values of $\lambda\in[10^{-10},10^{-9}]$ and $\xi\in[1,10]$, such that their ratio is always approximately $10^{-10}$, while setting $v=M_P/\sqrt{\xi}$ at each value of $\xi$. The behaviour recognised in that figure is similar to the one for varying $\alpha$, i.e. the values of $n_s$ do not change for varying $\xi$ and $\lambda$, but are dependent on the ratio $\lambda/\xi$ (similar to $\mathcal{A}_s$). However, increasing values of $\lambda$ and $\xi$ tend to decrease the already small values of $r$~\cite{Antoniadis2018}.

\subsection{\emph{Higgs field}}

In a previous section we addressed the possibility that the Higgs boson playing the role of the inflaton coupled minimally to the $R+R^2$ model. In the conventional metric formulation in order for the Higgs to successfully drive inflation a nonminimal coupling with the Einstein-Hilbert term has to be introduced~\cite{Bezrukov2008,DeSimone2009,Bezrukov2009a,Barbon2009,Rubio2019}. Initial interest of nonminimal Higgs inflation was targeted in the differences between the two formulations and it is then important to first review the simpler scenario, in which the $R^2$ term is absent.

\subsubsection*{\emph{Absence of the $R^2$ term -- an overview}}

Let us introduce once again the Higgs field coupled nonminimally with gravity; in the unitary gauge we obtain the following Lagrangian~\cite{Bauer2008}
\begin{equation}
    \mathscr{L}=\frac{1}{2}\left(M^2+\xi h^2\right)g^{\mu\nu}R_{\mu\nu}(\Gamma)-\frac{1}{2}(\nabla h)^2-V(h)\,,
\end{equation}
where $h(x)$ denotes the Higgs scalar and $M^2$ is an energy scale that together with the nonminimal coupling $\xi\phi^2$ identify the Planck scale. Next, we can eliminate the nonminimal coupling via a Weyl rescaling
\begin{equation}
    \overline{g}_{\mu\nu}(x)=\frac{M^2+\xi h^2}{M_P^2}\,g_{\mu\nu}(x)\,,
\end{equation}
to obtain the action in the Einstein frame
\begin{equation}
    \mathcal{S}=\int\!\mathrm{d}^4x\,\sqrt{-\overline{g}}\left\{\frac{M_P^2}{2}\,\overline{g}^{\mu\nu}R_{\mu\nu}(\Gamma)-\frac{M_P^2}{2(M^2+\xi h^2)}(\overline{\nabla}h)^2-\frac{\lambda h^4}{4(M^2+\xi h^2)^2}\right\}\,,
\end{equation}
where we ignored the Higgs VEV in the potential as we expect the field to assume values close to the Planck scale.

Let us normalise the kinetic term of the inflaton via a field redefinition of the form
\begin{equation}
    \frac{\mathrm{d}\varphi}{\mathrm{d}h}=\frac{1}{\sqrt{M^2+\xi h^2}}\qquad\implies\qquad\varphi=\frac{1}{\sqrt{\xi}}\sinh^{-1}(\sqrt{\xi}h)\,,
\end{equation}
where we identified the scale $M\equiv1$. Under this field redefinition the potential becomes
\begin{equation}\label{Cond:EFHiggsR2Pot}
    \bar{V}(\varphi)=\frac{\lambda}{4\xi^2}\,\tanh^2(\sqrt{\xi}\varphi)\,.
\end{equation}
At this point the theory is reminiscent of the free massive scalar field that is coupled minimally to the $R+R^2$ term, discussed in a previous section. Applying then the slow-roll approximation we can obtain the duration of inflation encoded in the formula 
\begin{equation}
    N=\int_{\varphi_f}^{\varphi_i}\!\frac{\bar{V}(\varphi)}{\bar{V}'(\varphi)}\,\mathrm{d}\varphi\approx\frac{1}{16\xi}\cosh{(2\sqrt{\xi}\varphi_i)}\approx\frac{h_i^2}{8}\,,
\end{equation}
where we assumed that $\varphi_i\gg\varphi_f$, or at least it dominates in the expression of $N=f(\varphi)$. Therefore, we can express the slow-roll parameters in terms of the approximate formula for the $e$-foldings $N$ as follows
\begin{equation}
    \epsilon_V\approx\frac{1}{8\xi N^2}\,\qquad\qquad \eta_V\approx-\frac{1}{N}\,.
\end{equation}
Assuming then an expansion of the inflationary observables around large numbers of $N$ we obtain the following approximate expressions~\cite{Bauer2008}
\begin{equation}\label{Cond:HiggswoutR2-rnsAs}
    n_s\approx1-\frac{2}{N}\,,\qquad r\approx\frac{2}{\xi N^2}\,\qquad \mathcal{A}_s\approx\frac{\lambda N^2}{12\pi^2\xi}\,.
\end{equation}
Clearly, since the value of the power spectrum depends on the ratio of the two free parameters $\lambda/\xi$ we can interchange one for the other using its observed value~\cite{Tenkanen2020,Gialamas2020}
\begin{equation}
    \xi\approx4\times10^6\, N^2\lambda\,.
\end{equation}
Thus, we are left effectively with only one free parameter, $\xi$ or $\lambda$. Then, in order for the observables eq.~\eqref{Cond:HiggswoutR2-rnsAs} to reside within the allowed region of observations for $N\!\sim\!55$ $e$-folds, the model parameters assume values in the range
\begin{equation}
    \xi\in[10^5,10^9]\,\implies\,\lambda\in[10^{-5},10^{-1}]\,,
\end{equation}
and vice verca.

Even the most conservative values of $\xi$ lead to highly suppressed values of $r\sim10^{-12}$, contrary to the usual metric formulation in which $r\sim10^{-3}$. Other than that the predicted values of $\xi$ in the Palatini formalism seem to be close to five magnitudes larger than those in the metric formalism, in which $\xi\sim(10^2-10^5)$. 

This encapsulates what is currently cited in the literature, however it does not constitute an appropriate comparison of the two formulations. A common link between the two can be found by invoking arguments based on the phenomenology of the proposed model, for example it is currently assumed that $\lambda\sim10^{-12}$ at energy scales close to the Planck scale~\cite{Rubio2019} and without any other BSM degrees of freedom present to stabilise the potential we are led to take it at face value, or at least assume some kind of tiny value of $\lambda$. Following the same procedure one can derive a similar relation between $\xi$ and $\lambda$ in the metric formulation of the model, reading~\cite{Bezrukov2008}
\begin{equation}
    \xi_{(2)}\approx4\times10^4\sqrt{\lambda_{(2)}}\,,
\end{equation}
where the subscript $(2)$ here denotes the metric (second-order) formalism. It is trivial then to show that in the case that $\lambda\sim10^{-12}$ both formulations demand values of $\xi\sim10^{-2}$, which in fact is in agreement with QFT wisdom that dimensionless couplings coming from perturbation theory (which is the main motivation of the nonminimal coupling) should be $\ll1$. If that is the case then the predicted value of the tensor-to-scalar ratio in the Palatini formalism is close to the current $2\sigma$ cut-off value of $r\sim0.05$, while the conventional metric formulation leads to $r_{(2)}\sim10^{-3}$ that depends only on the value of $N$ and not on the model parameters. This is an advantage of the \emph{Palatini-Higgs} model; since $r$ is dependent on $\xi$ any possible disagreement with future experiments can be remedied by placing stricter constraints on the allowed values of $\xi$, constraining in turn the values of $\lambda$ through $\mathcal{A}_s$.

\subsubsection*{\emph{Return of the $R^2$ term}}

Let us include the $R^2$ term in the gravitational sector, and while still under the assumption of the Palatini formalism, examine how the inflationary predictions of the model change. We noticed from earlier investigations that the $R^2$ term induces an asymptotic flatness to the inflaton potential in the Einstein frame. The total action functional is~\cite{Antoniadis2018}
\begin{equation}
    \mathcal{S}=\int\!\mathrm{d}^4x\,\sqrt{-g}\left\{\frac{1}{2}(M^2+2\xi|H|^2)R(g,\Gamma)+\frac{\alpha}{4}R^2(g,\Gamma)-|DH|^2-V(|H|)\right\}\,,
\end{equation}
where $H$ would be the Higgs field. Next, we may introduce the auxiliary field $\chi$ as we did before in order to eliminate the $R^2$; we obtain the following action in the scalar representation
\begin{equation}
    \mathcal{S}=\int\!\mathrm{d}^4x\,\sqrt{-g}\left\{\frac{1}{2}(M^2+2\xi|H|^2+\alpha\chi^2)R(g,\Gamma)-|DH|^2-V(|H|)-\frac{\alpha}{4}\,\chi^4\right\}\,.
\end{equation}
Then after a Weyl rescaling of the metric with the assumption that $M^2\approx M_P^2$ 
\begin{equation}
    \overline{g}_{\mu\nu}(x)=\left(1+\frac{\alpha\chi^2}{M_P^2}+2\xi\,\frac{|H|^2}{M_P^2}\right)g_{\mu\nu}(x)\,,
\end{equation}
we obtain the action in the Einstein frame
\begin{equation}
    \mathcal{S}=\int\!\mathrm{d}^4x\,\sqrt{-\overline{g}}\left\{\frac{M_P^2}{2}\overline{g}^{\mu\nu}R_{\mu\nu}(\Gamma)-\frac{|\overline{D}H|^2}{\left(1+\frac{\alpha\chi^2}{M_P^2}+2\xi\,\frac{|H|^2}{M_P^2}\right)}-\frac{V(|H|)+\frac{\alpha}{4}\chi^4}{\left(1+\frac{\alpha\chi^2}{M_P^2}+2\xi\,\frac{|H|^2}{M_P^2}\right)^2}\right\}\,.
\end{equation}
Variation of the action with respect to the auxiliary field $\chi$ leads to the following constraint equation~\cite{Antoniadis2018}
\begin{equation}
    \delta_\chi\mathcal{S}=0\ \implies\ \chi^2=\frac{\displaystyle{\frac{4V(H)}{M^2+2\xi|H|^2}}+\displaystyle{\frac{2|\overline{\nabla}H|^2}{M_P^2}}}{1-\displaystyle{\frac{2\alpha|\overline{\nabla}H|^2}{M_P^2(M^2+2\xi|\overline{\nabla}H|^2)}}}\,.
\end{equation}

If we adopt the unitary gauge $H=\frac{1}{\sqrt{2}}(0\ h)^\text{T}$ the Higgs potential reads
\begin{equation}
    V(H)=\lambda\left(|H|^2-\frac{v^2}{2}\right)^2=\frac{\lambda}{4}(h^2-v^2)^2\approx\frac{\lambda}{4}\,h^4\,,
\end{equation}
where in the last equality we assumed that in order for the Higgs field to play the role of the inflaton it has to be far away from the EW scale. Then we can substitute the expression of $\chi^2$ back into the action to obtain
\begin{equation}
    \mathcal{S}=\int\!\mathrm{d}^4x\,\sqrt{-\overline{g}}\left\{\frac{M_P^2}{2}\overline{g}^{\mu\nu}R_{\mu\nu}(\Gamma)-\frac{1}{2}(\overline{\nabla}h)^2\left(\frac{M_P^2\,\xi\, h^2}{\xi^2 h^4+4\alpha V(h)}\right)-V(h)\,\frac{M_P^4}{\xi^2h^4+4\alpha V(h)}+\mathcal{O}((\overline{\nabla}h)^4)\right\}\,.
\end{equation}

It is possible to reformulate the action functional in terms of a canonically normalised scalar field under the assumption that the Higgs field with its coupling satisfy the condition $\xi h^2\gg M_P^2$, at least during the first stages of inflation. Then we obtain
\begin{equation}
    \varphi=M_P\sqrt{\frac{\xi}{\xi^2+\alpha\lambda}}\,\sinh^{-1}{\left(\frac{h}{M_P}\sqrt{\xi\,\frac{\xi^2+\alpha\lambda}{\xi^2-\alpha\lambda}}\right)}\,,\qquad\text{iff }\ \xi h^2\gg M_P^2\,.
\end{equation}
Therefore the Lagrangian can be brought to its canonical form of
\begin{equation}
    \mathscr{L}\supset-\frac{1}{2}(\overline{\nabla}\varphi)^2-\bar{V}(\varphi)\,,
\end{equation}
where the scalar potential reads
\begin{equation}
    \bar{V}(\varphi)=\frac{\lambda}{4}\,\frac{M_P^4}{\xi^2+\alpha\lambda}\,\frac{\displaystyle{\sinh^2{\left(\frac{\varphi}{M_P}\sqrt{\frac{\xi^2+\alpha\lambda}{\xi}}\right)}}}{\displaystyle{\frac{2\xi^2}{\xi^2-\alpha\lambda}}+\sinh^2{\left(\frac{\varphi}{M_P}\sqrt{\frac{\xi^2+\alpha\lambda}{\xi}}\right)}}\,.
\end{equation}
Notice that at the limit of $\xi^2\gg \alpha\lambda$ we recover the potential eq.~\eqref{Cond:EFHiggsR2Pot} derived in the case that the $R^2$ term is absent, i.e. $\alpha=0$. Clearly this is not a surprising result, in fact it is expected that if the nonminimal coupling dominates the $R^2$ effectively does not contribute. On the other hand, if $\xi^2\ll\alpha\lambda$, or simply $\xi\ll1$, the potential takes the form of eq.~\eqref{Cond:MonomialPot-EF} for $n=2$ and $x=\sqrt{\xi}\varphi/M_P$. This is once again expected since in the limit of $\xi\to0$ we obtain the simple quartic potential minimally coupled $R+R^2$ term. In that case a larger amount of inflation is required as was noted in that section.

As we discussed earlier at the level of equations of motion the Levi-Civita condition is satisfied meaning that the connection coefficients are the Christoffel symbols with respect to the metric $\bar{\text{g}}$. Therefore, by assuming a flat FRW background and spatially homogeneous field $\varphi(\textbf{x},t)=\varphi(t)$ the equations of motion of the system read
\begin{equation}
    \ddot{\varphi}+3H\dot{\varphi}+\bar{V}'(\varphi)=0\,,\qquad\qquad 3H^2=\frac{1}{2}\dot{\varphi}^2+\bar{V}(\varphi)\,,
\end{equation}
where we set $M_P\equiv1$. Using the slow-roll approximation we can rewrite the system of equations as
\begin{equation}
    3H\dot{\varphi}+\bar{V}'(\varphi)\approx0\,,\qquad\qquad 3H^2\approx\bar{V}(\varphi)\,.
\end{equation}
Then, we can directly apply the mechanism of single-field slow-roll inflation via the slow-roll parameters, which in this case are given by~\cite{Antoniadis2018}
\begin{equation}
    \epsilon_V(\varphi)=8\xi^3(\alpha\lambda+\xi^2)\,\frac{\coth^2{\left(\frac{\varphi}{M_P}\sqrt{\frac{\alpha\lambda+\xi^2}{\xi}}\right)}\,\text{csch}^4{\left(\frac{\varphi}{M_P}\sqrt{\frac{\alpha\lambda+\xi^2}{\xi}}\right)}}{\left(M_P^2(\xi^2-\alpha\lambda)+2\xi^2\,\text{csch}^2{\left(\frac{\varphi}{M_P}\sqrt{\frac{\alpha\lambda+\xi^2}{\xi}}\right)}\right)^2}\,,
\end{equation}
where $\text{csch}(z)=1/\sinh{(z)}$ is the hyperbolic cosecant and $\coth{(z)}$ is the hyperbolic cotangent. The second slow-roll parameter reads~\cite{Antoniadis2018}
\begin{align}
    \eta_V(\varphi)&=\frac{4\xi(\alpha\lambda+\xi^2)\,\text{csch}^4{\left(\frac{\varphi}{M_P}\sqrt{\frac{\alpha\lambda+\xi^2}{\xi}}\right)}}{\left(2\xi^2\,\text{csch}^2{\left(\frac{\varphi}{M_P}\sqrt{\frac{\alpha\lambda+\xi^2}{\xi}}\right)}+M_P^2(\xi^2-\alpha\lambda)\right)^2}\times\\
    &\qquad\qquad\times\left[M_P^2(\alpha\lambda-\xi^2)\left(\cosh{\left(\frac{\varphi}{M_P}\sqrt{\frac{\alpha\lambda+\xi^2}{\xi}}\right)}+2\right)+2\xi^2\left(\coth^2{\left(\frac{\varphi}{M_P}\sqrt{\frac{\alpha\lambda+\xi^2}{\xi}}\right)}+1\right)\right]\,.\nonumber
\end{align}
Likewise, we obtain an expression for the scalar power spectrum given below
\begin{align}
    \mathcal{A}_s(\varphi)&=\frac{\lambda(\xi^2-\alpha\lambda)^2}{768M_P^6\pi^2\xi^3(\xi^2+\alpha\lambda)^5}\sinh^4{\left(\frac{\varphi}{M_P}\sqrt{\frac{\xi^2+\alpha\lambda}{\xi}}\right)}\tanh^2{\left(\frac{\varphi}{M_P}\sqrt{\frac{\xi^2+\alpha\lambda}{\xi}}\right)}\times\nonumber\\
    &\quad\qquad\times\left[M_P^4(\xi^3+\alpha\lambda\xi)^3+\alpha\lambda(\xi^2-\alpha\lambda)^2\sinh^4{\left(\frac{\varphi}{M_P}\sqrt{\frac{\xi^2+\alpha\lambda}{\xi}}\right)}\right]\,.
\end{align}

\begin{figure}
    \centering
    \includegraphics[scale=0.35]{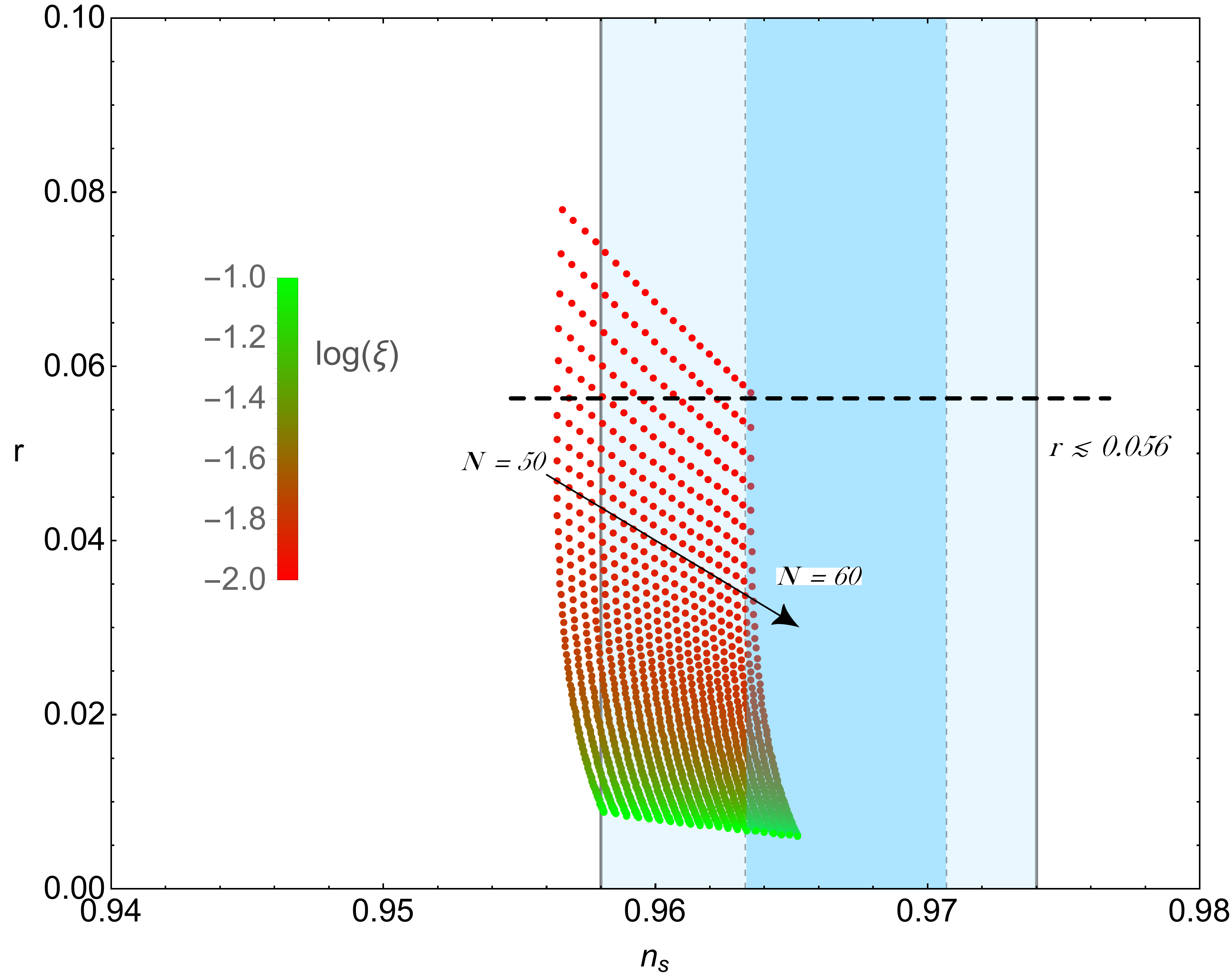}
    \caption{A plot of the $r-n_s$ region for various values of $\xi\in[10^{-2},10^{-1}]$, $\alpha=10^{8}$, $\lambda=10^{-12}$ and $M_P=1$. The light and dark blue region denote the $2\sigma$ and $1\sigma$ allowed region from the Planck2018 collaboration, while the red dashed line denotes the bound on $r\lesssim0.056$; following the notation of previous figures.}
    \label{fig:NonminHiggs-RNsxi}
\end{figure}

In fig.~\ref{fig:NonminHiggs-RNsxi} we employ numerical methods in order to calculate the predictions concerning the inflationary observables $r$ and $n_s$. Similarly to what was done in previous models, we obtain the field value at the end of inflation via the condition $\epsilon_V(\varphi_f)=1$ and for $N$ in the conventional range of $N\in[50,60]$ $e$-foldings we derive the field value at the start of inflation, $\varphi_i$. Then, in the figure we plot the values obtained for different values of the $\xi$ parameter and fixed values of $\alpha=10^8$ and $\lambda=10^{-12}$. Predictions close to values of $\xi\sim10^{-2}$ also break our initial condition that $\xi h^2\gg M_P^2$ and are not to be trusted entirely~\cite{Antoniadis2018}, they reside anyway outside the $2\sigma$ range of $r$. It is evident that values $\xi\gtrsim10^{-1}$ can result in successful inflation, with an ample amount of $N$ and observables in the $1\sigma$ range.

 It might prove useful to approach it also in a semi-analytic way, which is more conveniently presented if we reintroduce the original scalar field $h(t)$. For starters, the slow-roll parameters become
 \begin{align}
     \epsilon_V(h)&=\frac{8M_P^6(M_P^2+\xi h^2)}{h^2\left(\alpha\lambda h^4+(M_P^2+\xi h^2)^2\right)}\,,\\
     \eta_V(h)&=\frac{24M_P^6(M_P^2+\xi h^2)}{h^2\left(\alpha\lambda h^4+(M_P^2+\xi h^2)^2\right)}-\frac{4M_P^6}{h^2(M_P^2+\xi h^2)}-\frac{8M_P^4}{h^2}\,.
 \end{align}
 Next, we may solve for the number of $e$-folds $N$ to obtain
 \begin{equation}
     N=\left.\frac{h^2}{8M_P^4}\right|_{h_f}^{h_i}\approx\frac{h_i^2}{8M_P^4}\,.
 \end{equation}
The field value at the start of inflation, used in the expression of the observables, can be approximately expressed as $h_i^2\approx8M_P^2 N$. Therefore, the expressions of the spectral index and the tensor-to-scalar ratio become
 \begin{align}
     n_s&=1-\frac{16M_P^4}{h^2}-\frac{8M_P^6}{h^2(M_P^2+\xi h^2)}\approx1-\frac{2}{N}-\frac{1}{N(1+8\,\xi NM_P^2)}\stackrel{N\gg1}{\approx}1-\frac{2}{N}-\frac{1}{8M_P^2\xi N^2}+\mathcal{O}(N^{-3})\,,\\
     r&=\frac{128M_P^6(M_P^2+\xi h^2)}{h^2\left(\alpha\lambda h^4+(M_P^2+\xi h^2)^2\right)}\approx\frac{16}{N}\left(\frac{1+8\,\xi N M_P^2}{8\alpha\lambda N+(1+8\,\xi N M_P^2)^2}\right)\stackrel{N\gg1}{\approx}\frac{2\xi}{M_P^2(\alpha\lambda+\xi^2)N^2}+\mathcal{O}(N^{-3})\,.
 \end{align}
Depending on its magnitude, the parameter $\alpha$ can play a role in suppressing the values of $r$, however it does not affect the $n_s$ as noted by the numerical results. Likewise, the power spectrum of scalar perturbations can be expressed as
 \begin{equation}
     \mathcal{A}_s=\frac{\lambda h^6}{768M_P^4\pi^2(M_P^2+\xi h^2)}\approx\frac{2\lambda M_P^6N^3}{3\pi^2(1+8\xi N M_P^2)}\stackrel{N\gg1}{\approx}\frac{\lambda N^2 M_P^4}{12\pi^2\xi}-\frac{\lambda M_P^2N}{96\pi^2\xi^2}+\mathcal{O}(N^{-1})\,.
 \end{equation}
 
Since the power spectrum is independent from $\alpha$ (at all orders~\cite{Enckell2019,Gialamas2020b}), it is possible to place the same condition on the ratio of $\lambda/\xi$ as we did before, meaning that $\lambda/\xi\sim10^{-10}$ in order for $\mathcal{A}_s\approx10^{-9}$. The approximate formula derived for $r$ then leads to a bound on $\alpha$ depending on $\xi$, for example if $\xi\ll1$ then $\alpha\gtrsim10^8$, however if $\xi\gg1$ we are led to $\alpha<10^{12}$ and $\xi\gtrsim10^8$. The discussion agrees with our numerical results~\cite{Antoniadis2018} and with results presented in similar studies in refs.~\cite{Rasanen2019,Gialamas2020b,Tenkanen2020,Tenkanen2020a}.

\section{Extended interactions of gravity and matter}

During inflation it is expected that quantum effects can play a significant role. In previous sections we analysed certain ramifications of that statement, initially with the inclusion of the $R^2$ term and the nonminimal coupling of the inflaton field to the Einstein-Hilbert term, and in section~\ref{subsec:CWmodel} by considering the one-loop effective potential. In this section we are particularly interested in the type of coupling between the inflaton field and the gravitational sector. Currently, as we discussed previously in this work, our description of a quantum theory of gravity is lackluster at best and as such comments regarding what it entails should be treated with suspicion. Following the line of thought presented in this work it is interesting to consider the case in which the $\alpha$ parameter of the $\alpha R^2$ term is also dependent on the inflaton field, i.e. $\alpha\to\alpha(\phi)$. This type of coupling, alongside the $\xi\phi^2R$, are expected to be generated by quantum corrections of the $\phi$ field in a curved background, even if they are absent in the bare (tree-level) Lagrangian,\footnote{There are other local terms that are generated in the process, schematically reading as $R\,\nabla_\mu\phi\,\nabla^\mu\phi$ and $R^{\mu\nu}\,\phi\nabla_\mu\nabla_\nu\phi$ etc, which can be grouped for example in the Einstein tensor $G_{\mu\nu}\equiv R_{\mu\nu}-\frac{1}{2}g_{\mu\nu}R$, as $G^{\mu\nu}\nabla_\mu\phi\,\nabla_\nu\phi$. In ref.~\cite{Gialamas2020} it was studied specifically this type of combination in the Palatini formalism with a special interest given towards inflation.} see for example ref.~\cite{Parker2009}.

Let us consider a fundamental scalar field $\phi(x)$ that is nonminimally coupled to gravity in the way described above, then a general action describing its interactions would be~\cite{Lykkas2021}
\begin{equation}
    \mathcal{S}=\int\!\mathrm{d}^4x\,\sqrt{-g}\left\{\,\frac{1}{2}\left(1+f(\phi)\right)g^{\mu\nu}R_{\mu\nu}(\Gamma)+\frac{\alpha(\phi)}{4}\,R^2(\Gamma)-\frac{1}{2}\left(\nabla\phi\right)^2-V(\phi)\right\},
\end{equation}
where we fixed $M_P\equiv1$. Equivalently the action can be brought to its scalar representation via the introduction of an auxiliary field $\chi$ as follows
\begin{equation}
    \mathcal{S}=\int\!\mathrm{d}^4x\,\sqrt{-g}\left\{\frac{1}{2}\left(1+f(\phi)+\alpha(\phi)\chi\right)R-\frac{1}{2}\left(\nabla\phi\right)^2-V(\phi)-\frac{1}{4}\alpha(\phi)\chi^2\right\}\,.
\end{equation}
In a way similar to what was done in previous sections we can rescale the metric as
\begin{equation}
\overline{g}_{ \mu\nu}(x)\,=\,\left(1+f(\phi)+\alpha(\phi)\chi\right)g_{ \mu\nu}(x)\,,
\end{equation} 
to obtain the action in the Einstein frame
\begin{equation}\label{Action:ExtendedR2-withChi}
    \mathcal{S}[\overline{\text{g}},\Gamma,\phi,\chi]=\int\!\mathrm{d}^4x\,\sqrt{-\overline{g}}\left\{\,\frac{1}{2}\,\overline{g}^{\mu\nu}R_{\mu\nu}(\Gamma)\,-\frac{1}{2}\frac{\left(\overline{\nabla}\phi\right)^2}{\left(1+f(\phi)+\alpha(\phi)\chi\right)}\,-\frac{\left(\,V(\phi)+\frac{1}{4}\alpha(\phi)\chi^2\right)}{\left(1+f(\phi)+\alpha(\phi)\chi\right)^{2}}\right\}\,.
\end{equation}
In contrast to the metric formalism, terms involving derivatives of the $\phi$ and $\chi$ field are not generated through the Weyl rescaling of the metric and the action is schematically similar to the one of constant $\alpha$.

\subsection{\emph{Field equations \& Equations of motion}}

The action \eqref{Action:ExtendedR2-withChi} serves as the starting point in our analysis. In that case, variation of eq.~\eqref{Action:ExtendedR2-withChi} with respect to the ${\Gamma^\rho}_{\mu\nu}$ leads to the standard Levi-Civita condition, following exactly what was presented in previous sections and discussed in detail in ch.~\ref{Ch3:FirstOrder}; that is
\begin{equation}
\frac{\delta{\mathcal{S}}}{\delta{\Gamma^\rho}_{ \mu\nu}}=0\qquad\implies\qquad{\Gamma^\rho}_{ \mu\nu}=\frac{1}{2}\overline{g}^{ \rho\lambda}\left(\partial_{ \mu}\overline{g}_{ \lambda\nu}+\partial_{ \nu}\overline{g}_{ \mu\lambda}-\partial_{ \lambda}\overline{g}_{ \mu\nu}\right)\,.
\end{equation}
Next, variation with respect to the auxiliary field $\chi$ leads to the constraint
\begin{equation}
\frac{\delta{\mathcal{S}}}{\delta\chi}=0\qquad\implies\qquad\chi=\frac{4V(\phi)+\left(1+f(\phi)\right)\left(\overline{\nabla}\phi\right)^2}{\left(1+f(\phi)\right)-\alpha(\phi)\left(\overline{\nabla}\phi\right)^2}\,.
\end{equation}
Substitution of the above equations back to the level of the action \eqref{Action:ExtendedR2-withChi} leads to the final form of the action functional reading
\begin{equation}\label{Action:ExtendedR2-withoutChi}
\mathcal{S}[\overline{\text{g}},\phi]=\int\!\mathrm{d}^4x\,\sqrt{-\overline{g}}\left\{\frac{1}{2}\overline{R}-\frac{1}{2}K_0(\phi)\left(\overline{\nabla}\phi\right)^2+\frac{1}{4}K_2(\phi)\left(\overline{\nabla}\phi\right)^4-U(\phi)\right\}\,,
\end{equation}
where the functions encoding the noncanonical nature of the kinetic terms $K_0(\phi)$ and $K_2(\phi)$, alongside the scalar potential in the Einstein frame, are defined as follows
\begin{align}
    K_0(\phi)&\equiv\frac{1+f(\phi)}{\left[\left(1+f(\phi)\right)^2+4\alpha(\phi)V(\phi)\right]}\,,\\
    K_2(\phi)&\equiv\frac{\alpha(\phi)}{\left[\left(1+f(\phi)\right)^2+4\alpha(\phi)V(\phi)\right]}\,,\\
    U(\phi)&\equiv\frac{V(\phi)}{\left[\left(1+f(\phi)\right)^2+4\alpha(\phi)V(\phi)\right]}\,.
\end{align}
Notice that the action \eqref{Action:ExtendedR2-withChi} is exact, meaning that all terms of $\nabla\phi$ are present at this stage.

It is then straightforward to show that variation of the action \eqref{Action:ExtendedR2-withoutChi} with respect to the rescaled metric $\overline{\text{g}}$ leads to the following generalised Einstein field equations~\cite{Lykkas2021}
\begin{align}
\overline{G}_{ \mu\nu}\equiv\overline{R}_{ \mu\nu}-\frac{1}{2}\overline{g}_{ \mu\nu}\overline{R}=&\left(K_0(\phi)-K_2(\phi)\left(\overline{\nabla}\phi\right)^2\right)\partial_{ \mu}\phi\,\partial_{ \nu}\phi-\overline{g}_{ \mu\nu}\left(\frac{1}{2}K_0(\phi)\left(\overline{\nabla}\phi\right)^2-\frac{1}{4}K_2(\phi)\left(\overline{\nabla}\phi\right)^4+U(\phi)\right)\,,
\end{align}
and variation with respect to the scalar field $\phi(x)$ gives rise to a generalised Klein-Gordon equation of the following form
\begin{align}
\left(K_0(\phi)-K_2(\phi)\left(\overline{\nabla}\phi\right)^2\right)\overline{\square}\, \phi-K_2(\phi)\left(\partial_{ \mu}\left(\overline{\nabla}\phi\right)^2\right)\overline{g}^{ \mu\nu}\partial_{ \nu}\phi+\frac{1}{2}K'_0(\phi)\left(\overline{\nabla}\phi\right)^2-\frac{3}{4}K'_2(\phi)\left(\overline{\nabla}\phi\right)^4-U'(\phi)=0\,,
\end{align}
where the d' Alembertian operator is defined here as $\overline{\Box}\phi=\frac{1}{\sqrt{-\overline{g}}}\partial_\mu\left(\sqrt{-\overline{g}}\,\overline{g}^{\mu\nu}\partial_\nu\phi\right)$. The pair of these two field equations govern, in essence, the dynamics of the model.

Since we are interested primarily in understanding the predictions concerning the inflationary observables we assume a spatially homogeneous scalar field $\phi(x)=\phi(t)$ in a curved spacetime endowed with a flat FRW metric $\mathrm{d}s^2\,=\,-\mathrm{d}t^2\,+\,a^2(t)\,\delta_{ij}\,\mathrm{d}x^i\,\mathrm{d}x^j$. The equations of motion are then reduced to the generalised Friedmann equation
\begin{equation}
3\left(\frac{\dot{a}}{a}\right)^2=3H^2=\frac{1}{2}K_0(\phi)\dot{\phi}^2+\frac{3}{4}K_2(\phi)\dot{\phi}^4+U(\phi)=\rho\,,
\end{equation}
and the generalised Klein-Gordon equation becomes
\begin{equation}\label{Condition:ExtendedR2-GenKG}
\left(K_0(\phi)+3K_2(\phi)\dot{\phi}^2\right)\ddot{\phi}+3H\left(K_0(\phi)+K_2(\phi)\dot{\phi}^2\right)\dot{\phi}+\frac{1}{2}K'_0(\phi)\dot{\phi}^2+\frac{3}{4}K'_2(\phi)\dot{\phi}^4+U'(\phi)=0\,.
\end{equation}

As far as slow-roll inflation is concerned it is safe to assume that the kinetic terms obey
\begin{equation}
\frac{3}{4}\,K_2(\phi)\,\dot{\phi}^4\ll \frac{1}{2}\,K_0(\phi)\,\dot{\phi}^2\ll U(\phi)\,,
\end{equation}
at least during the initial stages of inflation, with the possibility that the condition is violated towards the end of inflation, in line with single-field slow-roll. Generally, corrections to the energy density due to the higher-order kinetic term $K_2(\phi)\dot{\phi}^4$ are encoded in deviations of the effective sound speed from unity.\footnote{In the present section we check these deviations numerically for each set of the parameters considered below and have found deviations smaller than $10^{-5}$, meaning that $c_s^2\approx1$ throughout inflation and we can safely neglect the higher-order kinetic terms altogether.} We also assume that 
\begin{equation}
|\ddot{\phi}|\ll |3H\dot{\phi}|\,.
\end{equation}
Therefore, at inflationary scales the action effectively reduces to
\begin{equation}
    \mathcal{S}=\int\!\mathrm{d}^4x\,\sqrt{-\overline{g}}\left\{\frac{1}{2}\overline{R}-\frac{1}{2}K(\phi)\left(\overline{\nabla}\phi\right)^2\,-U(\phi)\right\}\,,
\end{equation}
which after a redefinition of the scalar field 
\begin{equation}
    \Phi=\pm\int\!\mathrm{d}\phi\,\sqrt{K_0(\phi)}\,,
\end{equation}
can be brought into its canonical form yielding
\begin{equation}
    \mathscr{L}\supset-\frac{1}{2}\left(\overline{\nabla}\Phi\right)^2-U(\phi(\Phi))\,.
\end{equation}

The first-order slow-roll parameters containing the information of slow-roll inflation can be expressed in terms of the original field $\phi$ as follows
\begin{align}
    \epsilon_V&=\frac{1}{2}\left(\frac{U'(\Phi)}{U(\Phi)}\right)^2\,=\,\frac{1}{2K_0(\phi)}\left(\frac{{U}'(\phi)}{U(\phi)}\right)^2\,,\\
    \eta_V&=\frac{U''(\Phi)}{U(\Phi)}\,=\,\frac{1}{K_0(\phi)}\left(\frac{{U}''(\phi)}{U(\phi)}\right)-\frac{1}{2}\frac{{K_0}'(\phi)}{{K_0}^2(\phi)}\left(\frac{{U}'(\phi)}{U(\phi)}\right)\,,
\end{align}
and similarly the duration of inflation is given by
\begin{equation}
N=\int_{\Phi_i}^{\Phi_f}\!\mathrm{d}\Phi\left(\frac{U(\Phi)}{U'(\Phi)}\right)\,=\,\int_{\phi_*}^{\phi_f}\!\mathrm{d}\phi\,K_0(\phi)\left(\frac{U(\phi)}{U'(\phi)}\right)\,,
\end{equation}
where, following the notation of previous sections, the field values at end and start of inflation are denoted by $\Phi_i$ (or $\phi_i$) and $\Phi_f$ (or $\phi_f$), respectively. All of the above are related to observable quantities through their usual approximate expressions.

\subsection{\emph{Specifying the nonminimal couplings}}

The setup described up to this point has been purposefully quite general and involved the inflaton potential $V(\phi)$ and the two nonminimal coupling functions $f(\phi)$ and $\alpha(\phi)$ between the scalar field and the $R+R^2$ term. In general, since $R$ is dimensionful it suggests that the function $f(\phi)$ is a monomial in $\phi$, which, in the case that it also respects an internal $\mathbb{Z}_2$ symmetry it can be generally expressed as
\begin{equation}
    f(\phi)=\xi \phi^2\,,
\end{equation}
where the constant parameter $\xi$ is dimensionless. This was actually the type of coupling considered prior to this discussion.

In the case of the inflaton potential we suppose that only the renormalisable self-interaction terms of the inflaton would contribute and as such we assume that the potential is given by a quartic monomial in $\phi$, reading:
\begin{equation}
    V(\phi)=\frac{\lambda}{4!}\,\phi^4\,.
\end{equation}
As we discussed in the section~\ref{subsec:CWmodel}, the potential can be possibly enhanced by radiative corrections and obtain a logarithmic dependence on $\phi$ in the form of $\phi^4\ln{(\phi^2/\mu^2)}$, where $\mu$ denotes the would-be renormalisation scale. At this point we can parallelise the potential with the Higgs potential far away from the EW scale and identify the $\lambda$ parameter with the Higgs self-coupling. In what follows, values of the free parameters of the model are chosen such that they are phenomenologically consistent, at least approximately.

The other parametric function $\alpha(\phi)$ corresponds to a \emph{generalisation} of the Starobinsky constant\footnote{See refs.~\cite{Gundhi2021,Gundhi2021a} for a similar discussion in the metric formalism.} (or the mass term $\propto M_P^2\, R^2/M^2(\phi)$) that includes a \emph{dimensionless} dependence on the inflaton field $\phi$. Since the $R^2$ term is invariant under a Weyl rescaling (in the Palatini formalism) a direct relation between these two nonminimal couplings $f(\phi)$ and $\alpha(\phi)$ cannot be assumed ad hoc.\footnote{However, in ref.~\cite{Das2021} the contrary was assumed.} If we were to assume a UV completion of the theory the coupling $\alpha(\phi)$ would obtain a logarithmic correction from its running which could be represented here as
\begin{equation}
    \alpha(\phi)=\alpha_0+\beta_0\ln{\left(\frac{\phi^2}{\mu^2}\right)}\,,
\end{equation}
where $\alpha_0$ and $\beta_0$ are constant dimensionless parameters. In the following analysis we accompany the values of $\alpha_0$ and $\beta_0$ with a factor of at least $\alpha_0/\beta_0\gtrapprox\mathcal{O}(10)$ to account for the perturbative nature of $\alpha(\phi)$, as well as maintain the overall positivity of the $R^2$ term. Since $\alpha(\phi)$ depends on the field values $\phi$ we expect the model to differ substantially compared to the usual Palatini-$R^2$ models analysed earlier, in which the constant parameter $\alpha$ does not even affect the observables $n_s$ and $\mathcal{A}_s$~\cite{Enckell2019,Antoniadis2018}, and it should contribute drastically to the tensor-to-scalar ratio $r$ and the total number of $e$-folds $N$ since both depend manifestly on $\alpha$. Naturally, this depends on the magnitude of the value of $\alpha$ compared to the rest of the model parameters, as demonstrated in what follows. Notice that at field values close to the would-be renormalisation scale $\phi\to\mu$ the model is asymptotically scale-invariant resembling the undeformed Starobinsky model, which is recovered with corrections $\delta\phi$ around $\mu$ given by $\propto R+\left(\alpha_0+\beta_0(\delta\phi/\mu)^2\right)R^2$.

Considering that specific form of the parametric functions they give rise to the following expressions for the functions $K_0(\phi)$, $K_2(\phi)$ and $U(\phi)$ appearing in the final action~\cite{Lykkas2021}
\begin{align}
    K_0(\phi)&=\frac{1+\xi\phi^2}{(1+\xi\phi^2)^2+\displaystyle{\frac{\lambda}{6}}\phi^4\left(\alpha_0+\beta_0\ln(\phi^2/\mu^2)\right)}\,,\\
    K_2(\phi)&=\frac{\alpha_0+\beta_0\ln(\phi^2/\mu^2)}{(1+\xi\phi^2)^2+\displaystyle{\frac{\lambda}{6}}\phi^4\left(\alpha_0+\beta_0\ln(\phi^2/\mu^2)\right)}\,,\\
    U(\phi)&=\frac{\displaystyle{\frac{\lambda}{4!}}\phi^4}{(1+\xi\phi^2)^2+\displaystyle{\frac{\lambda}{6}}\phi^4\left(\alpha_0+\beta_0\ln(\phi^2/\mu^2)\right)}\,.\label{Condition:ExtendedR2-PotU}
\end{align}

Instinctively, a logarithmic correction to the scalar self-coupling can be added and actually, following the reasoning for the logarithmic form of $\alpha(\phi)$, it should. For this particular model such corrections would in principle affect the denominator of the above functions, however they can be absorbed in the definition of $\alpha_0$ and $\beta_0$ schematically as $\lambda(\phi)\alpha(\phi)\propto\lambda_0\alpha_0+\left(\lambda_1\alpha_0+\lambda_0\beta_1\right)\ln(\phi^2/\mu^2)$. As such, only the numerator of the Einstein-frame scalar potential would obtain a contribution, which could further improve the inflationary plateau even though it is subleading due to the smallness of the self-coupling $\lambda$. Specifically for inflation, the potential enters in the formulae of the slow-roll parameters in the form of $U'/U,\,(U'/U)'$, meaning that derivatives of the logarithmic corrections amount to inverse powers of the inflaton that are subleading in the large field limit~\cite{Lykkas2021}. The perturbative nature of the term $\beta_0\ln{(\phi^2/\mu^2)}$ ensures also that the plateau of the potential $U(\phi)$ would remain unaffected, violated only logarithmically at Planckian scales $\phi \gtrsim \mu\sim \mathcal{O}(1)$ as
\begin{equation}
 U(\phi)\,\approx\,\frac{\lambda}{4!\,\xi^2+4\lambda\left(\alpha_0+\beta_0\ln(\phi^2/\mu^2)\right)}\,.
\end{equation}
In terms of the canonically normalised scalar field $\Phi$ given by
\begin{equation}
\Phi=\int\!\mathrm{d}\phi\,\sqrt{K(\phi)}\,\stackrel{\xi\sqrt{\phi}\gg M_P}{\approx}\,\int\frac{\mathrm{d}\ln(\phi/\mu)}{\sqrt{\xi+\displaystyle{\frac{\lambda}{6\xi}}\left(\alpha_0+2\beta_0\ln(\phi/\mu)\right)}}\,=\,\frac{6\xi}{\lambda\beta_0}\sqrt{\xi+\frac{\lambda}{6\xi}\left(\alpha_0+2\beta_0\ln(\phi/\mu)\right)}\,,
\end{equation}
we obtain the following expression for the Einstein-frame potential
\begin{equation}
U(\Phi)\,\approx\,\frac{3\xi}{2\lambda\beta_0^2}\left(\frac{1}{\Phi^2+\ldots}\right)\,,
\end{equation}
 where the dots denote exponentially small corrections of $\mathcal{O}\left(e^{-(\lambda\beta_0/6\xi)\Phi^2}\right)$.
 
\begin{figure}
    \centering
    \includegraphics[width=0.55\textwidth]{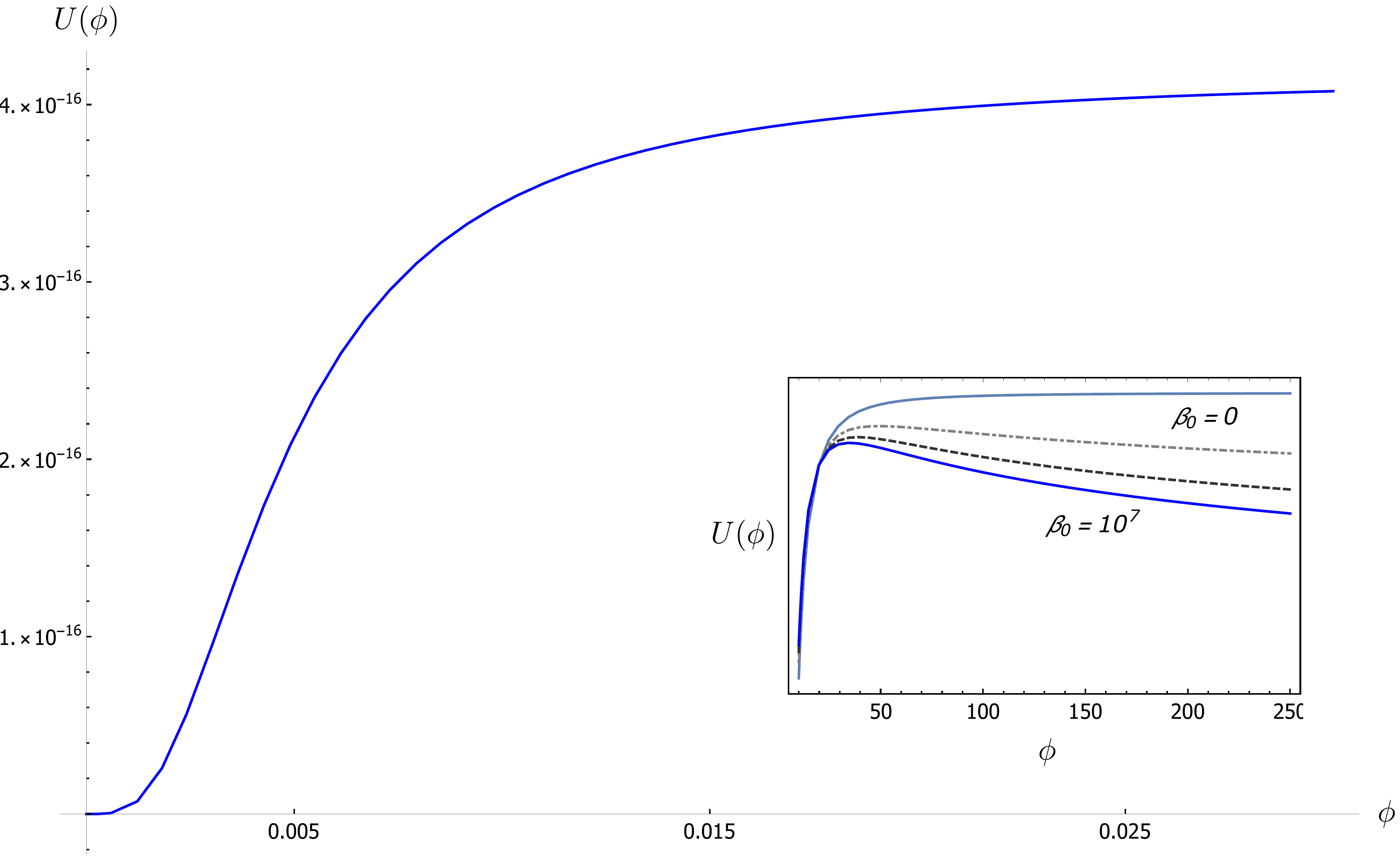}
    \caption{Plot of the scalar potential $U(\phi)$ in terms of the original field $\phi$, as given in eq.~\eqref{Condition:ExtendedR2-PotU}. In the main part of the figure the values of the free parameters are $\left\{\xi=10^5,\,\lambda=10^{-4},\,\mu=20,\,\alpha_0=10^8,\,\beta_0=10^7\right\}$. In the secondary figure presented in the bottom right corner, we showcase the asymptotic behaviour of the potential in the large field limit for varying values of $\beta_0$, from $\beta_0\approx10^7$ to $\beta_0=0$. The local maximum of the potential is not relevant to inflation, since, as we show later on, inflation occurs for field values way before the maximum. It is also important to note that for values above the scale $\mu$, which is always assumed to be $\simeq\phi_*$, the form of the potential cannot be trusted.}
    %\label{figpot}
\end{figure}

Let us proceed directly to the calculation of the SRPs in terms of the parametric functions assumed in the previous section. Without any additional assumptions the first and second SRPs are given as follows
\begin{align}
     \epsilon_V(\phi)&=\frac{1}{2\phi^2(1+\xi\phi^2)}\frac{\left(4(1+\xi\phi^2)-\displaystyle{\frac{\lambda\beta_0}{3}\phi^4}\right)^2}{\left(\,(1+\xi\phi^2)^2+\displaystyle{\frac{\lambda}{6}}\phi^4\left(\alpha_0+\beta_0\ln(\phi^2/\mu^2)\,\right)\,\right)}\,,\\
     \eta_V(\phi)&=3\,\epsilon_V(\phi)-\left(\frac{8}{\phi^2}+\frac{4}{\phi^2(1+\xi\phi^2)}+\frac{\lambda\beta_0}{3}\frac{\phi^2}{(1+\xi\phi^2)^2}\right)\,.
 \end{align}
 Then, in the slow-roll approximation the usual expressions for the inflationary observables lead to the following formulae
\begin{align}
     \mathcal{A}_s&\approx\frac{U(\phi)}{24\pi^2\epsilon_V}\,=\,\frac{1}{12\pi^2}\frac{\frac{\lambda}{4!}\phi^6(1+\xi\phi^2)}{\left(4(1+\xi\phi^2)-\frac{\lambda\beta_0}{3}\phi^4\right)^2}\label{Cond:ExtendedR2-As}\,\\
     n_s&=1-\left(\frac{16}{\phi^2}+\frac{8}{\phi^2(1+\xi\phi^2)}+\frac{2\lambda\beta_0}{3}\frac{\phi^2}{(1+\xi\phi^2)^2}\right)\label{Cond:ExtendedR2-Ns}\,\\
     r&\approx 16\epsilon_V=
 \frac{8}{\phi^2(1+\xi\phi^2)}\frac{\left(4(1+\xi\phi^2)-\frac{\lambda\beta_0}{3}\phi^4\right)^2}{\left[\,(1+\xi\phi^2)^2+\frac{\lambda}{6}\phi^4\left(\alpha_0+\beta_0\ln(\phi^2/\mu^2)\,\right)\,\right]}\label{Cond:ExtendedR2-r}\,.
 \end{align}
 The integral of the number of $e$-folds $N$ is exactly integrable after direct substitution of the kinetic function $K_0(\phi)$ and the potential $U(\phi)$ in its integrand, leading to the expression 
  \begin{equation}
     N=\left.\frac{b}{8\sqrt{4b+(\xi b)^2}}\left\{\left(1+\frac{1}{2}\xi^2b\right)\ln\left|\frac{\phi^2-\frac{1}{2}\xi b-\frac{1}{2}\sqrt{4b+(\xi b)^2}}{\phi^2-\frac{1}{2}\xi b+\frac{1}{2}\sqrt{4b+(\xi b)^2}}\right|\,+\,\frac{1}{2}\xi\sqrt{4b+(\xi b)^2}\ln\left|\phi^4-\xi b\phi^2-b\right|\,\right\}\right|_{\phi_f}^{\phi_*}
 \end{equation}
 where we made the definition of $b\equiv 12/(\lambda\beta_0)$ for the sake of brevity.
 
\subsection{\emph{Semi-analytic approach and numerical results}}

Before we move on with the predictions for the observables, let us entertain the possibility of $\xi=0$, namely the original minimally coupled model.\footnote{Of course the field $\phi$ is still coupled (nonminimally) with the $R^2$ term through the coupling $\alpha(\phi)$, so labeling it the ``minimal coupling'' is used just in comparison with previously considered minimal models.} The order of magnitude of $\mathcal{A}_s$ requires a specific power-play of $\lambda{\beta_0}^2\phi^2\sim\mathcal{O}(10^6)$ between the coupling constants, which in turn implies a very large value of the $\beta_0$ parameter. In what was considered the minimal case, implying $\xi\!=\!0\!=\!\beta_0$ in this particular case, it can successfully describe the inflationary era with appropriate inflationary observables, however it requires unusually large number of $e$-foldings $N\approx75$ $e$-folds, as was demonstrated in section~\ref{subsec:MinHiggs} (also see refs.~\cite{Antoniadis2019,Tenkanen2019}). Thus, if $\beta_0\neq0$ it is natural to expect large values of the parameter $\beta_0$ (not unlike the ones cited for $\alpha_0$ in the previous case~\cite{Antoniadis2019,Tenkanen2019,Gialamas2020b,Tenkanen2020}), since the self-coupling is approximately $\lambda_\text{max}\sim\mathcal{O}(10^{-4})$ when the inflaton $\phi$ assumes values around the Planck scale. However, large values of the parameter $\beta_0$ can lead to inconsistencies primarily conserning possible violation of the subleading nature of the kinetic terms, specifically their noncanonical functions $K_0(\phi)$ and $K_2(\phi)$. Pathologies of that nature can be detected in the values of the effective sound speed $c_s^2=\left(K_0(\phi)+K_2(\phi)\dot{\phi}^2\right)/\left(K_0(\phi)+3K_2(\phi)\dot{\phi}^2\right)$, namely in possible deviations from unity, which are especially alarming at field values close to the start of inflation. Additionally it can assume negative values hinting at instabilities or unphysical states.

A simple calculation via power counting shows that values of the parameters~\cite{Lykkas2021}
\begin{equation}
    \phi_i\sim 20 \qquad \text{and} \qquad \lambda\beta_0/\xi\sim\mathcal{O}(10^{-9})
\end{equation}
can indeed satisfy the observable quantities. The specific values presented above are of no particular interest, even though they can possibly reflect the high-energy phenomenology of the model and they are chosen primarily in order to satisfy the observation bounds. Although the value of $\alpha_0$ is accompanied by a factor of $\alpha_0/\beta_0\propto\mathcal{O}(10)$ based on perturbativity grounds, as shown later its specific value does not affect the predictions of the observables, which is in line with previous results of the Palatini-$R^2$ models. The expression of the spectral index $n_s$ in eq.~\eqref{Cond:ExtendedR2-Ns} shows that the field values $\phi_i$ are dominating at first order with contributions of $\phi_i^{-2}$ and ensuing corrections of $\propto\left(\lambda\beta_0/\xi\right)\phi_i^{-4}$, meaning that due to the smallness of the couplings we can approximately chose $\phi_i\sim20$ in order to satisfy the $1\sigma$ bound on $n_s$. Finally, the predictions are collectively presented below
\begin{equation}
    n_s\approx0.960\,,\qquad \mathcal{A}_s\sim 3.5\times10^{-9}\,,\qquad r\approx \frac{10^{-4}}{2\xi}\,,
\end{equation}
where the factor of $\mathcal{A}_s$ is not important at this point and can in principle be absorbed in the values of the parameters. The amount of inflation is encoded in $N$ which is approximately given by
\begin{equation}
    N\approx\frac{3\xi}{2\lambda\beta_0}\ln\left|\frac{1-\displaystyle{\frac{\lambda\beta_0}{12\xi}}\phi_i^2}{1-\displaystyle{\frac{\lambda\beta_0}{12\xi}}\phi_f^2}\right|\,.
\end{equation}
It is straightforward to show that assuming a conservative value of $N\sim 50$ $e$-folds the field value at the end of inflation is $\phi_f^2\approx\mathcal{O}(\sqrt{8/\xi})$.

In order to obtain the complete behaviour of the model concerning the inflationary period it is useful to employ numerical methods similar to the ones used in previous sections. In doing so we are able to produce the following table (table~\ref{table:ExtendedR2-T1}) including some characteristic values of the parameter space and the predictions regarding the inflationary observables. All of the entries are generated through numerical solution of the exact formulae in combination with the known conditions for inflation, for example $\epsilon_V(\phi_f)\equiv1$ being the condition for inflation to end. It is important to note that the parameters in table~\ref{table:ExtendedR2-T1} also lead to the desired value of the power spectrum $\mathcal{A}_s$, even if it is not presented.

\begin{table}
\centering
\begin{tabular}{cccccc}
& & \multicolumn{2}{c}{$r$} & \multicolumn{2}{c}{$n_s$}\\
\cmidrule(r{2pt}){3-4} \cmidrule(l){5-6}
    $\xi$ & $\lambda$ & $N=50$ & $N=60$ &  $N=50$ & $N=60$  \\ \midrule
    $10^{5}$&$10^{-4}$&$8\times10^{-9}$&$5.5\times10^{-9}$&$0.9600$&$0.9667$\\
    $10^{3}$&$10^{-6}$&$8\times10^{-8}$&$5.5\times10^{-8}$&$0.9600$&$0.9667$\\
    $10^3$&$10^{-5}$&$8\times10^{-7}$&$5.5\times10^{-7}$&$0.9600$&$0.9667$\\
    $10^2$&$10^{-7}$&$8\times10^{-6}$&$5.5\times10^{-6}$&$0.9600$&$0.9667$\\
    \bottomrule
    \end{tabular}
    \caption{Table including results of the numerical study of the exact expressions regarding inflationary observables. Assumed constant values of $\alpha_0=10$, $\beta_0=1$ and $\mu=20\,M_P\sim\phi_*$ that, together with $\xi$ and $\lambda$, we are able to reproduce the appropriate value of the scalar amplitude $\mathcal{A}_s$ within a field excursion $\Delta\phi\equiv\phi_i-\phi_f$ of $\phi_f\approx 10^{-1}M_P$ and $\phi_*\sim20M_P$.}
    \label{table:ExtendedR2-T1}
\end{table}

\begin{figure}
    \centering
    \includegraphics[scale=0.48]{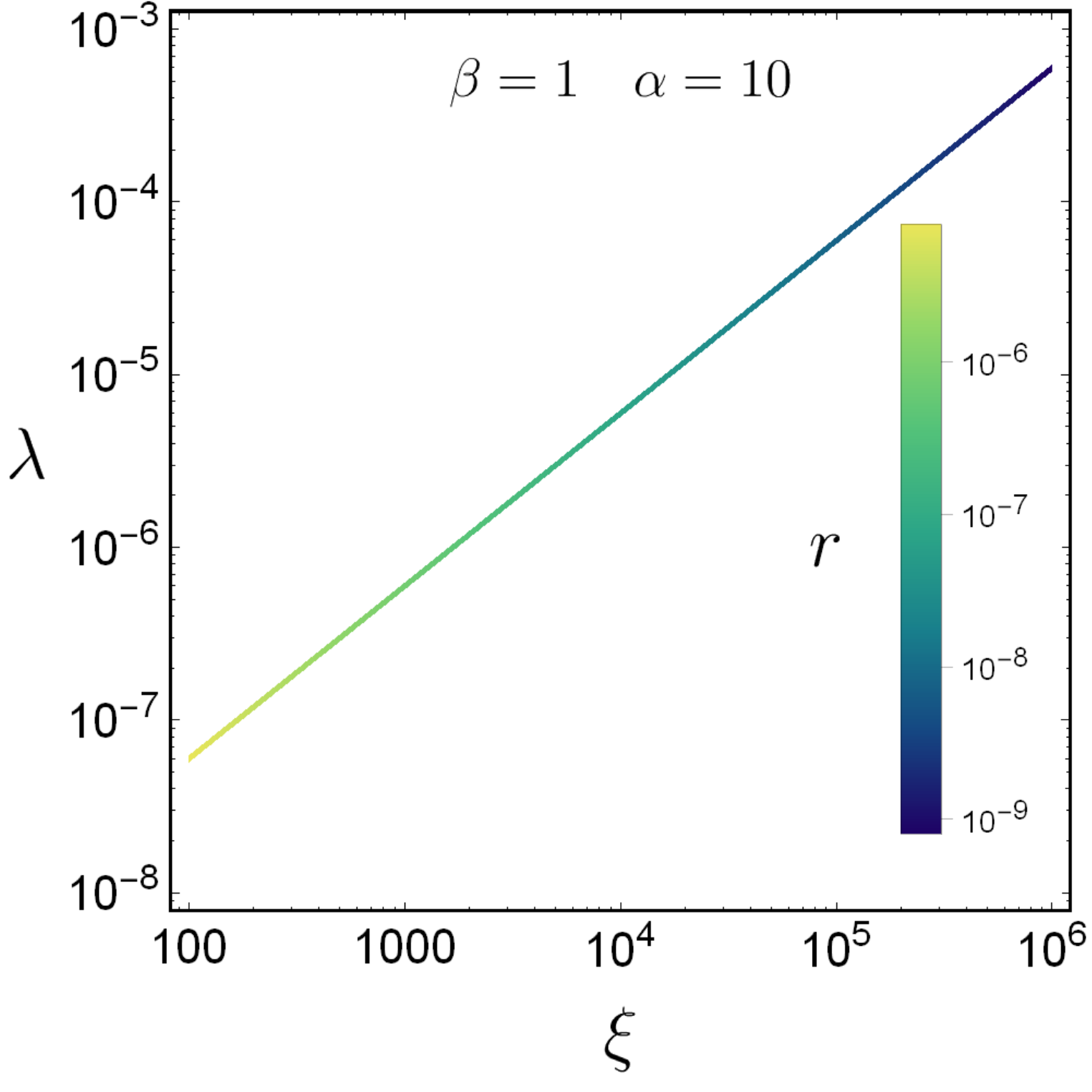}
    \caption{Plot of $\lambda=f(\xi)$ and the associated values of $r$ corresponding to the color grading displayed in the bar of the figure. These values are calculated at exactly $N=50$ $e$-folds and $\mathcal{A}_s\approx2.1\times10^{-9}$ is maintained throughout the displayed curve. In fact the figure is part of a larger contour plot, but only a small part of it leads to the appropriate value of the scalar power spectrum $\mathcal{A}_s$.}
    \label{fig:ExtendedR2-Aslambdaxi}
\end{figure}

As expected from eq.~\eqref{Cond:ExtendedR2-Ns} different values of $\left\{\xi,\,\lambda,\,\alpha_0,\,\beta_0\right\}$ presented in table~\ref{table:ExtendedR2-T1} lead to absolutely identical values of $n_s$~\cite{Lykkas2021}. This is a known feature of the Palatini--$R^2$ models (e.g. see refs.~\cite{Antoniadis2018, Antoniadis2019, Gialamas2020}) accompanied also by highly suppressed values of the tensor-to-scalar ratio $r$, which is effectively undetectable. In fig.~\ref{fig:ExtendedR2-Aslambdaxi} the values of $r$ are presented in terms of varying values of $\lambda$ and $\xi$ in a more comprehensive manner.

In previously considered models where $\beta_0=0$ larger values of $\alpha_0$ are assumed, namely $\alpha_0\sim10^8$, in order to satisfy the observational bounds. Similar values of $\alpha_0$ were also reported in ref.~\cite{LloydStubbs2020} capable of retaining the canonically normalised field $\Phi$ at sub-Planckian values during inflation. Therefore, in order to make contact with previous results we assume larger values of $\beta_0$ leading to larger values of $\alpha_0$ due to our condition $\alpha_0/\beta_0\sim\mathcal{O}(10)$. Then, from eqs.~\eqref{Cond:ExtendedR2-As}-\eqref{Cond:ExtendedR2-r} it is immediately noticeable that large values of $\beta_0$ can impact negatively the prediction of the observables. In table~\ref{table:ExtendedR2-T2}, following the same numerical algorithm employed in table~\ref{table:ExtendedR2-T1}, we present the predictions of the inflationary observables in the limit of large $\beta_0$ values.

\begin{table}
\centering
\begin{tabular}{cccccc}
& & \multicolumn{2}{c}{$r$} & \multicolumn{2}{c}{$n_s$}\\
\cmidrule(r{2pt}){3-4} \cmidrule(l){5-6}
    $\alpha_0$ & $\beta_0$ & $N=50$ & $N=60$ &  $N=50$ & $N=60$  \\\midrule
    $\left(10\sim10^{4}\right)$&$\left(1\sim10^{3}\right)$&$8\times10^{-9}$&$5.5\times10^{-9}$&$0.9600$&$0.9667$\\
    $10^{5}$&$10^{4}$&$7.7\times10^{-9}$&$5.3\times10^{-9}$&$0.9600$&$0.9666$\\
    $10^8$&$10^{7}$&$5.7\times10^{-9}$&$3.7\times10^{-9}$&$0.9530$&$0.9596$\\
    \bottomrule
    \end{tabular}
    \caption{Similar to table~\ref{table:ExtendedR2-T1} we present the numerical study of the exact expressions of the inflationary observables under the assumption of $\xi=10^5$, $\lambda=10^{-4}$ and $\mu=20\,M_P\sim\phi_*$. Together with the values of $\alpha_0$ and $\beta_0$ displayed in the table they are able to correctly reproduce the appropriate value of the scalar amplitude $\mathcal{A}_s$. The field excursion $\Delta\phi$ associated with the values in the table is approximated by $\phi_f\approx 10^{-2}M_P$ and $\phi_i\lesssim 20M_P$.}
    \label{table:ExtendedR2-T2}
\end{table}

\begin{figure}
    \centering
    \includegraphics[scale=0.5]{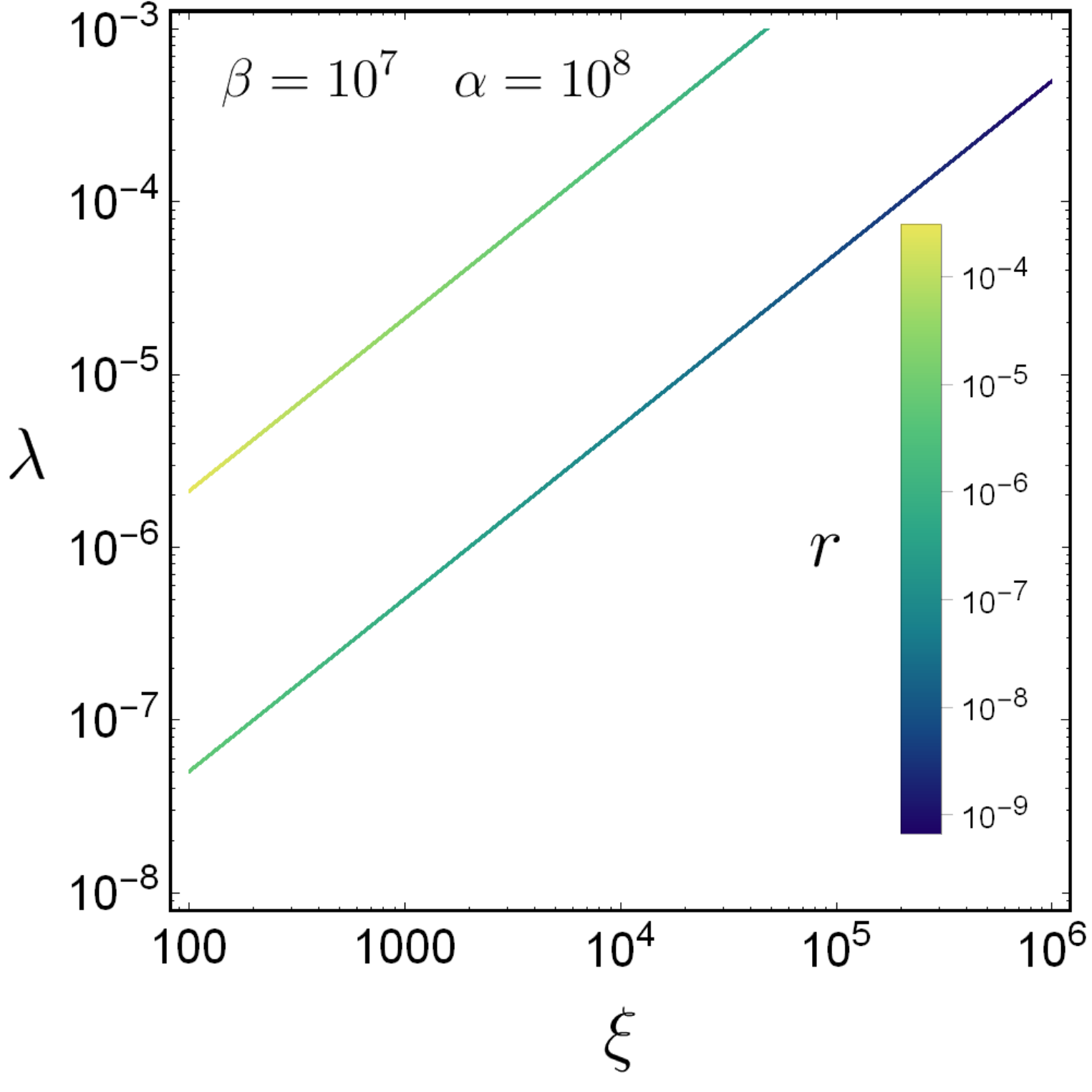}
    \caption{Plot of $\lambda=f(\xi)$ in terms of their predicted value of the tensor-to-scalar ratio $r$ following the colour grading of the legend in the figure. Similarly to the previous figure, fig.~\ref{fig:ExtendedR2-Aslambdaxi}, the values are presented at the point of $N=50$ $e$-folds and $\mathcal{A}_s\approx2.1\times10^{-9}$ is maintained throughout the curve(s). }
    \label{fig:ExtendedR2-Aslambdaxi-bigbeta}
\end{figure}

In fig.~\ref{fig:ExtendedR2-Aslambdaxi-bigbeta} larger values of the total coupling $\alpha(\phi)$ leads to larger values of the tensor-to-scalar ratio $r$ and surprisingly some of the predicted values for large $\alpha_0$ and $\beta_0$ reside in the projected accuracy of future experiments $r\sim10^{-4}$~~\cite{Matsumura2016,Kogut2011,Sutin2018}, meaning that they can possibly be differentiated. The two distinct curves that can reproduce the desired value of $\mathcal{A}_s$ are attributed to the power interplay between the parameters $\beta_0\lambda$ and $\xi^2$. Depending on their values one of the terms in the expression of the tensor-to-scalar ratio $r$~\eqref{Cond:ExtendedR2-r}, can dominate over the other, which in turn leads to large(r) values of $r$~\cite{Lykkas2021}, contrary to fig.~\ref{fig:ExtendedR2-Aslambdaxi}.

\begin{figure}
    \centering
    \includegraphics[scale=0.4]{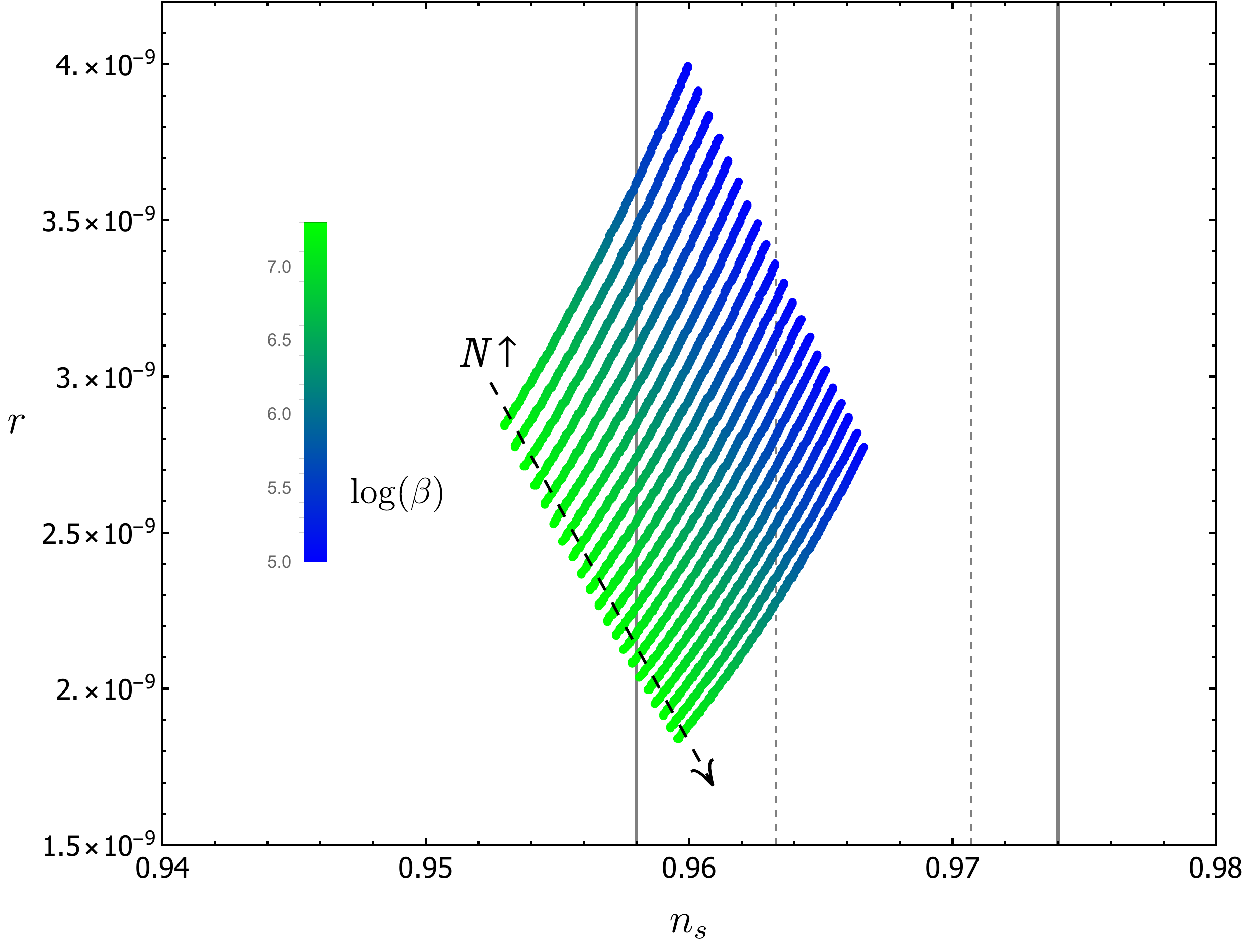}
    \caption{The predictions of the inflationary observables $r$ and $n_s$. Following the notation of previous figure, the dotted and solid grid lines denote the $1\sigma$ and $2\sigma$ allowed range of $n_s$. The parameter values assumed in the numerical analysis are $\alpha_0=10^8$, $\xi=2\times10^5$, $\lambda=10^{-4}$ and varying values of $\beta_0\in\left[10^5,10^7\right]$, as presented in the colour grading in the bar of the figure. The system is solved in the usual range of $N\in\left[50,60\right]$ values of which increase along the arrow displayed in the figure.}
    \label{fig:ExtendedR2-RNs}
\end{figure}

As seen from the entries in table~\ref{table:ExtendedR2-T2} large values of $\beta_0$ tend to decrease the predicted values of $n_s$, discernible also from its formula \eqref{Cond:ExtendedR2-Ns}. It seems that values of $\beta_0\lessapprox10^{-5}$ suggest that the value of the spectral index $n_s$ is determined primarily through the field value $\phi_i$. This behaviour is examined in fig.~\ref{fig:ExtendedR2-RNs}, in which we plot the $r$-$n_s$ for varying values of the parameter $\beta_0$ and a representative value of $\alpha_0$ since it affects only the values of $r$ which are already tiny.

\begin{figure}
    \centering
    \includegraphics[scale=0.6]{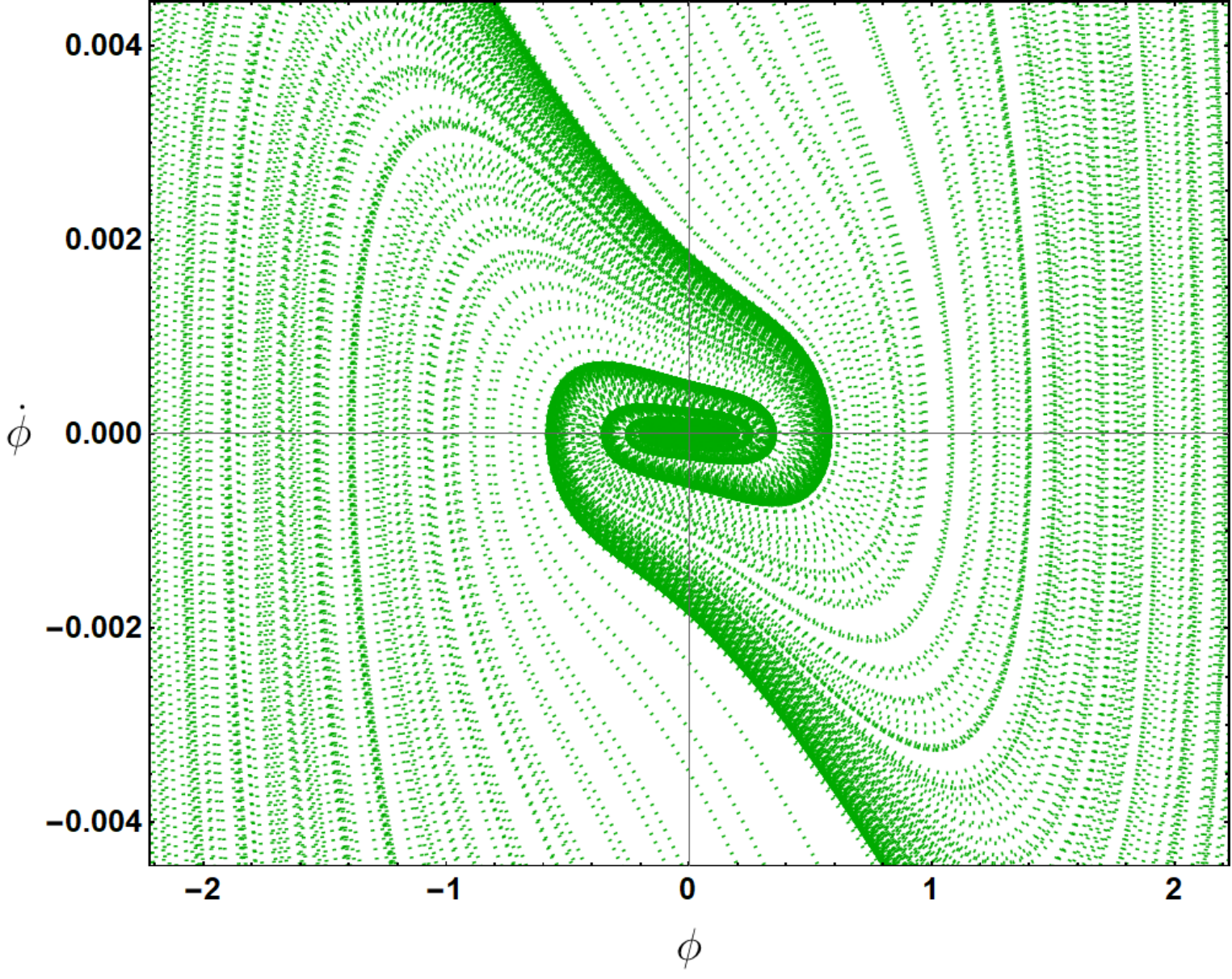}
    \caption{Plot of the phase-space trajectories $\dot{\phi}$-$\phi$ illustrating the attractor point of the potential for a particular part of the parameter space  $\left\{\alpha_0=10^8,\,\beta_0=10^7,\,\xi=2\times10^5,\,\lambda=10^{-4}\right\}$. The bolder line corresponds to the slow-roll trajectory resulting to the minimum of the potential at the center of the figure.}
    \label{fig:ExtendedR2-attractor}
\end{figure}

As we alluded to earlier large values of $\beta_0$ can destabilise the inflaton field out of its slow-roll inflation trajectory. In fig.~\ref{fig:ExtendedR2-attractor} we present a numerical study of the phase-space flow of numerical solutions of the equation of motion eq.~\eqref{Condition:ExtendedR2-GenKG}. For a plethora of initial conditions the trajectories $\dot{\phi}-\phi$ of the inflaton field fall into the slow-roll trajectory concluding at the potential minimum and oscillate around it~\cite{Lykkas2021}. Then, the attractive behaviour of the potential is retained even at large values of $\beta_0$. Note that only fine-tuned trajectories end up directly to the oscillatory phase without any prior amount of inflation. It should be mentioned that smaller values of $\beta_0$, as considered earlier in table~\ref{table:ExtendedR2-T1}, do not spoil the attractor behaviour of the model and also reproduce the results of refs.~\cite{Antoniadis2019, Tenkanen2020a}. The same is true in the case of $K_2(\phi)=0$, in which the higher-order kinetic term $\propto\dot{\phi}^4$ is completely disregarded.

\subsection{\emph{Reheating}}

In this section we examine if the proposed model is consistent with a period of reheating ensuing the inflationary era. As discussed in section~\ref{sec:ParReheating} of chapter~\ref{Ch2:Inflation} following the results of refs.~\cite{Dodelson2003,Liddle2003a,Dai2014,Munoz2014,Gong2015,Cook2015a} the quantities parametrising reheating, such as its number of $e$-foldings $N_\text{R}$ and the equation of state parameter $w_\text{R}$, in terms of parameters of inflation. This is done without discussing a particular reheating mechanism and in principle can restrict the allowed parameter space of an inflationary model or at least demonstrate if the model is consistent with a reheating era.

Without loss of generality we assume that the transition between the different eras, for example from inflation to reheating, is abrupt, in other words there is an instantaneous transition from $w=-1/3$ at the end of inflation to $w\to w_\text{R}$ at the start of reheating and so on. Additionally, we consider values of $w_\text{R}$ constant for the entirety of the reheating era. Therefore, following section~\ref{sec:ParReheating} one can derive expressions of the reheating temperature $T_\text{R}$ and the $e$-foldings $N_\text{R}$ in terms of inflationary parameters of a canonically normalised inflaton $\Phi$ with its potential $U(\Phi)$ as follows:
\begin{equation}
    T_\text{R}=\left(\frac{T_\gamma a_0}{k}\right)\left(\frac{43}{11g_\text{R}}\right)^{1/3}H_*e^{-N}e^{-N_\text{R}}\,,
\end{equation}
\begin{equation}
    N_\text{R}=\frac{4}{3(1+w_\text{R})}\left\{N+N_\text{R}+\ln{\frac{k}{a_0T_\gamma}}+\ln{\frac{U_f^{1/4}}{H_*}}+\ln{\left[\left(\frac{45}{\pi^2}\right)^{1/4}\left(\frac{11}{3}\right)^{1/3}g_\text{R}^{1/12}\right]}\right\}\,,
\end{equation}
with the identification of $U_f\equiv U(\Phi_f)$ and $g_\text{R}$ denoting the relativistic degrees of freedom at the point of reheating.

As discussed it is possible to directly solve for the number of $e$-foldings $N_\text{R}$ in case of instantaneous reheating $w_\text{R}=\sfrac{1}{3}$ leading to the following constraint of the number of $e$-folding $N$ during inflation
\begin{equation}
    N=61.6-\ln{\frac{U_f^{1/4}}{H_*}}\,.
\end{equation}
Specifically for the model at hand a direct substitution of the parameters $\alpha_0=10$, $\beta_0=1$, $\lambda=10^{-4}$, $\xi=10^5$ and $\mu\sim20 M_P$ leads to the maximum allowed value of $N$~\cite{Lykkas2021}
\begin{equation}\label{Result:Ninst}
    N^\text{inst}_\text{max}\,\approx\, 52\ e\text{-folds}\,,
\end{equation}
such that the model is consistent with the case of instantaneous reheating. The prediction is relatively robust to variations of the free parameters, tending to $N\approx51$ $e$-folds at very large values of the parameters $\alpha_0$ and $\beta_0$.

\begin{figure}
    \centering
    \includegraphics[scale=0.4]{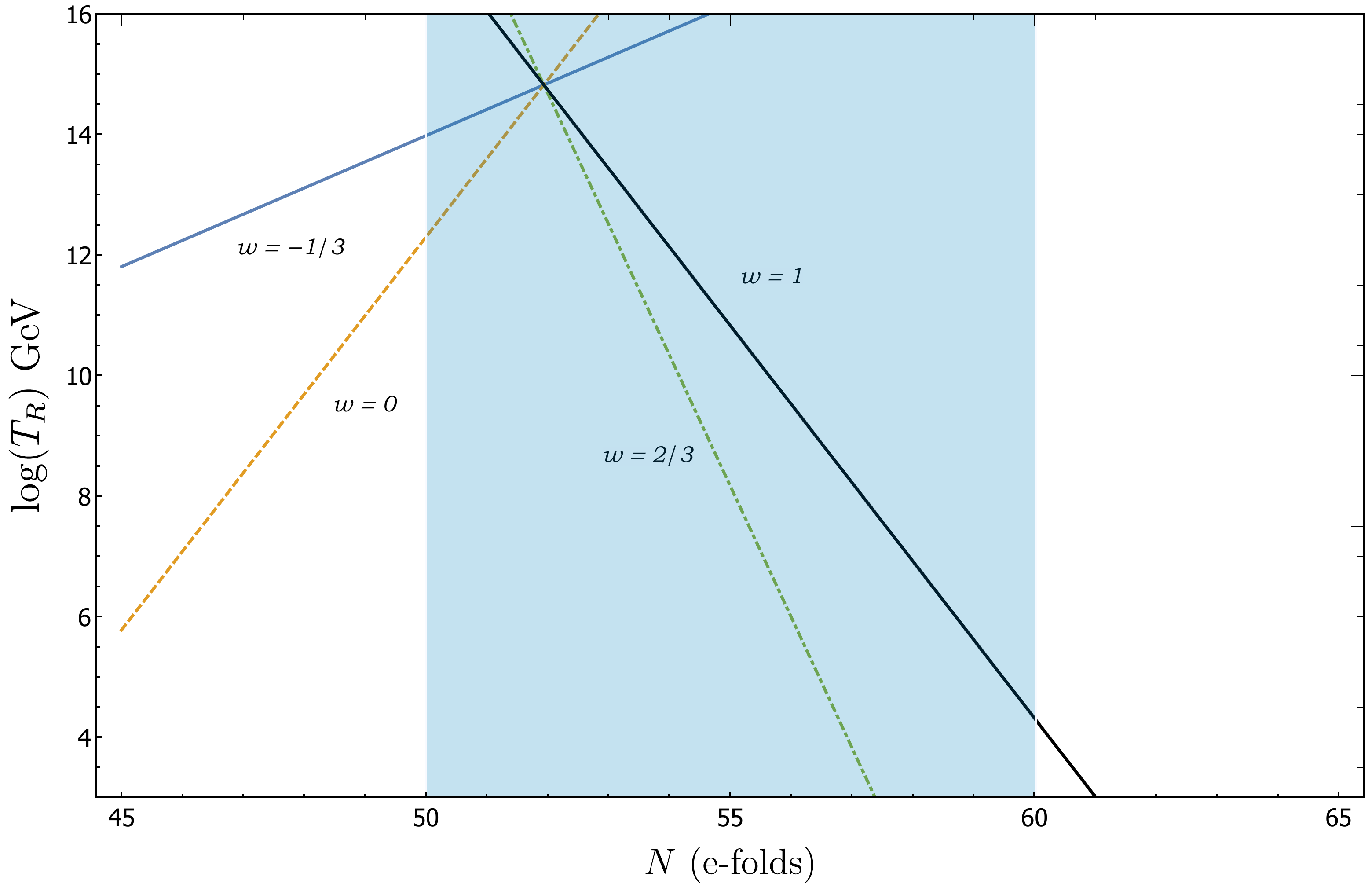}
    \caption{A plot of the reheating temperature $T_\text{R}$ in terms of the number of $e$-foldings assumed during the inflationary era, $N$. The coloured area represent the amount of $e$-foldings usually assumed in order to address the issues of early universe cosmology. The blue, dashing-orange, dotted green and black lines represent different values of $w_R\in\left\{\displaystyle{-\sfrac{1}{3},0,\sfrac{2}{3},1}\right\}$ respectively.}
    \label{fig:ExtendedR2-reheating}
\end{figure}

Different values of $w_\text{R}$ with $w_\text{R}\neq 1/3$ lead to a varying reheating temperature best described by the following formula
\begin{equation}
    T_\text{R}=\left\{\rho_f\left(\frac{30}{\pi^2g_R}\right)\right\}^{1/4}e^{-\frac{3}{4}(1+w_R)N_R}\equiv T_\text{R,max}\,e^{-\frac{3}{4}(1+w_R)N_R}\,,
\end{equation}
manifestly dependent on value of the state parameter $w_\text{R}$ and the number of $e$-foldings $N_\text{R}$. Then, we are able to present the values of $T_\text{R}$ in terms of the amount of inflation $N$ for different values of the parameter $w_\text{R}$, as shown in fig.~\ref{fig:ExtendedR2-reheating}. All the values of $w_\text{R}$ are consistent with reheating leading to a characteristic value of the temperature $T_\text{R}$, always in terms of $N$~\cite{Lykkas2021}. All the curves converge, as expected, to the point of instantaneous reheating ($w_\text{R}=1/3$) at a temperature of $T_\text{R}\approx10^{15}\,\text{GeV}$.

\subsection{\emph{Prospects of different coupling functions}}

It is worth mentioning some other forms of the coupling $\alpha(\phi)$. A general feature of the Palatini-$R^2$ models is that it provides the Einstein-frame rescaled inflaton potential a flat-enough region in the large field limit. For a general function $\alpha(\phi)$ this still holds, meaning that in the case of quartic potential $V(\phi)$ with a nonminimal coupling $\xi\phi^2 R$ we obtain
\begin{equation}
    U(\varphi)\stackrel{\varphi\to\infty}{\approx}\frac{\lambda}{4!}\,\frac{1}{\xi^2+\displaystyle{\frac{\lambda}{6}\alpha(\phi)}+\displaystyle{\frac{2\xi}{\phi^2}}+\displaystyle{\frac{1}{\phi^4}}+\ldots}\,,
\end{equation}
where we assumed for ease of notation that $M_P\equiv1$. Therefore, it is possible to obtain a plateau for any function $\alpha(\phi)$ that is well-behaved at large field values or at least within the field excursion required for inflation. Notice that in a previous section the parametric function had the form $\alpha(\phi)\propto\text{const.}\,+\ln{(\phi/\mu)}$ which ``blows up'' at large field values, however the field space in that theory is truncated, namely field values of the potential with $\phi\gg\mu$ are not to considered valid.

A similar behaviour is obtained if one assumes a string-inspired parametric function reading
\begin{equation}
    \alpha(\phi)=\alpha_0\, e^{\beta_0\,\phi^2/\mu^2}\,,
\end{equation}
where $\alpha_0$ and $\beta_0$ are constant free parameters and $\mu$ is once again some mass scale. Depending on the value of $\beta_0$ the scalar potential $U(\varphi)$ behaves at large field values as
\begin{equation}
    \left.U(\varphi)\right|_{\varphi\to\infty}\simeq\left\{\begin{matrix}0\,,&\beta_0>0\\ \displaystyle{\frac{\lambda}{24\xi^2}}\,,&\beta_0<0 \end{matrix}\right.
\end{equation}
In the case of positive $\beta_0>0$, the predicted values of the inflationary observables do not change much compared to what was discussed earlier, for example for a specific set of the parameters $\beta_0=\lambda=\alpha_0=10^{-5}$ and $\xi\sim2\cdot10^4$, at $N\!=\!55$ $e$-foldings we obtain the following expressions for the inflationary observables
\begin{equation}
    n_s(N=55)\simeq0.9637,\qquad\&\qquad r(N=55)\simeq3.4\times10^{-8}\,.
\end{equation}
Note that the parameters are capable of reproducing the appropriate value for the power spectrum of scalar perturbations $\mathcal{A}_s\approx10^{-9}$.

\begin{figure}
    \centering
    \includegraphics[scale=0.5]{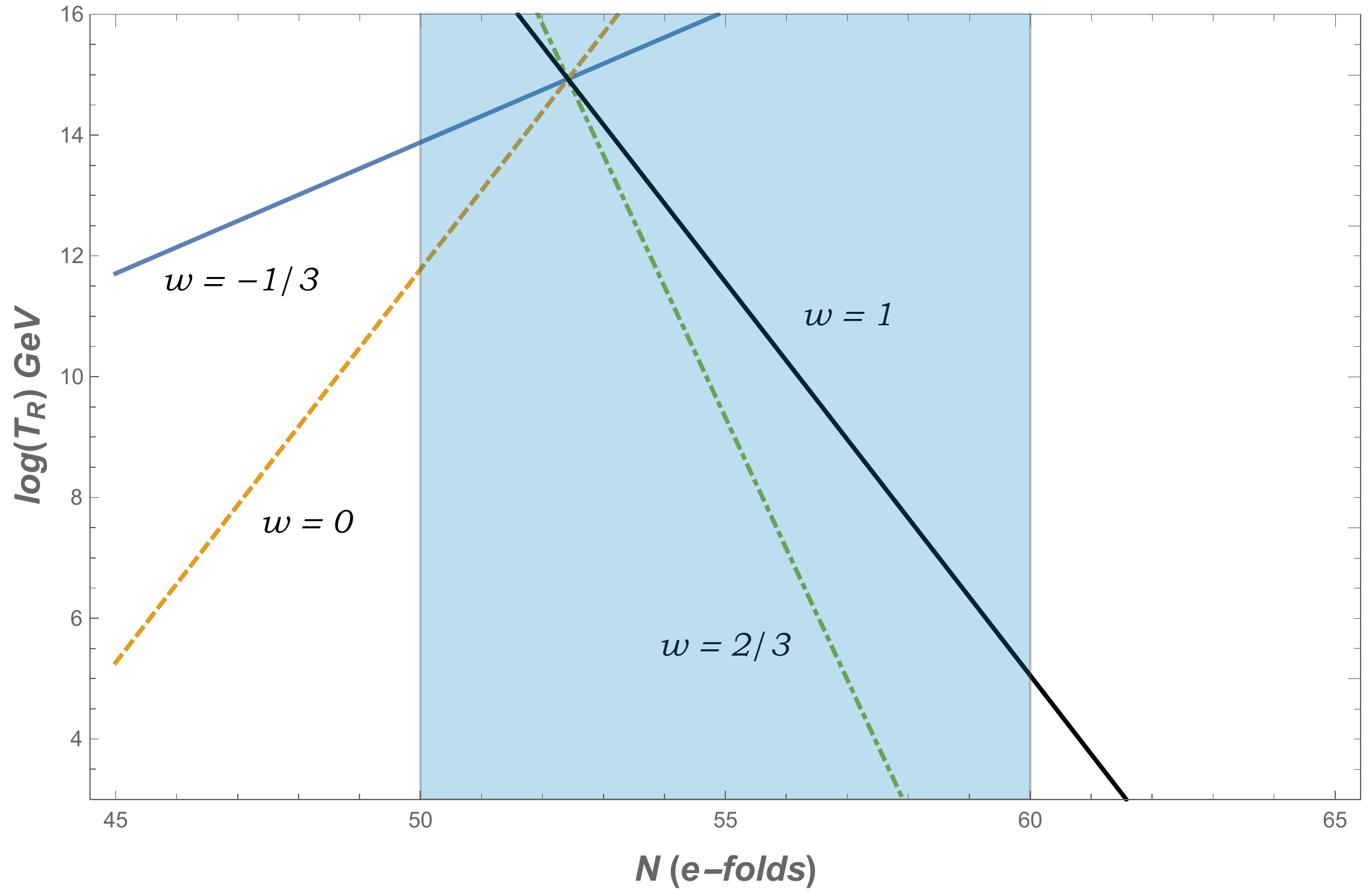}
    \caption{Plot of the reheating temperature as a function of the $e$-foldings $N$. The lines converge at the point of instant reheating at $N\approx52$ $e$-folds. The colour coding of the curves follows the one presented in previous figures.}
    \label{fig:ExtendedR2-reheatingString}
\end{figure}

If the condition of instant reheating is to be satisfied the maximum number of $e$-folds allowed during inflation is given by $N\approx52$ $e$-folds for $\xi\sim1.9\cdot10^4$ and $\alpha_0=\beta_0=\lambda=10^{-5}$. In fig.~\ref{fig:ExtendedR2-reheatingString} we present also case studies of the reheating temperature $T_\text{R}$ in terms of $N$ for varying values of the state parameter $w_\text{R}$. Similarly to the previous figure, fig.~\ref{fig:ExtendedR2-reheating}, the curves converge to the point of instantaneous reheating at $N\approx52$ $e$-foldings with a temperature of $T_\text{R}\sim10^{15}\,\text{GeV}$.

Let us consider the following dimensionless function of the $\alpha(\phi)R^2$ term, given by
\begin{equation}
    \alpha(\phi)=\alpha_0\tanh{\left(\frac{\phi^2}{\mu^2}\right)}\,,
\end{equation}
motivated primarily from mathematical amusement, however it is capable of reproducing some intuitive results. In the far UV region we obtain
\begin{equation}
    \left(\frac{\phi}{\mu}\right)\to\infty\qquad\implies\qquad \alpha(\phi)\to\alpha_0\,,
\end{equation}
meaning that the Starobinsky model is recovered alongside an asymptotic scale invariance. On the other hand at the small field limit we obtain
\begin{equation}
    \left(\frac{\phi}{\mu}\right)\to0\qquad\implies\qquad \alpha(\phi)\sim\alpha_0\,\frac{\phi^2}{\mu^2}+\mathcal{O}(\varphi^6)\,,
\end{equation}
which dynamically turns off the contribution of the $R^2$ term as the inflaton field approaches the minimum of the potential at $\phi\to 0$, and after doing so its condensate starts to dissipate.

The scalar potential $U(\varphi)$ in the Einstein frame obtains the desired plateau at large field values
\begin{equation}
\lim_{\phi\to\infty}U(\varphi)=\frac{\lambda}{4\alpha\lambda+24\xi^2}\,.
\end{equation}
Let us include also a note on the numerical results obtained following the same procedure as in previous sections. Results show that the free parameters are once again close to irrelevant as far as values of $n_s$ go. For example, if $\alpha_0=1$, $\xi=10^2$, $\lambda=10^{-7}$ we obtain, at $N=55$ $e$-foldings, 
\begin{equation}
    n_s(N=55)\simeq 0.9637\,,\qquad\qquad r(N=55)\simeq10^{-5}\,,
\end{equation}
where the tensor-to-scalar ratio chiefly depends on values of $\xi$. An important feature of the model is that relatively small values of the parameter space are able to generate the desired values for the inflationary observables. More importantly, the tensor-to-scalar ratio $r$ can assume large-enough values close to $r\sim10^{-4}$ with the possibility of detection by future missions.

\begin{figure}
    \centering
    \includegraphics[scale=0.5]{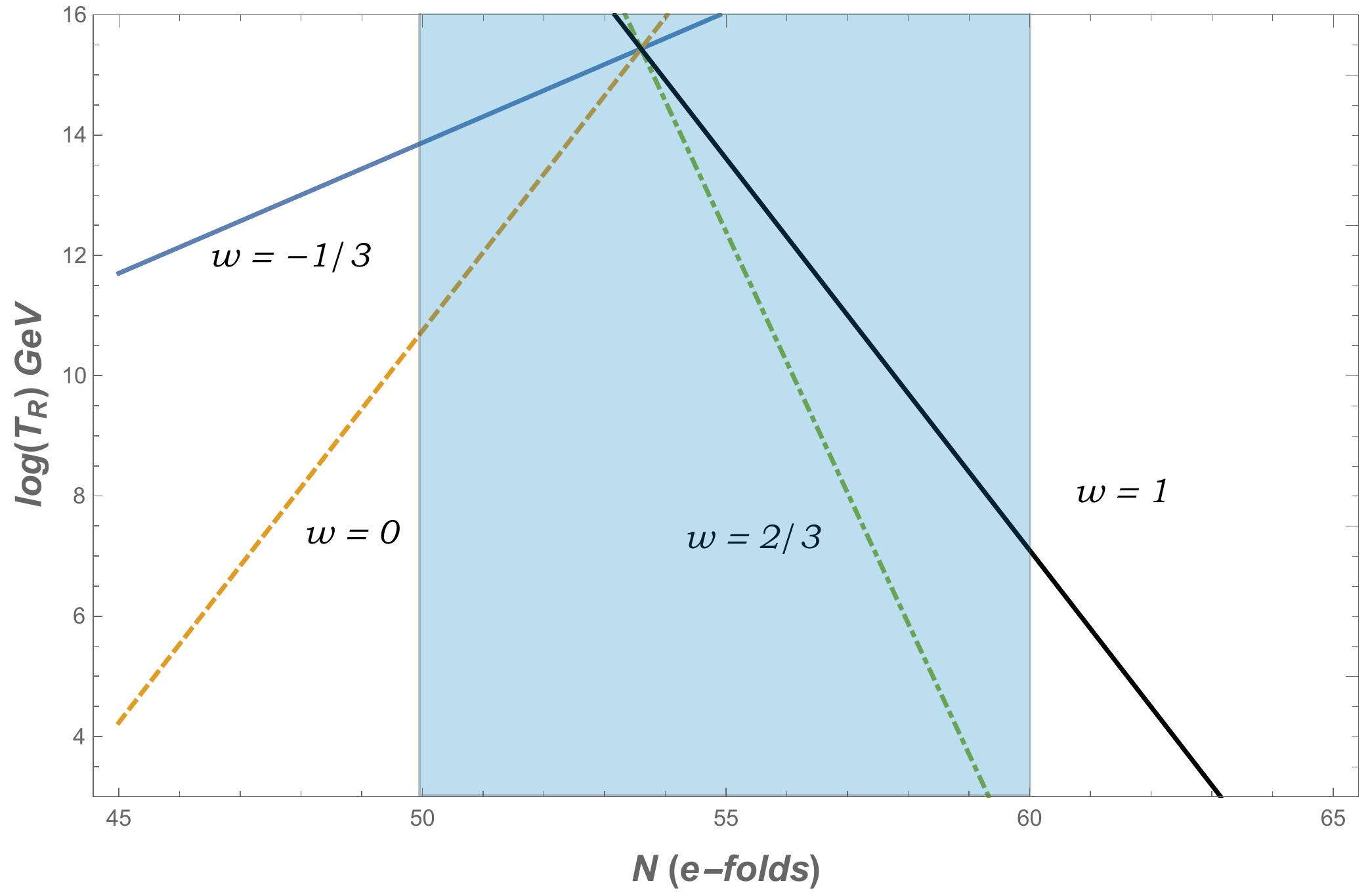}
    \caption{Plot of the reheating temperature in terms of the $e$-foldings $N$. The lines converge at the point of instant reheating at $N\approx54$ $e$-folds. The colour coding of the curves follows the one presented in previous figures.}
    \label{fig:ExtendedR2-reheatingTanh}
\end{figure}

The model is also consistent with a reheating phase succeeding inflation. For example for the particular values of the free parameters $\xi=1.8\times 10^2$, $\lambda\approx10^{-7}$ and $\alpha_0=10$, under the assumption of instant reheating an upper bound of $N\approx 54$ $e$-folds is obtained. Then in fig.~\ref{fig:ExtendedR2-reheatingTanh} the relation of the reheating temperature $T_\text{R}$ with $N$ is plotted for varying values of the state parameter $w_\text{R}$. As expected, the curves converge at the point of instantaneous reheating with a higher than usual reheating temperature of $T_\text{R}\lessapprox10^{16}\,\text{GeV}$.

\section{Constant-roll application}

The previous sections were devoted to understanding the inflationary models under the assumption of the slow-roll approximation, however we met in sec.~\ref{subsec:CRapprox} a different type of approximation that is capable of describing inflation, namely the constant-roll approximation. In the present section we are interested in examining the Palatini-$R^2$ models under the assumption of constant-roll.

\subsection{\emph{Preliminary considerations}}

Before we delve into more details of the approximation it is useful to cast the previous Palatini-$R^2$ models under a different light. We showed that if one considers as a starting point an action functional of the form 
\begin{equation}
    \mathcal{S}=\int\!\mathrm{d}^4x\,\sqrt{-g}\left\{\frac{1}{2}(M_P^2+\xi\phi^2)g^{\mu\nu}R_{\mu\nu}(\Gamma)+\frac{\alpha}{4}\left(g^{\mu\nu}R_{\mu\nu}(\Gamma)\right)^2-\frac{1}{2}(\nabla\phi)^2-V(\phi)\right\}\,,
\end{equation}
it can be equivalently represented in terms of a rescaled metric $\overline{\text{g}}$ in the Einstein frame where the scalar field $\phi$ obtains a noncanonical kinetic factor as well as a higher-order kinetic term $\propto(\nabla\phi)^4$. All of the above can be grouped schematically in the following way ($M_P\equiv 1$ henceforth)
\begin{equation}
    \mathcal{S}=\int\!\mathrm{d}^4x\,\sqrt{-\overline{g}}\left(\frac{1}{2}\overline{g}^{\mu\nu}\,R_{\mu\nu}(\Gamma)+\mathscr{L}(\phi,X)\right),
\end{equation}
expressed in terms of an effective Lagrangian
\begin{equation}
    \mathscr{L}(\phi,X) \equiv A(\phi)X+B(\phi)X^2-U(\phi)\,,
\end{equation}
where we defined $X\,\equiv\,\frac{1}{2}(\overline{\nabla}\phi)^2$  and the model functions
\begin{align}
    A(\phi)&\equiv-\left(1+\xi\phi^2+4\alpha\frac{V(\phi)}{(1+\xi\phi^2)}\right)^{-1}\,,\\
    B(\phi)&\equiv\alpha\left(\,(1+\xi\phi^2)^2+4\alpha V(\phi)\,\right)^{-1}\,=\,-\frac{\alpha A(\phi)}{(1+\xi\phi^2)}\,,\\
    U(\phi)&\equiv\frac{V(\phi)}{(1+\xi\phi^2)^2+4\alpha V(\phi)}=-\frac{A(\phi)\,V(\phi)}{1+\xi\phi^2}\,.
\end{align}
Note that the Lagrangian belongs to a generalised class of the so-called $k$-inflation models and, as previously stated, in their original formulation the models assumed a vanishing potential, however, since then generalisations of them were proposed that included a scalar potential, similar to the model at hand.

The energy-momentum tensor governing the dynamics of the source field $\phi$ is given by
\begin{equation}
    T_{\mu\nu}\equiv \frac{2}{\sqrt{-\overline{g}}}\,\frac{\delta\mathcal{S}}{\delta \overline{g}^{\mu\nu}}=-\frac{\partial\mathscr{L}}{\partial X}\,\left(\overline{\nabla}_{\mu}\phi\right)\left(\,\overline{\nabla}_{\nu}\phi\right)+\overline{g}_{\mu\nu}\,\mathscr{L}
\end{equation}
or, expressed in terms of the model functions
\begin{equation}\label{Cond:ConstRoll-GenStressEnergy}
T_{ \mu\nu}=-\left(A(\phi)+2B(\phi)X\right)\left(\overline{\nabla}_{ \mu}\phi\right)\left(\overline{\nabla}_{ \nu}\phi\right)+\overline{g}_{ \mu\nu}\left(A(\phi)X+B(\phi)X^2-U(\phi)\right)\,.
\end{equation}
Therefore, assuming that the inflaton field is spatially homogeneous, dependent only on time, the energy density $\rho=T_{00}$ and the pressure $T_{ij}=p\,\overline{g}_{ij}=\mathscr{L}\,\overline{g}_{ij}$ are obtained as
\begin{align}
    \rho&=A(\phi)X+3B(\phi)X^2+U(\phi)\,,\label{Cond:ConstRoll-Energy}\\
    p&=A(\phi)X+B(\phi)X^2-U(\phi)\,.\label{Cond:ConstRoll-Pressure}
\end{align}
In order to describe inflation we assume a flat FRW metric $\overline{\text{g}}$ that gives rise to the following equations of motion
\begin{align}
    &3H^2=\rho\,,\\
    &\dot{\rho}+3H(\rho+p)=0\,,
\end{align}
which can be combined into
\begin{equation}
    2\dot{H}+3H^2=-p\,.
\end{equation}
Likewise, the equation of motion for the scalar field, obtained by variation of the action $\delta_\phi\mathcal{S}=0$, is given by
\begin{equation}\label{Eq:ConstRoll-GenKG}
    \ddot{\phi}(A(\phi)+6B(\phi)X)+3H\dot{\phi}(A(\phi)+2B(\phi)X)-A'(\phi)X-3B'(\phi)X^2=U'(\phi)\,.
\end{equation}
Note that in our conventions the kinetic term $X=-(\sfrac{1}{2})\dot{\phi}^2$ is negative.

\subsubsection*{\emph{Digression on energy conditions}}

The state parameter $w=p/\rho$ during inflation must satisfy
\begin{equation}
    w=\frac{p}{\rho}<-\frac{1}{3}\,,
\end{equation}
leading to the following inequality
\begin{align}
    3B(\phi)X^2+2A(\phi)X-U(\phi)<0&\,\implies\\
    -A(\phi)\left(3\alpha X^2-2(1+\xi\phi^2)X-V(\phi)\right)<0&\,\implies\nonumber\\
    \frac{3\alpha}{4}\,\dot{\phi}^4+(1+\xi\phi^2)\dot{\phi}^2<V(\phi)\,,\label{Cond:ConstRoll-ForInflation}
\end{align}
which in the slow-roll regime is trivially satisfied for any potential bounded from below $V(\phi)>0$ (at least locally). Notice that in order to obtain the last inequality we assumed that $A(\phi)<0$ which is exactly the condition such that $AX$ is a \emph{ghost-free} term and is trivially satisfied $\forall\alpha>0$. As expected, eq.~\eqref{Cond:ConstRoll-ForInflation} is also obtained if we start from the equivalent condition that $\epsilon_H<1$.

Since the formulation of Einstein's field equations $G_{\mu\nu}\!\propto\!T_{\mu\nu}$ the fact that $T_{\mu\nu}$ is not a universal function similar to $G_{\mu\nu}$, but is instead dependent on the type of matter and its interaction, has led the scientific community to come up with general \emph{rules}, known as the ``\emph{energy conditions}'' (of GR)~\cite{Hawking1973}, that express in a mathematical way the notion of \emph{locally} positive energy densities. The conditions seem to be violated one after the other by quantum effects and some of them are even abandoned altogether~\cite{Visser2000}. Those that are still relevant in the literature are the null, weak, dominant and the strong conditions, with the null energy condition being the weakest of them and as such it is expected that any reasonable theory should satisfy that one at least. In what follows we discuss possible implications arising from the energy conditions in terms of the stress-energy tensor provided in eq.~\eqref{Cond:ConstRoll-GenStressEnergy}.

So the \emph{null energy condition} (NEC) is the statement that for any future-pointing null vector $\vec{k}$ it should hold that $T_{\mu\nu}k_{\mu}k_{\nu}\geq0$, which leads to
\begin{equation}
    \rho+p\geq0\,.
\end{equation}
Therefore in terms of the energy density and pressure derived earlier in eq.~\eqref{Cond:ConstRoll-Energy} and eq.~\eqref{Cond:ConstRoll-Pressure} respectively, it becomes
\begin{equation}
    2X(A(\phi)+2B(\phi)X)\geq0\implies1+\xi\phi^2+\alpha\dot{\phi}^2\geq0\,,
\end{equation}
satisfied for $\xi>0$ and $\alpha>0$, as per standard wisdom. In fact, at the slow-roll regime it tends to $1+\xi\phi^2>0$ which is simply the condition that gravity remains attractive for all $\phi$.

The \emph{weak energy condition} (WEC) states that for every timelike vector $\vec{Y}$ the matter density has to be nonnegative, meaning that $T_{\mu\nu}Y^\mu Y^\nu\geq0$. The condition then has an overlap with the NEC since it demands that $\rho\geq0$ as well as $\rho+p\geq0$. Assuming that the latter is covered by the NEC let us focus on the first part amounting to
\begin{equation}
    A(\phi)X+3B(\phi)X^2+U(\phi)\geq0\,\implies\,\frac{1}{2}\dot{\phi}^2+\frac{3}{4}\alpha\dot{\phi}^4+V\geq0\,,
\end{equation}
which is similar to the condition for inflation \eqref{Cond:ConstRoll-ForInflation} and as such is trivially satisfied.

The \emph{dominant energy condition} (DEC) says that additionally to the WEC for every future-pointing null or timelike vector $\vec{W}$ the vector field ${T^\mu}_\nu W^\nu$ should also be future-pointing and causal. For a perfect fluid the condition simply reads $\rho\geq|p|$; in the case that $p<0$, which also makes sense for inflation (see eq.~\eqref{Cond:ConstRoll-ForInflation}) the condition collapses to the NEC. So for $p>0$ we obtain
\begin{equation}
    B(\phi)X^2+U(\phi)\geq0\,\implies\,\frac{\alpha}{4}\dot{\phi}^4+V(\phi)\geq0\,,
\end{equation}
which is also satisfied for any $\alpha>0$ and especially at the slow-roll limit for any potential $V(\phi)>0$ $\forall\phi$. The DEC then encapsulates the condition that the original metric $\text{g}$ is rescaled into $\overline{\text{g}}$ by a positive factor. Let us remind the reader that the Weyl rescaling is done through
\begin{equation}
    \overline{g}_{\mu\nu}=(1+\xi\phi^2+\alpha\chi^2)g_{\mu\nu}\stackrel{!}{=}\left(\frac{4\alpha V(\phi)+(1+\xi\phi^2)^2}{1+\xi\phi^2+\alpha\dot{\phi}^2}\right)g_{\mu\nu}\,,
\end{equation}
where the last equality is obtained by substituting the on-shell relation of the auxiliary field $\chi^2$ in terms of $\phi$. Assuming that the overall factor is positive, it leads to the condition best summarised by DEC.

The \emph{strong energy condition} (SEC) demands that for any timelike vector $\vec{Y}$ the trace of the tidal tensor measured by observers is always nonnegative, meaning $(T_{\mu\nu}-\frac{1}{2}Tg_{\mu\nu})Y^\mu Y^\nu\geq0$, where $T\equiv {T^\mu}_\mu$, leading to the following conditions in the case of a perfect fluid
\begin{equation}
    \rho+p\geq0\,,\qquad\quad\qquad\rho+3p\geq0\,.
\end{equation}
The first part is simply the NEC while the second part directly violates our assumption for inflation \eqref{Cond:ConstRoll-ForInflation}. Nevertheless, it leads to the following inequality
\begin{equation}
    2A(\phi)X+3B(\phi)X^2-U(\phi)\geq0\,\implies\,\dot{\phi}^2(1+\xi\phi^2)+\frac{3}{4}\alpha\dot{\phi}^4-V(\phi)\geq0\,,
\end{equation}
which is violated in the slow-roll regime.

We avoided discussing the energy conditions of GR up to this point, however the addition of higher-order terms lead to nontrivial contributions to the energies densities, especially close to and after the end of inflation. Notice that some parts if not all of the energy conditions can also be derived from different considerations, such as the no-ghost condition $A(\phi)<0$ and others, as stated above.

\subsection{\emph{Parameters \& observables}}

With the $\phi$ field being the sole scalar degree of freedom capable of driving inflation we assume that it satisfies the constant-roll condition
\begin{equation}\label{Cond:ConstRoll-Condition}
    \ddot{\phi}=\beta H\dot{\phi}\,,
\end{equation}
where $\beta$ is some undetermined constant dimensionless parameter. As we discussed in sec.~\ref{subsec:CRapprox} the condition approaches the slow-roll approximation in the limit of $\beta\ll1$ in which $\ddot{\phi}\approx0$. Let us introduce the following slow-roll parameters encoding inflation~\cite{Odintsov2020}
\begin{equation}\label{Cond:ConstRoll-SRPs}
    \epsilon_1=-\frac{\dot{H}}{H^2}\,,\qquad\epsilon_2=-\frac{\ddot{\phi}}{H\dot{\phi}}\,,\qquad\epsilon_3=\frac{\dot{F}}{2HF}\,,\qquad\epsilon_4=\frac{\dot{E}}{2HE}\,,
\end{equation}
in terms of the quantities $F$ and $E$ defined by
\begin{equation}
    F=\frac{\partial \mathscr{L}}{\partial R}, \qquad E=-\frac{F}{2X}\left(X\,\frac{\partial\mathscr{L}}{\partial X}+2X^2\,\frac{\partial^2\mathscr{L}}{\partial X^2}\right).
\end{equation}

In order for the SRPs $\epsilon_i$ to make sense we shall also assume that $(\sfrac{1}{2})\dot{\phi}^2\ll U(\phi)$ at least at the very initial stages of inflation. The magnitude of the SRPs is checked numerically later on for each particular part of the parameter space that is able to provide us with a successful inflation. Under the constant-roll condition \eqref{Cond:ConstRoll-Condition} the second SRP becomes $\epsilon_2=-\beta$ and since we are in the Einstein frame with $F=\sfrac{1}{2}$ we obtain $\epsilon_3=0$. Nevertheless, we can express the inflationary observables $n_s$ and $r$ in terms of the SRPs $\epsilon_i$ as follows~\cite{Odintsov2020}
\begin{align}
    n_s&=1-2\,\frac{2\epsilon_1-\epsilon_2-\epsilon_3+\epsilon_4}{1-\epsilon_1}\,,\\
    r&=4\,|\epsilon_1|\,c_s\,,
\end{align}
where $c_s$ represents the effective sound speed of propagation of primordial perturbations given by
\begin{equation}
    c_s^2=\frac{\mathscr{L}_X}{\mathscr{L}_X+2X\mathscr{L}_{XX}}=\frac{A(\phi)-B(\phi)\dot{\phi}^2}{A(\phi)-3B(\phi)\dot{\phi}^2}\,=\,\frac{1+\xi\phi^2+\alpha\dot{\phi}^2}{1+\xi\phi^2+3\alpha\dot{\phi}^2}\,.
\end{equation}
Notice that in the present case the sound speed is bounded by $0<{c_s}^2<1$ as it should. Then, the corresponding power spectrum of scalar perturbations is
\begin{equation}\label{Cond:ConstRoll-PowSpec}
\mathcal{A}_s\approx\frac{H^2}{8\pi^2\epsilon_1(\phi)}=\frac{1}{72\pi^2}\frac{\left(A(\phi)X+3B(\phi)X^2+U(\phi)\right)^2}{\left(A(\phi)X+2B(\phi)X^2\right)}\,.
\end{equation}
Evaluated at the horizon crossing $\phi=\phi_i$ has to yield the observed value of $\mathcal{A}_s\sim10^{-9}$.

One of the advantages of the constant-roll approximation is that the generalised Klein-Gordon equation of motion for the inflaton can be solved analytically, which is not the case in general for such complicated systems, as was demonstrated in previous sections in which we studied the same models in the slow-roll regime. Supposing then that such a solution exists we substitute the constant-roll condition \eqref{Cond:ConstRoll-Condition} into the equation of motion \eqref{Eq:ConstRoll-GenKG} yielding~\cite{Antoniadis2020}
\begin{equation}
\dot{\phi}\,H\left[(\beta+3)A(\phi)+6(\beta+1)B(\phi)X\right]-A'(\phi)\,X-3B'(\phi)\,X^2=U'(\phi)\,.
\end{equation}
Next we can solve the above equation for the Hubble parameter $H$ to obtain the expression
\begin{equation}
H=\sqrt{\frac{U}{3}\left(1-A\left(\frac{\dot{\phi}^2}{2U}\right)+3B\left(\frac{\dot{\phi}^2}{2U}\right)^2\right)}\approx\sqrt{\frac{U}{3}} \left(1-\frac{A}{4U}\,\dot{\phi}^2+\frac{1}{8}\left(\frac{3B}{U}-\frac{A^2}{4U^2}\right)\dot{\phi}^4\right)\,,
\end{equation}
where the approximate formula is derived by expanding around powers of $\dot{\phi}^2/U(\phi)$ and keeping terms up to $\mathcal{O}\left((\dot{\phi}^2/2U)^2\right)$. Substitution of the last expression back into the initial equation leads to a cubic polynomial in terms of $\dot{\phi}$ reading
\begin{equation}
    \sqrt{\frac{U}{3}}\left(3B(\beta+1)+\frac{A^2}{4U}(\beta+3)\right)\dot{\phi}^3-\frac{1}{2}A'\,\dot{\phi}^2-A\sqrt{\frac{U}{3}}(\beta+3)\dot{\phi}+U'=0\,,
\end{equation}
where terms $\mathcal{O}(\dot{\phi}^4)$ are neglected. In principle even if these terms are present the resulting quartic equation is solvable, however due to the assumed tiny values of $\dot{\phi}$ leading to insignificant effects the added complications are not justified. It is known that a cubic equation can be rewritten as the depressed cubic in the form of
\begin{equation}
    x^3+\nu_1\,x+\nu_0=0\,,\qquad\qquad\text{with}\quad x\,\equiv\,\dot{\phi}-\frac{A'}{6\gamma}\,,
\end{equation}
where the coefficients are defined as~\cite{Antoniadis2020}
\begin{align}
    \gamma&=\sqrt{\frac{U}{3}}\,\left(3B(\beta+1)+\frac{A^2}{4U}(\beta+3)\right),\\
    \nu_1&=-\frac{1}{\gamma}\left[(\beta+3)A\sqrt{\frac{U}{3}}+\frac{(A')^2}{12\gamma}\right],\\
    \nu_0&=\frac{1}{\gamma}\left[U'-A(\beta+3)\,\frac{A'}{6\gamma}\sqrt{\frac{U}{3}}-\frac{(A')^3}{108\gamma^2}\right]
\end{align}
Then a real solution to the depressed cubic equation reads
\begin{equation}
    x=\frac{\left(-9\nu_0+\sqrt{3}\sqrt{4{\nu_1}^2+27{\nu_0}^2}\right)^{1/3}}{2^{1/3}\,3^{2/3}}-\frac{\left(\frac{2}{3}\right)^{1/3}\nu_1}{\left(-9\nu_0+\sqrt{3}\sqrt{4{\nu_1}^2+27{\nu_0}^2}\right)^{1/3}}\,,
\end{equation}
at which point we say that we have a solution of $\dot{\phi}$ in terms of $\phi$. Therefore, all of the expressions of the SRPs and subsequently the ones for the observables can be rephrased purely in terms of the inflaton field $\phi$, which is also the case for the number of $e$-folds defined as
\begin{equation}
N=\int_{\phi_i}^{\phi_f}\frac{d\phi}{\dot{\phi}}H=\frac{1}{\sqrt{3}}\int_{\phi_i}^{\phi_f}\frac{d\phi}{\dot{\phi}}\sqrt{A(\phi)X+3B(\phi)X^2+U(\phi)}\,.
\end{equation}

\subsection{\emph{Higgs field}}

Under the assumption that the inflaton field $\phi$ is a fundamental scalar that interacts with the rest of the matter fields, interactions which may prove important at the stage of reheating, the self-interacting potential $V(\phi)$ can be restricted to a renormalisable form of $V(\phi)=m^2\phi^2/2+\lambda\phi^4/4$, which in the large field limit is best approximated by a quartic monomial
\begin{equation}
    V(\phi)=\frac{\lambda}{4}\,\phi^4\,,
\end{equation}
even though in principle the higher-order terms cannot be ruled out completely. A quartic potential is appealing in particular since it can be identified with the Higgs potential far away from the EW scale, namely $V(H)=\lambda\,(|H|^2-v^2/2)^2$ for $|H|\gg v$, driven also by a nonminimal coupling to the Einstein-Hilbert term in the form of $\xi|H|^2$. As noted ealier in the thesis, the subject of Higgs inflation is studied extensively especially in the metric formalism with limited studies spent on the Higgs-$R^2$ models due to their complexity ($2$-dimensional field space; see however refs.~\cite{Ema2017,Wang2017,He2018}). It is then interesting to compare results obtained in the slow-roll regime from our previous studies of the Palatini-$R^2$ Higgs models with ones in the constant-roll approximation.

\subsubsection*{\emph{Minimally coupled Higgs field}}

Let us consider first a simpler model in which the Higgs field is coupled minimally to gravity, corresponding to $\xi=0$. Then, the model functions become
\begin{equation}
    A=-(1+4\alpha V)^{-1},\qquad B=-\alpha A,\qquad U=-V\,A
\end{equation}
and their derivatives with respect to the inflaton read
\begin{equation}
    A'=4\alpha A^2 V',\qquad B'=-4\alpha^2A^2V', \qquad U'=-V'A^2\,,
\end{equation}
where the initial Jordan-frame potential is the quartic potential. By substituting the above relations we can simplify the expressions of the SRPs~\cite{Antoniadis2020}
\begin{align}
    \epsilon_1&=3\,\frac{AX+2BX^2}{AX+3BX^2+U}=3\,\frac{\dot{\phi}^2+\alpha\dot{\phi}^4}{\dot{\phi}^2+\frac{3}{2}\alpha\dot{\phi}^4+2V},\\
    \epsilon_2&=-\beta,\\
    \epsilon_3&=0,\\
    \epsilon_4&=\frac{\sqrt{3}}{2}\,\frac{\dot{\phi}(A'+6B'X)+12\beta BHX}{(A+6BX)\sqrt{AX+3BX^2+U}}=\left(\frac{3\alpha\beta\dot{\phi}^2}{1+3\alpha\dot{\phi}^2}-\frac{2\sqrt{3}\,\alpha}{\sqrt{1+4\alpha V}}\,\frac{V'\,\dot{\phi}}{\sqrt{\frac{1}{2}\dot{\phi}^2+\frac{3}{4}\alpha\dot{\phi}^4+V}}\right).
\end{align}
Therefore the power spectrum $\mathcal{A}_s$ is given by
\begin{equation}
    \mathcal{A}_s=\frac{1}{72\pi^2}\,\frac{\left(AX+3BX^2+U\right)^2}{X(A+2BX)}=\frac{1}{36\pi^2}\frac{\left(\frac{1}{2}\dot{\phi}^2+\frac{3}{4}\alpha\dot{\phi}^4+V\right)^2}{\dot{\phi}^2\left(1+\alpha\dot{\phi}^2\right)\left(1+4\alpha V\right)}\,,
\end{equation}
and a similar albeit more involved expression can be reached for the tensor-to-scalar ratio $r$ and the spectral index $n_s$.

In what follows we study the system numerically by first substituting the real solution of $\phi(\dot{\phi})$ into the expressions of the SRPs. Then, by demanding that inflation ends at some point defined by $\epsilon_1(\phi_f)\equiv1$ we obtain the field value $\phi_f$ at the end of inflation. Allowing for a conservative range of $e$-folds between $N\in[50,60]$ $e$-folds we are able to obtain the value at horizon crossing $\phi_i$. In table~\ref{table:ConstantRoll-T1} we present our findings for specific values of $\alpha=10^7$ and $\lambda=10^{-13}$, while varying small values of $\beta$.

\begin{table}[H]
    \centering
    \begin{tabular}{cc}
    \toprule
         $\beta$& $n_s\ (N=55)$ \\\midrule
         $0.019$   & $0.9761$     \\
         $0.020$   & $0.9721$       \\
         $0.021$   & $0.9681$       \\
         $0.022$   & $0.9641$        \\
         $0.023$   & $0.9601$       \\
         $0.024$   & $0.9561$       \\\bottomrule
    \end{tabular}
    \caption{Values of the spectral tilt $n_s$ for $\alpha=10^{7}$, $\lambda=10^{-13}$ and varying values of $\beta$. Note that as $\beta$ increases the spectral tilt $n_s$ decreases rapidly, while the tensor-to-scalar ratio $r$ is largely unaffected, being $r_{N=55}\sim5\times10^{-3}$.}
    \label{table:ConstantRoll-T1}
\end{table}

The values of $\alpha=10^7$ and $\lambda=10^{-13}$ are directly linked to the power spectrum and as such they are chosen so that the correct value $\mathcal{A}_s\approx2.1\times10^{-9}$ is reproduced. Interestingly, if we allow for various values of $\alpha$ and keep the other parameters $\beta$ and $\lambda$ constant we obtain the following table~\ref{table:ConstantRoll-T2} using the same procedure as with table~\ref{table:ConstantRoll-T1}.

\begin{table}[H]
    \centering
    \begin{tabular}{ccc}
    \toprule
         $\alpha$& $n_s\ (N=55)$ & $r\ (N=55)$\\\midrule
         $5\times10^{6}$ & $0.9589$    	& $9.7\times10^{-3}$	\\
         $7\times10^6$   & $0.9639$      & $7.4\times10^{-3}$	 \\
         $9\times10^6$   & $0.9670$      & $5.9\times10^{-3}$	\\
         $10^7$   		& $0.9681$      & $5\times10^{-3}$ \\
         $2\times10^7$   & $0.9739$      & $3\times10^{-3}$		\\
         $3\times10^7$   & $0.9763$      & $2.1\times10^{-3}$				\\\bottomrule
    \end{tabular}
    \caption{Values of the spectral tilt $n_s$ and the tensor-to-scalar ratio $r$ for $\beta=0.021$, $\lambda=10^{-13}$ and varying values of $\alpha\in\left\{5,30\right\}\times10^6$.}
    \label{table:ConstantRoll-T2}
\end{table}

Then it is noticeable that increasing values of $\alpha$ lead to an increase of $n_s$ as well as a decrease in $r$. It was demonstrated earlier that in the slow-roll approximation, namely the limit of $\beta\to0$, the spectral index is manifestly independent of $\alpha$, however in the present case of the constant-roll the parameter $\alpha$ and therefore the $R^2$ term plays an important role in determining the value of $n_s$, even if $\beta$ takes up small values~\cite{Antoniadis2020}. On the other hand, the dependence of $r$ on the parameter $\alpha$ is already known, however, the actual values of $r$ are, in this case, possibly detectable by future missions contrary to the case of the slow-roll regime, in which they are effectively undetectable, $r\lesssim10^{-10}$~\cite{Antoniadis2018,Antoniadis2019}.

The predictions for the inflationary observables are within the $1\sigma$ of the allowed region of observations. Nevertheless, we noticed in a previous section that the minimal Higgs model in the Palatini-$R^2$ requires an unusually large amount of inflation of $N\sim75$ $e$-folds~\cite{Antoniadis2019} when considered in the slow-roll regime. However, in the present case we notice that a conservative amount of $N\sim55$ $e$-folds suffices~\cite{Antoniadis2020} in order to obtain the desired values of $n_s$ and $r$.

The hierarchy of the kinetic terms $AX/BX^2\sim1$ at $N=60$ $e$-folds for $\alpha=10^7$, $\lambda=10^{-13}$ and $\beta=2.1\times10^{-2}$ suggests that the higher-order kinetic term can have a considerable contribution to the inflaton field dynamics in the constant-roll regime. Therefore, it is important to ensure that the SRPs remain small, $\epsilon_i\ll1$, for the duration of inflation so that the assumed approximations remain valid. For example, assuming the same values for the parameters we obtain $\epsilon_1\approx 10^{-3}$, $\left|\epsilon_4\right|\sim 10^{-2}$ and obviously $|\epsilon_2|\approx10^{-2}$. It should be noted that it was also checked numerically that the solution of $\phi(\dot{\phi})$ does indeed satisfy the equation of motion, where small deviations were found at field values well after the end of inflation, $\phi<\phi_f$, which is to be expected since violations of the approximation tend to appear at field values approaching $\phi_f$.

Inflation takes place in the large field domain with the exact scale of inflation determined in terms of the canonically normalised scalar field $\Phi$ defined through
\begin{equation}
-\frac{1}{2}\left(\overline{\nabla}\Phi\right)^2=\frac{1}{2}\,A(\phi)\left(\overline{\nabla}\phi\right)^2+\frac{1}{4}\,B(\phi)\left(\overline{\nabla}\phi\right)^4\,,
\end{equation}
which in this particular case of the minimally coupled Higgs it becomes
\begin{equation}
\left(\frac{\mathrm{d}\Phi}{\mathrm{d}\phi}\right)^2=-A(\phi)\left(1+\frac{\alpha}{2}\,\dot{\phi}^2\right)=\frac{1}{1+\alpha\lambda\phi^4}\left(1+\frac{\alpha}{2}\,\dot{\phi}^2\right).
\end{equation}
Then we can substitute the solution of $\phi(\dot{\phi})$ into the above formula and expand the overall expression for large field values $\dot{\phi}\simeq\vartheta_0+\vartheta_1/\phi$, where the constants $\vartheta_i(\alpha,\lambda,\beta)$ are depending solely on the values of the model parameters~\cite{Antoniadis2020}. This remains consistent with our results that the inflaton field $\phi$ resides in the transPlanckian region throughout inflation. Finally, the relation of the two fields can be approximated in the large field limit by
\begin{equation}
\Phi\approx \mathcal{C}\mp \left(\sqrt{\frac{1+\frac{\alpha}{2}\,\vartheta_0^2}{\alpha\lambda}}\right)\frac{1}{\phi}\mp\left(\frac{\alpha\,\vartheta_0\,\vartheta_1}{4\sqrt{\alpha\lambda}\sqrt{1+\frac{\alpha}{2}\vartheta_0^2}}\right)\frac{1}{\phi^2}\,,
\end{equation}
where $\mathcal{C}$ is an integration constant. Independently of $\mathcal{C}$ the exact excursion of the field $\Phi$ can be calculated directly from the formula above, yielding $\Delta\Phi\equiv\Phi_f-\Phi_i=14\,M_P$ if we assume that $\alpha=10^7$, $\lambda=10^{-13}$ and $\beta=0.021$. Note that it was verified once again that the equation of motion expressed in terms of the normalised field $\Phi$ is satisfied, where similar deviations were found in the region of $\Phi<\Phi_f$~\cite{Antoniadis2020}. Therefore, under the assumption of constant-roll we are able to approximately rewrite the higher-order and noncanonical kinetic terms in terms of a canonically normalised field $\Phi$ with a complicated self-interacting potential $U(\Phi)$.

\begin{figure}[h!t]
    \centering
    \includegraphics[scale=0.5]{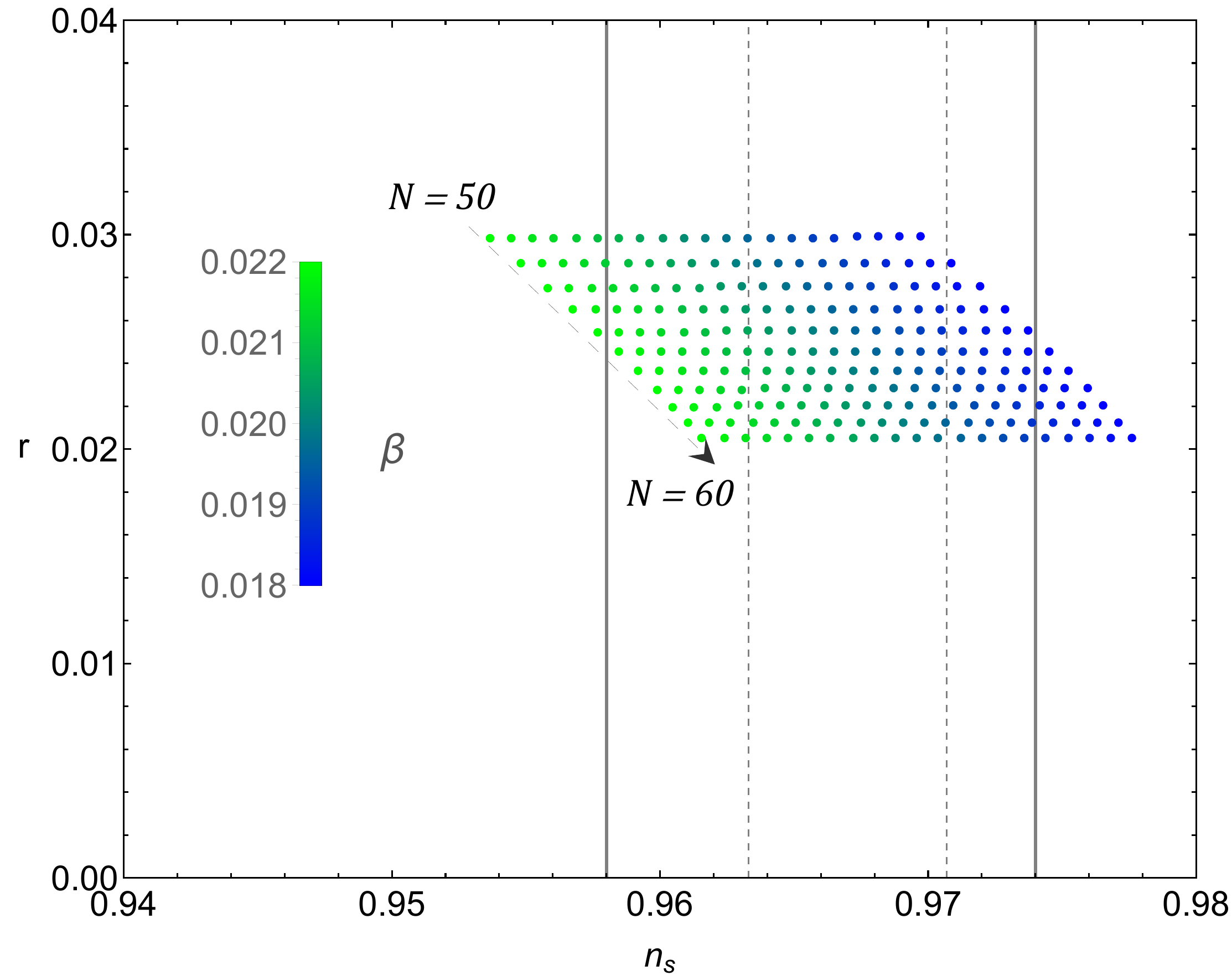}
    \caption{A plot of the $r$-$n_s$ plane for $\alpha=10^7$, $\lambda=10^{-13}$, $\xi=10^{-6}$ and varying values of $\beta\in\left\{0.018,0.022\right\}$. The dashed and solid lines represent the $1\sigma$ and $2\sigma$ allowed range of the $n_s$, respectively. All of the values of the tensor-to-scalar ratio $r$ are within the allowed region of observations $r<0.06$. Once again, as $|\beta|$ increases, the spectral tilt $n_s$ decreases, while the effect on the tensor-to-scalar ratio $r$ is minimal, $r_{N=50}=0.03$ and $r_{N=60}=0.02$.}
    \label{fig:ConstantRoll-RNs}
\end{figure}

\subsubsection*{\emph{Nonminimally coupled Higgs field}}

Let us now assume a nonminimal coupling between the inflaton and gravity in the form of $\xi\phi^2R$, which is known that in the slow-roll framework the interplay of this coupling and the $R^2$ term yields appropriate values of the inflationary observables. 

It is then straightforward to follow the same procedure with the assumption that $\xi\neq0$. In fig.~\ref{fig:ConstantRoll-RNs} we present the results coming from that exact numerical analysis, meaning that the field excursion is obtained by the condition that inflation ends at $\epsilon_1(\phi_f)\equiv1$ and lasts some amount $N\in[50,60]$ $e$-folds. The figure is obtained for characteristic values of the parameters, $\alpha=10^7$, $\lambda=10^{-13}$, $\xi=10^{-6}$ and varying values of $\beta\in\left\{0.018,0.022\right\}$. It is immediately noticeable that larger values of $\beta$ tend to decrease the values of $n_s$ and minimally affect $r$, as expected from the results we obtained in the minimal case. The parameters are chosen such that the power spectrum of scalar primordial perturbations takes up its observed value of $\mathcal{A}_s\approx2.1\times10^{-9}$. The model is able to provide a successful inflation with appropriate values for the observables, similarly to its slow-roll counterpart, however it requires smaller values of $\xi$ compared to the slow-roll paradigm. Another thing to note is that the SRPs remain small during inflation reporting values at $\epsilon_1\lesssim10^{-2}$ and $|\epsilon_4|\sim10^{-2}$, as well as $|\epsilon_2|\approx10^{-2}$ by definition.

The numerical values of the parameter ${c_s}^2$ both in the minimal and the nonminimal scenario are approximately ${c_s}^2\sim0.4$ and close to the start of inflation are approximately unity as $\phi$ tends to $M_P$. As it should, the values are ${c_s}^2<1$ and possible instabilities due to negative values ${c_s}<0$ are avoided.

\chapter{Summary \& Conclusions}
\label{Ch5:conclusions}

The theory of cosmic inflation is currently the best candidate that provides a natural solution to puzzles from the Big Bang singularity and, most importantly, the seeds for large structure formation in our universe, with the latter being a prediction of the theory. There exist numerous models that can describe inflation, meaning that they can reproduce the correct values of the inflationary observable quantities, with varying success. Advances in inflation are connected with developments in (future) experiments and our ability to probe the region of observables with higher precision. A large part of the simplest models has already been ruled out or they are at marginal contact with observational data (e.g. see fig.~\ref{fig:Planck2018}), however the Starobinsky model ($R+R^2$) resides persistently within the $1\sigma$ allowed region of observations since its inception. The continued success of the model has led to a plethora of other models that attempt to modify or extend the initial model, while also retaining some of its attractive features.

Predictions of the theory of inflation rely on the interplay of gravity and matter, and our understanding of it. Therefore, it is naturally connected with the \emph{parametrisation} of the gravitational degrees of freedom. The so-called \emph{Palatini} or \emph{first-order formulation} of gravity offers a different way to parametrise the gravitational DOFs, by generalising the connection on the manifold to a \emph{metric-affine} one assuming no a priori dependence on the metric. As far as GR is concerned, it was shown that it is \emph{equivalent} to the conventional metric or \emph{second-order formulation} (at least at the classical level)~\cite{Palatini:1919}, and the Palatini formalism was lost to obscurity for the most part. However, recent developments in modified theories of gravity and a rising interest in non-Riemannian geometries elevated the Palatini formalism to a fundamental question on the gravitational degrees of freedom. Since extended theories of gravity are prevalent in the inflationary paradigm it was promptly shown that they tend to have different descriptions between the two formalisms~\cite{Bauer2008}, even leading to contrasting results in some cases. This observation is the stepping-stone of the main part of the thesis as we explain later on.

Based on the previous discussion it seemed important to review some aspects of single-field inflation, which was done in chapter~\ref{Ch2:Inflation}. Starting with a brief overview of modern cosmology we are led to the puzzles of Big Bang, which in turn lead us straight to the central idea of inflation. Most importantly, we review the \emph{simplest} mechanism of single-field \emph{slow-roll inflation} and obtain the (approximate) expressions of the \emph{observable quantities} in terms of the slow-roll parameters. In the same direction, we study the \emph{constant-roll approximation} noting some of its features and differences with the slow-roll one. Towards the end of the chapter we offer a discussion on reheating, i.e. the era right after the end of inflation. First, we begin by briefly analysing some of the proposed mechanisms of reheating, such as the well-known \emph{perturbative reheating} and \emph{preheating}. Then, trailing the results of refs.~\cite{Dodelson2003,Liddle2003a,Dai2014,Munoz2014,Gong2015,Cook2015a}, we review a different way to parametrise the reheating parameters in terms of the inflationary ones that allows us to possibly place stricter constraints on the model parameters.

In the following chapter, ch.~\ref{Ch3:FirstOrder}, we comment on different features of the first-order formalism, paying close attention to the ones that are further highlighted in inflationary models. After a short historical review we revisit the conventional metric formalism and derive the Einstein field equations of GR. In the following section we introduce the notion of the Palatini variation, starting with an example of the first-order and second-order of Electromagnetism. Then, a detailed summary of \emph{metric-affine spaces} is needed in order to properly introduce the \emph{metric-affine connection} (especially the nonmetricity and torsion) in GR. After that, the \emph{on-shell equivalence} between the metric and the Palatini formulations of GR is directly derived, concluding the chapter with a discussion on their \emph{possible} equivalence at the quantum level.

In the last chapter, ch.~\ref{Ch4:QuadGrav}, we present the main results of the thesis, in which the \emph{Starobinsky inflationary model} is the focal point. As such, we begin by first reviewing the model in the metric formalism highlighting how the scalar degree of freedom (the \emph{scalaron} field) sourced by the $R^2$ term arises in the \emph{scalar representation} of the theory. Then, it is straightforward to derive the \emph{Starobinsky potential} in the Einstein frame, following a Weyl rescaling of the metric and a field redefinition, and apply the mechanism of single-field slow-roll inflation, as described in ch.~\ref{Ch2:Inflation}, in order to obtain the famous predictions of the model. We continue by considering a coupling of the Starobinsky model with a real scalar field and its self-interacting potential, first in a \emph{minimal} way (the global term $\sqrt{-g}$) and later in a \emph{nonminimal} via a coupling to the EH term of the form of $\xi\phi^2 R$. It is evident that in both cases the theory effectively contains \emph{two} scalar degrees of freedom, the scalaron $\chi$ and the original scalar field $\phi$. There we notice that an application of the models to inflation is highly complicated since both fields can in principle contribute in driving inflation, and especially in the case of the nonminimal coupling, the kinetic terms of the scalar fields in the Einstein frame mix \emph{nontrivially} further complicating the analysis.

Considering the Starobinsky model in the Palatini formalism we show that the $R^2$ does \emph{not} actually lead to a dynamical scalaron, $\chi$. Since in the first-order formulation the connection and the metric do not dependent on each other a priori, a Weyl rescaling of the metric leaves the Ricci tensor \emph{invariant} being purely a function of the connection, i.e. $R_{\mu\nu}(\Gamma)$. Therefore, there is no way for the scalaron to obtain a kinetic term in the Einstein frame. Then, higher-order curvature invariants are \emph{unable} to contribute a scalar DOF, so in order for the model to describe inflation the inflaton field has to be manifestly included into the action in the form of a \emph{fundamental} scalar field $\phi$. Our starting point is then an action of that form. In the Einstein frame we show that the scalaron $\chi$ is included in multiplicative factors of the kinetic term of $\phi$ and its potential $V(\phi)$. Then, after a variation of the final action with respect to $\chi$ we obtain its \emph{constraint} equation, which we then substitute back into the action. There we find that the action obtains \emph{higher-order kinetic terms} of the original scalar field, $\propto(\nabla\phi)^4$, as well as complicated expressions for the \emph{noncanonical kinetic function} and scalar potential. Next, we are able to obtain the generalised Einstein field equations for the system after variation with respect to the metric and the scalar field and show that the equation of motion of the connection leads to the \emph{Levi-Civita condition} with respect to the Weyl rescaled metric $\bar{\text{g}}$. By considering the path integral formulation of the same theory it is evident (even though not conclusive) that the nondynamical nature of the scalaron is not a figment of the classical action, but it remains at the quantum level even when other matter fields are included in the background of the action, leading to local terms that we can safely ignore. In closing of that section, we offer a brief discussion on the \emph{issue of frames},namely the Einstein and Jordan, since the transition from the Jordan frame to the Einstein frame is at the \emph{heart} of our analysis.

In the next section we focus in a number of inflationary models that are already \emph{ruled out} from observations in their metric version, and assume a minimal coupling with the $R+R^2$ gravitational sector. First, we consider the so-called \emph{natural inflation model}, where we showed that the $R^2$ term has a considerable contribution to the inflaton potential by inducing a flat region in the large field limit of the inflaton $\phi$. In fact, the result generally holds for any sensible form of an inflaton potential $V(\phi)$, leading to a plateau for \emph{quite general} conditions. In the particular case of natural inflation the flattening of the potential is evident in fig.~\ref{fig:NaturalInf-PotU}. Since we are interested in slow-roll inflation we neglect the contribution of the higher-order kinetic terms $\propto\dot{\phi}^4$ and apply the conventional mechanism of single-field inflation. There we find that when the natural inflation model is minimally coupled to the Starobinsky model in the Palatini formalism we are able to obtain acceptable predictions for the inflationary observable quantities, therefore allowing for the possibility for the model to describe inflation, \emph{contrary} to their metric formulation in which they fall short.

The simplest scenario of a free massive scalar field also shares the same fate as the natural inflation, however considered here in the Palatini formalism coupled minimally to the Starobinsky model we obtain values of the inflationary observables within the allowed $1\sigma$ region for values of the mass term around $m\sim 10^{13}\,\text{GeV}$. The scale of inflation, defined as the field values of the canonically normalised inflaton field, is \emph{slightly} above the Planck scale.

Following the success of the previous models we are interested in realising the \emph{Higgs inflation} scenario with a minimal coupling to gravity. It is known that in the metric formalism a nonminimal coupling is \emph{required}, however, considered in this specific context we show that the effect of the $R^2$ term allows for a successful inflation with the caveat that the number of $e$-folds required is \emph{larger than usual}, close to $N\sim75$ $e$-foldings.

Having analysed the spectrum of minimally coupled models, the wildly popular models that are \emph{nonminimally} coupled to the EH term are also discussed. The general feature of the flattening of the Einstein-frame scalar potential still remains. Implementing this program to the CW and the induced gravity model we obtain acceptable inflationary observables for both models for a large part of the model parameter space. Of particular importance is the scenario of \emph{nonminimal Higgs inflation}, in which we show that the nonminimal coupling $\xi$ between Higgs and the EH term can assume \emph{small} values compared to the ones obtained in absence of the $R^2$ term. Contrary to its minimal formulation the model leads to appropriate inflationary observables for values of $N\in[50,60]$ $e$-folds. 

Motivated from the success of the previous models we investigated the scenario in which the Starobinsky is nonminimally coupled to a quartic potential, and \emph{promoted} the Starobinsky constant $\alpha$ to include \emph{logarithmic} corrections $\propto\log{(\phi^2/\mu^2)}$ of the fundamental scalar field, $\alpha\mapsto\alpha(\phi)$. The resulting action in the Einstein frame has a similar form to the previously considered Palatini-$R^2$ models, however the plateau of the Einstein-frame inflaton potential is violated logarithmically at field values $\phi>\mu$. We find that prediction of the model regarding the inflationary observables are in good agreement with recent observational data, notably a large coupling $\alpha(\phi)$ can actually \emph{influence} the values of the observables (see fig.~\ref{fig:ExtendedR2-RNs}), \emph{contrary} to the previous model in which only the tensor-to-scalar ratio $r$ was dependent on $\alpha$. This suggests that other models that reside outside the $(+)2\sigma$ allowed region can in principle be in agreement with observational bounds by introducing a field dependent constant $\alpha(\phi)$. In addition the values of $r$ can range from \emph{tiny}, which is a general feature of the Palatini-$R^2$ models, to rather \emph{large} approaching the bound on $r$, meaning that they can be in contact with future experiments of expected precision $10^{-3}$ or even $10^{-4}$. After the end of inflation the model undergoes a process of reheating. Through the mechanism highlighted in section~\ref{sec:ParReheating} it is possible to parametrise the reheating parameters in terms of the inflation ones, which for different values of the reheating state parameter $w_\text{R}$ we showed that the model is indeed capable of supporting a reheating era for the specific values of the model parameters assumed during inflation with maximum reheating temperature $T_\text{R}\sim 10^{15}\,\text{GeV}$, fig.~\ref{fig:ExtendedR2-reheating}. Under the assumption of instantaneous reheating we obtained an upper bound on $N\approx52$ $e$-folds. Concluding this section, other forms of field dependence of $\alpha(\phi)$ were considered that also retain the desired plateau of the inflaton potential. For a specific part of the model parameter space we were able to find agreement with observational bounds even in these cases.

In the last section we shift our focus and analyse the inflationary phenomenology of the Higgs field coupled to the quadratic gravity under the assumption of the \emph{constant-roll approximation}. Similarly to previous models the resulting Einstein-frame action is of the form of a generalised $k$-inflation type and we analyse the cases of \emph{minimal} and \emph{nonminimal} coupling of the Higgs with gravity, by assuming that the constant-roll condition $\ddot{\phi}\sim\beta H\dot{\phi}$ holds, where $\beta$ is a constant parameter. In both cases the predictions of the inflationary observables show a significant dependence on the higher-order kinetic terms, contrary to their slow-roll counterparts. Particularly for the minimal scenario we found that in the constant-roll regime acceptable values for the observables are obtained for $N\in[50,60]$ $e$-folds, which is in contrast with the slow-roll case where it was shown that large values of $N\gtrsim70$ $e$-folds are required in order for $n_s$ to reside within the $2\sigma$ allowed region.

\bibliography{thesis_refs}
\bibliographystyle{jhep}

\end{document}